\documentclass{emulateapj}
\usepackage{graphicx}
\usepackage{subfigure}
\usepackage{rotating}
\usepackage{lscape}
\usepackage{natbib}
\def\gae{\lower 2pt \hbox{$\, \buildrel {\scriptstyle >}\over {\scriptstyle \sim}\,$}}
\def\lae{\lower 2pt \hbox{$\, \buildrel {\scriptstyle <}\over {\scriptstyle \sim}\,$}} 

\received{3/30/09}
\revised{8/20/09}
\accepted{9/4/09}

\begin{document}

\title{58 Radio Sources Near Bright Natural Guide Stars\footnotemark[1]\ $^,$\footnotemark[2]}
\footnotetext[1]{
Based on observations obtained at the Canada-France-Hawaii Telescope
(CFHT) which is operated by the National Research Council of Canada,
the Institut National des Sciences de l'Univers of the Centre National
de la Recherche Scientifique of France,  and the University of
Hawai`i.}
\footnotetext[2]{Based on observations obtained at the Gemini
  Observatory, which is operated by the Association of Universities
  for Research in Astronomy, Inc., under a cooperative agreement with
  the NSF on behalf of the Gemini partnership: the National Science
  Foundation (United States), the Science and Technology Facilities
  Council (United Kingdom), the National Research Council (Canada),
  CONICYT (Chile), the Australian Research Council (Australia), CNPq
  (Brazil) and SECYT (Argentina) 
}
\author{B. Stalder\footnotemark[3], K. C. Chambers}\footnotetext[3]{
Current Address: Department of Physics, Harvard University, 17 Oxford
Street, Cambridge, MA 02138} 
\affil{Institute for Astronomy, University of Hawai`i \\
2680 Woodlawn Drive, Honolulu, HI 96822}
\author{and \\ William D. Vacca}
\affil{SOFIA-USRA, NASA Ames Research Center \\
MS N211-3, Moffett Field, CA 94035-1000}
\email{bstalder@physics.harvard.edu (B.S.); chambers@ifa.hawaii.edu
  (K.C.C.); wvacca@sofia.usra.edu (W.D.V.)}

\begin{abstract}

We present a preliminary survey of 58 radio sources within the isoplanatic 
patches ($r < 25''$) of bright ($11<R<12$) stars suitable for use as
natural guide stars with high-order adaptive optics (AO).
An optical and near-infrared imaging survey was conducted
utilizing tip-tilt corrections in the optical and AO in the
near-infrared.   
Spectral Energy Distributions (SEDs) were fit to the multi-band
data for the purpose of obtaining photometric 
redshifts using the Hyperz code \citep{bolzonella2000}.  Several of
these photometric redshifts were confirmed with spectroscopy, a result
that gives more confidence to the redshift distribution for the whole sample.
Additional long-wavelength data from Spitzer, SCUBA, SHARC2, and VLA
supplement the optical and near-infrared data. 
We find the sample generally follows and extends the
magnitude-redshift relation found for more powerful local radio
galaxies.  The survey has identified several reasonably bright
($H=19-20$) objects at significant redshifts ($z>1$) that are now
within the capabilities of the current generation of AO-fed integral-field
spectrographs. 
These objects constitute a unique sample 
that can be used for detailed ground-based AO
studies of galactic structure, evolution, and AGN formation at high
redshift.   

\end{abstract}


\section{Introduction}

One of the most active debates in extragalactic astronomy is on the
nature of spheroid formation at high redshifts.
The hierarchical galaxy formation scenario is based on the conclusions
of semi-analytic $\Lambda$CDM models
\citep{thomas1999,kauffmann1999,kauffmann2000}
 as well as observations of galaxies 
\citep{miyazaki2003,dickinson2003,van-dokkum2008}.
It is typically described as a building up of  
a large stellar system from smaller ones, and predicts that massive early-type
galaxies should have undergone final assembly at relatively late
epochs ($1 < z < 2$). 
Conversely, the monolithic formation scenario models, supported by
other observations \citep{thomas2002a,brodie2006,mcgrath2008},
require that massive spheroids form at much higher redshifts directly
from primordial density fluctuations (see \citet{peebles2002} for a
review and discussion) and evolve passively to the present epoch. 

In order to test the predictions of these scenarios, and thereby
determine which provides the more accurate model of galaxy formation,
we have attempted to study massive galaxies at high redshift by
observing the host galaxies of radio sources.  Though these objects
are not classified as normal galaxies, this technique has certain
advantages over other high redshift galaxy search methods
(e.g. Lyman-break, and NIR-selection).  High redshift radio galaxies
(HzRGs) tend to have large stellar masses and high near-infrared
luminosities making them accessible to ground-based observations.
Furthermore, radio galaxies follow a magnitude-redshift relation (the
near-IR Hubble diagram, see \citet{van-breugel1998}), which allows a
selection of high redshift objects based on their
apparent brightnesses.  \citet{lilly1984} also found that the K-band
magnitude-redshift (K-z) relation
of the 3CR catalog of powerful radio galaxies could be well fit by a
passively evolving old stellar population, similar to present-day
elliptical galaxies.  The tendency for HzRGs to reside in over-dense
regions in the early universe, makes them likely progenitors of the
brightest cluster galaxies in the present epoch \citep{best1998}.  

There are a few disadvantages to observing HzRGs.  First, the
``alignment effect''
\citep{chambers1987,mccarthy1987,chambers1990,pentericci2001}
was identified when it was discovered that these objects' optical
morphologies at
high redshift were aligned with structures in the radio, and therefore
their optical/infrared magnitudes could not be treated as independent
of their radio properties.  
Several mechanisms have been proposed to explain the alignment effect
(e.g.,  anisotropic interactions between cluster members,
jet-triggered star formation, optical synchrotron emission, inverse
Compton scattering of the CMB photons, and thermal continuum emission
from the plasma ionized by the AGN, see \citet{mccarthy1993}).  Each
of these mechanisms has been shown to be present in certain objects,
but because no single explanation is satisfactory
for all cases, it is generally thought that two or more may
be the source of the alignment effect.  Regardless of the causes of the
alignment effect, in order to obtain the fundamental parameters of the
underlying host galaxy, the contributions from the AGN and the host
galaxy must be disentangled.  

Studies of the relative contribution of evolved stellar population
and flat-spectrum (and presumably aligned) components in optical and
infrared images of 3CR radio galaxies revealed that the components
responsible for the alignment effect contribute only about $10\%$ 
to the total SED of the host galaxy
\citep{best1998} affecting both continuum and emission-line
morphologies \citep{mccarthy1987}.   
The alignment effect also generally diminishes at wavelengths longward
of the 4000\AA\  break \citep{rigler1992}, although there are some exceptions
\citep{eisenhardt1990,chambers1988}. 
When the alignment effect is examined in detail, only one-fourth to
one-third of low redshift radio galaxies have any detectable
morphological peculiarities (at the 
$\mu_V > 25$mag/arcsecond$^2$ level), and this fraction becomes
smaller in less powerful radio galaxies
\citep{heckman1986,dunlop1993}, as a weaker radio source has less of
an impact on the rest of the galaxy.   
This suggests the millijansky-level radio source population predominantly
consists of sources with radio fluxes sufficiently low that the
optical/near-IR morphologies and SEDs of host galaxies are probably completely
dominated by the stellar population of the host galaxy. 

An additional disadvantage of observing high redshift galaxies results from the
loss of angular and spatial resolution when using traditional ground-based 
observational techniques, which is needed to accurately
determine the galaxy's fundamental properties.  This can easily be achieved from
space using HST, though in the near-infrared (where these objects' SEDs are 
less-contaminated by their AGNs) HST is limited by both the diffraction 
 size (about 0.2''in K-band) and light-collecting aperture.  This makes a survey of 
these sources extremely time-consuming on a highly competitive telescope and spectroscopy
on these faint sources is almost impossible.  However, with the advent of ground-based AO
techniques, a large imaging survey can be conducted with a comparable spatial resolution and 
exposure time on a 3-4 meter class telescope, with the limitation that most AO systems
require a bright guide star in proximity to the galaxy on the sky.  
Luckily radio sources are common enough that a significant number meet
this criteria in wide-area radio survey data sets. 

The Faint Images of the Radio Sky at Twenty-centimeters (FIRST) Survey
\citep{white1997} at the Very Large Array (VLA) has 
80$\%$ completeness down to 1 mJy and better than 1$''$ astrometry
that can be used to search for HzRGs.   
The catalog is only minimally contaminated (about 10$\%$ by number,
see \citet{jackson2005}) by low redshift star forming galaxies which
quickly become more numerous at levels fainter than 0.1 mJy.  
More importantly, the
VLA's astrometry allows for mostly straightforward optical
counterpart identifications. 
And lastly, the density of sources (about 90 sources per square degree
over the survey's 10,000 square degrees) is sufficiently high that a
sizable sample of sources are within the isoplanatic patch (about
25$''$ in the NIR) of a V$<$12 star.   With these criteria, this
survey provides a highly efficient means to preselect likely
candidates for high redshift galaxies with undisturbed stellar
populations and light profiles that can be observed at high spatial
resolution from the ground initially using natural guide star AO.   
As laser guide star and multi-conjugate AO (which also require nearby 
natural guide stars) become more mature, this sample would also be 
well-suited for these instruments and as an initial target list for
JWST as it relates to several of its key science objectives.

In this paper, we present the first phase of a project, involving
an imaging survey plus a supplemental 
spectroscopic survey of the FIRST-BNGS sample.  The imaging survey
consists of optical, NIR, FIR, submillimeter, and radio observations.
However, the focus of the imaging survey was on the optical and NIR
wavelength regions  
(since we are mainly interested in the stellar populations of these
objects). 
The goal of this survey was to obtain multi-wavelength photometry to
provide identifications and probable redshifts for candidates for high
precision diagnostics of galaxies at high redshift to study  
galactic structure, evolution, and AGN formation at high redshift.
Supplemental spectroscopic observations were also obtained 
to refine redshifts or remove ambiguities in the photometric redshift
solutions.  In section 2, we introduce the sample we compiled to
search for high redshift radio galaxies. Section 3 describes our
observations and reduction process for the imaging survey.   
In section 4, we present our photometric measurement and redshift fitting
procedures and results.  Section 5 discusses the spectroscopic
observations, reduction process, and results, while section 6 briefly
summarizes the VLA radio data. 
Finally, in section 7, we summarize our findings and discuss the
potential of this sample. 
 
For this paper we adopt (unless otherwise stated) the
Friedmann-Lemaitre world cosmological model with $\Omega_0$=0.3,
$\Omega_\Lambda$=0.7, and $H_0$=70km/s/Mpc, giving the present age of
the Universe as 13.47 Gyr.  Also, the spectral index, $\alpha$, will
be defined such that the flux of a source, $S_\nu$,  is proportional
to $\nu^{\alpha}$.  

\section{Sample Selection}

A cross-correlation of the VLA FIRST survey and the USNO-A2.0 Catalog
\citep{monet1998} yields 58 sources with $S_{1.4GHz}\gae$1-mJy,
galactic latitude $|b_{II}| > 35$ located within an
annulus $15<r<25$ arcseconds around a $11 < R < 12$ star. These criteria ensure 
that the FIRST source lies within the isoplanatic patch of a sufficiently bright 
guide star for the NIR AO observations under most seeing conditions without
introducing additional observing or photometric complications (e.g., PSF wings 
of the guide star contaminating the sky background if the source is too close 
to the star or the core of the guide star saturating significant sections of the 
detectors if the star is too bright).
We expect near diffraction-limited performance 
longward of 1.2$\mu m$ \citep{roddier1999} for a typical high-order AO 
system at Mauna Kea.  In the optical, where only tip-tilt correction is 
available, the point-source sensitivity is increased by up to a factor
of four over non-compensated images.  Since we are probably not resolving 
the high redshift (z$>$1) FIRST sources, we would expect similar performance. 

Table 1 lists the coordinates of each radio source and corresponding
USNO star corrected for proper motion to 2005.0.  These 58 objects
comprise the FIRST-BNGS sample used in this paper.  About 90\% are expected to be FRI or FRII
galaxies \citep{jackson2005}, and based on scaling the 151MHz local radio luminosity
function to 1.4 GHz, probably half will be at significant redshift
($z>1$).

\section{Observations}

Here we describe our optical and near-infrared imaging data.  Figures 1-8 show
54$''$x54$''$ FIRST radio contour maps and 14$''$x14$''$ 
thumbnail NIR images for each source overlaid with FIRST
radio contours. 

\subsection{Optical Imaging}

The broadband optical imaging data of the FIRST-BNGS sample were obtained with the 
Orthogonal Parallel Transfer
Imaging Camera (OPTIC; \citet{tonry2002}) mounted at the f/10 focus of 
the University of Hawai`i 2.2-meter (UH2.2m) telescope and at the Nasmyth
focus of the Wisconsin, Indiana, Yale, NOAO (WIYN) 3-meter telescope at Kitt
Peak National Observatory.  This camera utilizes an effective
``tip-tilt'' correction feature of a specialized CCD, called an
orthogonal transfer array, 
with no physically moving parts. 
This technique moves 
the accumulated charge from an astronomical source around on the CCD
based on the centroid of a  
guide star read out at short intervals (20-100ms) in a nearby region
of the chip.  All of our sources are well within a BNGS isoplanatic patch for
tip-tilt correction (about 10$'$ for a 2 meter telescope).  The tip-tilt usually 
improves the point-source sensitivity by nearly a factor of four in most seeing 
conditions and slightly increases the achieved spatial resolution.  Our best measured FWHM
was 0.3$''$ for a 300 second exposure.  Since we expect the angular size
of these high redshift sources to be about this scale, tip-tilt 
correction only provides an increase in efficiency for flux measurements.

Our observations consisted of several sequential orthogonally
transferred (OT) 300-second exposures in various 
broadband filters (B, V, R, I, and z$'$).
Data were taken on about 50 nights between 2002 December and 2005
April (Table 2) under photometric conditions.  The 6$'$ field of view
was oriented strategically with the  
bright guide star in the guiding region of the chip and 
the object in the imaging region.  Due to the proximity of the science target to 
the guide star, diffraction spikes and PSF wings were sometimes a
problem.  In addition, those objects that were particularly close to
the guide star were at the edge of the detector's science region.  As 
the guiding regions are at the top and bottom of the detector, the 
instrument was occasionally rotated 90 degrees in order to accommodate
target objects to the east and west  
of the guide star.  Supplemental rough guiding
was also accomplished by OPTIC communicating frequent guide offsets to
the telescope control system. 
 
With the exception of the generation of flat-field frames (``flats''),
standard IRAF procedures were for the data reduction. 
The flats were constructed using a special program, kindly 
provided by John Tonry, called ``conflat'', which constructs
a flat field from any normal flat image (in this case, the flat was
created from the median of several high-signal dome flat images) by
convolving it with the sequence of OT 
shifts executed by OPTIC during the science exposure.  A separate
convolved flat is therefore needed 
for each exposure.  Each set of images was flat-fielded, background
subtracted, aligned using bright stars, then averaged together.  An
absolute position was derived from the location of the bright guide
star using the USNO-A2.0 catalog \citep{monet1998}.  A similar
procedure was used for sequences of short exposure on photometric
standard \citet{landolt1992} fields, but the frames were not stacked
so that the individual exposures could be used to estimate
uncertainties in the calibration.  Photometric zeropoints were derived
based on these Landolt standards in the Vega magnitude system.

\subsection{NIR Imaging}

For the NIR survey of 58 FIRST-BNGS sources, our strategy was to obtain at least 
H-band photometry for the entire sample.  Because the HzRGs in the
sample are relatively compact, this was most efficiently accomplished using
the 3.6-meter Canada-France-Hawai`i Telescope (CFHT) with the 
the Pueo AO bonnette and KIR infrared detector \citep{rigaut1998}.  Pueo-KIR 
incorporates a wavefront curvature sensor \citep{roddier1991} and a
19-electrode deformable mirror.  This provides near
diffraction-limited imaging in good conditions using guide stars similar to 
those in the FIRST-BNGS sample.  Since the diffraction limit of CFHT is around 0.2$''$ in 
H-band, most the high redshift sources are barely resolved or unresolved, and therefore 
we are presenting only flux measurement from these observations.  However, the increase in point-source 
sensitivity over non-AO NIR observations allowed us to rapidly observe all sources in 
the sample in a handful of nights.  Some additional non-AO J, H, and K
photometric data were taken with the Quick Infrared Camera (QUIRC;
\citet{hodapp1995}) on the UH 2.2m telescope and the SpeX imager
\citep{rayner2003} on the IRTF to supplement the CFHT H-band data.  A subsequent 
K-band AO imaging campaign to measure morphologies of these sources was carried 
out on a subsample of 18 FIRST-BNGS high redshift candidates (Stalder \& 
Chambers\nocite{stalder2009a}, in prep.).
These observations were made using the Subaru 8-meter telescope
and the Infrared Camera and Spectrograph (IRCS) \citep{kobayashi2000} mounted
behind the 36-element curvature-sensing AO system and the photometry
is included in this data set.   

\begin{deluxetable*}{rlrrrrrrrrrrrrrrrrrr}
\tabletypesize{\scriptsize}
\tablecaption{Object Positions (J2000.0).}
\tablehead{
\multicolumn{1}{r}{$\#$} & \multicolumn{1}{l}{Name} & 
\multicolumn{1}{c}{ID RA} & 
\multicolumn{1}{c}{ID Dec} & 
\multicolumn{1}{c}{USNO-A2.0} &
\multicolumn{1}{c}{R} & \multicolumn{1}{c}{B-R} & 
\multicolumn{1}{c}{RA offset} & \multicolumn{1}{c}{Dec offset} & \multicolumn{1}{c}{Tot offset} &
\multicolumn{1}{c}{$b_{II}$} & \multicolumn{1}{c}{$S_{1.4Ghz}$}
}
\startdata
1	& F0023-0904 	 & 00:23:57.043 & -09:04:43.02 & 0750-00092600 & 11.6 & 1.1 & -9.01 & 25.44 & 26.99 & -53 & 15.28 \\ 
2	& F0129-0140 	 & 01:29:42.917 & -01:40:40.03 & 0825-00339308 & 12.0 & 1.0 & 13.66 & 16.92 & 21.75 & -59 & 3.70 \\ 
3	& F0152-0029 	 & 01:52:00.671 & -00:29:13.53 & 0825-00426996 & 11.5 & 1.2 & 15.30 & 8.59 & 17.55 & -58 & 24.69 \\ 
4	& F0152+0052 	 & 01:52:16.147 & +00:52:16.35 & 0900-00437561 & 11.1 & 0.8 & -29.71 & -3.29 & 29.89 & -57 & 16.85 \\ 
5	& F0202-0021 	 & 02:02:37.247 & -00:21:00.82 & 0825-00472596 & 11.0 & 1.5 & -15.66 & 19.85 & 25.28 & -57 & 2.79 \\ 
6	& F0216+0038 	 & 02:16:46.186 & +00:39:00.90 & 0900-00529744 & 12.0 & 1.6 & -14.83 & 6.41 & 16.16 & -54 & 29.94 \\ 
7	& F0916+1134 	 & 09:16:08.084 & +11:34:23.04 & 0975-06196986 & 11.0 & 0.7 & 19.04 & 2.09 & 19.16 & 35 & 4.54 \\ 
8	& F0919+1007 	 & 09:19:34.330 & +10:07:22.55 & 0975-06212600 & 11.6 & 0.7 & -16.44 & -1.12 & 16.47 & 36 & 8.97 \\ 
9	& F0938+2326 	 & 09:38:39.209 & +23:26:43.90 & 1125-05944823 & 11.4 & 1.9 & 22.97 & 9.92 & 25.02 & 36 & 8.09 \\ 
10	& F0939-0128 	 & 09:39:43.995 & -01:28:03.09 & 0825-06948720 & 11.6 & 0.9 & -20.20 & 11.81 & 23.40 & 37 & 5.06 \\ 
11	& F0942+1520 	 & 09:42:58.821 & +15:20:28.01 & 1050-06112221 & 11.8 & 1.4 & 11.44 & 12.18 & 16.71 & 40 & 15.70 \\ 
12	& F0943-0327 	 & 09:43:15.625 & -03:27:03.82 & 0825-06969818 & 11.0 & 1.1 & 7.82 & -17.95 & 19.58 & 39 & 99.49 \\ 
13	& F0950+1619 	 & 09:50:36.928 & +16:19:53.27 & 1050-06140249 & 11.0 & 1.4 & 8.21 & -19.06 & 20.75 & 42 & 1.69 \\ 
14	& F0952+2405 	 & 09:52:20.644 & +24:05:53.87 & 1125-05996595 & 11.2 & 1.1 & 24.17 & 4.27 & 24.54 & 38 & 1.27 \\ 
15	& F0955+2951 	 & 09:55:12.289 & +29:51:30.83 & 1125-06008285 & 11.5 & 0.7 & -21.52 & -9.50 & 23.52 & 35 & 6.65 \\ 
16	& F0955+0113 	 & 09:55:18.949 & +01:13:37.24 & 0900-06543413 & 11.5 & 0.9 & 16.53 & 4.55 & 17.14 & 40 & 8.58 \\ 
17	& F0956-0533 	 & 09:56:12.923 & -05:33:20.54 & 0825-07043060 & 11.5 & 1.7 & 6.61 & -18.81 & 19.94 & 42 & 2.40 \\ 
18	& F0958+2721 	 & 09:58:46.919 & +27:21:17.78 & 1125-06020281 & 11.7 & 0.9 & -15.25 & 2.71 & 15.49 & 37 & 2.62 \\ 
19	& F1000-0636 	 & 10:00:03.474 & -06:36:38.53 & 0825-07062955 & 11.9 & 1.0 & -10.58 & -17.47 & 20.42 & 44 & 1.41 \\ 
20	& F1008-0605 	 & 10:08:34.084 & -06:05:29.94 & 0825-07106082 & 11.7 & 2.3 & 2.07 & -16.13 & 16.26 & 45 & 1.52 \\ 
21a	& F1010+2527N 	 & 10:10:09.833 & +25:27:58.94 & 1125-06057690 & 11.3 & 1.7 & -19.57 & -15.23 & 24.80 & 41 & 1.31 \\ 
21b	& F1010+2527S 	 & 10:10:09.854 & +25:27:58.26 & 1125-06057690 & 11.3 & 1.7 & -19.28 & -15.91 & 25.00 & 41 & 1.31 \\ 
22	& F1010+2727 	 & 10:10:17.095 & +27:27:39.14 & 1125-06058003 & 11.4 & 1.0 & -14.40 & 6.47 & 15.79 & 39 & 5.94 \\ 
23	& F1014+1438 	 & 10:14:30.351 & +14:38:55.90 & 0975-06445830 & 11.5 & 1.1 & 8.74 & 22.55 & 24.18 & 47 & 32.74 \\ 
24	& F1016+1513 	 & 10:16:53.648 & +15:13:02.45 & 1050-06244113 & 11.2 & 1.0 & -4.26 & 16.41 & 16.95 & 47 & 6.88 \\ 
25	& F1024-0031 	 & 10:24:23.499 & -00:31:21.86 & 0825-07189101 & 12.0 & 1.0 & 21.04 & -13.22 & 24.85 & 46 & 158.37 \\ 
26	& F1027+0520 	 & 10:27:51.341 & +05:20:51.45 & 0900-06695802 & 11.9 & 0.9 & 22.22 & 1.86 & 22.30 & 49 & 22.64 \\ 
27	& F1039+2602 	 & 10:39:57.545 & +26:02:12.17 & 1125-06156471 & 12.0 & 1.0 & -21.27 & 17.16 & 27.33 & 46 & 11.53 \\ 
28	& F1040+2323 	 & 10:40:53.550 & +23:23:31.69 & 1125-06159536 & 11.6 & 0.7 & -23.65 & -0.74 & 23.66 & 49 & 1.65 \\ 
29	& F1116+0235 	 & 11:16:10.460 & +02:35:46.26 & 0900-06888400 & 11.0 & 0.7 & 5.41 & 17.89 & 18.69 & 56 & 2.00 \\ 
30	& F1133+0312 	 & 11:33:01.502 & +03:12:20.78 & 0900-06953454 & 11.8 & 0.3 & 21.64 & 23.57 & 32.00 & 59 & 0.78 \\ 
31	& F1140+1316 	 & 11:40:48.122 & +13:16:19.10 & 0975-06779562 & 11.7 & 1.0 & 11.49 & 26.40 & 28.79 & 65 & 5.21 \\ 
32	& F1147+2647 	 & 11:47:47.329 & +26:47:46.69 & 1125-06384622 & 11.9 & 0.6 & -19.52 & 12.69 & 23.28 & 59 & 0.66 \\ 
33	& F1155+2620 	 & 11:55:18.857 & +26:20:04.88 & 1125-06409314 & 11.6 & 1.0 & 5.57 & 21.69 & 22.39 & 61 & 0.99 \\ 
34	& F1158+1716 	 & 11:58:45.785 & +17:16:53.16 & 1050-06622092 & 11.9 & 1.0 & -23.17 & 6.09 & 23.96 & 68 & 2.64 \\ 
35	& F1202+0654 	 & 12:02:35.855 & +06:54:58.22 & 0900-07060106 & 11.1 & 0.9 & -0.73 & -19.62 & 19.63 & 66 & 0.99 \\ 
36	& F1211+3616 	 & 12:11:36.405 & +36:16:31.00 & 1200-06860134 & 11.9 & 0.7 & -9.01 & -15.11 & 17.59 & 51 & 1.14 \\ 
37	& F1215+3242 	 & 12:15:51.397 & +32:43:14.57 & 1200-06870734 & 11.3 & 0.7 & -9.83 & 27.33 & 29.04 & 57 & 45.07 \\ 
38a	& F1217-0529E 	 & 12:17:19.755 & -05:29:18.87 & 0825-07696891 & 11.5 & 0.9 & 17.66 & 9.33 & 19.98 & 66 & 10.23 \\ 
38b	& F1217-0529W 	 & 12:17:19.657 & -05:29:18.36 & 0825-07696891 & 11.5 & 0.9 & 16.20 & 9.84 & 18.95 & 66 & 10.23 \\ 
39	& F1217+3810 	 & 12:17:35.870 & +38:10:51.30 & 1275-08065221 & 11.7 & 1.0 & -7.45 & -18.79 & 20.21 & 49 & 0.84 \\ 
40	& F1218-0625 	 & 12:18:44.115 & -06:25:36.97 & 0825-07703320 & 12.0 & 0.3 & -14.79 & -4.57 & 15.48 & 67 & 4.29 \\ 
41	& F1218-0716 	 & 12:18:53.524 & -07:16:18.85 & 0825-07703926 & 11.3 & 0.7 & 15.94 & -9.57 & 18.59 & 68 & 0.79 \\ 
42	& F1234+2001 	 & 12:34:32.770 & +20:01:33.04 & 1050-06743069 & 11.2 & 1.0 & 13.95 & -0.82 & 13.98 & 74 & 4.80 \\ 
43	& F1237+1141 	 & 12:37:17.118 & +11:41:15.10 & 0975-06997522 & 11.5 & 0.8 & -21.43 & 16.90 & 27.29 & 73 & 17.83 \\ 
44	& F1315+4438 	 & 13:15:41.557 & +44:38:21.80 & 1275-08221253 & 11.4 & 0.9 & 13.79 & 11.53 & 17.98 & 46 & 0.89 \\ 
45	& F1329+1748 	 & 13:29:10.570 & +17:48:10.35 & 1050-06936553 & 11.7 & 0.5 & -2.30 & -23.79 & 23.90 & 80 & 1.34 \\ 
46	& F1355+3607 	 & 13:55:29.855 & +36:07:12.61 & 1200-07189476 & 11.5 & 1.0 & 4.24 & 17.59 & 18.09 & 68 & 3.48 \\ 
47	& F1430+3557 	 & 14:30:08.889 & +35:57:19.20 & 1200-07317559 & 11.7 & 0.5 & 15.69 & 9.84 & 18.52 & 73 & 8.54 \\ 
48	& F1435-0019 	 & 14:35:28.042 & -00:19:51.33 & 0825-08378453 & 11.1 & 0.5 & -2.28 & -10.36 & 10.61 & 52 & 10.00 \\ 
49	& F1445+2702 	 & 14:45:36.895 & +27:02:27.19 & 1125-07063557 & 11.9 & 0.8 & 0.88 & 21.41 & 21.43 & 85 & 34.22 \\ 
50	& F1447+1217 	 & 14:47:33.593 & +12:17:11.51 & 0975-07595012 & 12.0 & 0.7 & 25.05 & -5.40 & 25.62 & 62 & 48.90 \\ 
51	& F1451+0556 	 & 14:51:15.153 & +05:56:43.97 & 0900-07812705 & 11.4 & 0.7 & 14.16 & -15.23 & 20.79 & 54 & 2.35 \\ 
52a	& F1458+4319NW 	 & 14:58:50.327 & +43:19:46.79 & 1275-08563676 & 11.9 & 0.9 & -11.33 & -14.85 & 18.68 & 61 & 11.11 \\ 
52b	& F1458+4319SE 	 & 14:58:50.410 & +43:19:44.66 & 1275-08563676 & 11.9 & 0.9 & -10.42 & -16.98 & 19.92 & 61 & 11.11 \\ 
52c	& F1458+4319E 	 & 14:58:50.823 & +43:19:43.44 & 1275-08563676 & 11.9 & 0.9 & -5.91 & -18.20 & 19.14 & 61 & 11.11 \\ 
53	& F1505+4457 	 & 15:05:11.548 & +44:57:31.91 & 1275-08584140 & 11.3 & 0.8 & 4.97 & 12.34 & 13.30 & 59 & 17.08 \\ 
54	& F1524+5122 	 & 15:24:06.270 & +51:22:15.02 & 1350-08618364 & 11.6 & 0.8 & 1.41 & 20.85 & 20.90 & 46 & 22.45 \\ 
55	& F1644+2554 	 & 16:44:34.831 & +25:54:37.12 & 1125-07786245 & 11.9 & 0.7 & 1.97 & 20.59 & 20.68 & 59 & 0.66 \\ 
56	& F2217-0837 	 & 22:17:18.793 & -08:37:41.25 & 0750-21161298 & 11.4 & 0.2 & 0.56 & -18.77 & 18.78 & -35 & 0.99 \\ 
57a	& F2217-0138E 	 & 22:17:47.228 & -01:38:45.55 & 0825-19614577 & 11.9 & 0.9 & -2.52 & 21.16 & 21.31 & -43 & 26.37 \\ 
57b	& F2217-0138W 	 & 22:17:47.112 & -01:38:45.78 & 0825-19614577 & 11.9 & 0.9 & -4.26 & 20.93 & 21.36 & -43 & 26.37 \\ 
58	& F2354-0055 	 & 23:54:42.286 & -00:55:29.38 & 0825-20052970 & 11.1 & 1.2 & 9.76 & 19.49 & 21.80 & -58 & 1.38 \\ 
\enddata
\end{deluxetable*}

\begin{figure*}[htp]
\vspace{+1.5in}
\begin{center}
\includegraphics[scale=0.8,angle=0,viewport= 20 150 600 700]{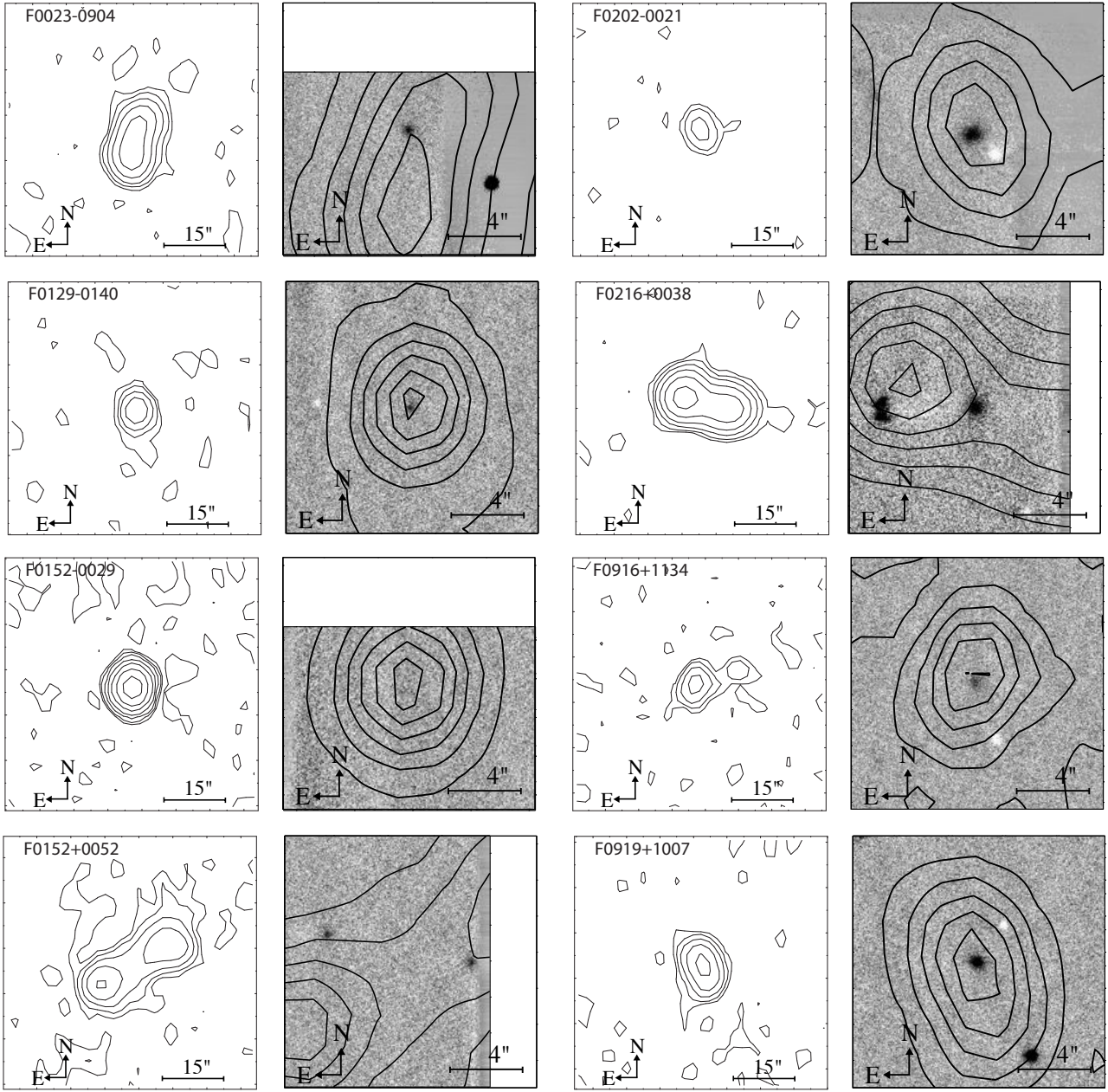}
\end{center}
\vspace{+0.5in}
\caption{\scriptsize 
 Images of 8 radio source fields from the BNGS sample. 
         Images are centered on the optical identification or center of the 
	 radio source if undetected.
         The left panel is the FIRST contour radio map.
         The right panel is H-band Pueo-KIR AO images.
         The FIRST radio map is 54$''$ on a side (the
         contours are set as  
         0.25, 0.5, 1, 2, 4, 8, 16, and 32 mJy)
	 and the H-band images are 14$''$ on a side.  The
         contours are set to be 5 equal linear levels 
	 from the peak flux in the field to 0 mJy flux density.
         At left, from top to bottom shows the optical/IR sources, F0023-0904,
         F0129-0140, F0152-0029, and F0152+0052.  At right, from top to bottom shows, 
         F0202-0021, F0216+0038, F0916+1134, and F0919+1007. 
	 Note that there is the
         contamination from negative (white) amplifier and positive (black hourglass) 
         ghosting artifacts (in
         Pueo-KIR images) from the saturated guide star.  The edges of
         the detectors are also visible in some frames.
         }
\end{figure*}

\begin{figure*}[htp]
\vspace{+0.9in}
\begin{center}
\includegraphics[scale=0.8,angle=0,viewport= 20 150 600 700]{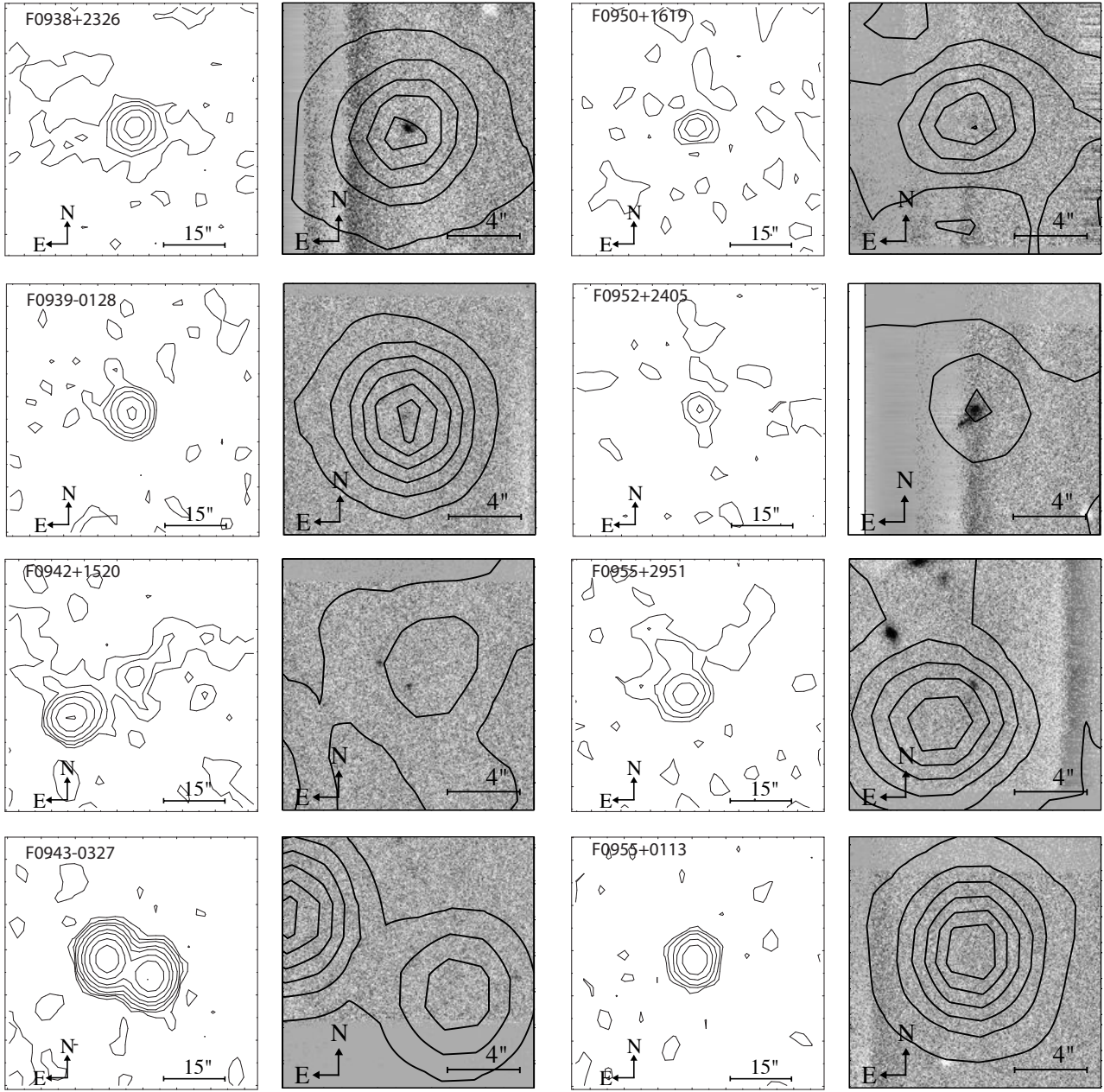}
\end{center}
\vspace{+0.5in}
\caption{\scriptsize
Same as Figure 1, for 8 additional
  objects.  This set at left, from top to bottom shows, 
F0938+2326, F0939-0128, F0942+1520, and F0943-0327.
This set at right, from top to bottom shows, 
	F0950+1619, F0952+2405, F0955+2951, and F0955+0113.
	 }
\end{figure*}

\begin{figure*}[htp]
\vspace{+0.9in}
\begin{center}
\includegraphics[scale=0.8,angle=0,viewport= 20 150 600 700]{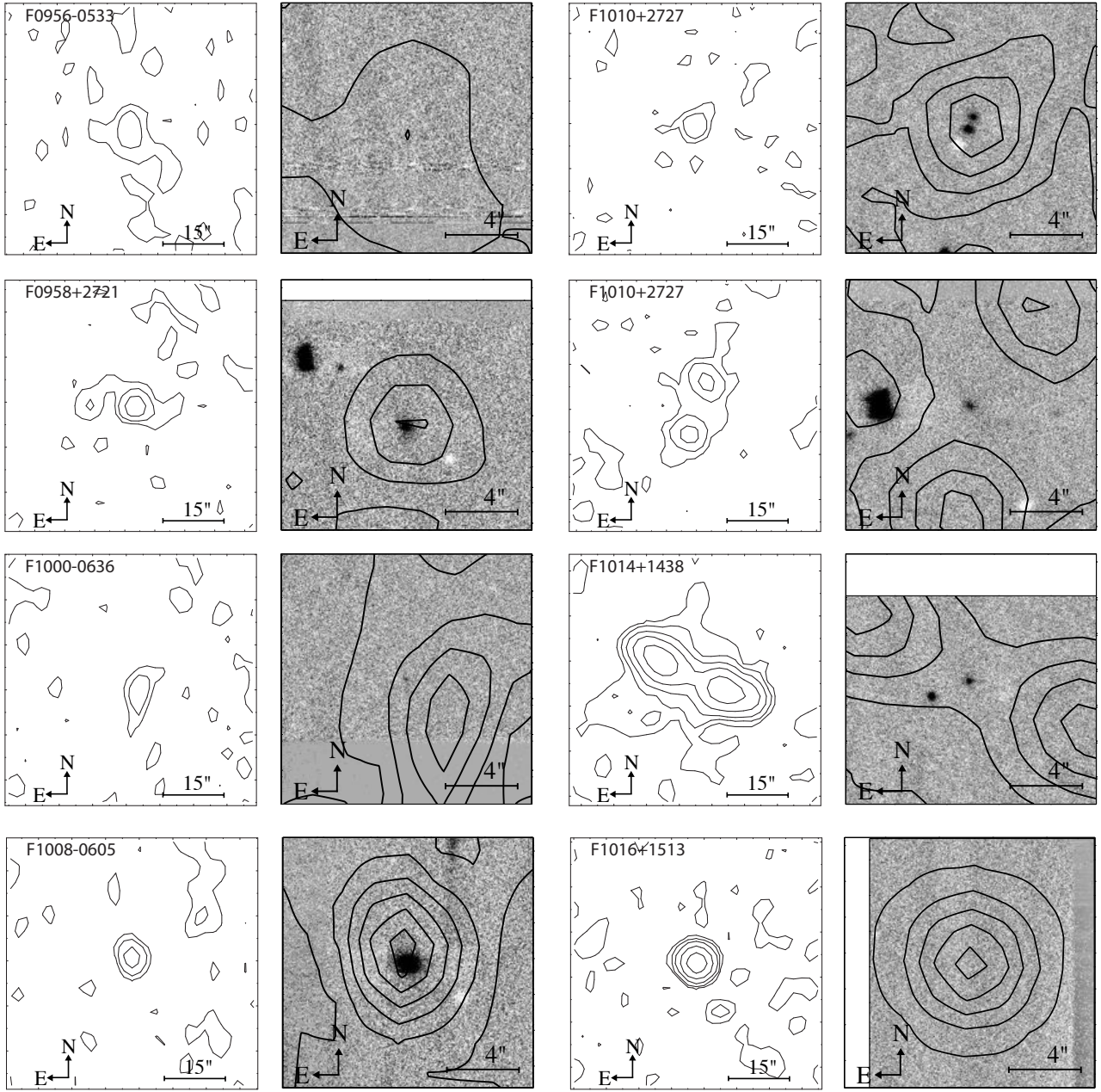}
\end{center}
\vspace{+0.5in}
\caption{\scriptsize
Same as Figure 1, for 8 additional
  objects.  This set at left, from top to bottom shows, 
F0956-0533, F0958+2721, F1000-0636, and
	 F1008-0605.
This set at right, from top to bottom shows, 
	 F1010+2527, F1010+2727, F1014+1438, and F1016+1513.
	 }
\end{figure*}

\begin{figure*}[htp]
\vspace{+0.9in}
\begin{center}
\includegraphics[scale=0.8,angle=0,viewport= 20 150 600 700]{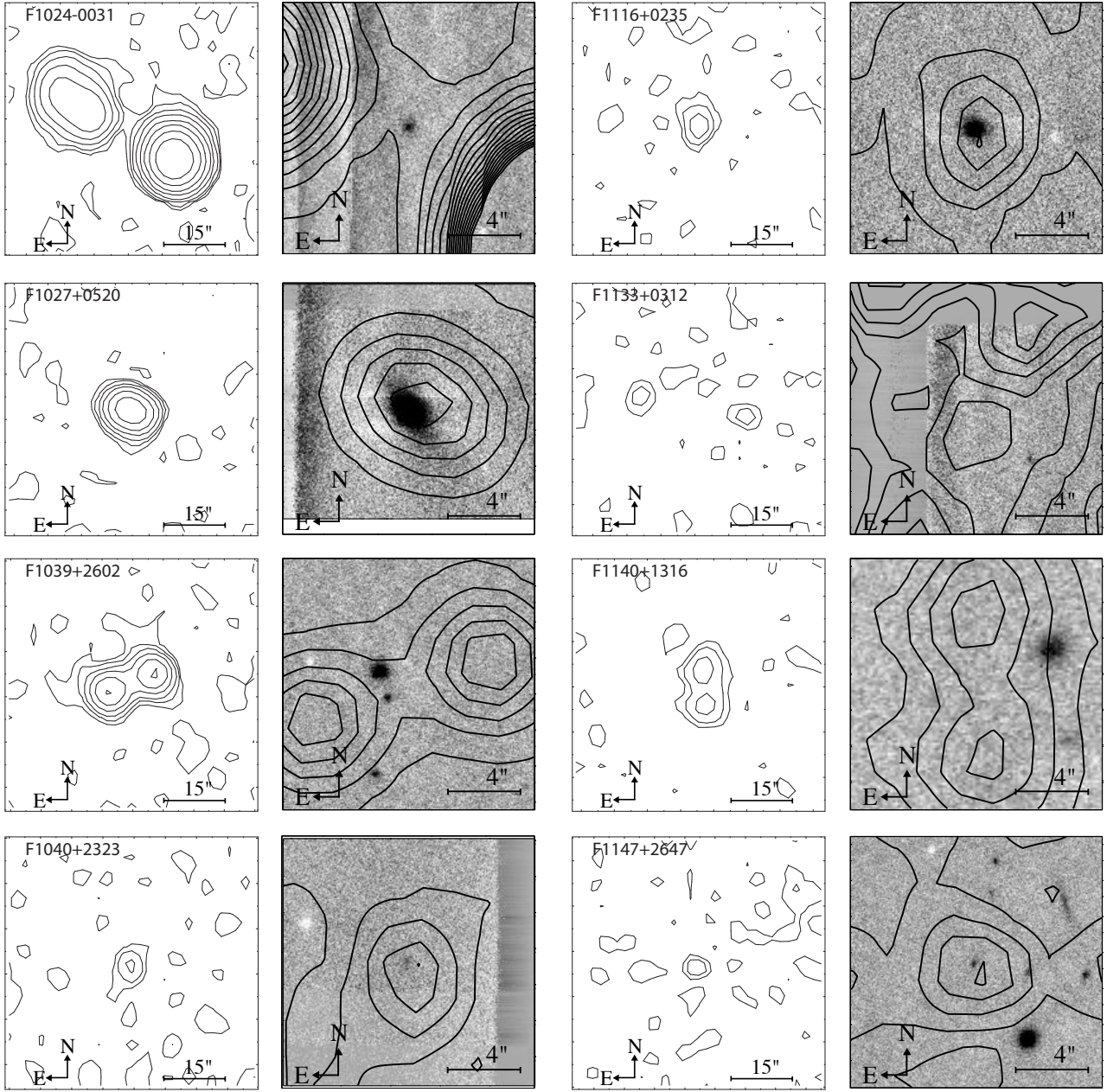}
\end{center}
\vspace{+0.5in}
\caption{\scriptsize
Same as Figure 1, for 8 additional
  objects with the exception that F1140+1316 is OPTIC I-band with the same spatial scale and FIRST contours as the H-band Pueo images.  This set at left, from top to bottom shows, 
	F1024-0031, F1027+0520, F1039+2602, and F1040+2323.
This set at right, from top to bottom shows, 
F1116+0235, F1133+0312, F1140+1316, and
	F1147+2647.
	 }
\end{figure*}

\begin{figure*}[htp]
\vspace{+0.9in}
\begin{center}
\includegraphics[scale=0.8,angle=0,viewport= 20 150 600 700]{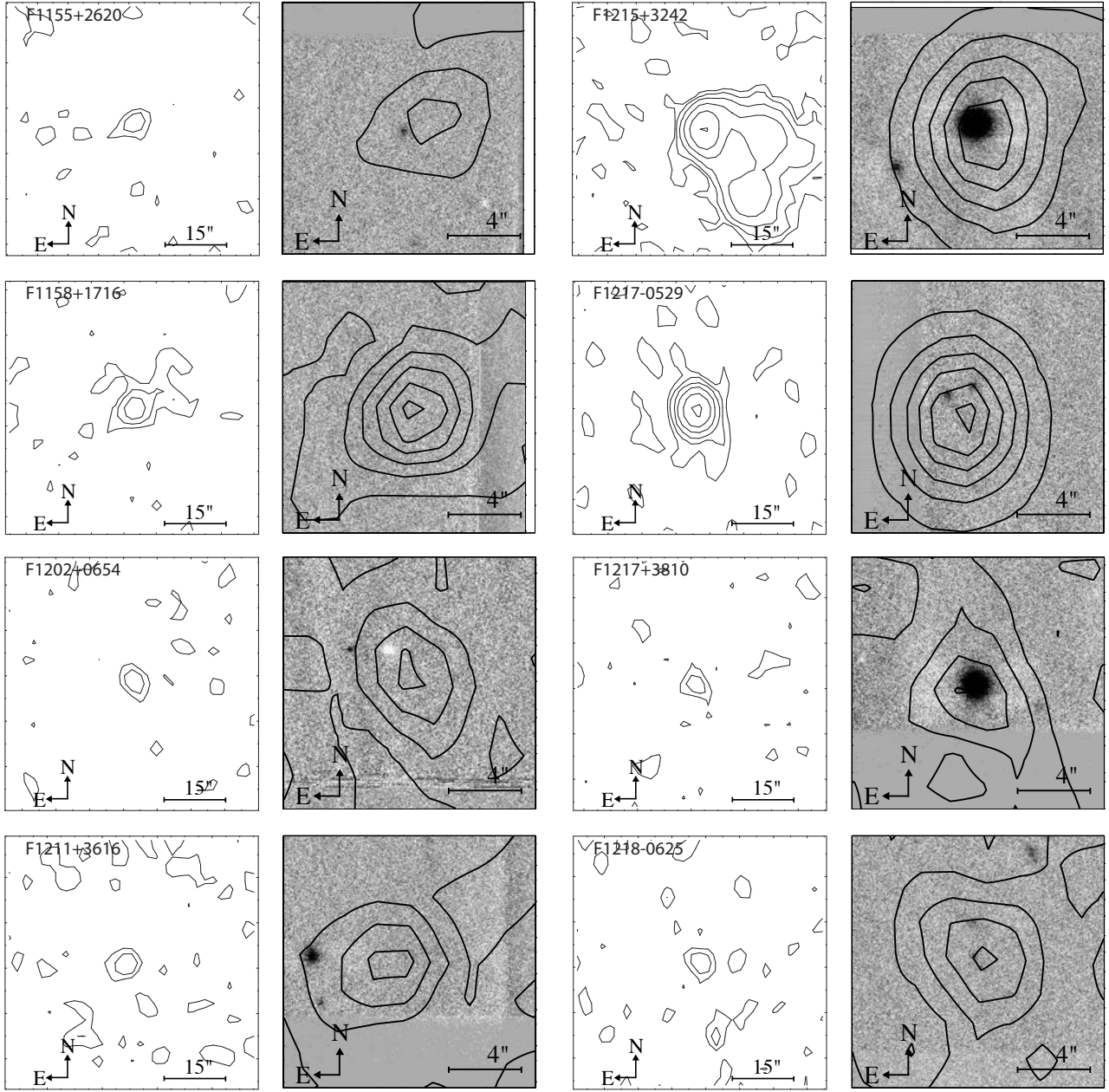}
\end{center}
\vspace{+0.5in}
\caption{\scriptsize
Same as Figure 1, for 8 additional
  objects.  This set at left, from top to bottom shows, 
	 F1155+2620, F1158+1716, F1202+0654, and F1211+3616.
This set at right, from top to bottom shows, 
F1215+3242, F1217-0529, F1217+3810, and F1218-0625.
	 }
\end{figure*}

\begin{figure*}[htp]
\vspace{+0.9in}
\begin{center}
\includegraphics[scale=0.8,angle=0,viewport= 20 150 600 700]{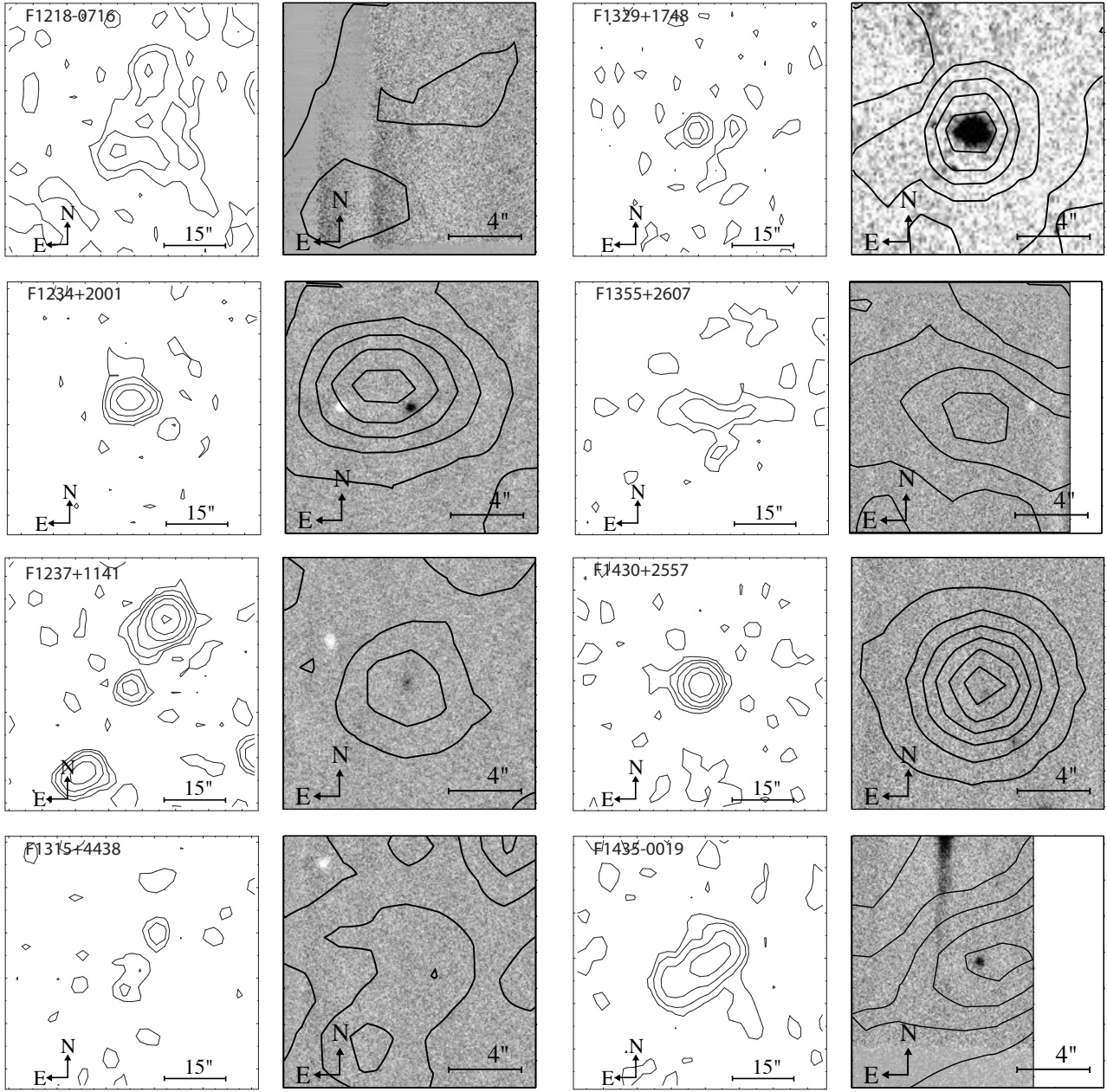}
\end{center}
\vspace{+0.5in}
\caption{\scriptsize
Same as Figure 1, for 8 additional
  objects with the exception that F1329+1748 is OPTIC I-band with the same spatial scale and FIRST contours as the H-band Pueo images.  This set at left, from top to bottom shows, 
	 F1218-0716, F1234+2001, F1237+1141, and F1315+4438.
This set at right, from top to bottom shows, 
	F1329+1748, F1355+3607, F1430+3557, and F1435-0019.
	 }
\end{figure*}

\begin{figure*}[htp]
\vspace{+0.9in}
\begin{center}
\includegraphics[scale=0.8,angle=0,viewport= 20 150 600 700]{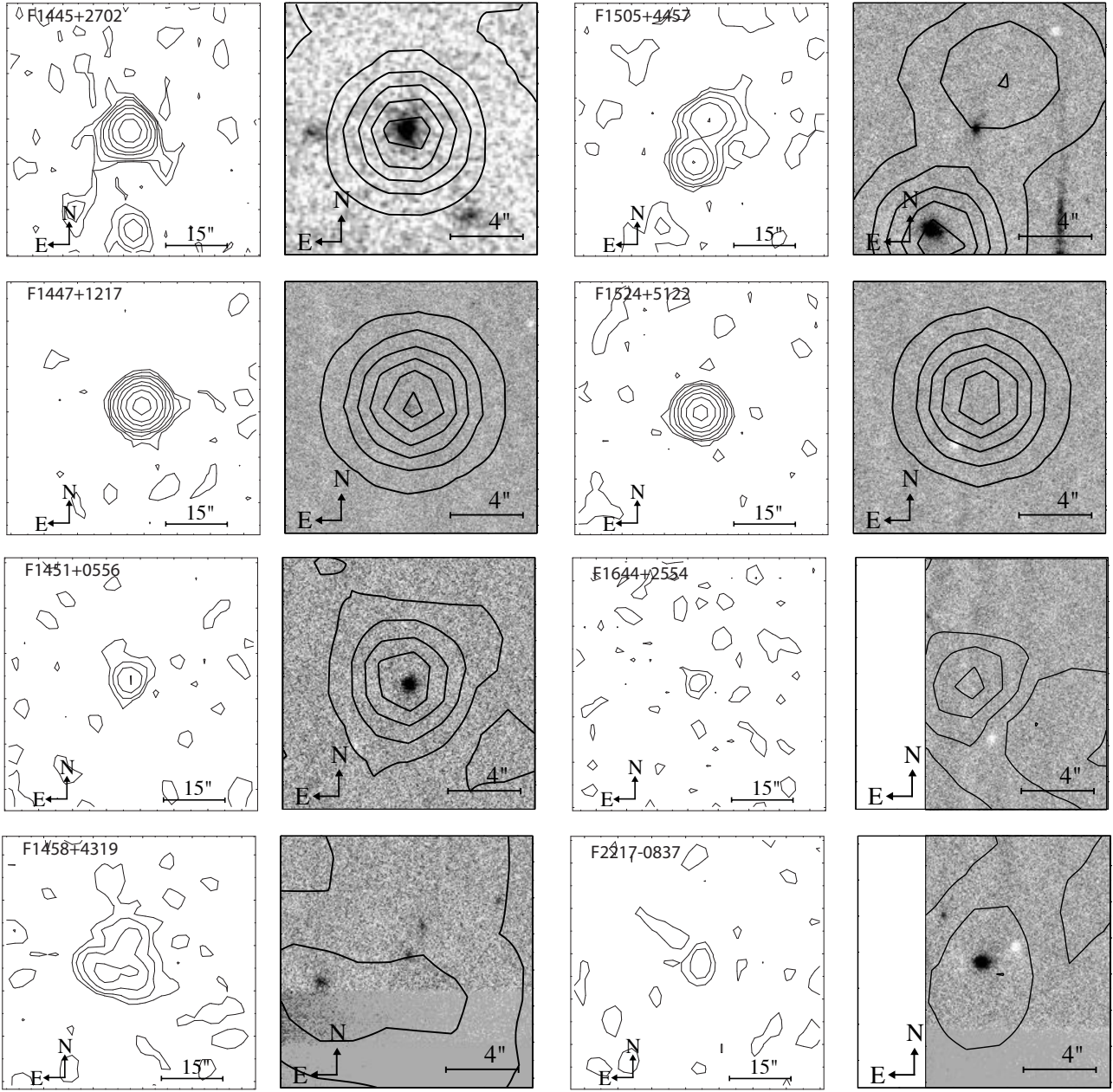}
\end{center}
\vspace{+0.5in}
\caption{\scriptsize
Same as Figure 1, for 8 additional
  objects with the exception that F1445+2702 is OPTIC I-band with the same spatial scale and FIRST contours as the H-band Pueo images.  This set at left, from top to bottom shows, 
F1445+2702, F1447+1217, F1451+0556, and F1458+4319.
This set at right, from top to bottom shows, 
	F1505+4457, F1524+5122, F1644+2554, and F2217-0837.
}
\end{figure*}

\begin{figure*}[htp]
\vspace{+1.2in}
\begin{center}
\includegraphics[scale=0.8,angle=0,viewport= 20 150 600 700]{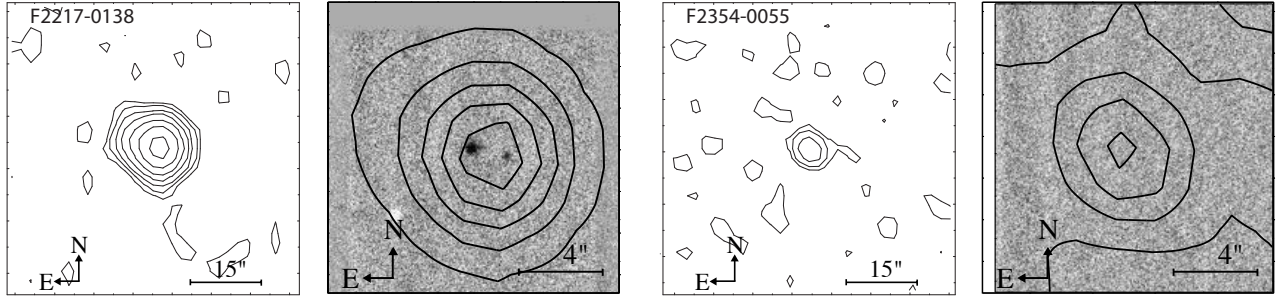}
\end{center}
\vspace{-5.0in}
\caption{\scriptsize
Same as Figure 1, for 2 additional
  objects.  At left is
	F2217-0138 and at right is F2354-0055.
}
\vspace{1.0in}
\end{figure*}

\begin{deluxetable*}{llllc}
\tabletypesize{\scriptsize}
\tablecaption{OPTIC and NIR observations}
\tablehead{
\multicolumn{1}{c}{Telescope} & 
\multicolumn{1}{c}{Instrument} & \multicolumn{1}{c}{Filters} & \multicolumn{1}{c}{Dates (UT)} & 
\multicolumn{1}{c}{FWHM\tablenotemark{a}}}
\startdata
UH 2.2-meter & OPTIC    & B,V,R,I,z$'$ & 2002 April 23-27 & 1.0$''$-1.5$''$\\
&&& 2002 December 7-10  & 0.7$''$-1.2$''$\\
&&& 2003 January 24-27  & 0.5$''$-0.9$''$\\
&&& 2003 March 20-25    & 0.6$''$-1.2$''$\\
&&& 2003 May 20         & 1.0$''$-1.2$''$\\
&&& 2003 August 1-5     & 0.7$''$-1.0$''$\\
&&& 2004 February 27    & 0.8$''$-1.0$''$\\
&&& 2004 July 28        & 0.5$''$\\
&&& 2004 August 21-24   & 0.8$''$-2.0$''$\\
&&& 2005 February 10-17 & 0.7$''$-2.5$''$\\
WIYN         & OPTIC    & B,V,R,I,z$'$ & 2003 October 14 & 0.6$''$-1.2$''$\\
&&& 2004 June 4         & 1.0$''$-1.2$''$\\
&&& 2004 November 9-10  & 1.2$''$-1.5$''$\\
&&& 2005 April 26-28    & 1.0$''$-1.2$''$\\
&&& 2006 January 1      & 1.0$''$ \\
&&& 2006 June 29-31     & 0.5$''$-2.0$''$ \\
UH 2.2-meter & QUIRC	& K'    & 2003 August 10-11 & 1.0$''$ \\
IRTF		 & SPEX		& J,H,K & 2006 August 14-15 & 0.7$''$-0.9$''$\\
CFHT         & Pueo-KIR & J,H   & 2003 March 14-16  & 0.5$''$-0.7$''$\\
&&& 2003 October 11-14  & 0.2$''$-0.3$''$\\
&&& 2004 April 3-4      & 0.3$''$-0.4$''$\\
&&& 2004 October 1-3    & 0.2$''$-0.4$''$\\
&&& 2005 April 17-23    & 0.2$''$-0.3$''$\\
Subaru		& IRCS	& K'	& 2004 February 2-3 & 0.2$''$-0.5$''$\\
&&& 2004 November 29    & 0.5$''$ \\
&&& 2005 January 16     & 0.4$''$\\
&&& 2005 February 18-19 & 0.3$''$-0.6$''$\\
\enddata
\tablenotetext{a}{Corrected FWHM range in observed band.}
\end{deluxetable*}
\normalsize

The J and H AO data were collected under excellent seeing and
photometric conditions over the course of 
19 nights (Table 2).  The K-band AO data were obtained during
photometric conditions, but seeing conditions were below average
(1.5-2.0$''$ natural seeing in V-band) on several of the nights,
unfortunately giving a mediocre corrected K-band FWHM (0.2-0.5$''$).  
The non-AO data were under photometric and varying seeing conditions.

Calculated Strehl ratios (the ratio of the maximum of the on-axis PSF
to the maximum of a theoretical diffraction-limited PSF) for Pueo-KIR
were around 10$\%$ in J and consistently above 40$\%$ and
reached as high as 80$\%$-90$\%$ in H, and remained stable throughout
these nights.  The IRCS K-band Strehls were around 20-25$\%$.  These
are typical Subaru Strehl ratios because the system has only 36
subapertures on an 8-meter diameter, whereas Pueo has 19 subapertures
over the 3.6-meter diameter that more finely samples the isoplanatic
regions over the aperture.  Both AO systems resulted in near
diffraction-limited performance in average to good natural seeing
(0.15$''$ on-axis corrected FWHM for CFHT in H and 0.08$''$ for Subaru
in K).  Unfortunately, due to worse than average seeing conditions,
little of the IRCS data achieved this.  The exposure times were
180-seconds in J, 120-seconds in H, and 90 in K.  A custom dithering
pattern was used that kept the bright guide star on the field for
aligning the images, but did not allow the regions saturated 
by the guide star to overlap with the object, to avoid persistence
problems in photometric measurements and in flat frame construction.  
For the Subaru AO data, an off-axis PSF star at the same distance to
the guide star and similar brightness to our science targets was
observed each night to represent the PSF in our fitting process  
for the K-band data.  

The Pueo-KIR detector suffers from additional amplifier related 
artifacts from the saturated guide star reflected in each quadrant, plus a
``ghost'' guide star due to internal reflections within the instrument.  
These were easily identified because of their consistency, but are 
remarked on here because of the effect on the overall
quality of the data.  The amplifier artifacts can be seen as negative
(white) signals in most H-band thumbnail images, for example, just
southwest of the F0202-0021 object at H-band in Figure 2.  The
``ghost'' image can plainly be seen as an hourglass shaped object to
the east of F0216+0038 in Figure 2 and also in the
H-band image of F1010+2727 in Figure 3. 
Some objects were also near the edge of the detector FOV, 
as seen in several images (a consequence of keeping both 
guide star and science object on the chip).

The non-AO data was taken over 5 nights in 2003 August using QUIRC on the UH2.2m
and 2006 August using SpeX on IRTF with similar exposure times and
dithering patterns as the CFHT Pueo-KIR and Subaru IRCS data.   

Standard IRAF procedures were used to reduce the near-infrared data.
Median sky flats were constructed from all of the exposures of each
field then normalized by the mode of the sky values.  For each
exposure, the data was divided by the normalized flat and sky
subtracted based on the mode.  Finally, the science frames were
stacked, using the bright guide star for registration.  World
coordinate system data was also entered into the file header based on
the position of the guide star.  The same procedure was used for short
exposure photometric standard fields (by using the science object's
median sky frame closest in time to when the sequence was taken), but
the fields were not stacked in order to obtain multiple measurements.
Photometric zeropoints were derived from the magnitudes in the UKIRT
Faint Standard catalog  \citep{leggett2006} for the Mauna Kea filter
set \citep{simons2002,tokunaga2002} on the Vega magnitude system. 

\subsection{Spitzer Archive Data}

A search of all 58 FIRST-BNGS sources in the Spitzer Archives
resulted in serendipitous observations of 2 objects (F1237+1141 and
F1430+3557).  The Spitzer IRAC 3.6/5.8$\mu$m and MIPS 24$\mu$m,
70$\mu$m, and 160$\mu$m mosaiced images were produced from basic
calibrated data (BCD) images using the Mosaicking and Point Source
Extraction tool, MOPEX \citep{makovoz2005}.  Some cleaning was done
for the MIPS 70 and 160$\mu$m data also using MOPEX.  Aperture
photometry was accomplished using the IRAF ``phot'' routines with an
appropriate aperture correction from the respective IRAC and MIPS
websites.  Sensitivity was measured through fitting the background
assuming Poisson statistics.  For F1237+1141 at 3.6$\mu$m, the guide star has 
some diffraction spikes near the object, but was mitigated 
by measuring the flux with a smaller aperture and separately at two observed rotation angles 
(with different spike orientations).
For 5.8$\mu$m and the MIPS channels, the nearby guide star 
was much dimmer at these wavelengths so the flux measurements were not 
affected by gradients or artifacts.

\begin{deluxetable*}{lccrrr}
\tablecaption{Spitzer, SHARC2 and SCUBA Fluxes and 3-$\sigma$ upper-limits.}
\tablehead{
\multicolumn{1}{c}{Name} & 
\multicolumn{1}{c}{instrument} & 
\multicolumn{1}{c}{dates(UT)} &
\multicolumn{1}{c}{$\lambda (\mu m)$} & 
\multicolumn{1}{c}{$F_{\nu}$(mJy)} & 
\multicolumn{1}{c}{$\sigma_{\nu}$(mJy)}}
\startdata
F0152-0029 & SHARC2 & Nov16-19,2005  &   350 & $<$24.6 & \\
F1116+0235 & SCUBA  & Nov22,2003  &   450 & $<$39.0 & \\
F1116+0235 & SCUBA  & Nov22,2003  &   850 &     4.1 & 2.1 \\
F1237+1141 & IRAC   & May27,2004$\&$Jun09,2004  &   3.6 &  0.0673 & 0.0013 \\
F1237+1141 & IRAC   & May27,2004$\&$Jun09,2004  &   5.8 &  0.1928 & 0.0017 \\
F1237+1141 & MIPS   & Jun22,2004$\&$Jun23,2004  &  24.0 &  0.725  & 0.073 \\
F1237+1141 & MIPS   & Jun22,2004$\&$Jun23,2004  &  70.0 &  3.21   & 0.64  \\
F1237+1141 & MIPS   & Jun22,2004$\&$Jun23,2004  & 160.0 & 12.5    & 2.5   \\
F1430+3557 & MIPS   & Feb01,2004  &  24.0 &  0.374  & 0.037 \\
F1430+3557 & MIPS   & Feb01,2004  & 160.0 & $<$33   & \\
F1644+2554 & SCUBA  & Aug27,2003  &   450 & $<$90   & \\
F1644+2554 & SCUBA  & Aug27,2003  &   850 & $<$6.6  & \\
F2217-0138 & SHARC2 & Nov19,2005  &   350 & $<$195  & \\
\enddata
\end{deluxetable*}

\subsection{Submillimeter Bolometer Observations}

In an attempt to detect submillimeter emission from the objects in the
sample, we obtained time on CSO and JCMT.  
We were able to observe 2 objects in the sample using the CSO
bolometer array, SHARC2 over the course of several nights.  SHARC2
data were reduced using the CRUSH pipeline (with faint and compact
flags enabled), and aperture photometry was measured with IRAF
routines.  Though no detections were made, the background noise was
measured to get sensitivity for each field.  Several hours of JCMT
SCUBA queue time was also awarded, though only a few observations were
actually executed on 2 sources.  The photometric data were reduced and
extracted using the ORAC-DR pipeline using the nearest calibrators.  A
weak detection was made of F1116+0235, which is a low-redshift source.
Table 3 shows the measurements from these data.

\subsection{Identification}

The reduced and registered optical and near-IR images were compared to
the FIRST survey cutouts at the coordinates of the radio source in the catalog.  
Optical identifications of single radio sources were straightforward.
We required the counterpart to be  
within 3$''$ of the peak of the radio source.  This allowed for the
FIRST positional accuracy (about 1$''$) and for slightly elongated
radio structures where the central source positions are not well
measured.  For double-lobed radio sources 
to be matched, detections had to be along the radio axis and near the
center of the two lobes.  For irregular radio morphologies, many
objects in the corresponding area were  marked as possible matches
(i.e., N, S, E, W, etc.). The objects nearest to the radio source peak
were usually selected as the most probable counterparts.

\subsection{Photometry}

Calibrated flux measurements were made using the 
$\chi^2$-minimization surface-brightness fitting routine GALFIT
\citep{peng2002} for each object for all optical and NIR bands
observed.  GALFIT was chosen due to the density of sources and
presence of artifacts in the data as it fits multiple objects
simultaneously and is better able to deal with artifacts than aperture
photometry.  For the B, V, R, I, z$'$, J, H and non-AO K no
PSF-convolution was used (the Pueo-KIR J and H data have relatively
low S/N and do not have an observed off-axis PSF star), so each object was
fitted to a pure Sersic ($I\sim r^{1/n}$) profile which spans the possible
profiles of unresolved Gaussian ($n=0.5$), exponential disk
($n=1.0$), and de Vaucouleurs ($n=4.0$) .  For the K-band Subaru AO data, 
the objects were fit as Sersic profiles convolved with a
normalized PSF derived from an off-axis 17th magnitude star imaged the
same night and if necessary, smoothed to the FWHM measured in the
science frame.   These stars were selected to be at the same distance to its
guide star as the science objects, and at a similar airmass, 
so should capture the long temporal scale  
off-axis PSF as long as conditions did not change much throughout the night.
This was verified with other point sources in our science fields.
Because the background is of critical importance in
these fits, tests were run using different background fitting
algorithms, using both aperture fitting and using the GALFIT
integrated algorithm.  The aperture-measured background was found to
be more accurate and reproducible; the same conclusion was reached by
\citet{haussler2007} using both real and simulated data.  Residuals
were checked to verify that the final fit was good.  Only total
magnitudes were used from these fits.  The magnitude errors reported
by GALFIT were larger than the 0.03 mag intrinsic scatter for S/N $>$
 2.5 found by \citet{haussler2007}, so the GALFIT photometric errors 
added in quadrature to this was used as an estimate of the error.  
The result for each object was total flux in each observed filter.
These measurements are aperture and atmospheric-seeing independent as
they are derived from a model fit.  This was confirmed by smoothing
sample data and GALFIT recovering essentially the same flux (within
measurement errors).  These photometry measurements were also robust against 
gradients and artifacts caused by the nearby bright guide star, this was 
verified by repeatability of the measurements of the same object at various 
position angles and with the guide star on and off the detector.  In cases where
an artifact is particularly close to the science object, they are modeled in the 
GALFIT fit along with the science object, producing excellent residuals.
Table 4 shows the best-fit magnitudes from GALFIT or 3-sigma
upper-limits based on background noise in the data and assuming
Poisson statistics. 

\section{SED fitting}

\subsection{Optical-NIR Photometric Redshifts}

The initial redshift determinations for each object with 4 or more
bands of photometric data (plus 1 object with a large break between 2
bands, F0942+1520) were derived using the public SED fitting code,
Hyperz \citep{bolzonella2000}.  The library of the
\citet{mannucci2001} near-infrared extensions to the
\citet{kinney1996} templates for E, S0, Sa, Sb, Sc, and SB1 to SB6
(starburst models over a range of reddening) were used as the range of
possible SEDs for this sample of radio sources  
(see Appendix A for a more detailed description and comparison plots
of the templates).  Extensions in the UV and mid to far IR wavelength
regions were also added (see Appendix). 

Hyperz uses a $\chi^2$-minimization procedure to determine the
best-fit SED and redshift after applying a correction for Lyman forest
absorption from \citet{madau1995}.  Within the limitations of the
code, the redshift range was set to between $0<z<7$ in steps of 0.05, and 
the extinction in the rest-frame V-band ($A_V$) was allowed to vary
from 0 to 10 magnitudes. Reddening was calculated using the
\citet{calzetti2000} extinction law. Figures 9-17 shows the best-fit
SEDs for these objects with the parameters listed in Table 5. 

Since only low redshift templates were used, reddening, which is fit
for by Hyperz, was relied on to find most likely solution for a given
object at each redshift.  This is reasonable given the degeneracies
between star formation history, age, metallicity and reddening.  It
should be pointed out that therefore no detailed information beyond
redshift, absolute magnitude, and general SED-type (early, late,
star-burst) about these parameters should be considered certain, so
were not given in our analysis. 

All fits were confirmed using our own software that carries out a
procedure similar to that of Hyperz.  The code performs a
$\chi^2$-minimization to fit the SED which first converts magnitudes
to flux, then creates a library of every SED template at every
redshift step in the given redshift range, applies the
\citet{madau1995} absorption curve to them, integrates each redshifted
model over the filter bandpasses, computes the least squares fit
between the observed fluxes and the model fluxes, and finally finds
the minimum $\chi^2$ among all the redshifts for each model and among
all the models.  The output includes the best template, redshift, and
scale factor to convert to absolute flux.  These fits were only used
to confirm the Hyperz results because reddening was not taken into
account in our code.

\tabletypesize{\scriptsize}
\begin{deluxetable*}{rlccrcrcrcrcrcrcrc}
\tablecaption{OPTIC and NIR Photometric Data.}
\tablehead{
\multicolumn{1}{c}{$\#$} &
\multicolumn{1}{c}{Name} & 
\multicolumn{1}{c}{B} & \multicolumn{1}{c}{$\sigma_B$} &
\multicolumn{1}{c}{V} & \multicolumn{1}{c}{$\sigma_V$} &
\multicolumn{1}{c}{R} & \multicolumn{1}{c}{$\sigma_R$} &
\multicolumn{1}{c}{I} & \multicolumn{1}{c}{$\sigma_I$} &
\multicolumn{1}{c}{z$'$} & \multicolumn{1}{c}{$\sigma_{z'}$} &
\multicolumn{1}{c}{J} & \multicolumn{1}{c}{$\sigma_J$} &
\multicolumn{1}{c}{H} & \multicolumn{1}{c}{$\sigma_H$} &
\multicolumn{1}{c}{K} & \multicolumn{1}{c}{$\sigma_K$} }
\startdata
1	& F0023-0904	& 24.44   	& .16	& 24.45   	& .17	& 23.41   	& .15	& 21.49   	& .10	& 21.69   	& .11	& 19.31   	& .25	& 18.83   	& .16	&       	&    \\
2	& F0129-0140	& 23.69   	& .15	& 23.09   	& .11	& 22.68   	& .04	& 21.93   	& .21	& 21.99   	& .08	& 20.84   	& .27	& 19.23   	& .15	&       	&     \\
3	& F0152-0029	& $>$25.25 	&    	& $>$24.75 	&    	& $>$24.17 	&    	& $>$23.51 	&    	& $>$23.73 	&    	&       	&    	& $>$20.05 	&    	& 19.34   	& .12 \\
4	& F0152+0052	& $>$25.07 	&    	& $>$24.87 	&    	& $>$25.11 	&    	& $>$24.99 	&    	& $>$24.84 	&    	& $>$18.47 	&    	& $>$19.00 	&    	&       	&     \\
5	& F0202-0021	& 25.45   	& .15	& 22.61   	& .07	& 21.64   	& .06	& 19.85   	& .07	& 20.05   	& .25	& 18.31   	& .20	& 18.15   	& .24	& 16.72   	& .23 \\
6	& F0216+0038	& $>$24.73 	&    	& 22.91   	& .11	& 21.78   	& .06	& 20.20   	& .07	& 19.74   	& .11	& 18.84   	& .20	& 17.65   	& .24	& 16.37   	& .23 \\
7	& F0916+1134	&       	&    	& 22.65   	& .11	& 22.50   	& .06	& 21.84   	& .07	& 21.33   	& .23	&       	&    	& 19.69   	& .15	&       	&     \\
8	& F0919+1007	& $>$25.11 	&    	& 23.61   	& .11	& 22.66   	& .06	& 20.48   	& .07	& 19.54   	& .25	&       	&    	& 17.85   	& .15	& 17.17   	& .09 \\
9	& F0938+2326	&       	&    	& 21.73   	& .11	& 20.88   	& .06	& 20.56   	& .08	& 20.05   	& .02	&       	&    	& 19.19   	& .15	&       	&     \\
10	& F0939-0128	&       	&    	& $>$24.91 	&    	& $>$25.15 	&    	&       	&    	& $>$24.54 	&    	&       	&    	& 20.98   	& .20	&       	&     \\
11	& F0942+1520	& $>$24.56 	&    	& $>$24.31 	&    	& $>$25.08 	&    	& $>$24.26 	&    	&       	&    	&       	&    	& 20.83   	& .17	& 18.13   	& .15 \\
12	& F0943-0327	& $>$25.05 	&    	& $>$24.89 	&    	& $>$25.14 	&    	& 23.57   	& .21	&       	&    	&       	&    	& $>$20.07 	&    	&       	&     \\
13	& F0950+1619	&       	&    	&       	&    	& 23.74   	& .10	& 22.49   	& .09	& $>$24.55 	&    	&       	&    	& $>$20.00 	&    	&       	&     \\
14	& F0952+2405	&       	&    	&       	&    	&       	&    	& 20.72   	& .07	&       	&    	&       	&    	& 17.26   	& .23	&       	&     \\
15	& F0955+2951	&       	&    	& 24.13   	& .13	& 23.07   	& .07	& 21.91   	& .08	& 21.43   	& .10	&       	&    	& 20.31   	& .14	&       	&     \\
16	& F0955+0113	&       	&    	& $>$24.88 	&    	& $>$25.16 	&    	& $>$25.02 	&    	& $>$24.54 	&    	&       	&    	& $>$20.02 	&    	&       	&     \\
17	& F0956-0533	&       	&    	& $>$24.91 	&    	& $>$25.15 	&    	& $>$24.97 	&    	&       	&    	&       	&    	& $>$19.06 	&    	&       	&     \\
18	& F0958+2721	&       	&    	&       	&    	&       	&    	& 18.96   	& .07	&       	&    	&       	&    	& 17.85   	& .14	&       	&     \\
19	& F1000-0636	&       	&    	&       	&    	&       	&    	& $>$25.00 	&    	&       	&    	&       	&    	& $>$20.02 	&    	&       	&     \\
20	& F1008-0605	& 21.71   	& .16	& 20.08   	& .12	& 18.94   	& .20	& 17.98   	& .18	&       	&    	&       	&    	& 16.32   	& .14	&       	&     \\
21a	& F1010+2527N\tablenotemark{a}	& $>$23.78   	&    	& $>$23.15 	&    	& 23.73   	& .16	& 22.23   	& .16	& 22.08   	& .18	&       	&    	& 20.49   	& .17	& 18.55   	& .22 \\
21b	& F1010+2527S\tablenotemark{a}	& $>$23.78   	& .23	& $>$23.15 	& .11	& 22.48   	& .06	& 21.00   	& .07	& 21.02   	& .08	&       	&    	& 20.02   	& .14	& 17.61   	& .25 \\
22	& F1010+2727	& $>$24.22 	&    	& $>$24.47 	&    	& 22.44   	& .25	& 21.74   	& .07	& 22.27   	& .17	&       	&    	& 19.00   	& .18	& 18.08   	& .15 \\
23	& F1014+1438	& $>$25.18 	&    	& $>$25.63 	&    	& 24.46   	& .11	& 22.81   	& .07	& 22.46   	& .08	&       	&    	& 19.88   	& .17	& 18.89   	& .13 \\
24	& F1016+1513	&       	&    	&       	&    	& $>$25.28 	&    	& $>$25.14 	&    	& $>$24.58 	&    	&       	&    	& $>$20.11 	&    	&       	&     \\
25	& F1024-0031	&       	&    	&       	&    	&       	&    	&       	&    	&       	&    	&       	&    	& 17.49   	& .33	&       	&     \\
26	& F1027+0520	&       	&    	& 18.17   	& .11	& 17.33   	& .06	& 16.67   	& .07	& 16.78   	& .08	&       	&    	& 15.14   	& .14	&       	&     \\
27	& F1039+2602	& $>$25.14 	&    	& 24.61   	& .34	& 23.85   	& .45	& 22.33   	& .28	& 22.29   	& .44	&       	&    	& 20.24   	& .15	& 18.33   	& .04 \\
28	& F1040+2323	&       	&    	&       	&    	& $>$25.20 	&    	& $>$25.09 	&    	& 23.27   	& .13	&       	&    	& 20.66   	& .17	&       	&     \\
29	& F1116+0235	& 21.06   	& .15	& 20.50   	& .12	& 19.52   	& .06	& 18.68   	& .07	& 18.83   	& .08	&       	&    	& 16.58   	& .14	&       	&     \\
30	& F1133+0312	&       	&    	&       	&    	& $>$25.20 	&    	& $>$25.07 	&    	& $>$24.55 	&    	&       	&    	& $>$20.12 	&    	&       	&     \\
31	& F1140+1316	& $>$25.14 	&    	& $>$25.74 	&    	& $>$25.20 	&    	& $>$25.04 	&    	& $>$24.55 	&    	&       	&    	&       	&    	&       	&     \\
32	& F1147+2647	& $>$24.99 	&    	& $>$24.89 	&    	& 25.24   	& .10	& 22.50   	& .20	& 22.82   	& .12	&       	&    	& 20.66   	& .16	&       	&     \\
33	& F1155+2620	& $>$25.00 	&    	& $>$24.91 	&    	& 23.88   	& .19	& 22.30   	& .08	& 21.98   	& .09	&       	&    	& 19.80   	& .17	&       	&     \\
34	& F1158+1716	&       	&    	&       	&    	&       	&    	&       	&    	&       	&    	&       	&    	& $>$19.06 	&    	&       	&     \\
35	& F1202+0654	&       	&    	&       	&    	&       	&    	&       	&    	&       	&    	&       	&    	& $>$19.02 	&    	&       	&     \\
36	& F1211+3616	&       	&    	& $>$25.72 	&    	& $>$25.98 	&    	& $>$25.60 	&    	&       	&    	&       	&    	& $>$20.13 	&    	&       	&     \\
37	& F1215+3242	& 20.36   	& .25	& 18.86   	& .21	& 17.98   	& .26	& 17.23   	& .27	& 16.33   	& .19	&       	&    	& 15.83   	& .14	&       	&     \\
38a	& F1217-0529E\tablenotemark{a}	&       	&    	& $>$24.88 	&    	& 24.66   	& .09	& 22.34   	& .11	& 22.82   	& .09	&       	&    	& 18.90   	& .26	&       	&     \\
38b	& F1217-0529W\tablenotemark{a}	&       	&    	& $>$24.88 	&    	& 24.83   	& .17	& 22.98   	& .09	& 22.95   	& .14	&       	&    	& 20.73   	& .17	&       	&     \\
38	& F1217-0529S\tablenotemark{a}	&       	&    	& $>$24.88 	&    	& 25.47   	& .19	& 23.58   	& .21	& $>$24.05 	&    	&       	&    	& $>$20.93 	&    	&       	&     \\
39	& F1217+3810	&       	&    	&       	&    	&       	&    	&       	&    	&       	&    	&       	&    	& 15.52   	& .14	&       	&     \\
40	& F1218-0625	&       	&    	& 24.13   	& .20	& 23.27   	& .20	& 21.91   	& .28	& 22.14   	& .42	&       	&    	& 19.48   	& .18	&       	&     \\
41	& F1218-0716	&       	&    	& 23.82   	& .14	&       	&    	&       	&    	&       	&    	&       	&    	& 21.54   	& .20	&       	&     \\
42	& F1234+2001	&       	&    	& $>$24.88 	&    	& $>$23.20 	&    	& 19.85   	& .28	& 19.58   	& .20	&       	&    	& 18.63   	& .16	&       	&     \\
43	& F1237+1141	& $>$24.02   	&    	& 23.99   	& .07	& 23.27   	& .06	& 22.47   	& .07	& 22.38   	& .04	&       	&    	& 19.64   	& .29	& 18.68   	& .14 \\
44	& F1315+4438	&       	&    	& 23.41   	& .30	& 23.37   	& .20	& 22.66   	& .66	& 22.83   	& .48	&       	&    	& $>$20.04 	&    	& 21.32   	& .15 \\
45	& F1329+1748	&       	&    	& 20.78   	& .11	& 20.32   	& .06	& 19.68   	& .07	&       	&    	&       	&    	&       	&    	&       	&     \\
46	& F1355+3607	&       	&    	&       	&    	& $>$23.42 	&    	& $>$23.87 	&    	& $>$23.68 	&    	&       	&    	& $>$21.07 	&    	& $>$20.86 	&     \\
47	& F1430+3557	& $>$25.18 	&    	& $>$24.14 	&    	& $>$23.52 	&    	& $>$23.14 	&    	& $>$23.59 	&    	&       	&    	& $>$19.89 	&    	& 18.86   	& .13 \\
48	& F1435-0029	& $>$25.47 	&    	& $>$25.86 	&    	& $>$26.11 	&    	& $>$25.00 	&    	& 22.91   	& .12	&       	&    	& 19.87   	& .15	&       	&     \\
49	& F1445+2702	&       	&    	&       	&    	&       	&    	& 20.69   	& .07	&       	&    	&       	&    	&       	&    	&       	&     \\
50	& F1447+1217	& $>$24.89 	&    	& $>$24.70 	&    	& 24.00   	& .11	& 22.60   	& .09	& 22.16   	& .11	&       	&    	& $>$20.25 	&    	& 20.20   	& .07 \\
51	& F1451+0556	& 19.80   	& .15	& 19.59   	& .11	& 19.01   	& .06	& 18.81   	& .07	& 19.07   	& .15	&       	&    	& 17.80   	& .14	& 17.06   	& .20 \\
52a	& F1458+4319NW\tablenotemark{b}	& $>$25.03 	&    	& $>$24.91 	&    	& 24.73   	& .14	& 22.30   	& .09	& 21.99   	& .12	&       	&    	& 21.45   	& .18	&       	&     \\
52b	& F1458+4319SE\tablenotemark{b}	& $>$25.03 	&    	& $>$24.91 	&    	& 24.41   	& .09	& 21.38   	& .10	& 21.77   	& .09	&       	&    	& 20.19   	& .16	&       	&     \\
52c	& F1458+4319E\tablenotemark{b}	& $>$25.03 	&    	&       	&    	&       	&    	& 21.17   	& .08	& 21.28   	& .09	&       	&    	& 19.29   	& .33	&       	&     \\
53	& F1505+4457	& $>$25.33 	& .15	& 23.98   	& .07	& 21.90   	& .11	& 21.18   	& .07	& 20.94   	& .08	&       	&    	& 18.46   	& .16	& 18.24   	& .28 \\
54	& F1524+5122	&       	&    	&       	&    	& $>$23.87 	&    	& $>$23.17 	&    	& $>$23.55 	&    	&       	&    	& $>$20.31 	&    	& $>$20.96 	&     \\
55	& F1644+2554	&       	&    	&       	&    	& $>$25.20 	&    	& $>$25.13 	&    	& $>$24.55 	&    	&       	&    	& $>$19.89 	&       &       	&     \\
56	& F2217-0837	& 23.14   	& .20	& 20.30   	& .13	& 20.10   	& .07	& 19.03   	& .07	& 19.16   	& .08	& 17.77   	& .24	& 17.31   	& .14	& 16.59   	& .03 \\
57a	& F2217-0138E\tablenotemark{a}	& $>$24.98 	&    	& 24.48   	& .16	& 22.42   	& .08	& 20.59   	& .07	& 20.88   	& .09	& 20.75   	& .28	& 18.64   	& .15	& 17.48   	& .10 \\
57b	& F2217-0138W\tablenotemark{a}	& $>$24.98 	&    	& $>$24.89 	&    	& 24.60   	& .15	& 22.32   	& .09	& 22.01   	& .14	& 22.00   	& .50	& 20.39   	& .16	& 19.29   	& .34 \\
58	& F2354-0055	& $>$24.99 	&    	& $>$24.89 	&    	& $>$25.16 	&    	& $>$25.00 	&    	& $>$24.45 	&    	&       	&    	& $>$19.93 	&    	&       	&     \\
\enddata
\tablenotetext{a}{Multiple source, possibly an interacting system.}
\tablenotetext{b}{ID unclear}
\end{deluxetable*}
\normalsize

\begin{deluxetable}{llllc}
\tabletypesize{\scriptsize}
\tablecaption{Hyperz Results}
\tablehead{
\multicolumn{1}{c}{Name} & 
\multicolumn{1}{c}{$z_{phot}$} & \multicolumn{1}{c}{$\sigma_{z}$} & \multicolumn{1}{c}{$M_V$} & 
\multicolumn{1}{c}{SED} }
\startdata
F0023-0904  	 & 1.49 & $^{+0.14}_{-0.07}$ & -24.10 & Sc    \\
F0129-0140  	 & 2.44 & $^{+0.40}_{-0.90}$ & -24.73 & SB4   \\
F0202-0021  	 & 0.58 & $^{+0.08}_{-0.02}$ & -22.13 & E     \\
F0216+0038  	 & 0.65 & $^{+0.30}_{-0.11}$ & -22.09 & S0    \\
F0916+1134  	 & 1.12 & $^{+0.07}_{-0.05}$ & -22.00 & SB3   \\
F0919+1007  	 & 0.78 & $^{+0.06}_{-0.03}$ & -23.08 & S0    \\
F0938+2326  	 & 3.88 & $^{+0.14}_{-0.07}$ & -26.42 & SB3   \\
F0942+1520\tablenotemark{a} 	 & 3.36 & $^{+3.56}_{-0.54}$ & -26.45 & S0   \\
F0955+2951  	 & 4.39 & $^{+0.14}_{-0.13}$ & -25.77 & SB4   \\
F1008-0605  	 & 0.27 & $^{+0.15}_{-0.08}$ & -21.32 & E     \\
F1010+2527N  	 & 4.41 & $^{+0.33}_{-0.38}$ & -26.57 & SB4   \\
F1010+2527S  	 & 4.63 & $^{+0.08}_{-0.06}$ & -27.32 & SB2   \\
F1010+2727  	 & 4.53 & $^{+0.20}_{-0.11}$ & -27.87 & Sb    \\
F1014+1438  	 & 4.47 & $^{+0.34}_{-0.22}$ & -26.58 & SB5   \\
F1027+0520  	 & 0.60 & $^{+0.07}_{-0.04}$ & -25.18 & SB1   \\
F1039+2602  	 & 3.62 & $^{+0.69}_{-0.26}$ & -26.18 & SB3   \\
F1116+0235  	 & 0.62 & $^{+0.04}_{-0.04}$ & -23.20 & SB2   \\
F1147+2647  	 & 4.75 & $^{+0.15}_{-0.11}$ & -25.98 & SB1   \\
F1155+2620  	 & 4.54 & $^{+0.69}_{-0.49}$ & -26.79 & SB3   \\
F1215+3242  	 & 0.11 & $^{+0.29}_{-0.11}$ & -19.86 & S0    \\
F1217-0529E  	 & 4.95 & $^{+0.21}_{-0.06}$ & -29.93 & S0    \\
F1217-0529W  	 & 4.97 & $^{+0.32}_{-0.20}$ & -28.96 & E     \\
F1217-0529S  	 & 4.82 & $^{+0.71}_{-0.80}$ & -23.50 & SB2   \\
F1218-0625  	 & 3.98 & $^{+0.53}_{-3.99}$ & -27.25 & Sb    \\ 
F1234+2001  	 & 5.40 & $^{+0.70}_{-0.41}$ & -27.35 & SB4   \\
F1237+1141  	 & 2.38 & $^{+0.49}_{-0.14}$ & -24.34 & SB2   \\
F1315+4438  	 & 2.77 & $^{+0.16}_{-0.28}$ & -22.49 & SB2   \\
F1447+1217  	 & 4.70 & $^{+0.13}_{-0.18}$ & -25.11 & SB3   \\
F1451+0556  	 & 2.71 & $^{+0.15}_{-0.05}$ & -26.70 & SB2   \\
F1458+4319NW  	 & 5.05 & $^{+0.25}_{-0.10}$ & -24.40 & SB2   \\
F1458+4319SE  	 & 5.03 & $^{+0.34}_{-0.19}$ & -26.06 & SB2   \\
F1505+4457  	 & 0.55 & $^{+0.10}_{-0.08}$ & -20.52 & Sa   \\
F2217-0837  	 & 0.26 & $^{+0.04}_{-0.02}$ & -20.16 & E     \\
F2217-0138E  	 & 4.55 & $^{+0.06}_{-0.21}$ & -27.81 & SB2    \\
F2217-0138W  	 & 4.99 & $^{+0.40}_{-0.25}$ & -26.35 & SB2   \\
\enddata
\tablenotetext{a}{Only 2 bands detected, Lyman break between H and K.}
\end{deluxetable}
\normalsize

\begin{figure*}[htp]
\vspace{+0.2in}
\begin{center}
\rotatebox{270}{\includegraphics[width=2.0in,height=3.2in]{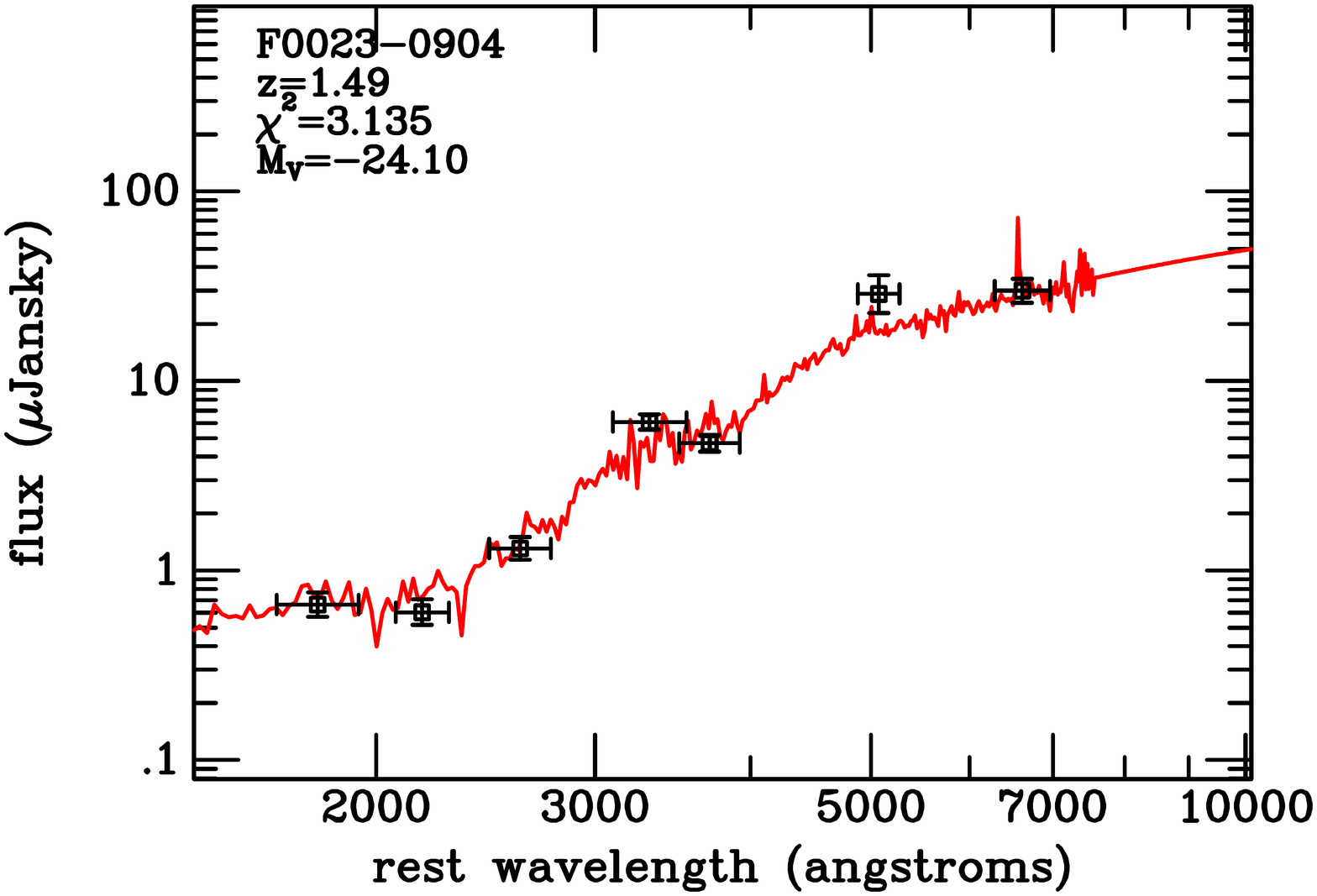}}
\rotatebox{270}{\includegraphics[width=2.0in,height=3.2in]{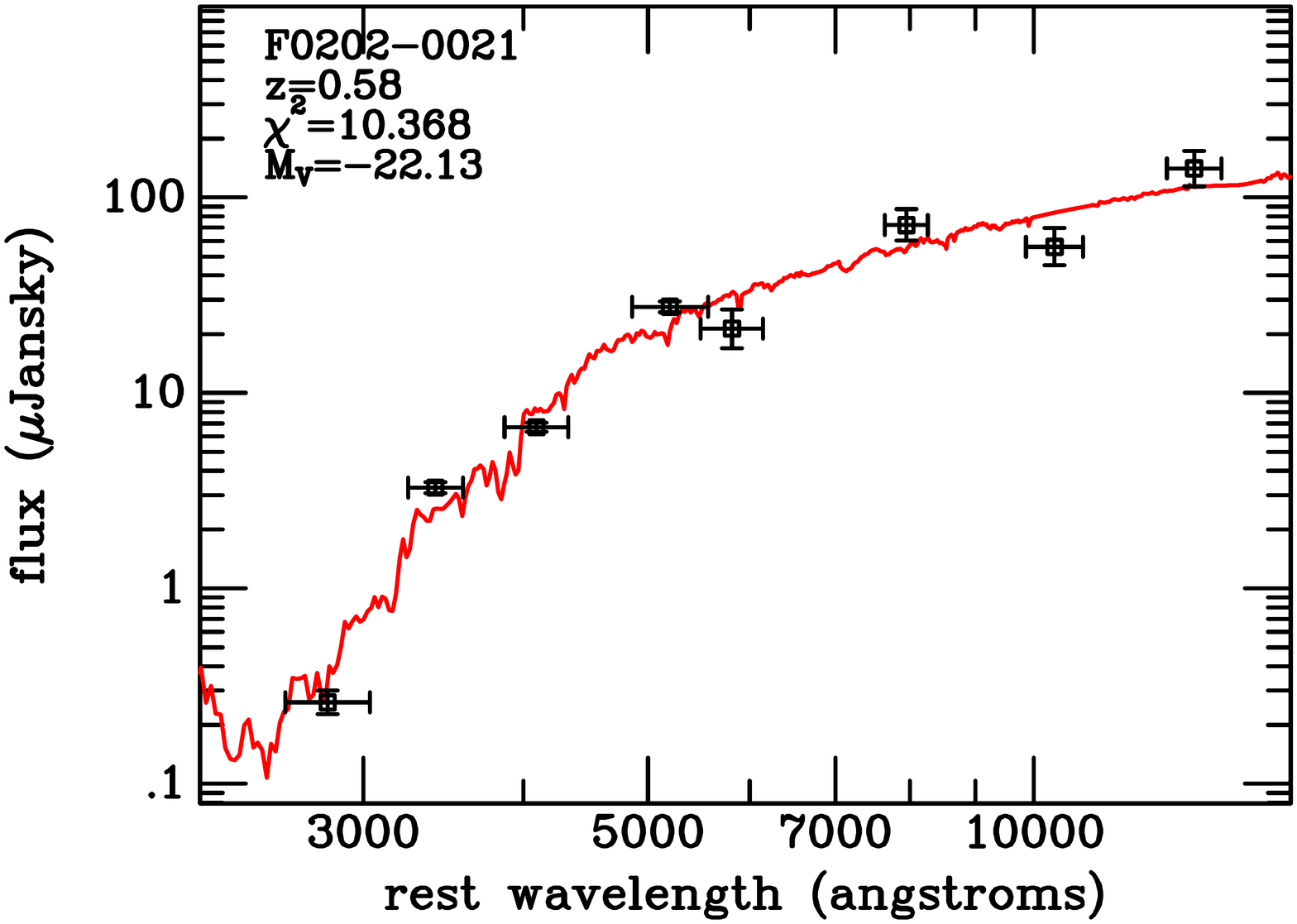}}
\end{center}
\vspace{-1.5in}
\begin{center}
\includegraphics[scale=0.42,angle=0,viewport= 120 50 500 575]{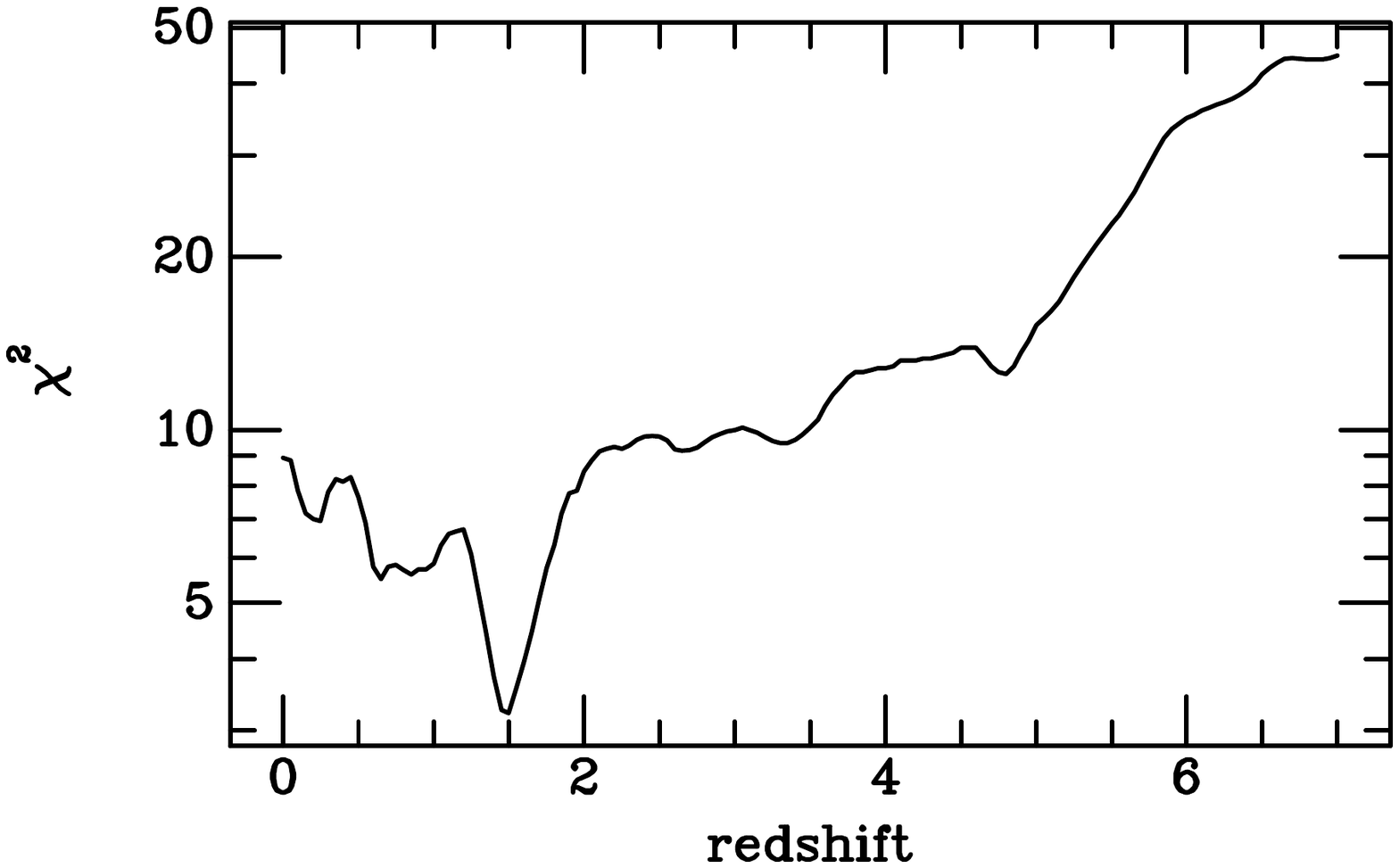}
\includegraphics[scale=0.42,angle=0,viewport= -50 50 500 575]{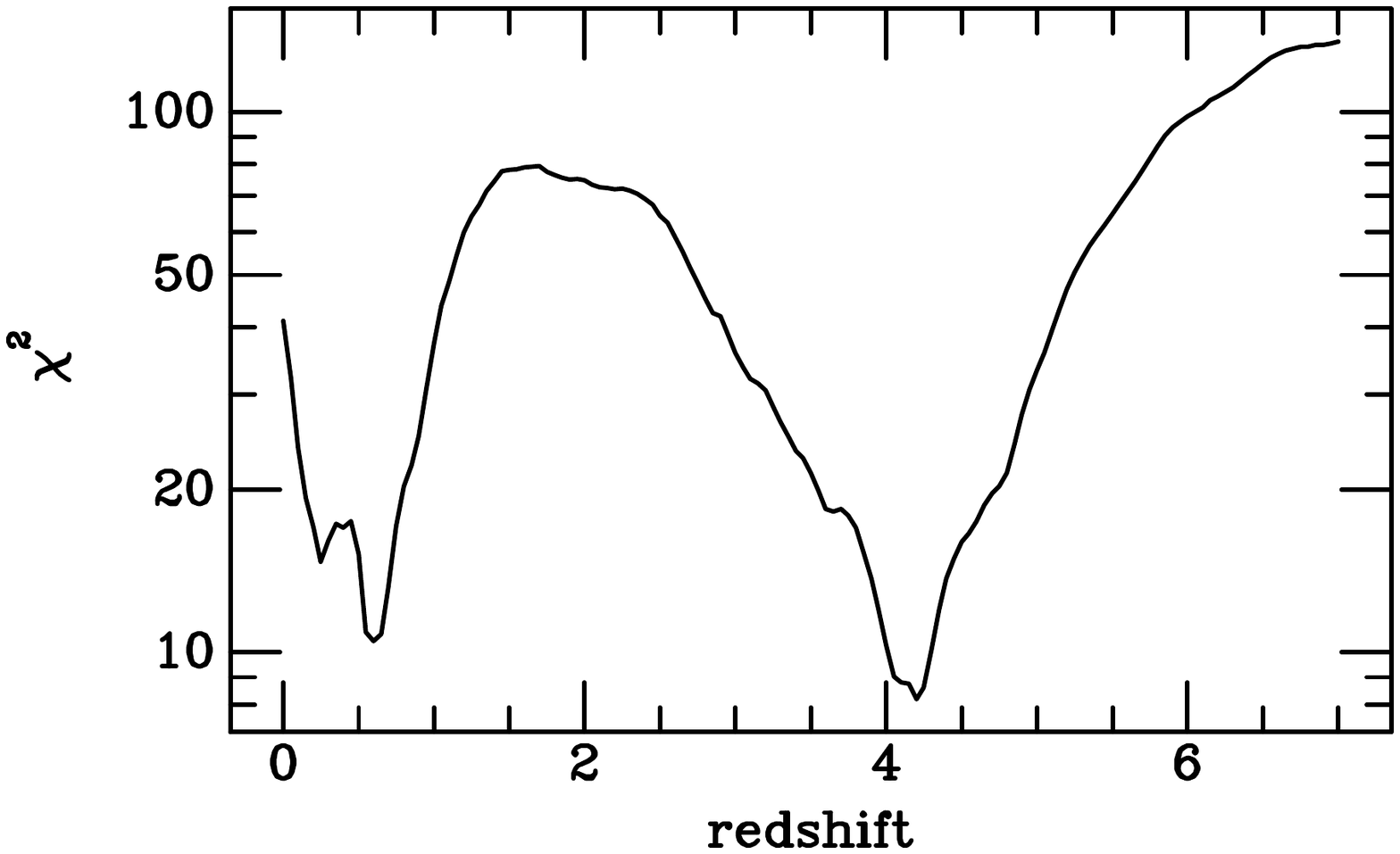}
\end{center}
\vspace{-0.75in}
\begin{center}
\rotatebox{270}{\includegraphics[width=2.0in,height=3.2in]{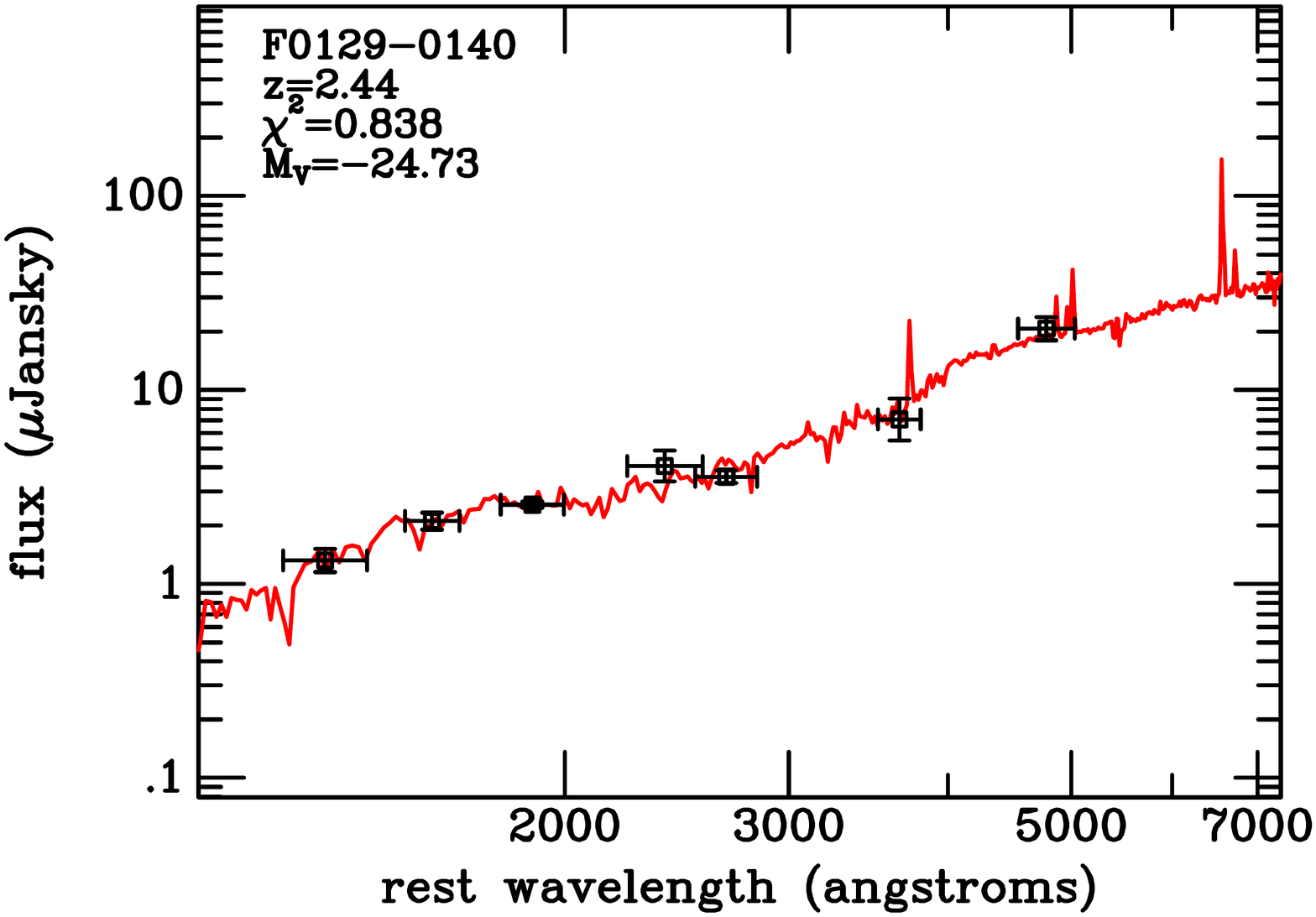}}
\rotatebox{270}{\includegraphics[width=2.0in,height=3.2in]{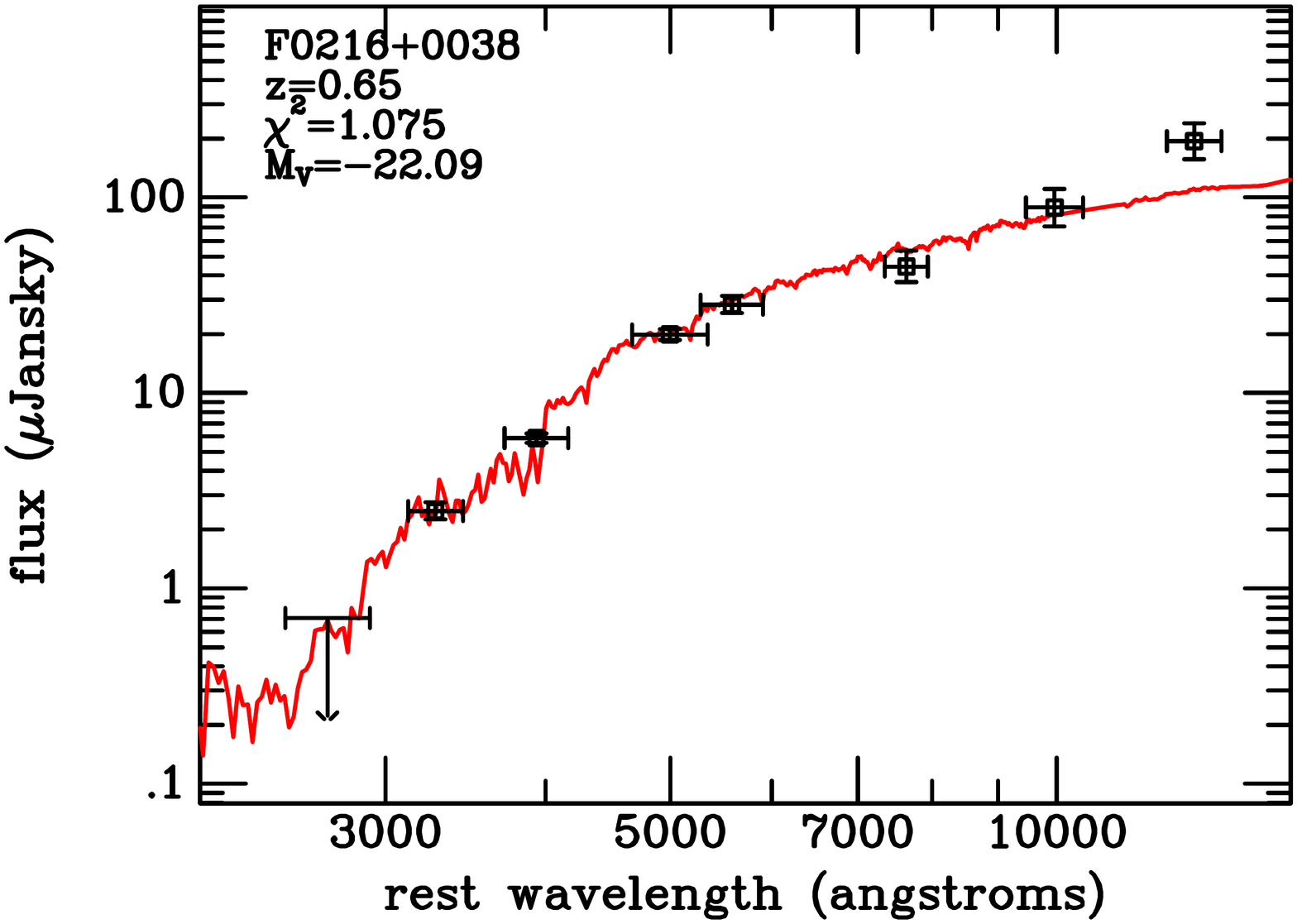}}
\end{center}
\vspace{-1.5in}
\begin{center}
\includegraphics[scale=0.42,angle=0,viewport= 120 50 500 575]{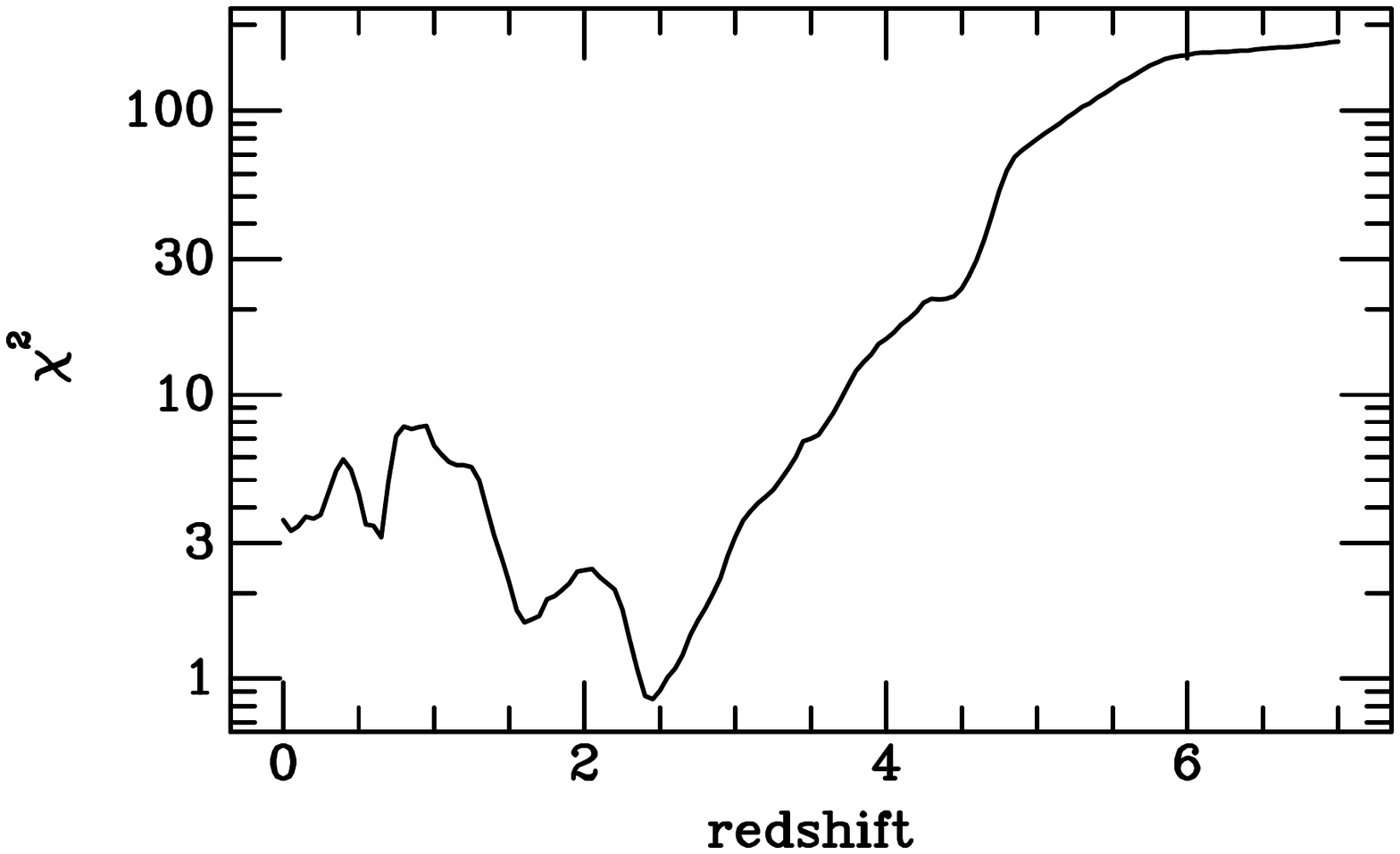}
\includegraphics[scale=0.42,angle=0,viewport= -50 50 500 575]{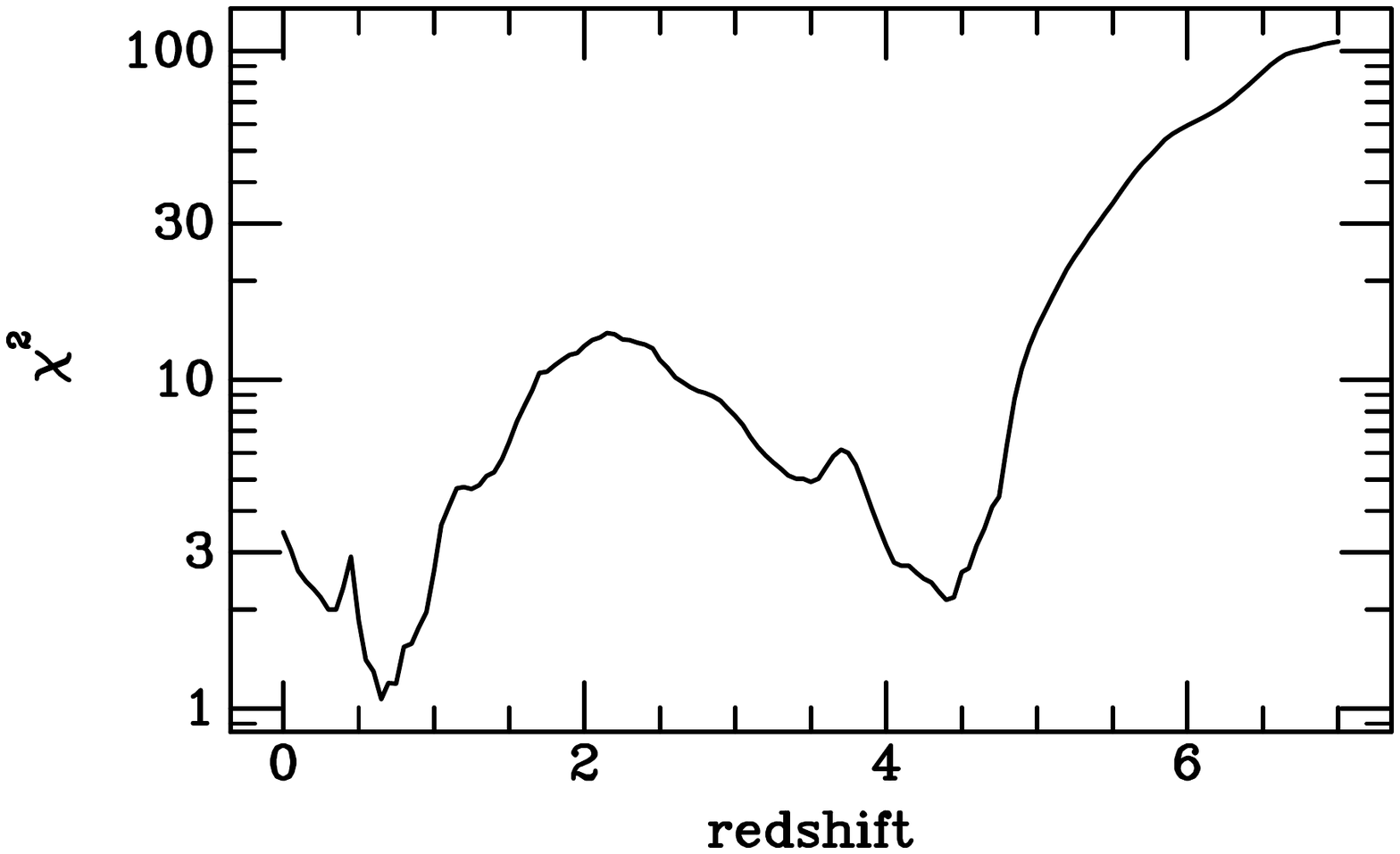}
\end{center}
\vspace{-0.15in}
\caption{\scriptsize
SED fits of 4 objects in the FIRST-BNGS sample.  For each object, the top panel shows the
best-fit SED and the actual broadband photometry data points with
1-sigma error bars in the rest frame of the galaxy.
The bottom panel shows the distribution of best $\chi^2$ for the 11 templates at each redshift.
}
\end{figure*}

\begin{figure*}[htp]
\vspace{+0.4in}
\begin{center}
\rotatebox{270}{\includegraphics[width=2.0in,height=3.2in]{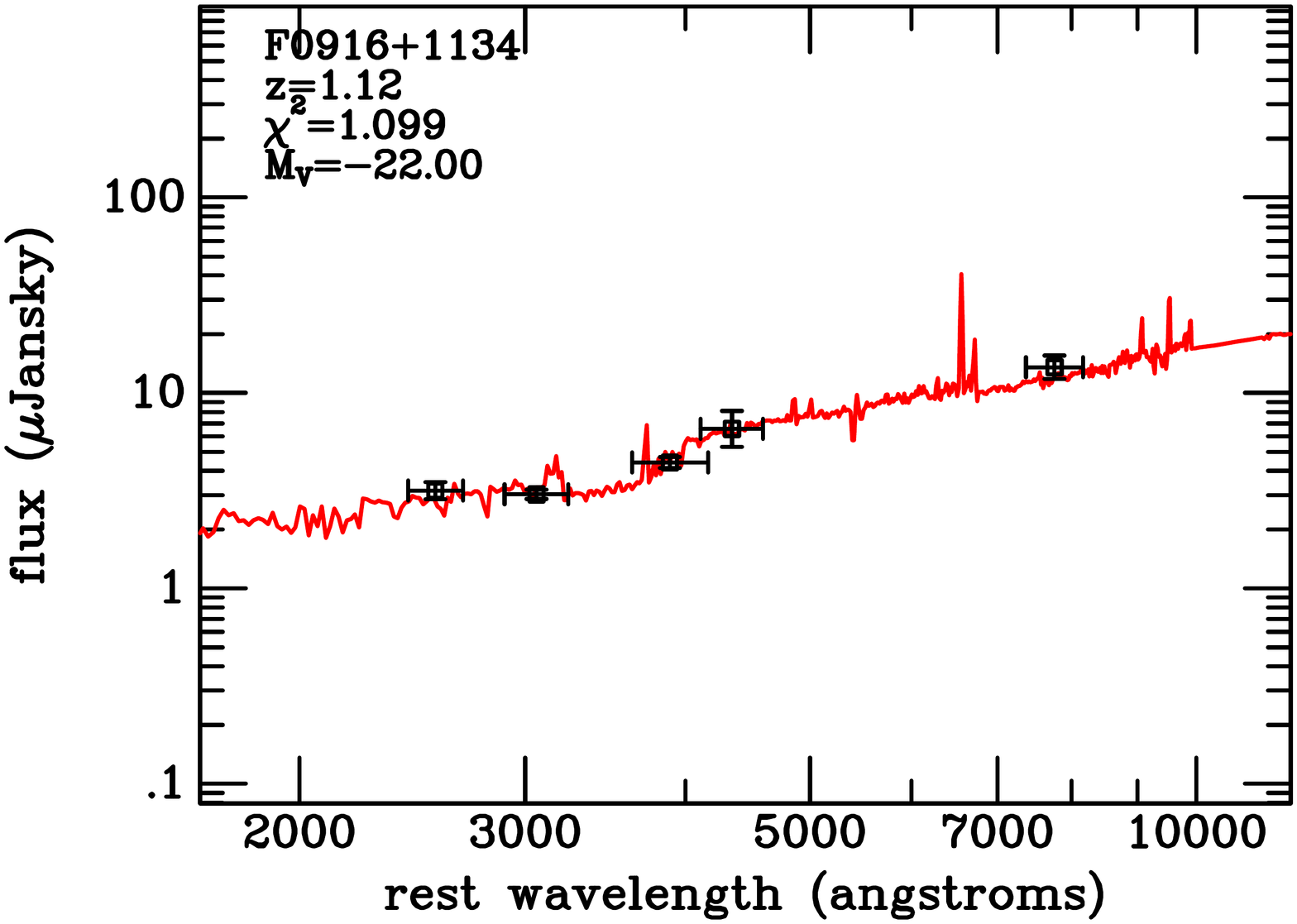}}
\rotatebox{270}{\includegraphics[width=2.0in,height=3.2in]{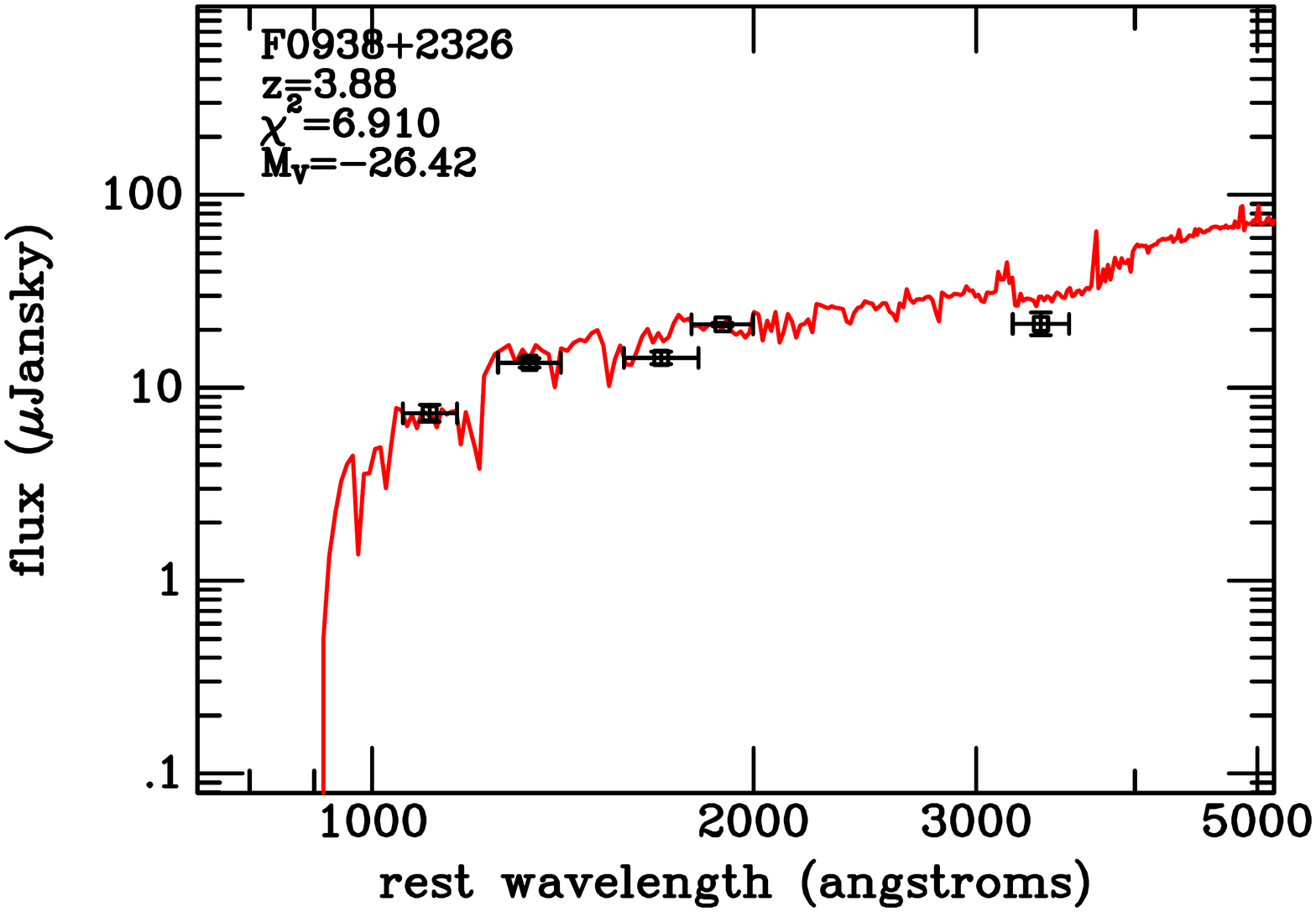}}
\end{center}
\vspace{-1.5in}
\begin{center}
\includegraphics[scale=0.42,angle=0,viewport= 120 50 500 575]{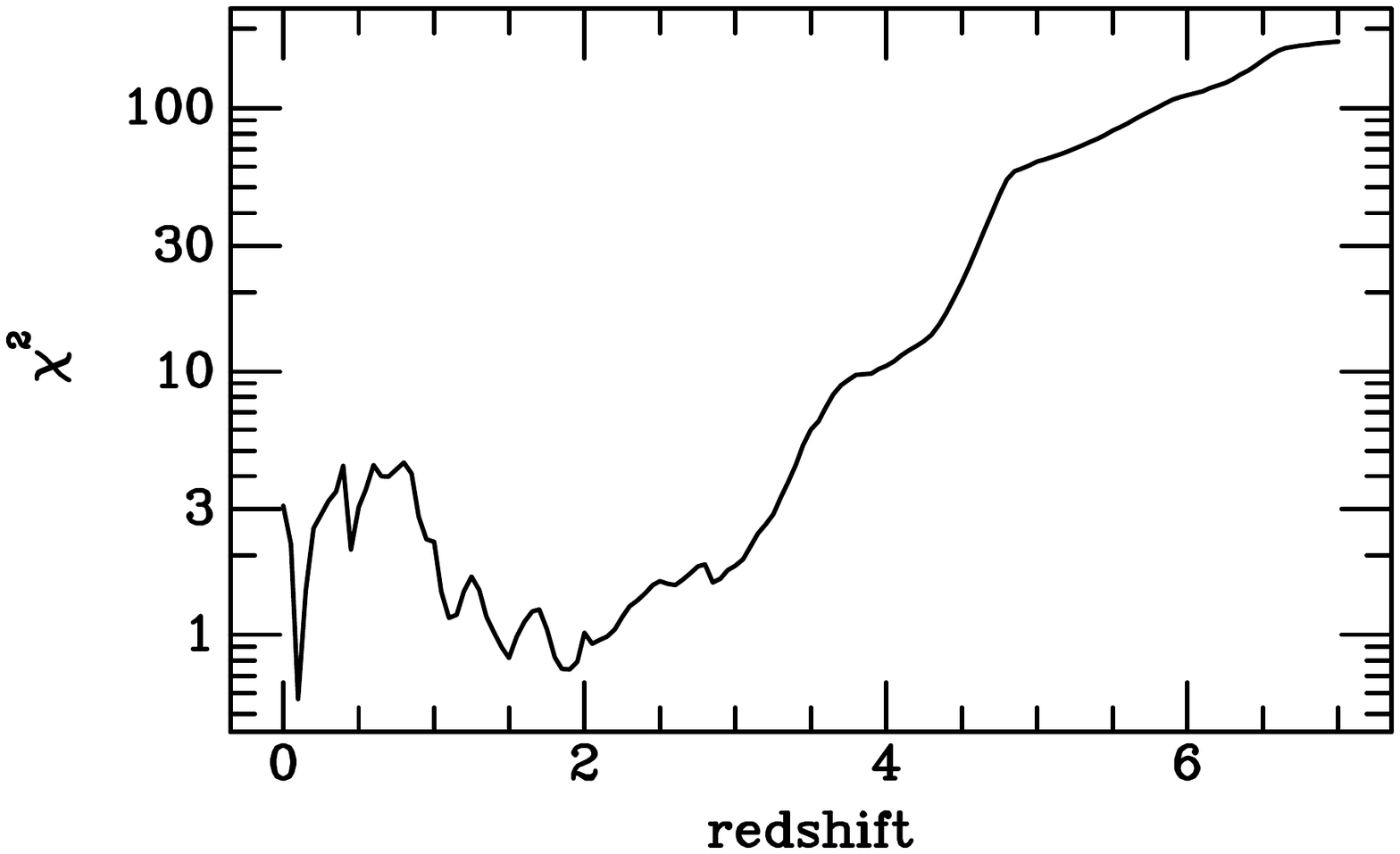}
\includegraphics[scale=0.42,angle=0,viewport= -50 50 500 575]{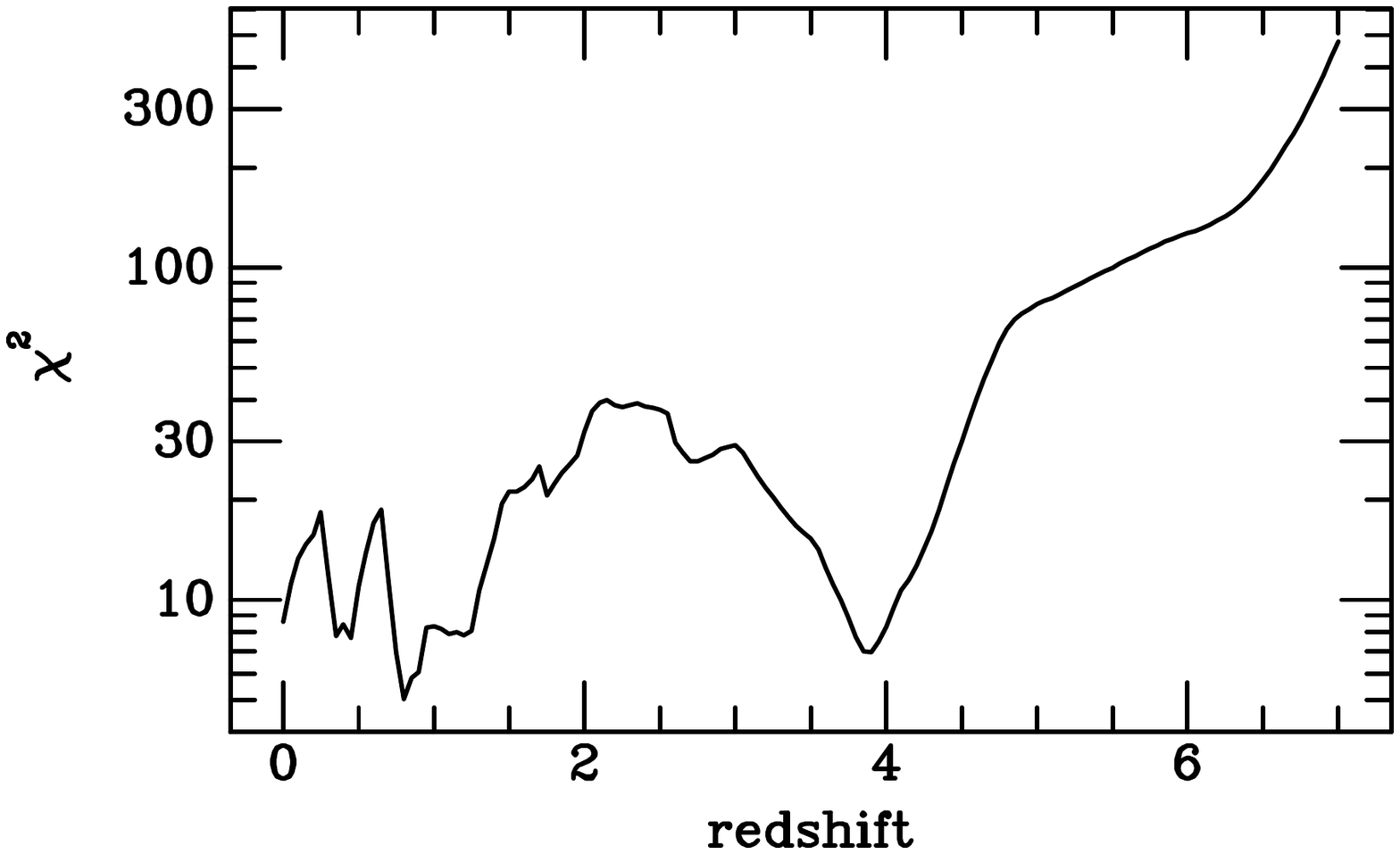}
\end{center}
\vspace{-0.75in}
\begin{center}
\rotatebox{270}{\includegraphics[width=2.0in,height=3.2in]{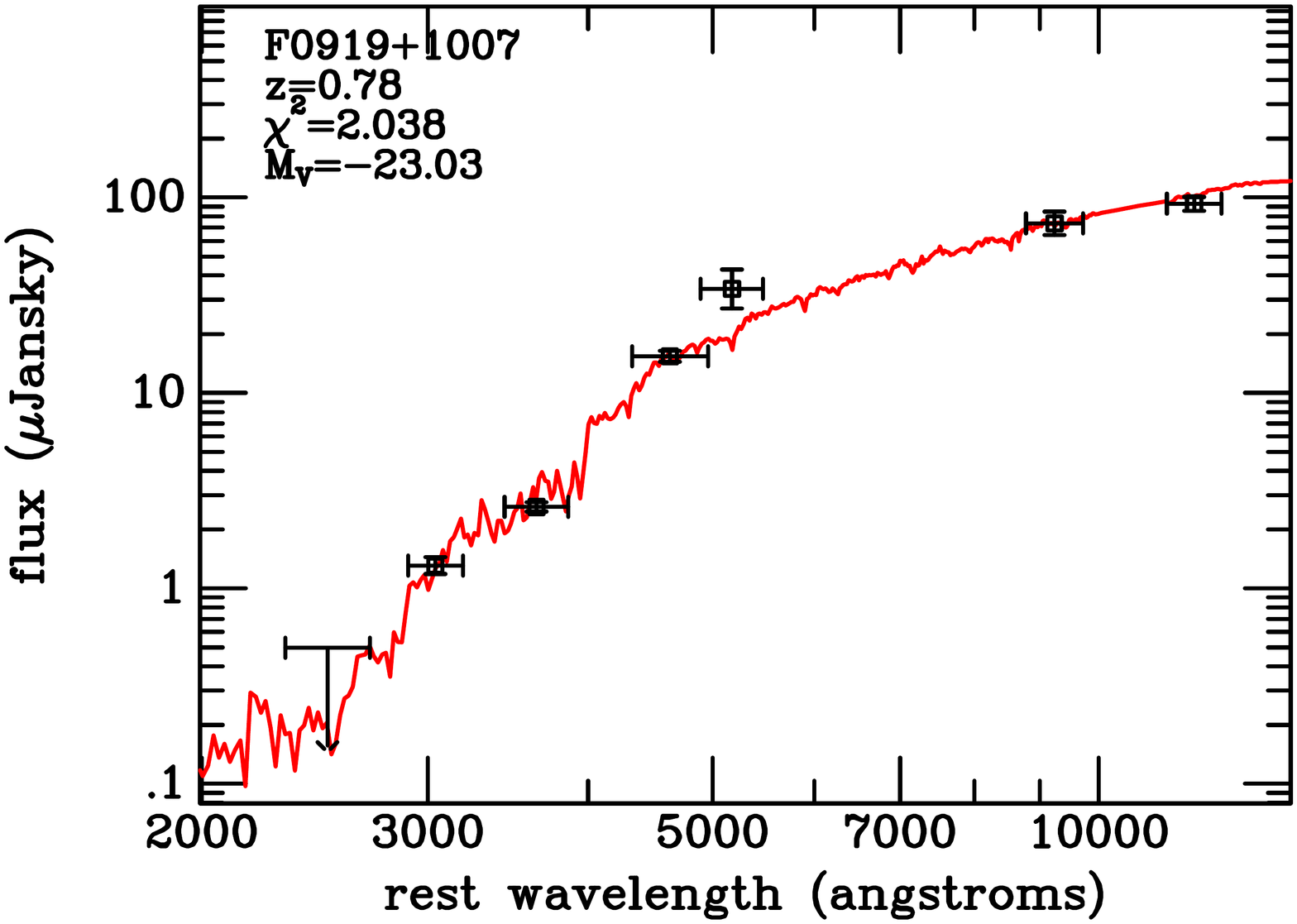}}
\rotatebox{270}{\includegraphics[width=2.0in,height=3.2in]{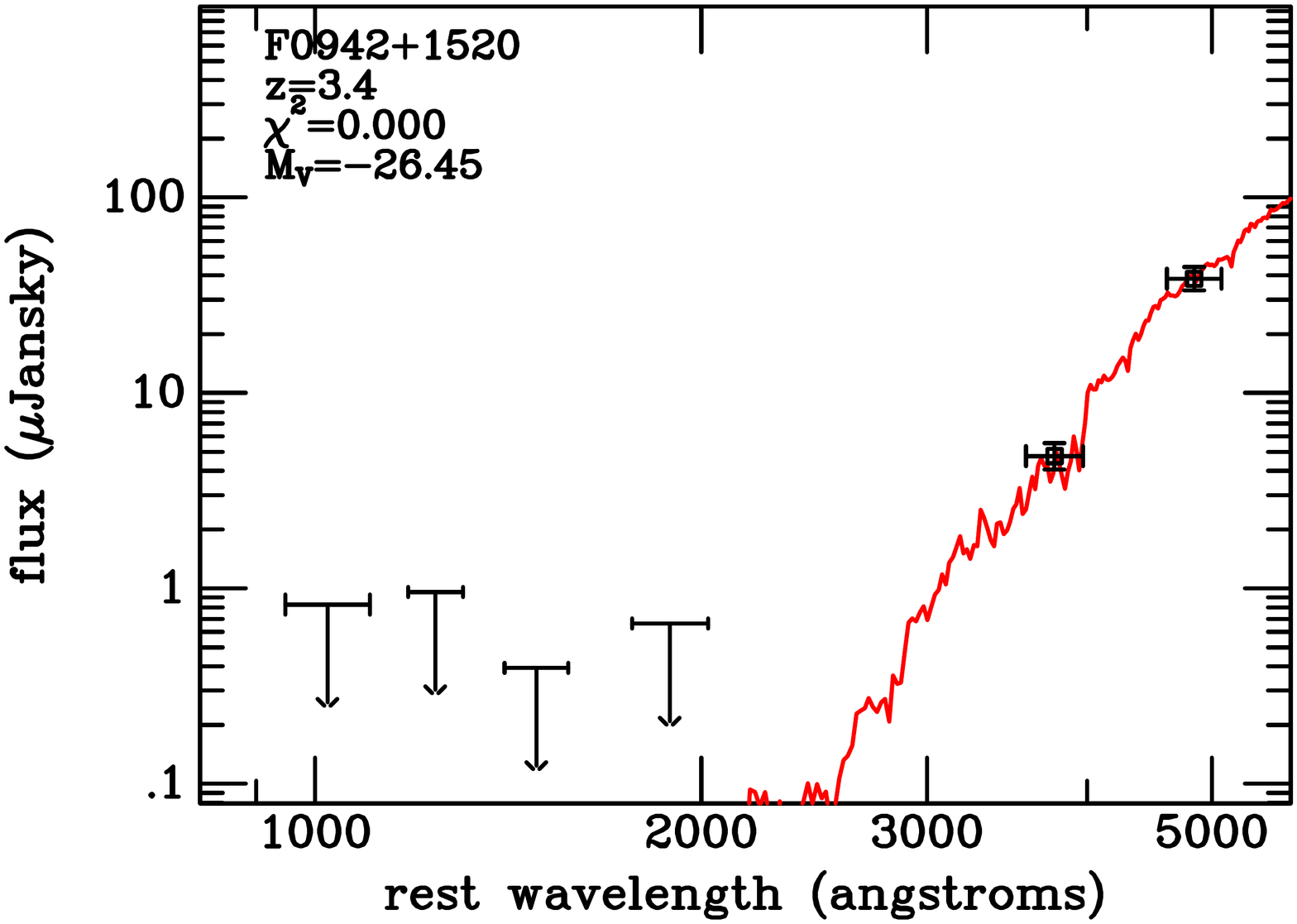}}
\end{center}
\vspace{-1.5in}
\begin{center}
\includegraphics[scale=0.42,angle=0,viewport= 120 50 500 575]{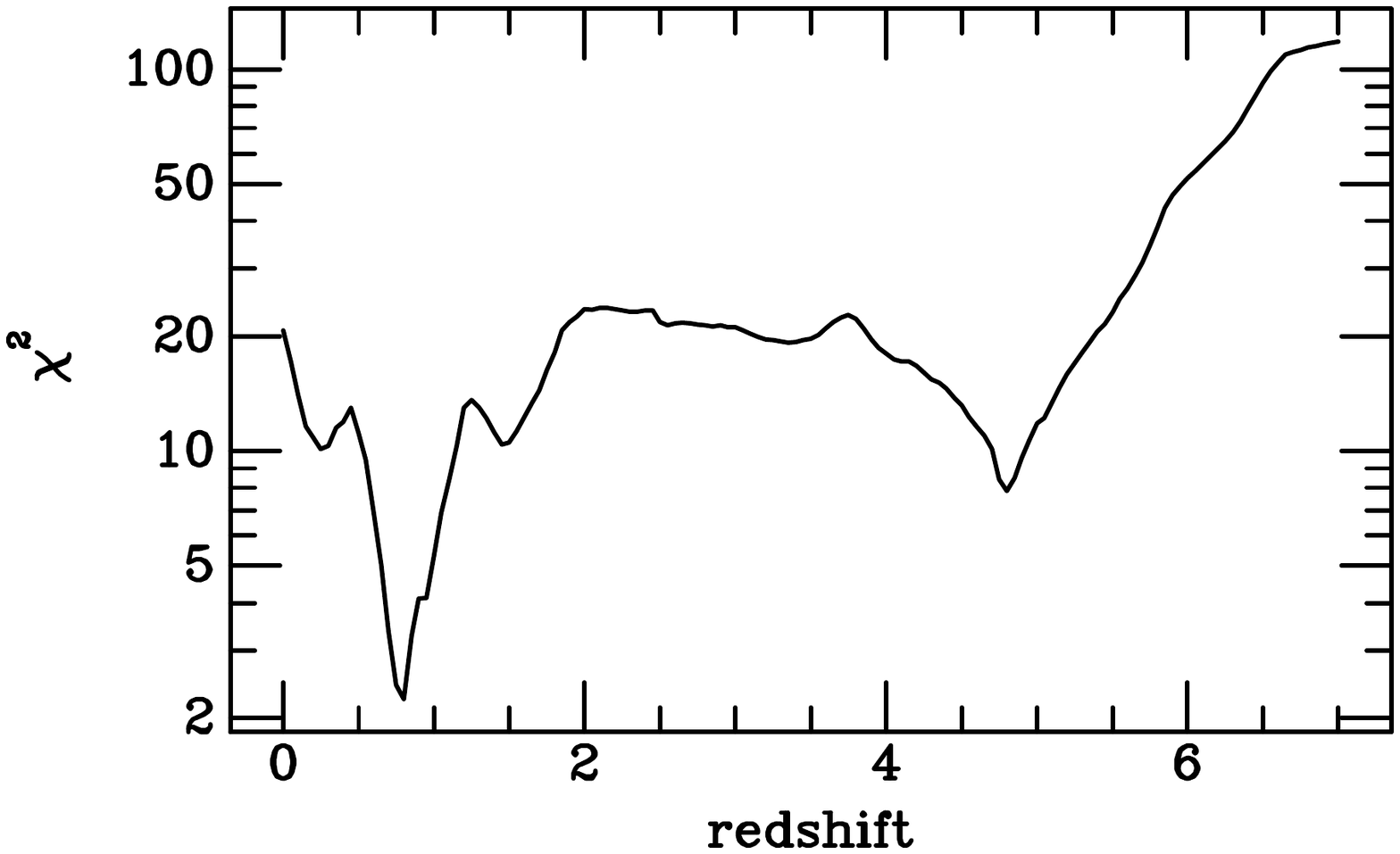}
\includegraphics[scale=0.42,angle=0,viewport= -50 50 500 575]{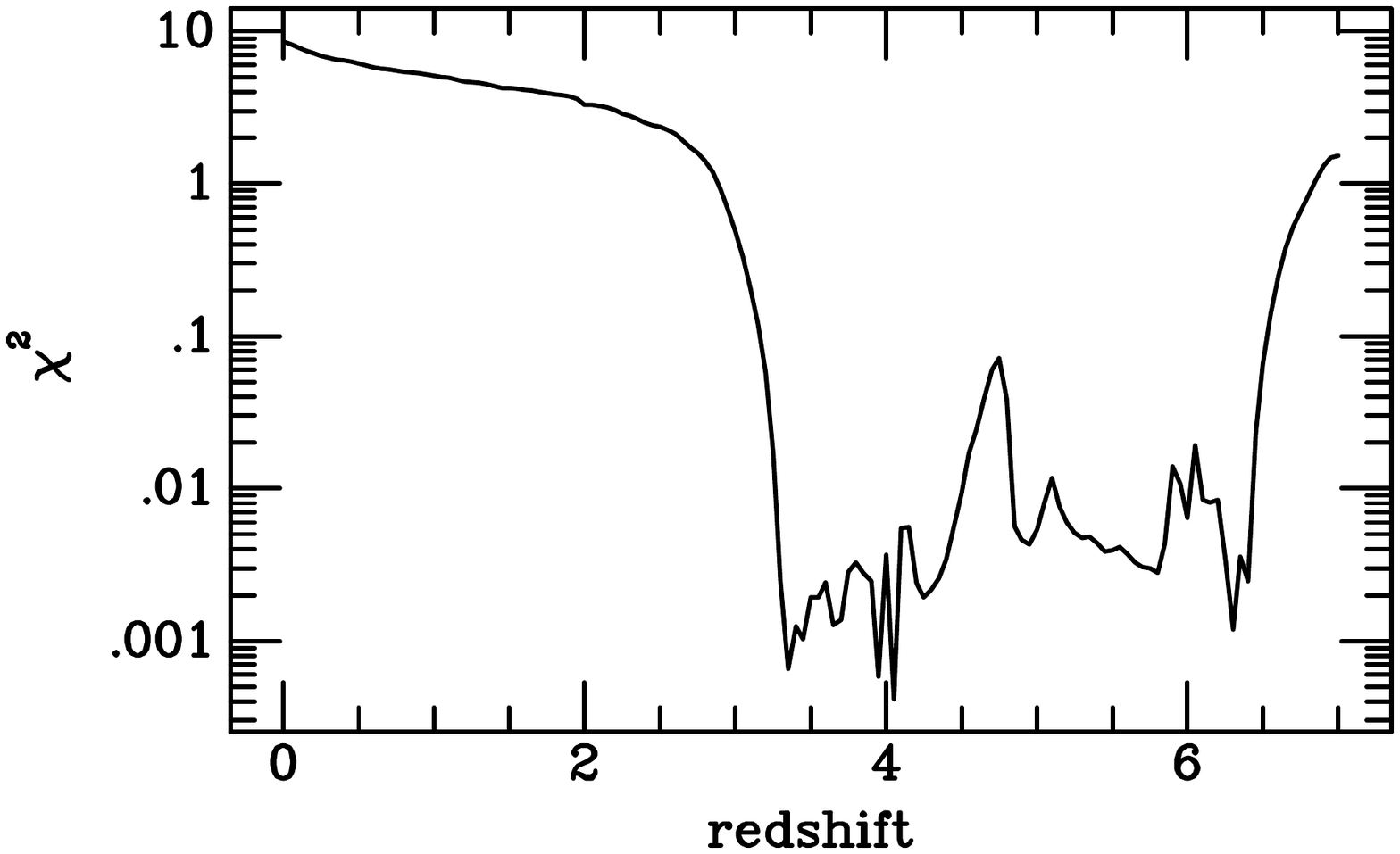}
\end{center}
\vspace{-0.15in}
\caption{\scriptsize
Same as Figure 9 for 4 additional FIRST-BNGS sources.
}
\end{figure*}

\begin{figure*}[htp]
\vspace{+0.4in}
\begin{center}
\rotatebox{270}{\includegraphics[width=2.0in,height=3.2in]{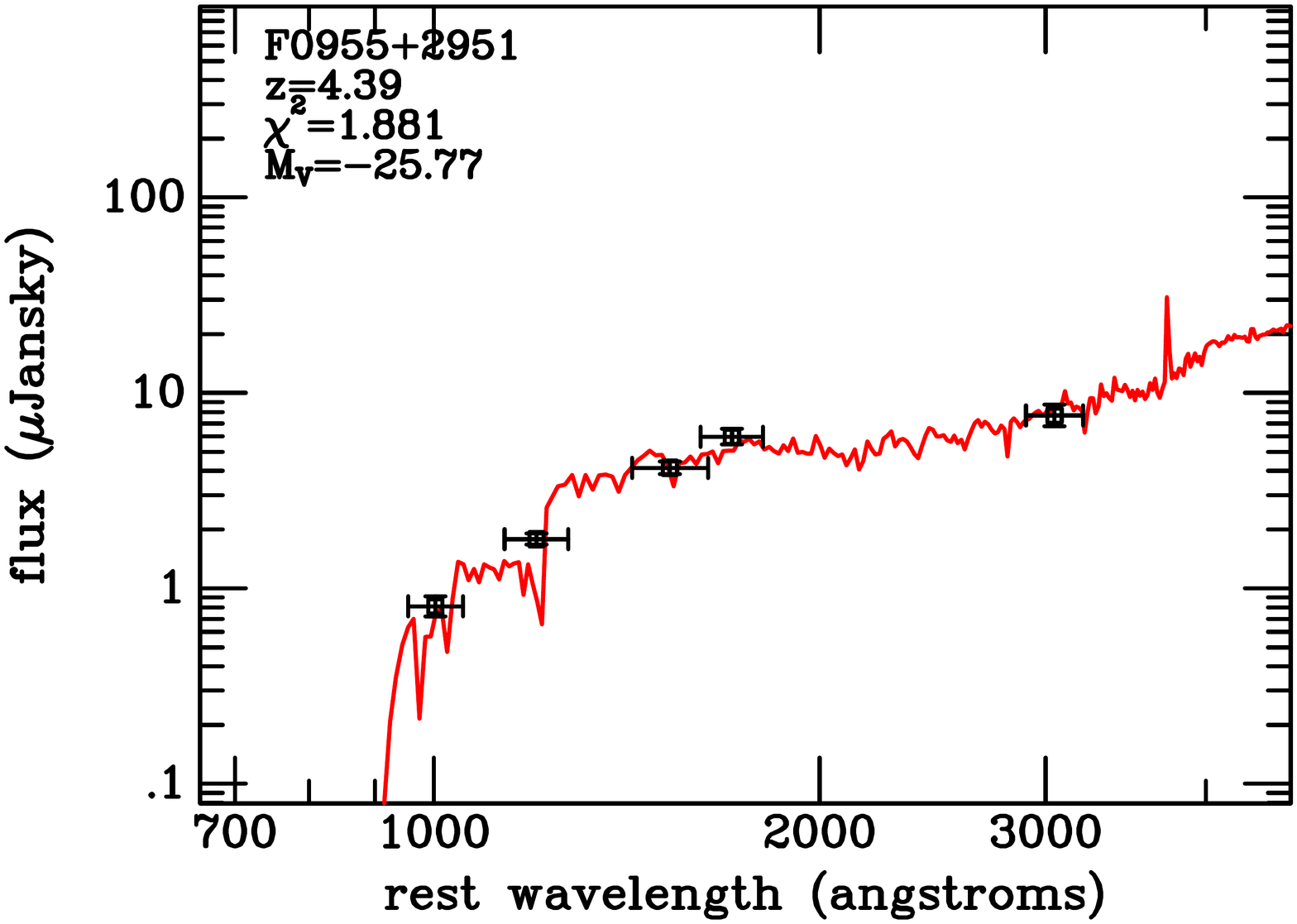}}
\rotatebox{270}{\includegraphics[width=2.0in,height=3.2in]{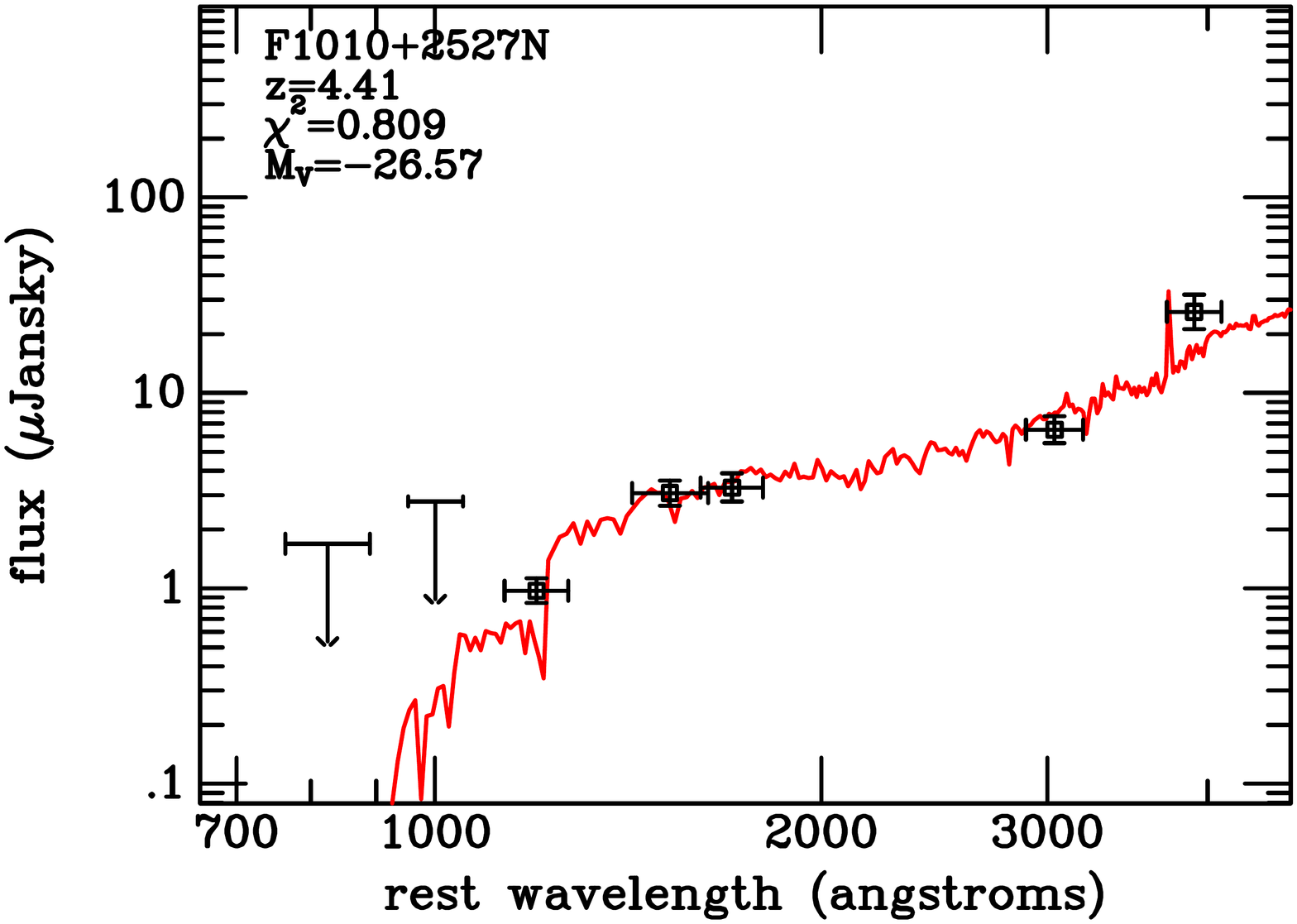}}
\end{center}
\vspace{-1.5in}
\begin{center}
\includegraphics[scale=0.42,angle=0,viewport= 120 50 500 575]{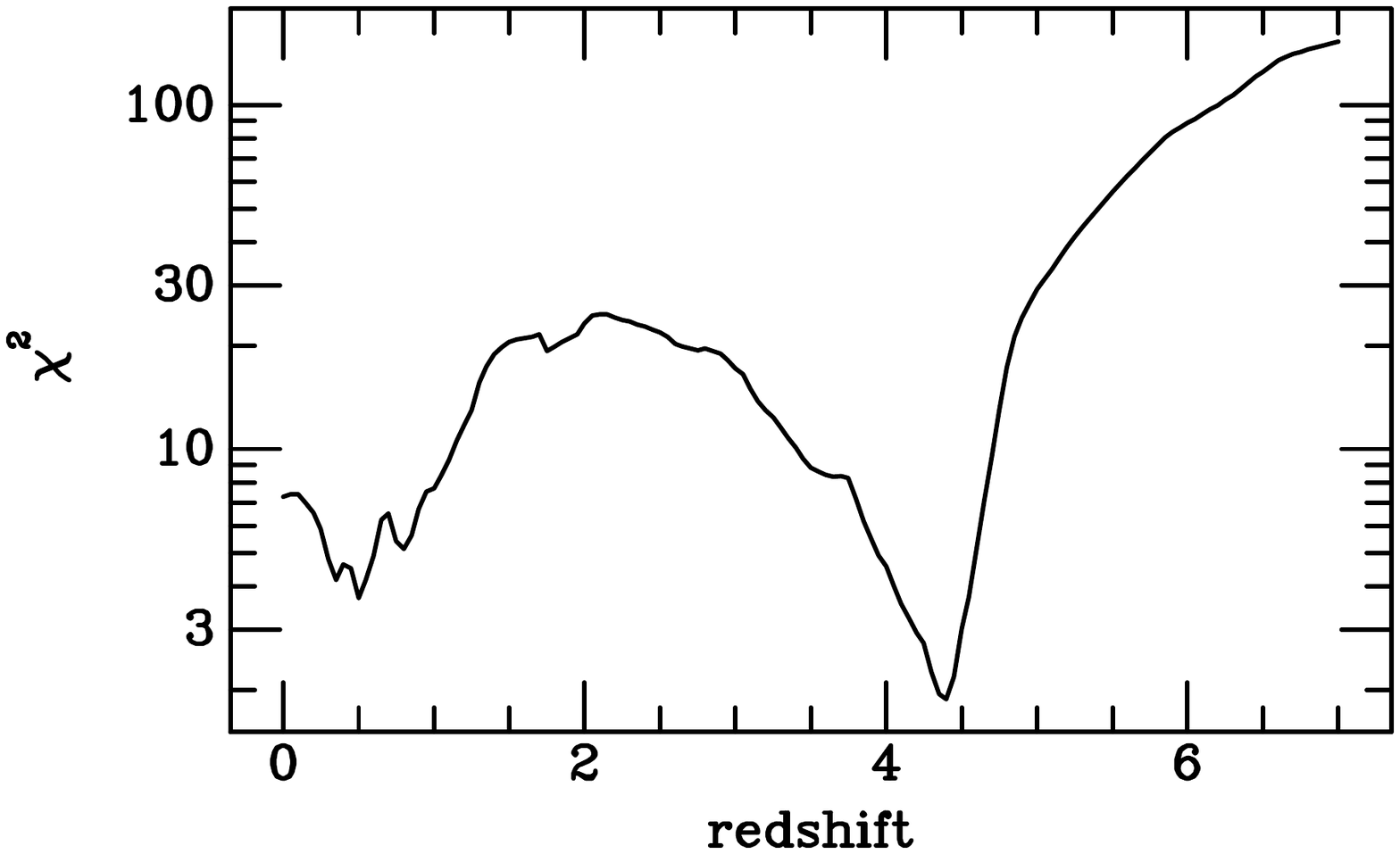}
\includegraphics[scale=0.42,angle=0,viewport= -50 50 500 575]{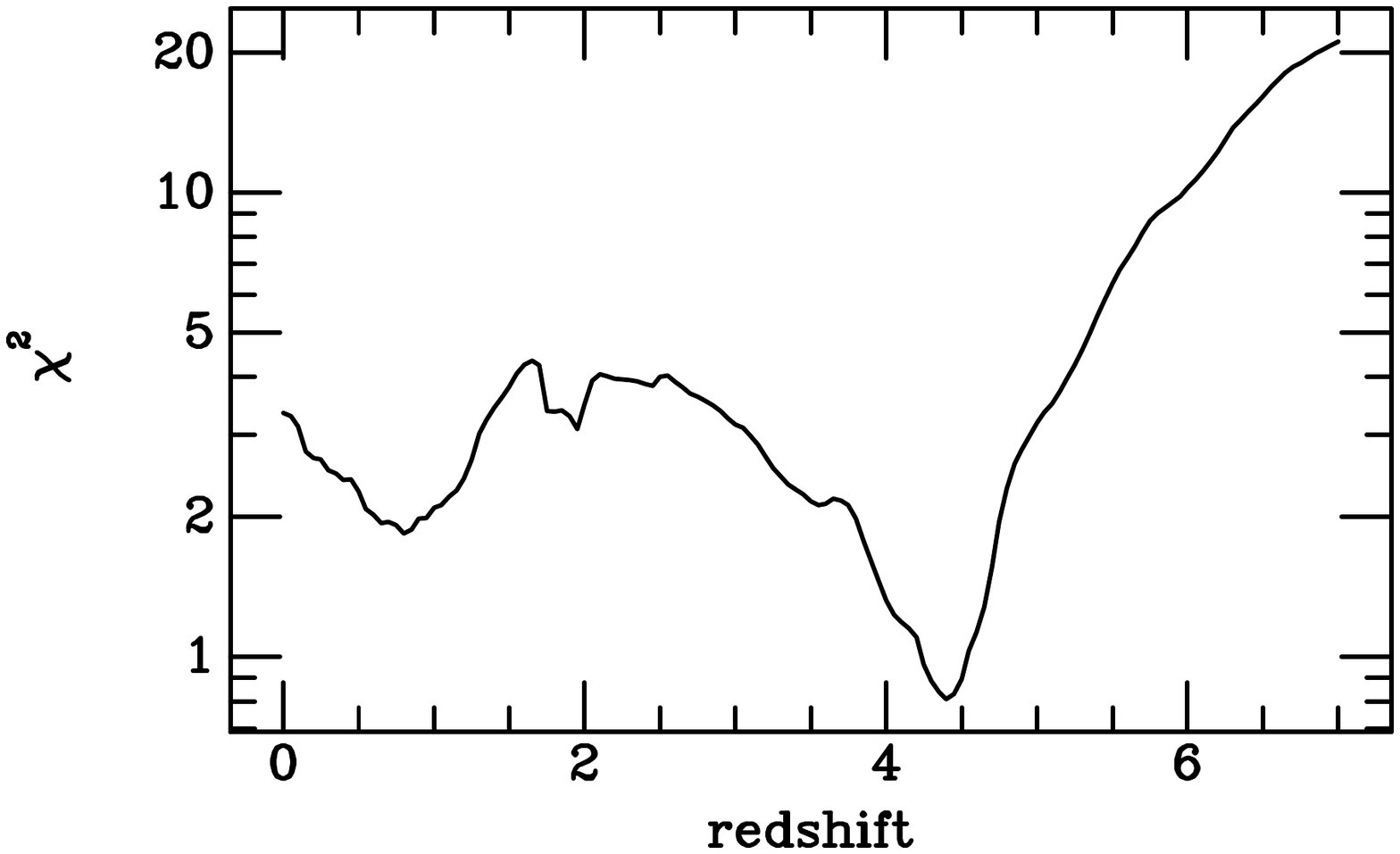}
\end{center}
\vspace{-0.75in}
\begin{center}
\rotatebox{270}{\includegraphics[width=2.0in,height=3.2in]{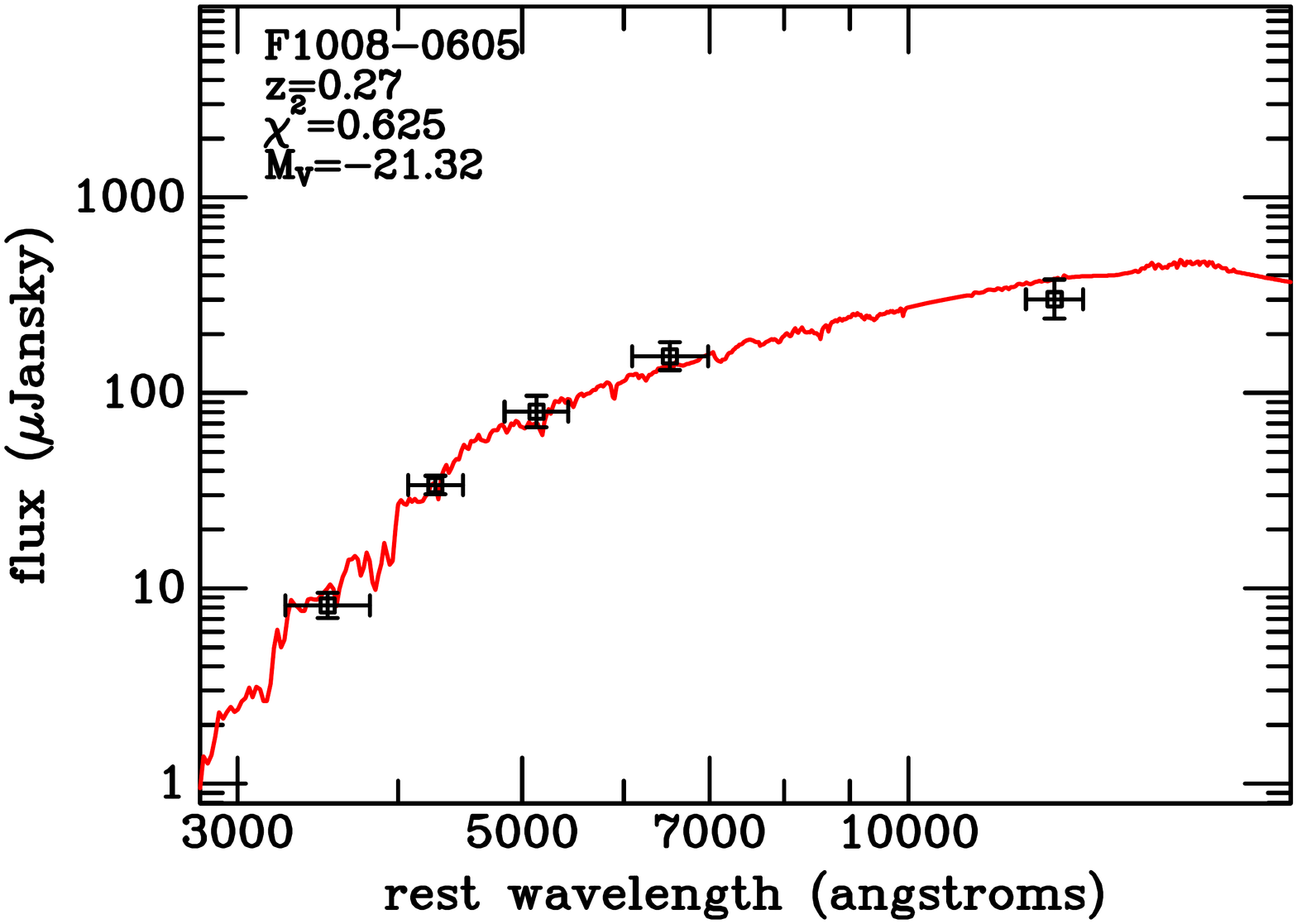}}
\rotatebox{270}{\includegraphics[width=2.0in,height=3.2in]{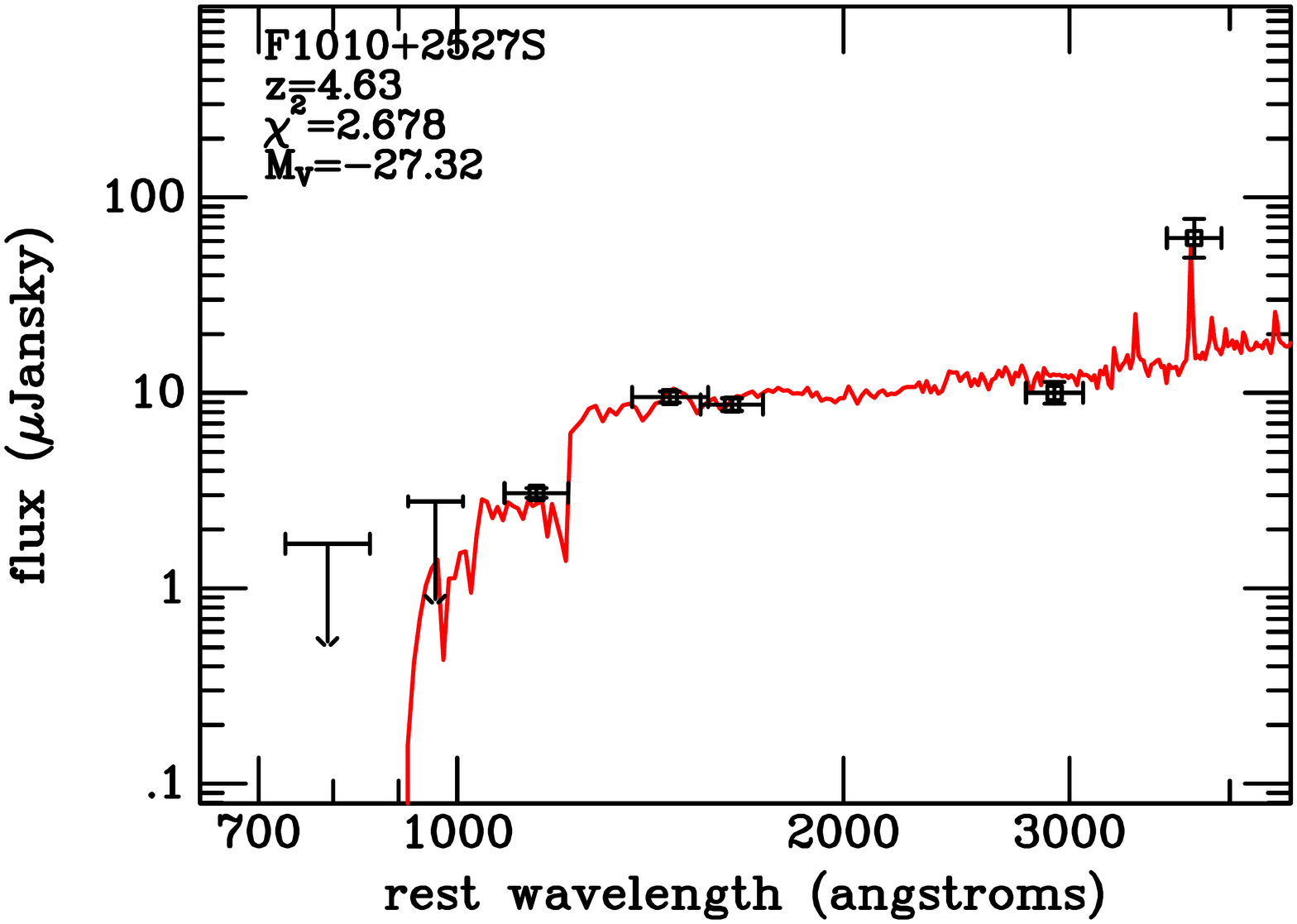}}
\end{center}
\vspace{-1.5in}
\begin{center}
\includegraphics[scale=0.42,angle=0,viewport= 120 50 500 575]{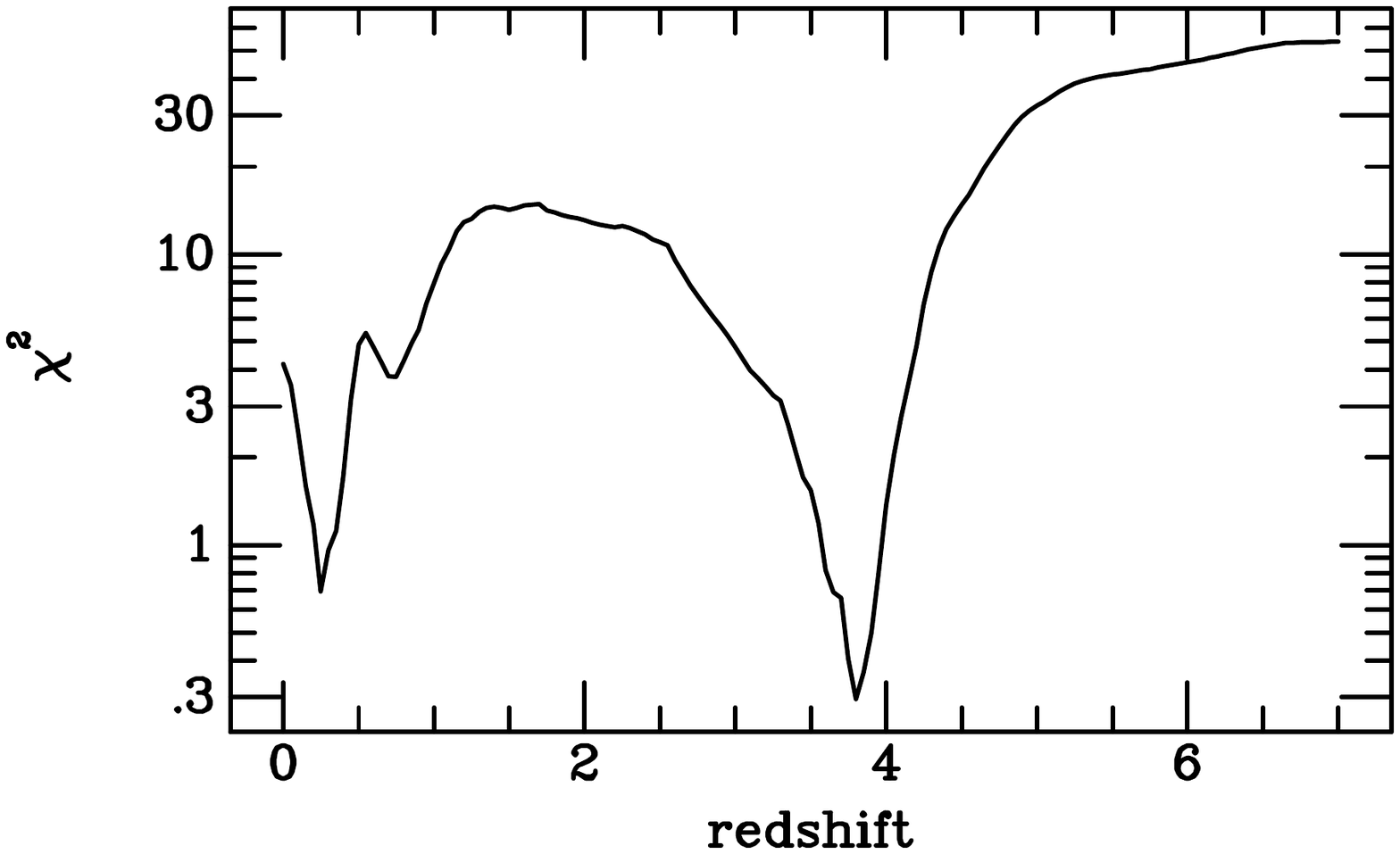}
\includegraphics[scale=0.42,angle=0,viewport= -50 50 500 575]{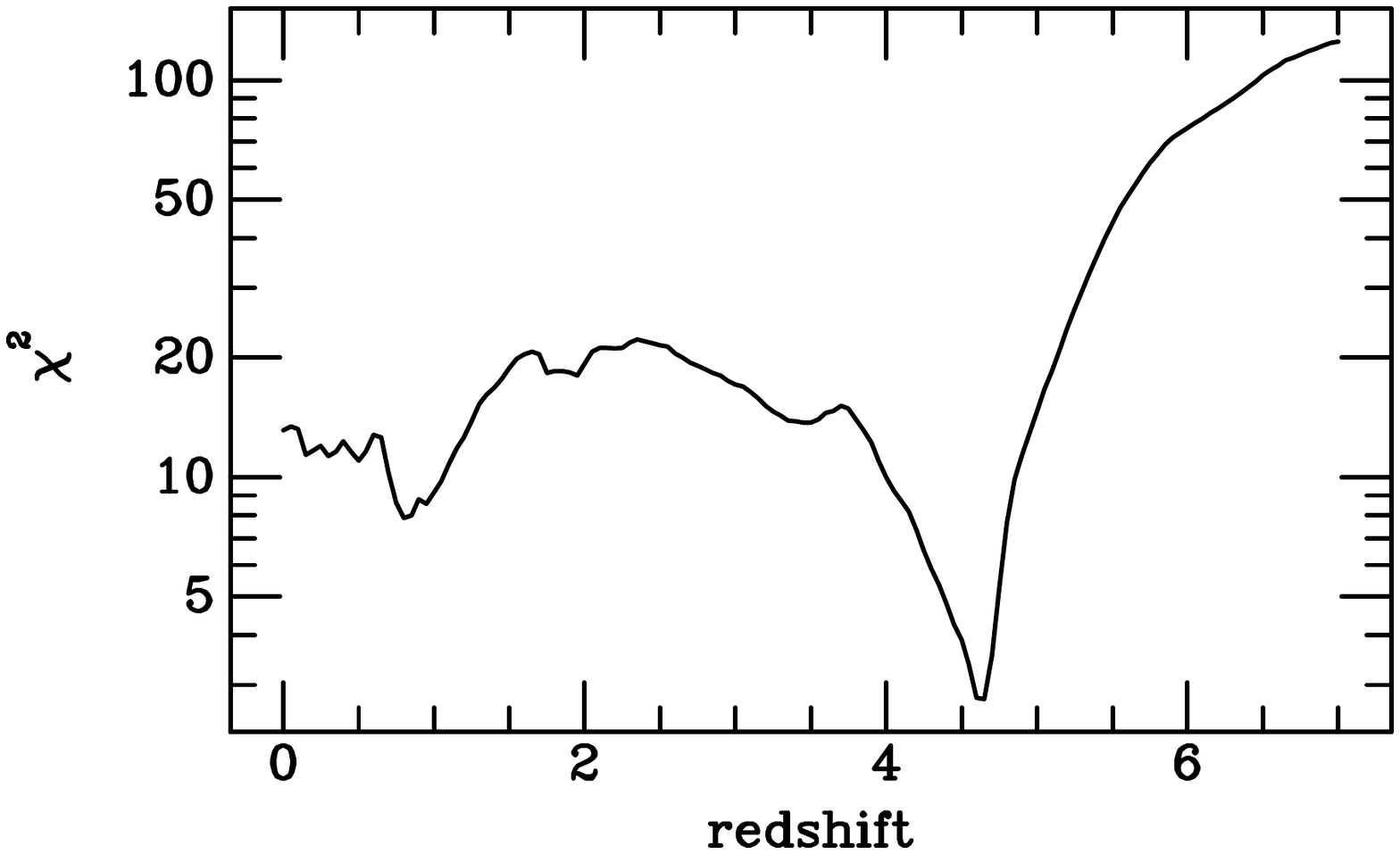}
\end{center}
\vspace{-0.15in}
\caption{\scriptsize
Same as Figure 9 for 4 additional FIRST-BNGS sources.
}
\end{figure*}

\begin{figure*}[htp]
\vspace{+0.4in}
\begin{center}
\rotatebox{270}{\includegraphics[width=2.0in,height=3.2in]{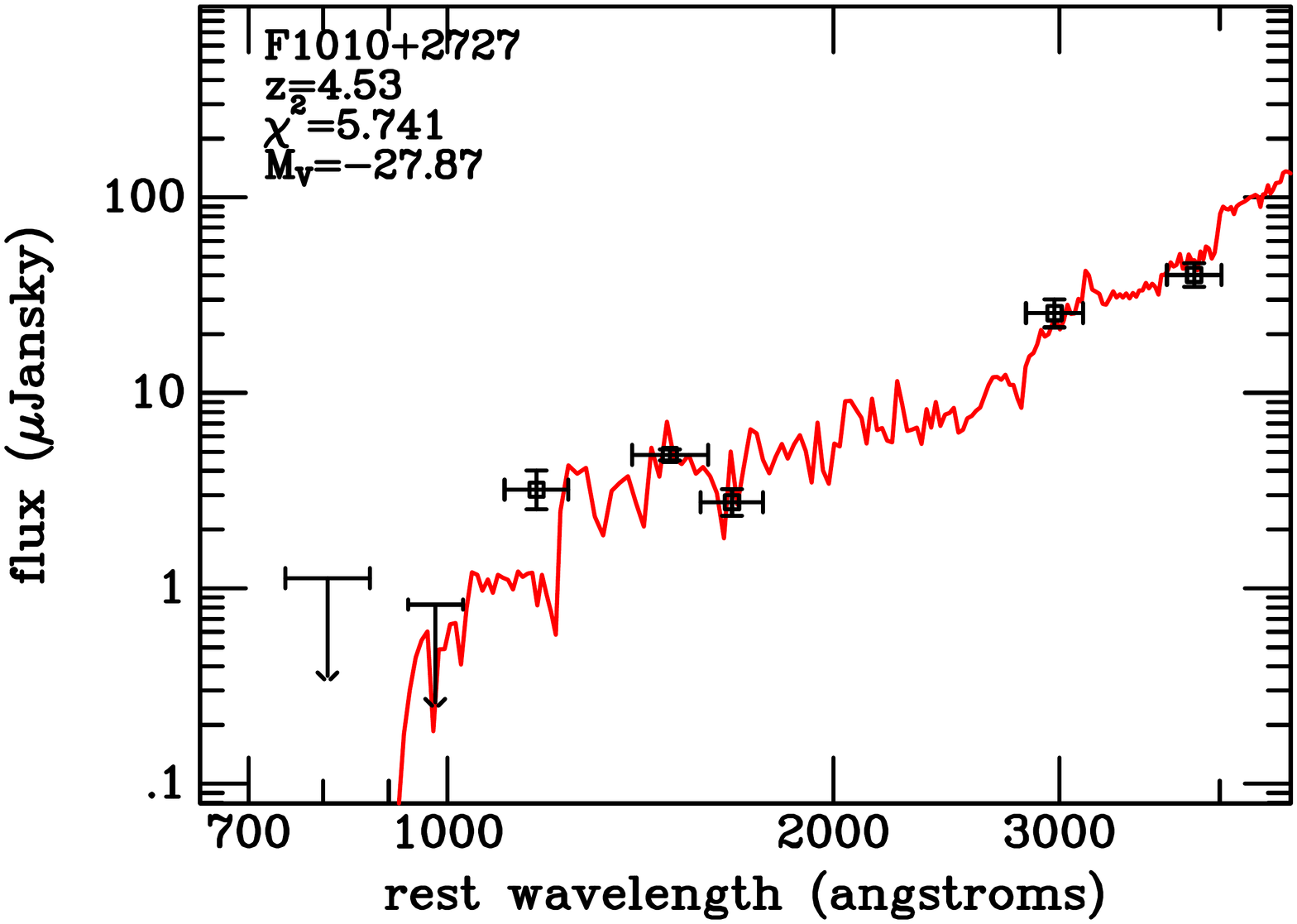}}
\rotatebox{270}{\includegraphics[width=2.0in,height=3.2in]{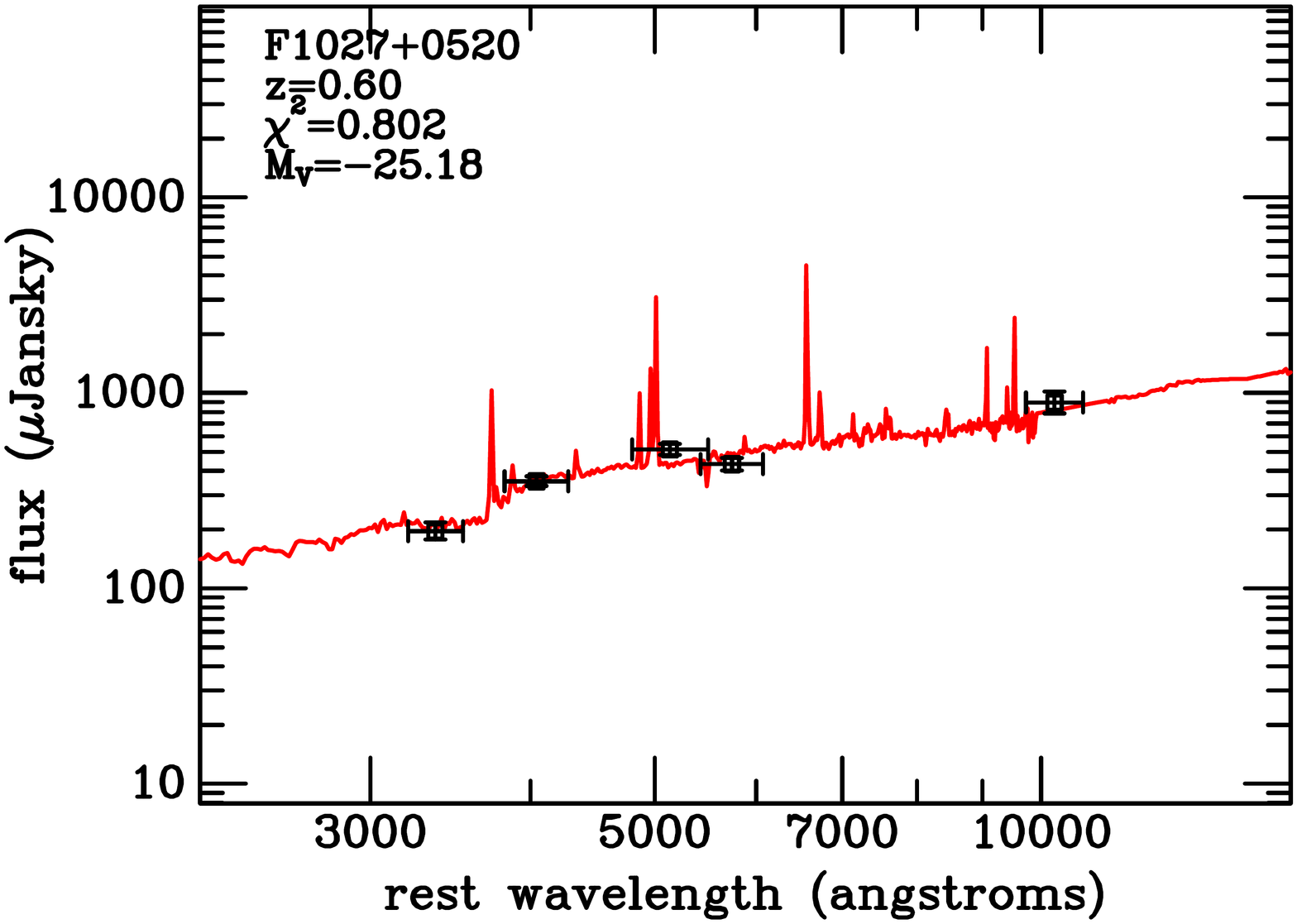}}
\end{center}
\vspace{-1.5in}
\begin{center}
\includegraphics[scale=0.42,angle=0,viewport= 120 50 500 575]{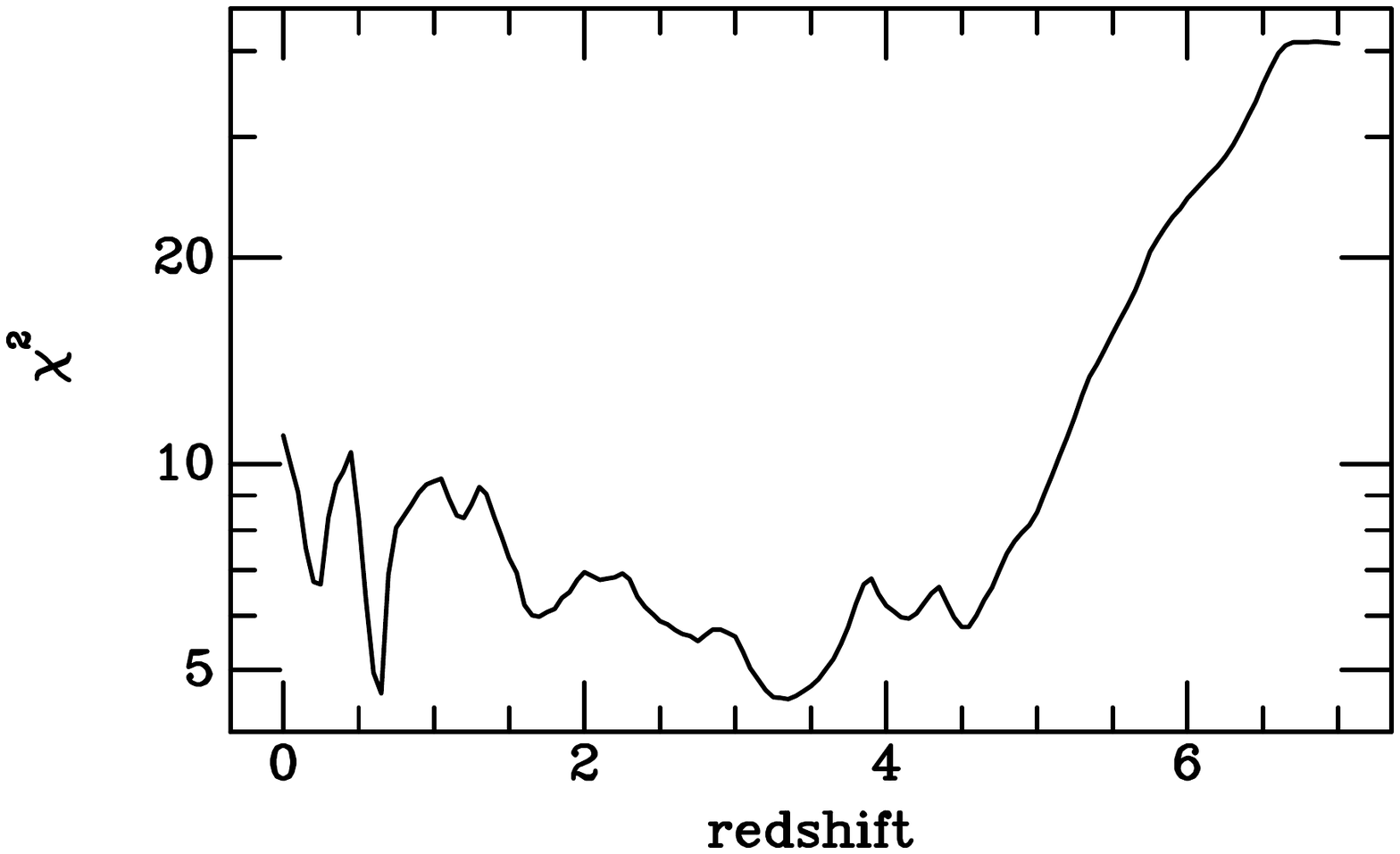}
\includegraphics[scale=0.42,angle=0,viewport= -50 50 500 575]{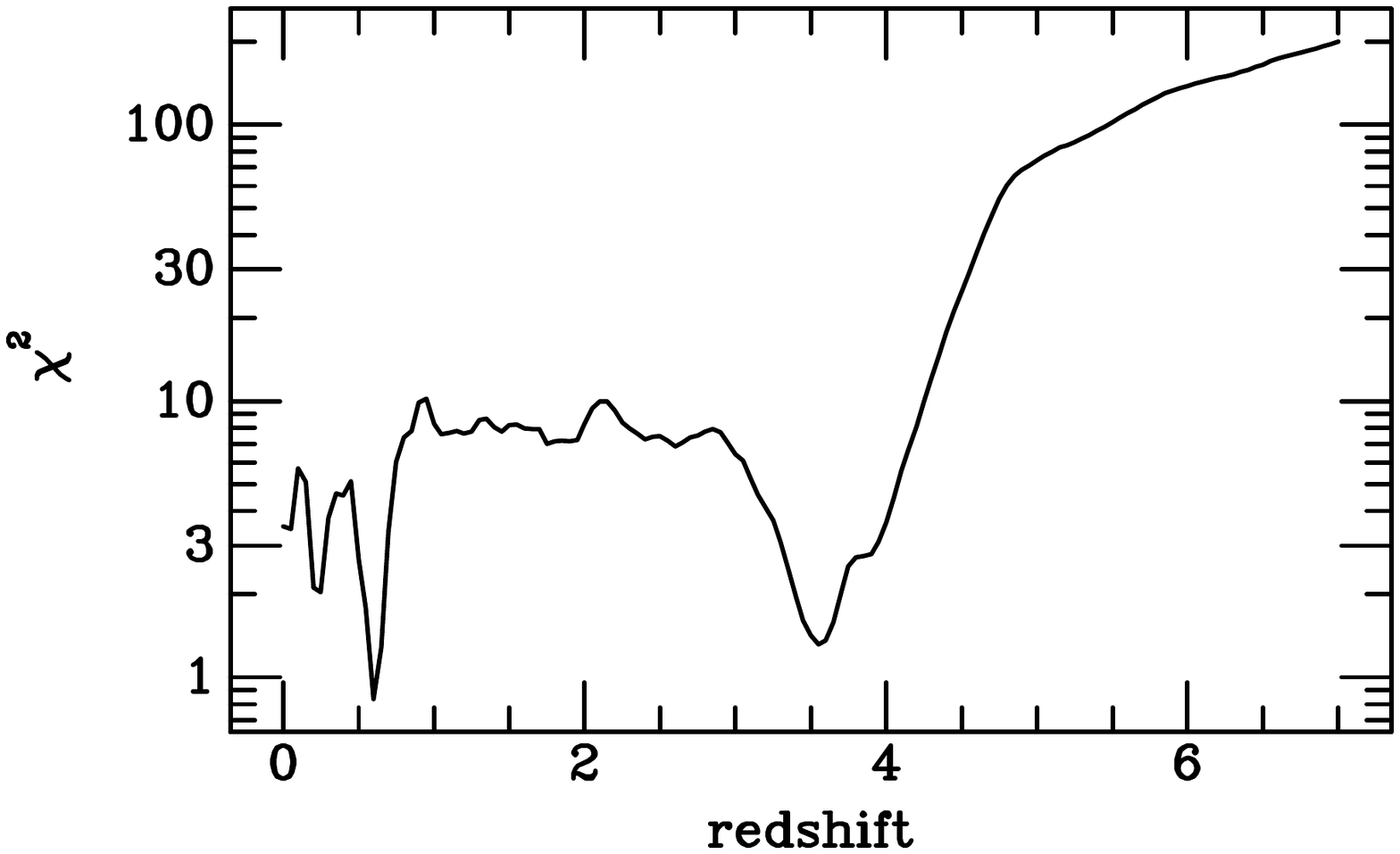}
\end{center}
\vspace{-0.75in}
\begin{center}
\rotatebox{270}{\includegraphics[width=2.0in,height=3.2in]{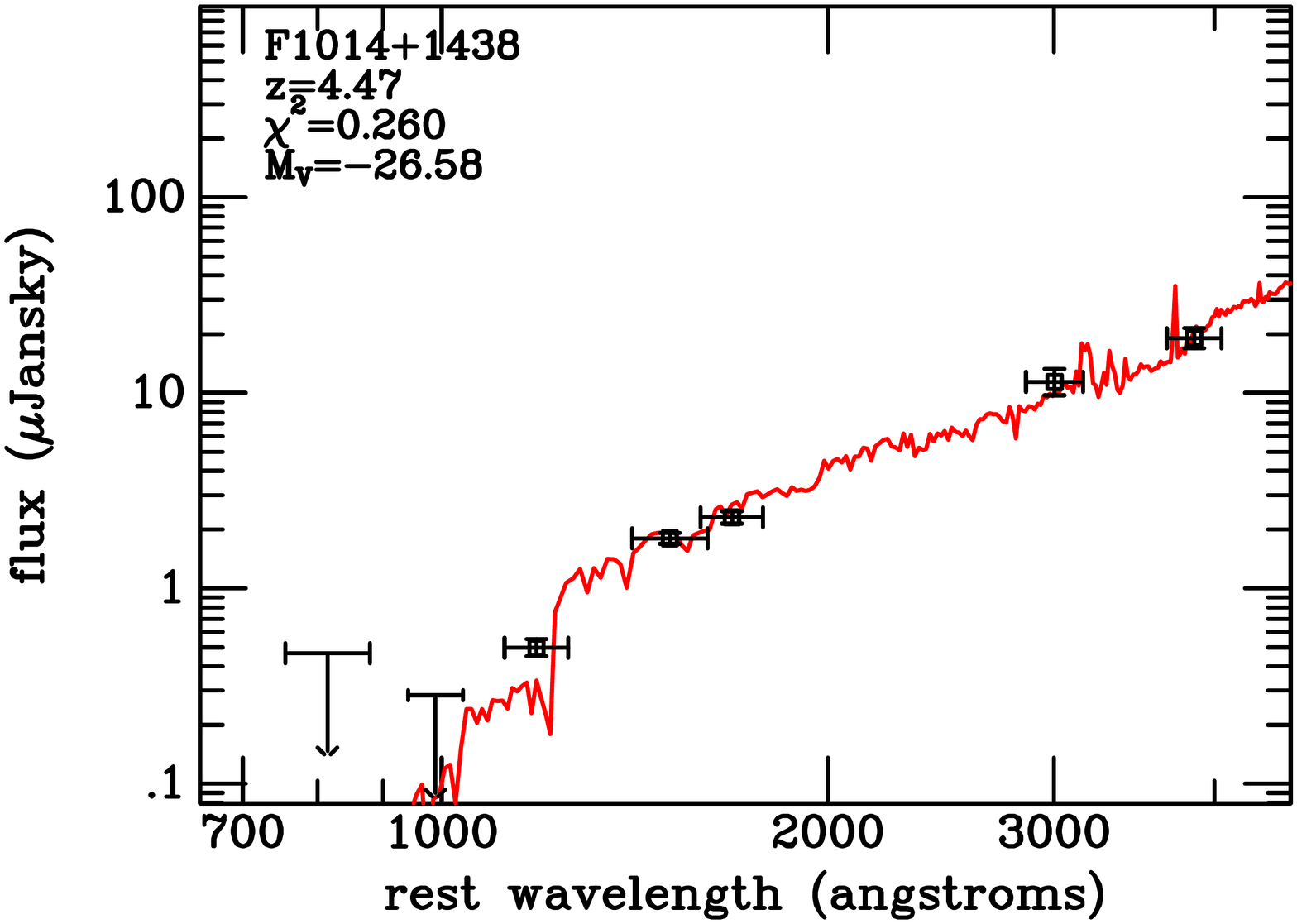}}
\rotatebox{270}{\includegraphics[width=2.0in,height=3.2in]{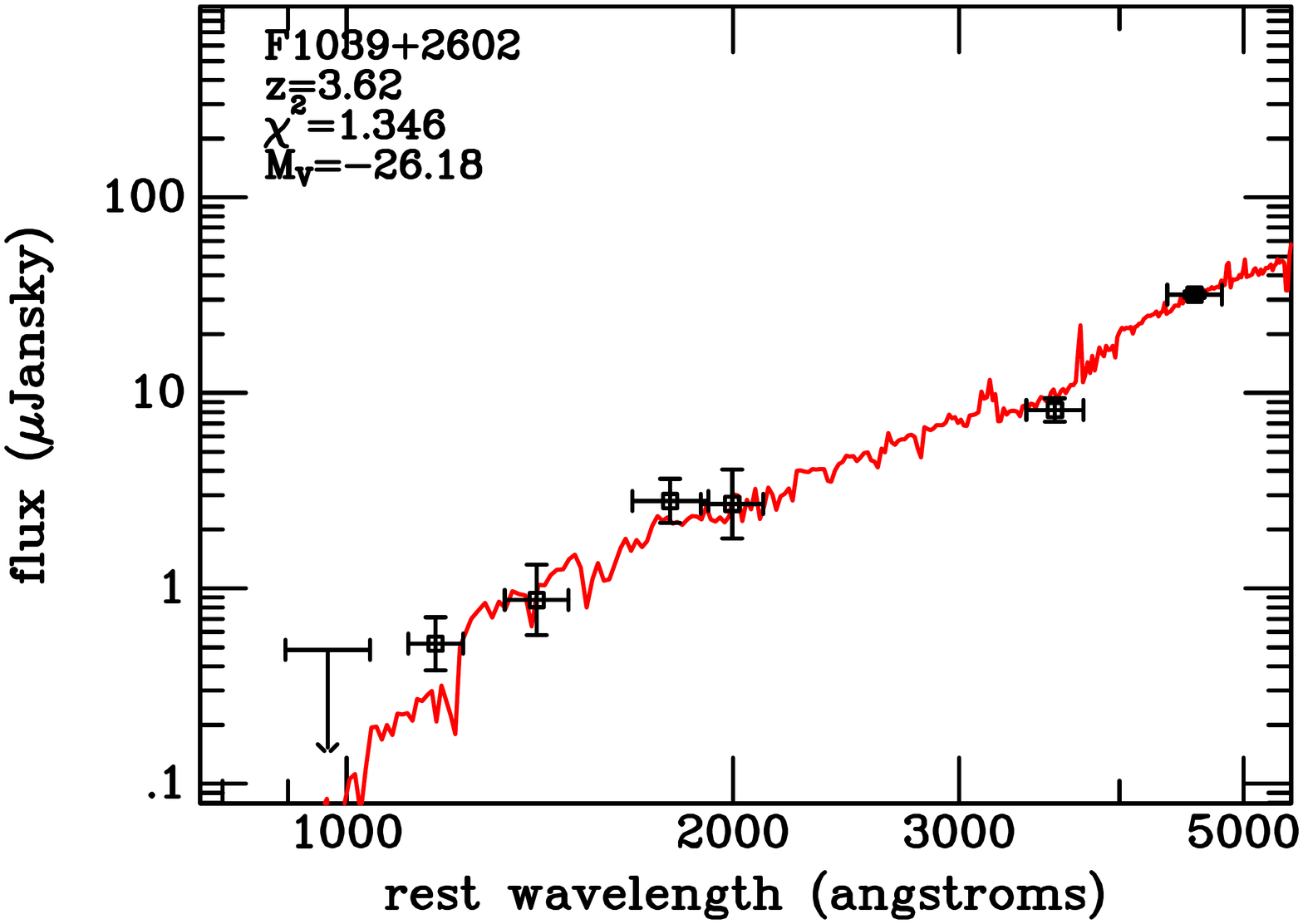}}
\end{center}
\vspace{-1.5in}
\begin{center}
\includegraphics[scale=0.42,angle=0,viewport= 120 50 500 575]{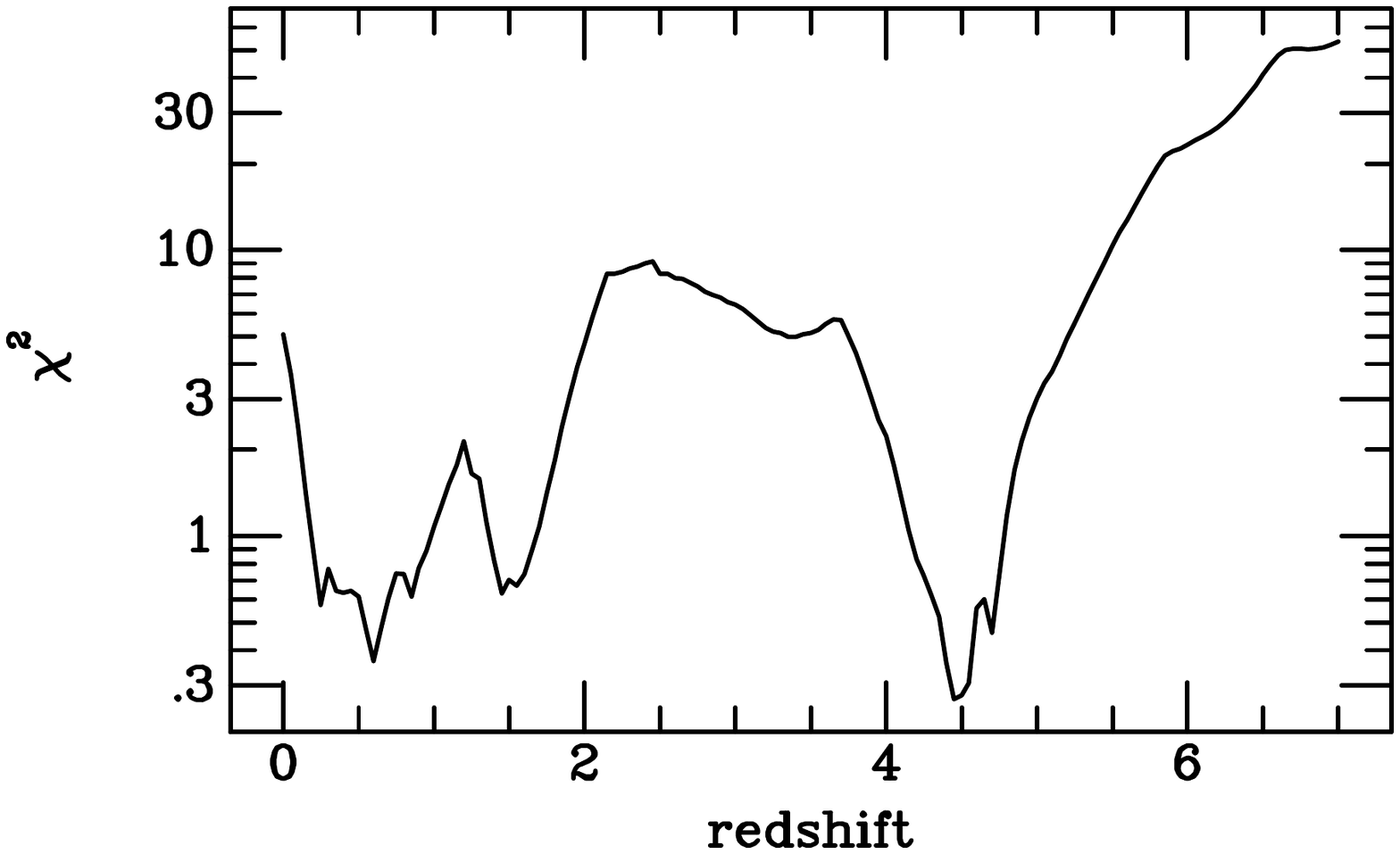}
\includegraphics[scale=0.42,angle=0,viewport= -50 50 500 575]{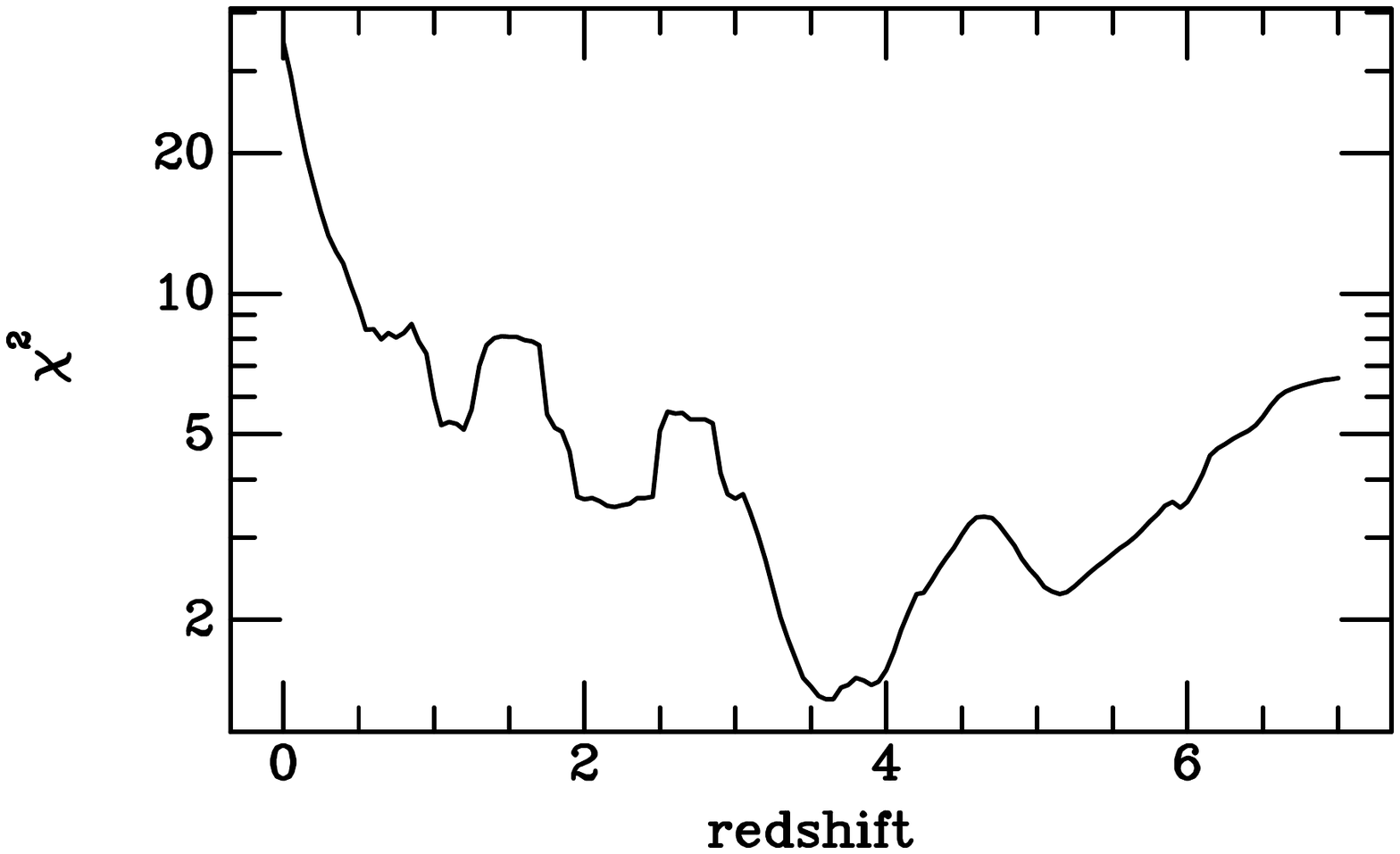}
\end{center}
\vspace{-0.15in}
\caption{\scriptsize
Same as Figure 9 for 4 additional FIRST-BNGS sources.
}
\end{figure*}

\begin{figure*}[htp]
\vspace{+0.4in}
\begin{center}
\rotatebox{270}{\includegraphics[width=2.0in,height=3.2in]{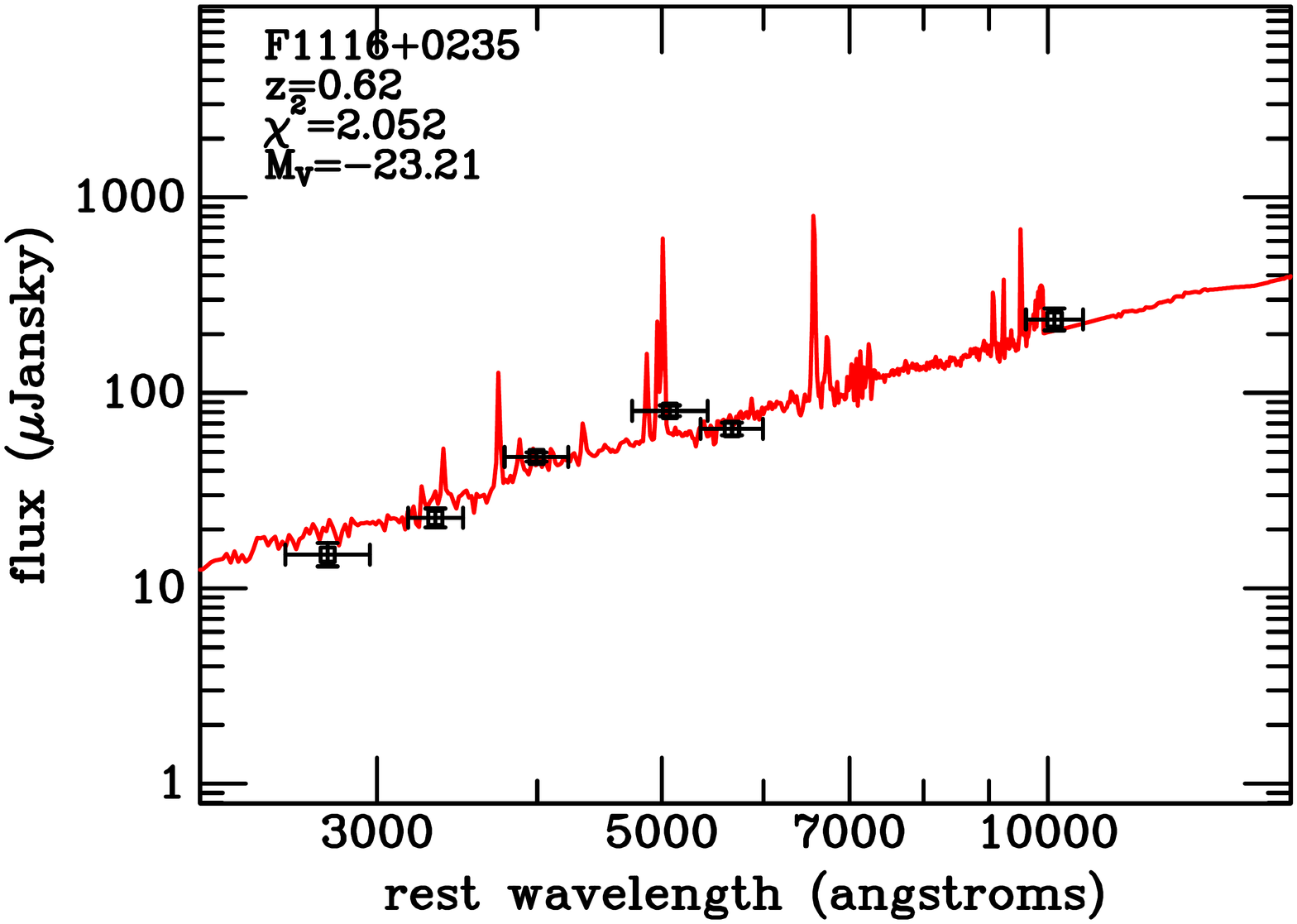}}
\rotatebox{270}{\includegraphics[width=2.0in,height=3.2in]{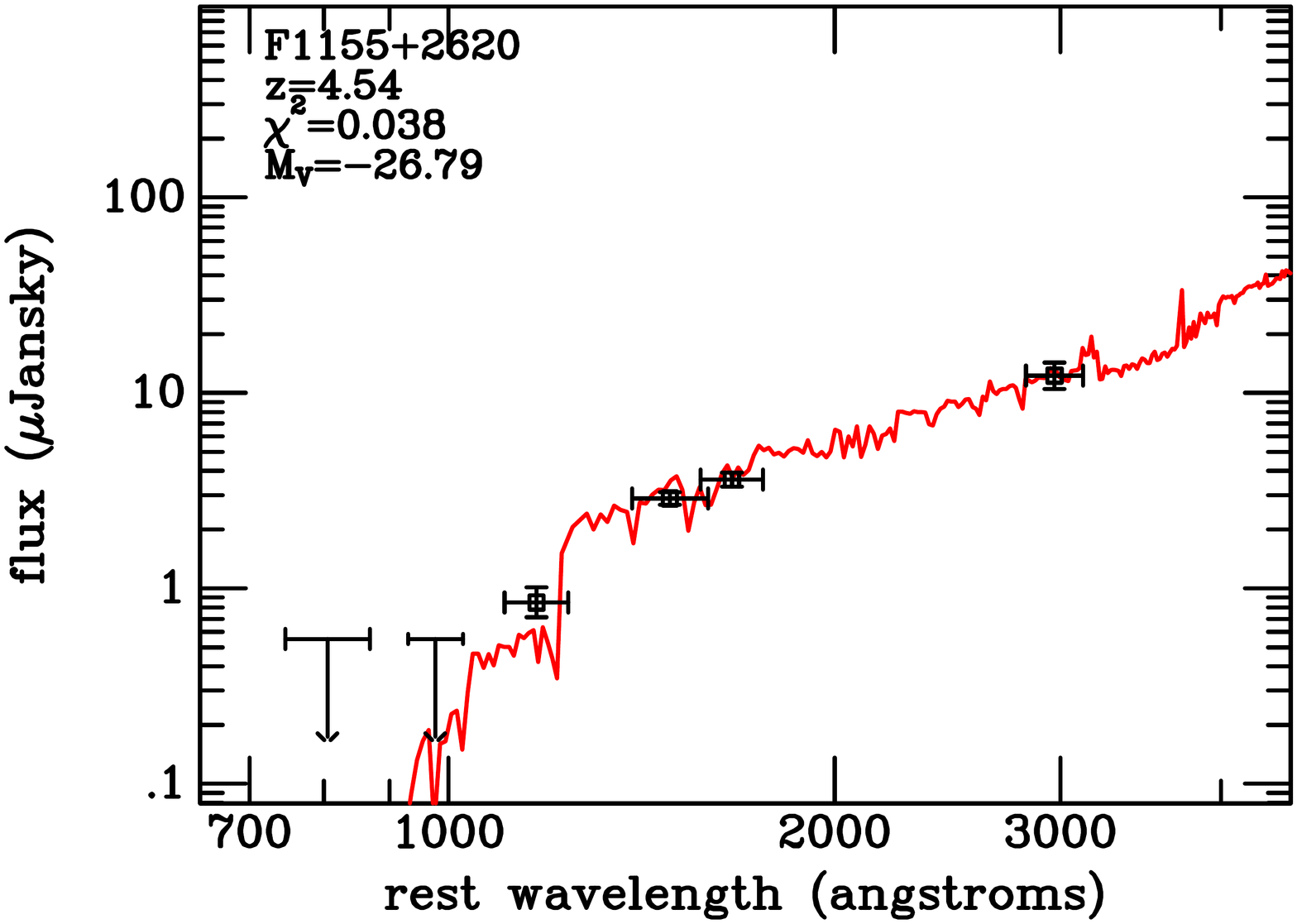}}
\end{center}
\vspace{-1.5in}
\begin{center}
\includegraphics[scale=0.42,angle=0,viewport= 120 50 500 575]{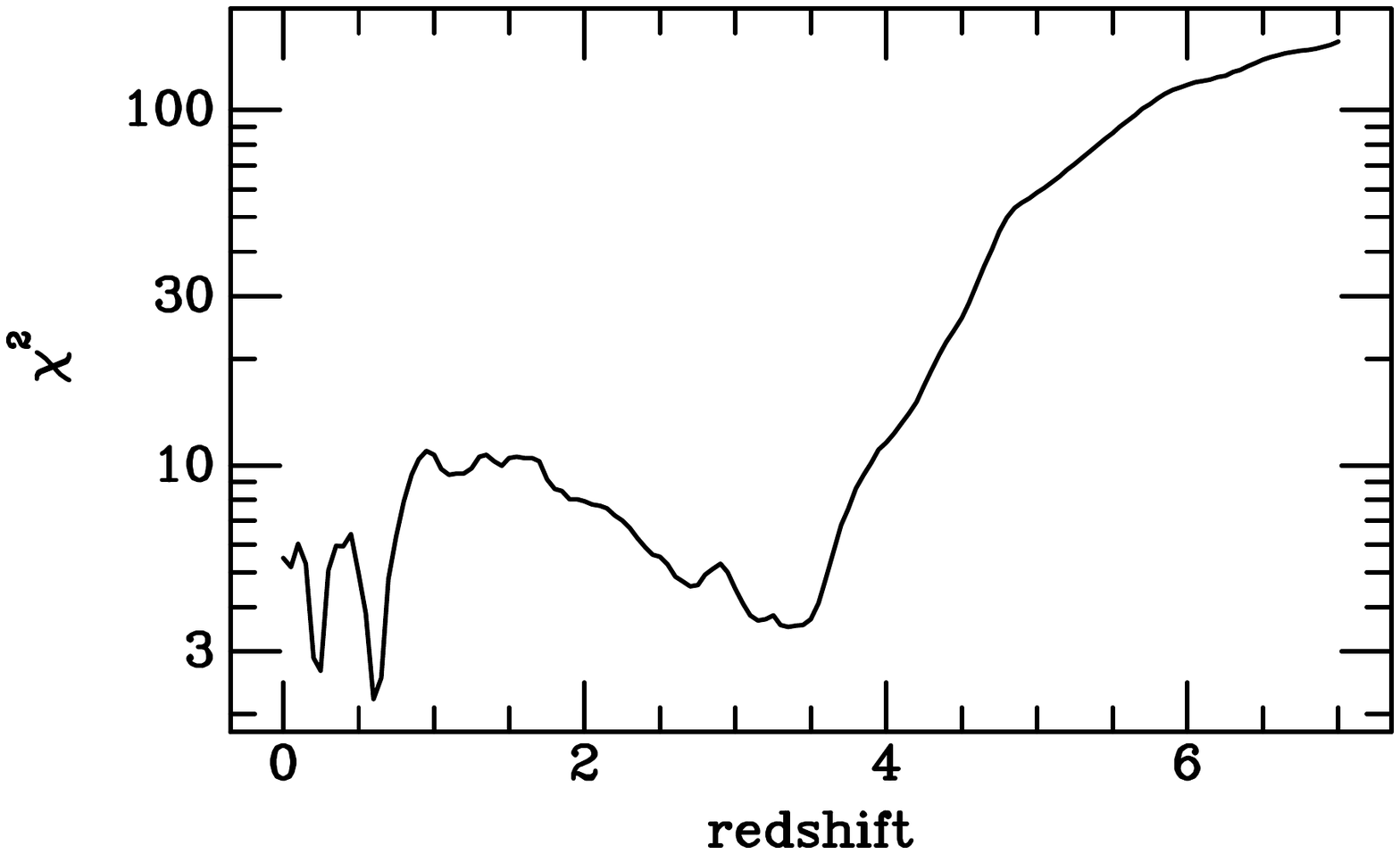}
\includegraphics[scale=0.42,angle=0,viewport= -50 50 500 575]{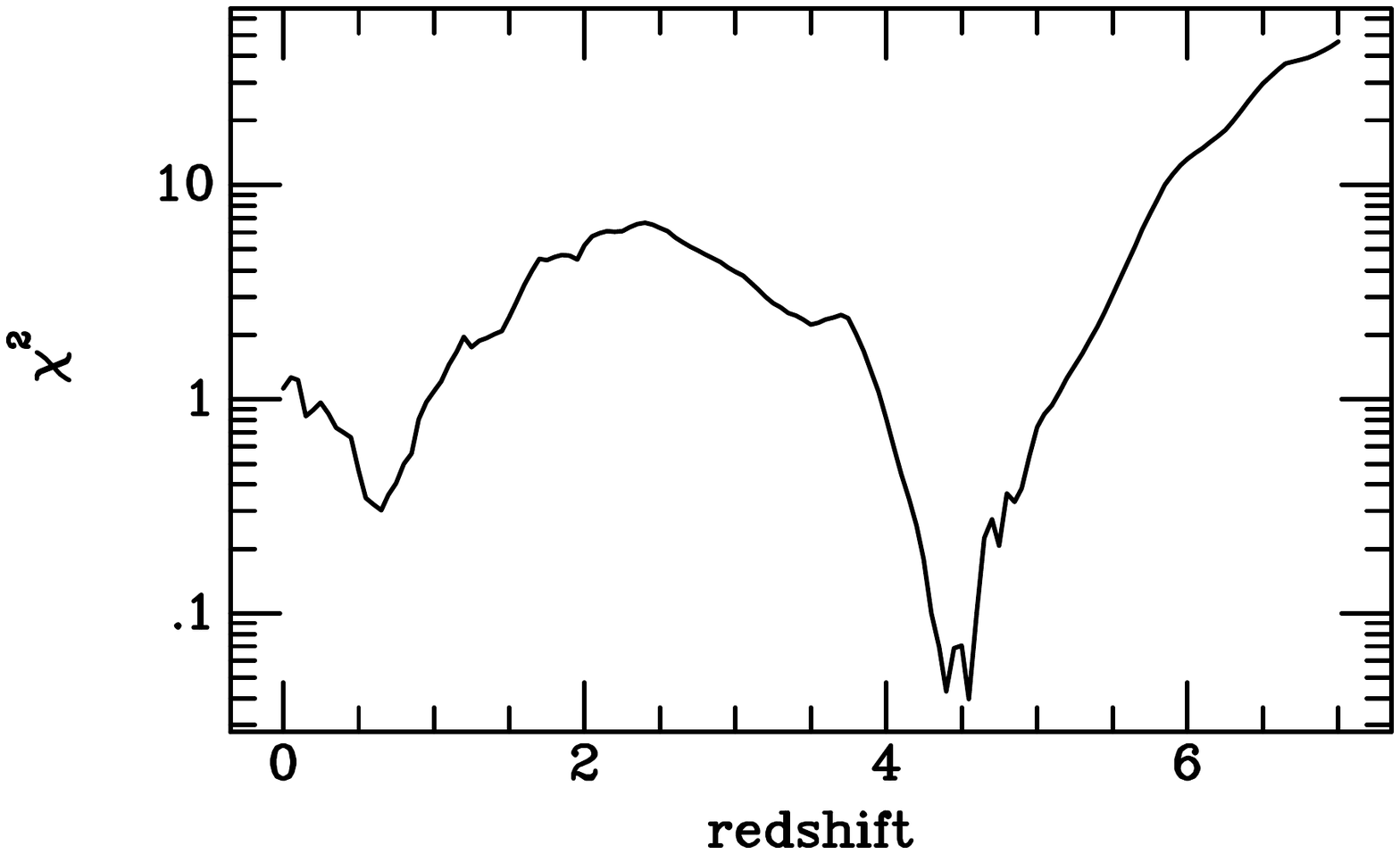}
\end{center}
\vspace{-0.75in}
\begin{center}
\rotatebox{270}{\includegraphics[width=2.0in,height=3.2in]{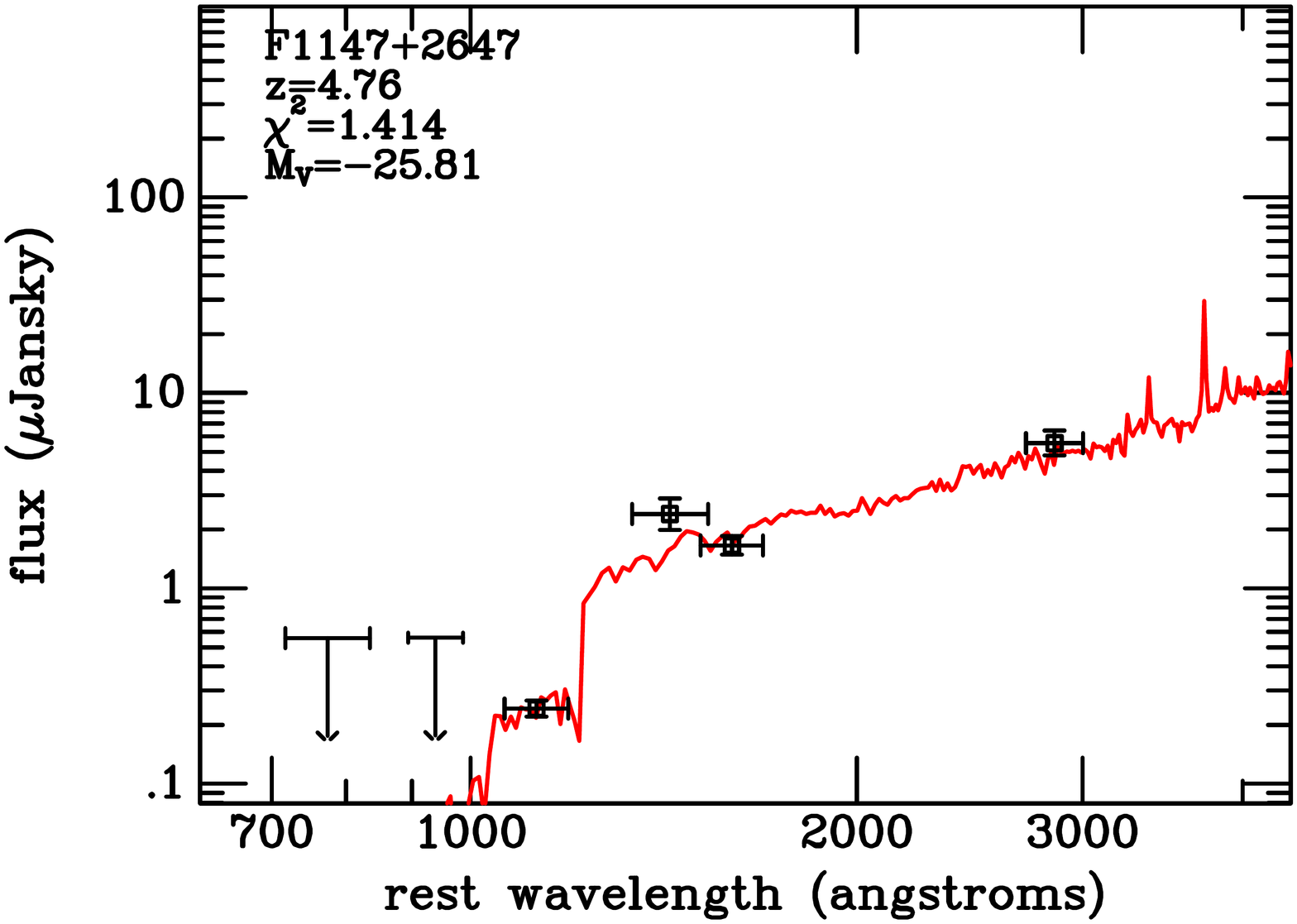}}
\rotatebox{270}{\includegraphics[width=2.0in,height=3.2in]{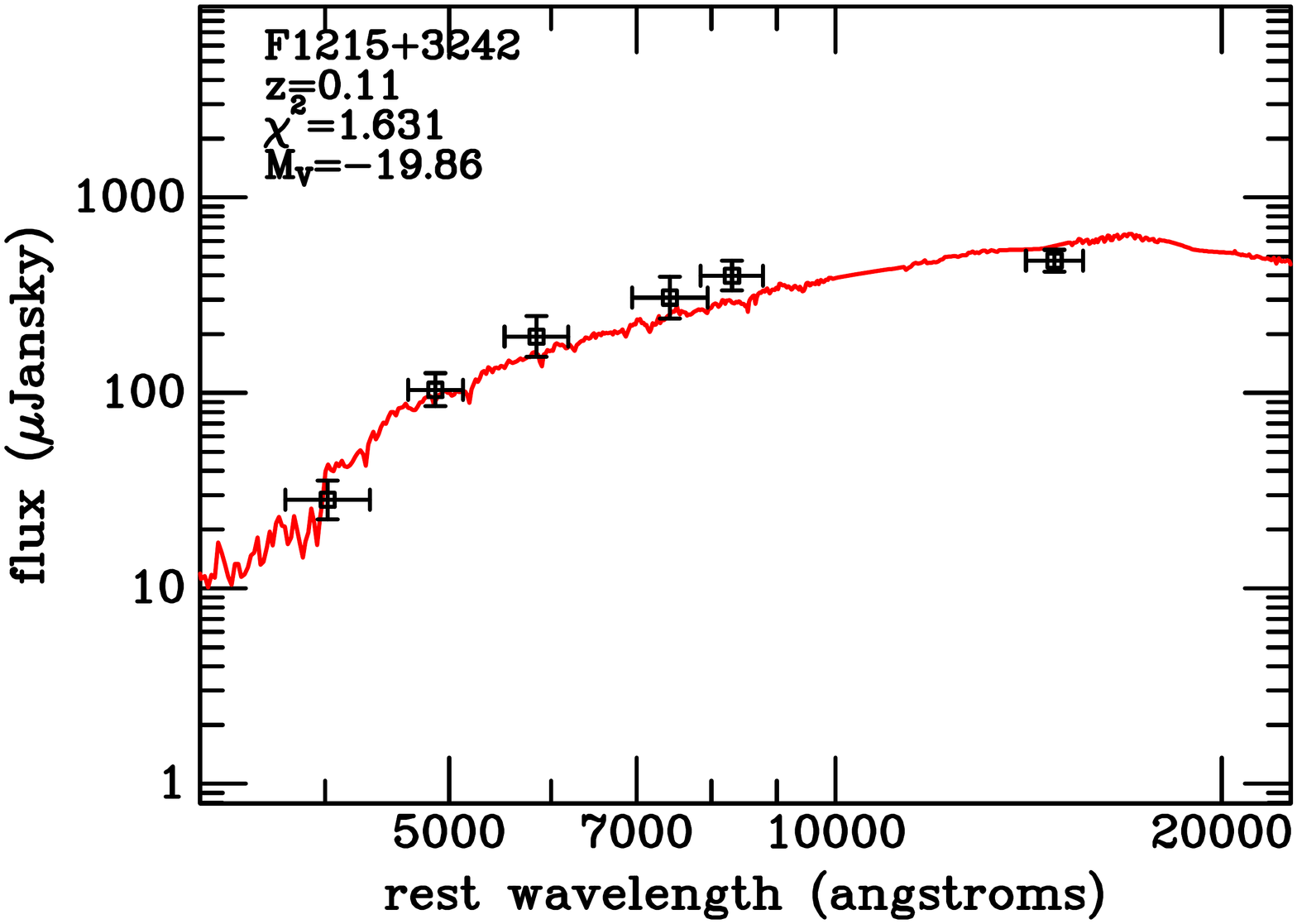}}
\end{center}
\vspace{-1.5in}
\begin{center}
\includegraphics[scale=0.42,angle=0,viewport= 120 50 500 575]{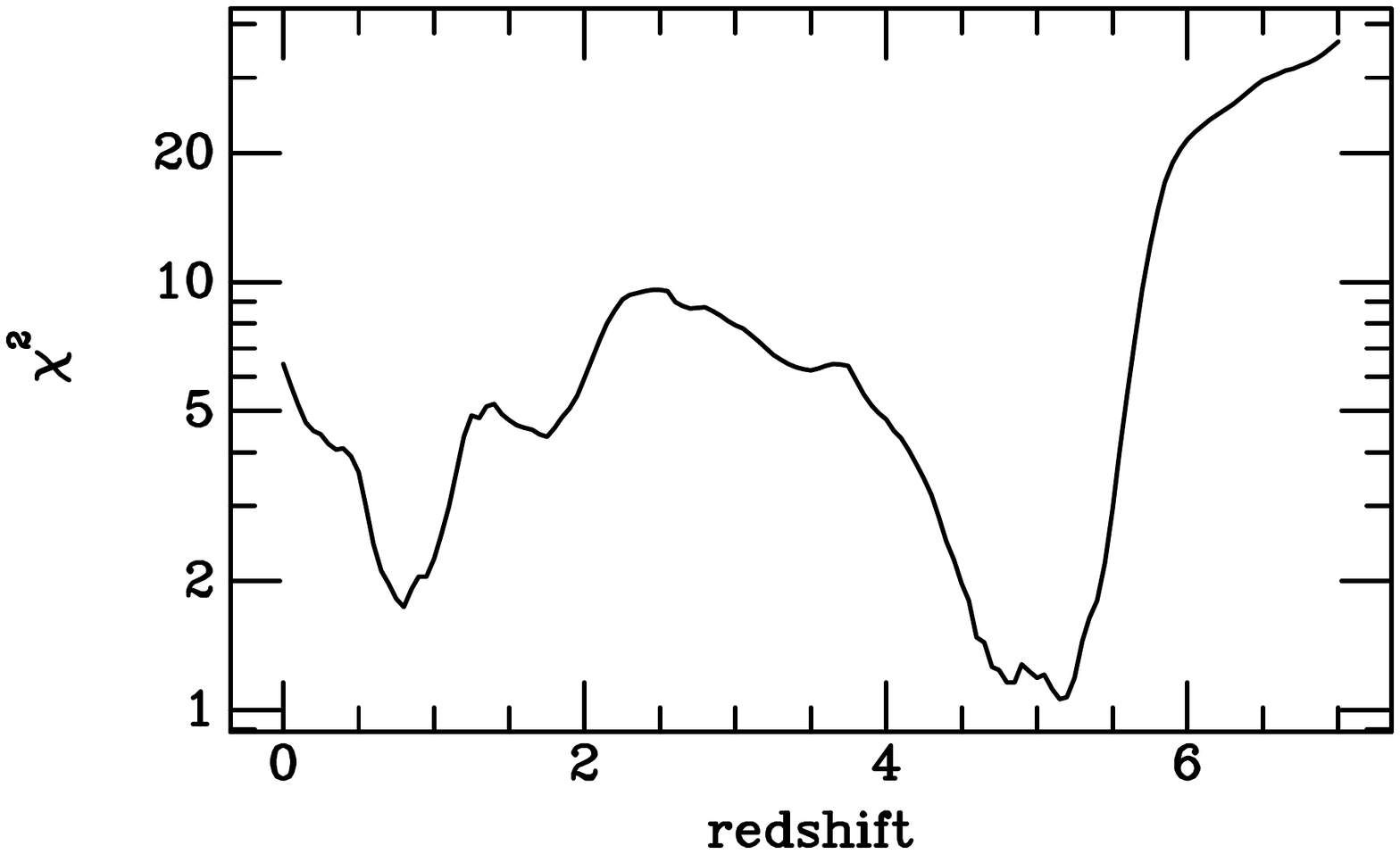}
\includegraphics[scale=0.42,angle=0,viewport= -50 50 500 575]{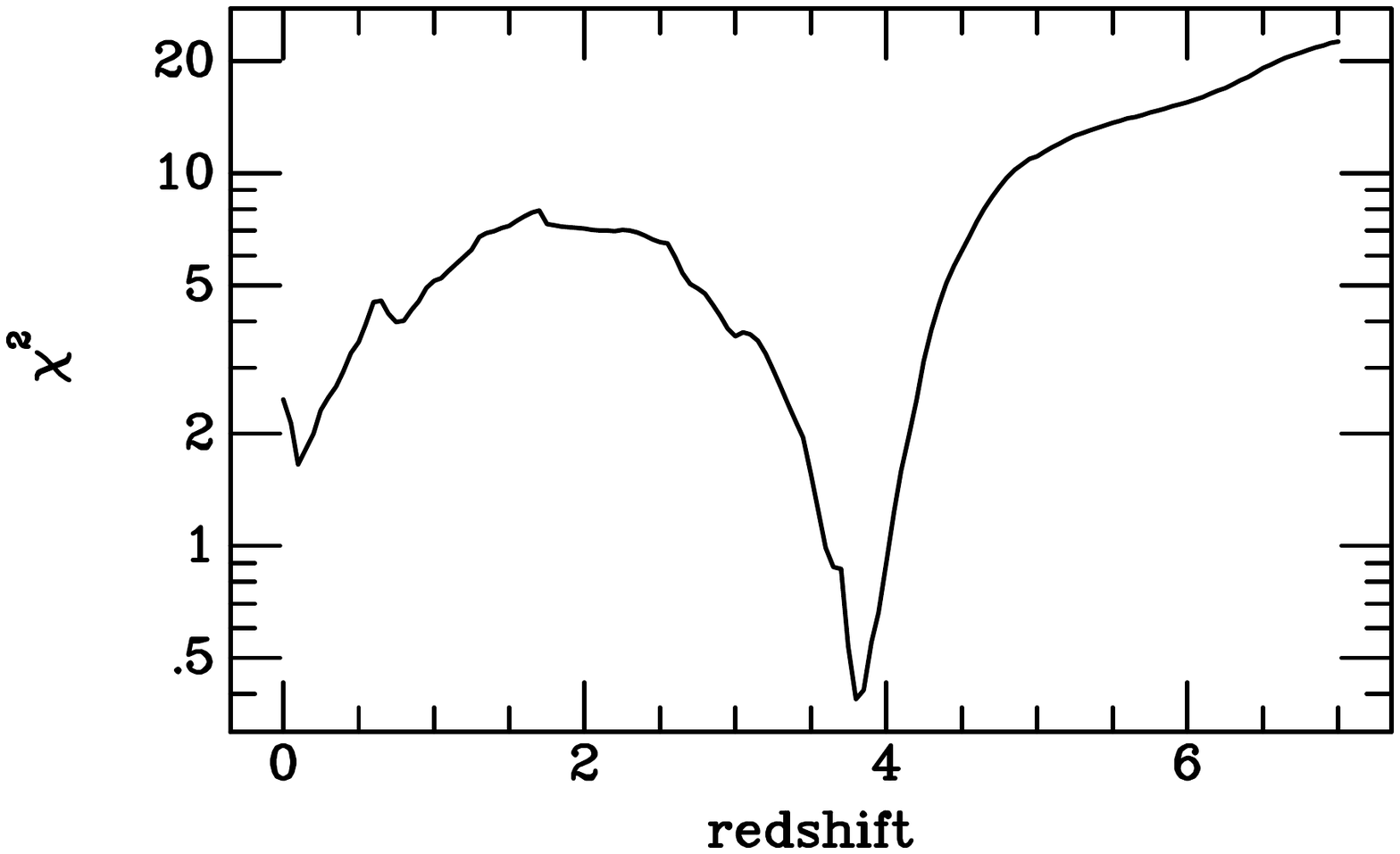}
\end{center}
\vspace{-0.15in}
\caption{\scriptsize
Same as Figure 9 for 4 additional FIRST-BNGS sources.
}
\end{figure*}

\begin{figure*}[htp]
\vspace{+0.4in}
\begin{center}
\rotatebox{270}{\includegraphics[width=2.0in,height=3.2in]{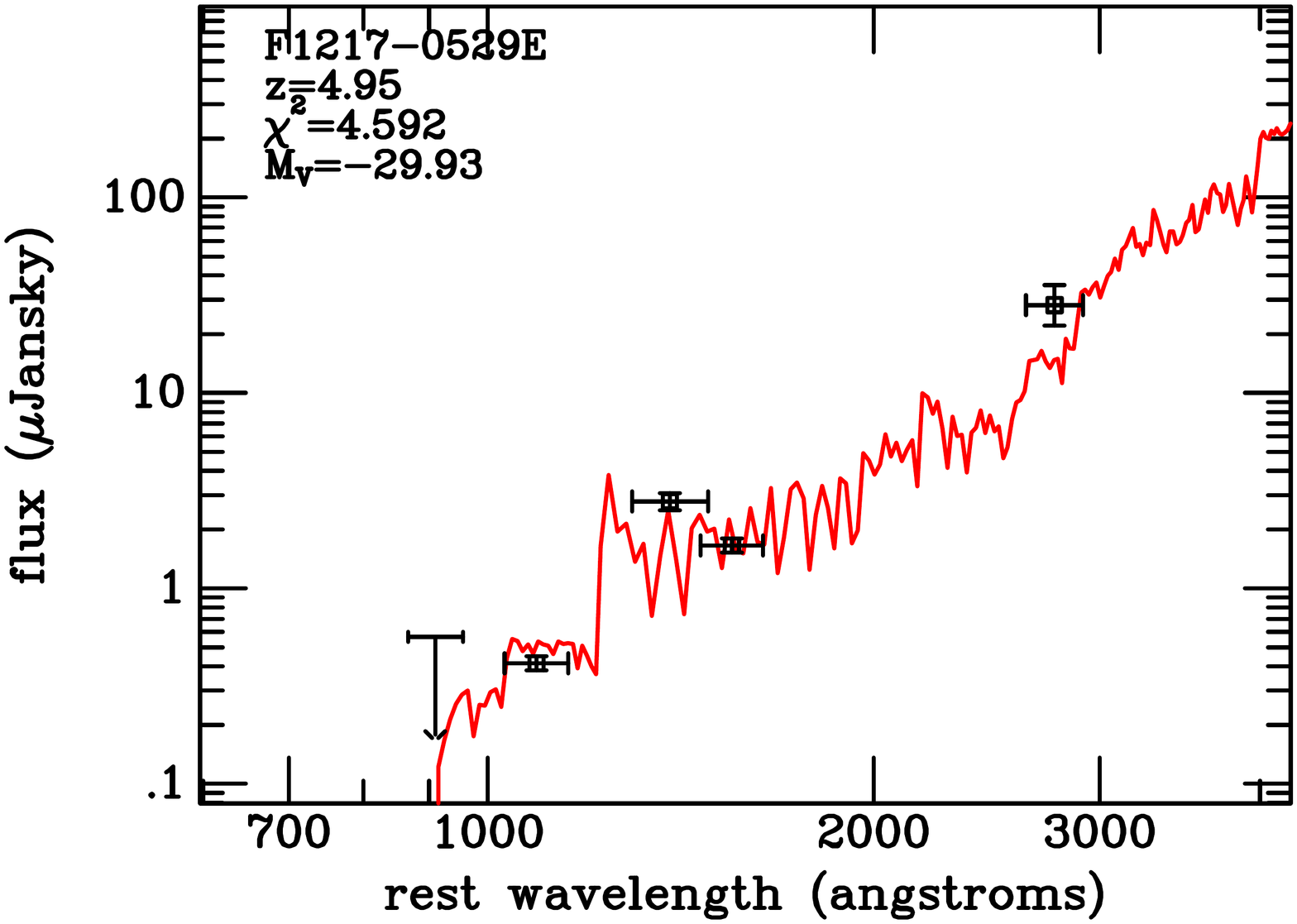}}
\rotatebox{270}{\includegraphics[width=2.0in,height=3.2in]{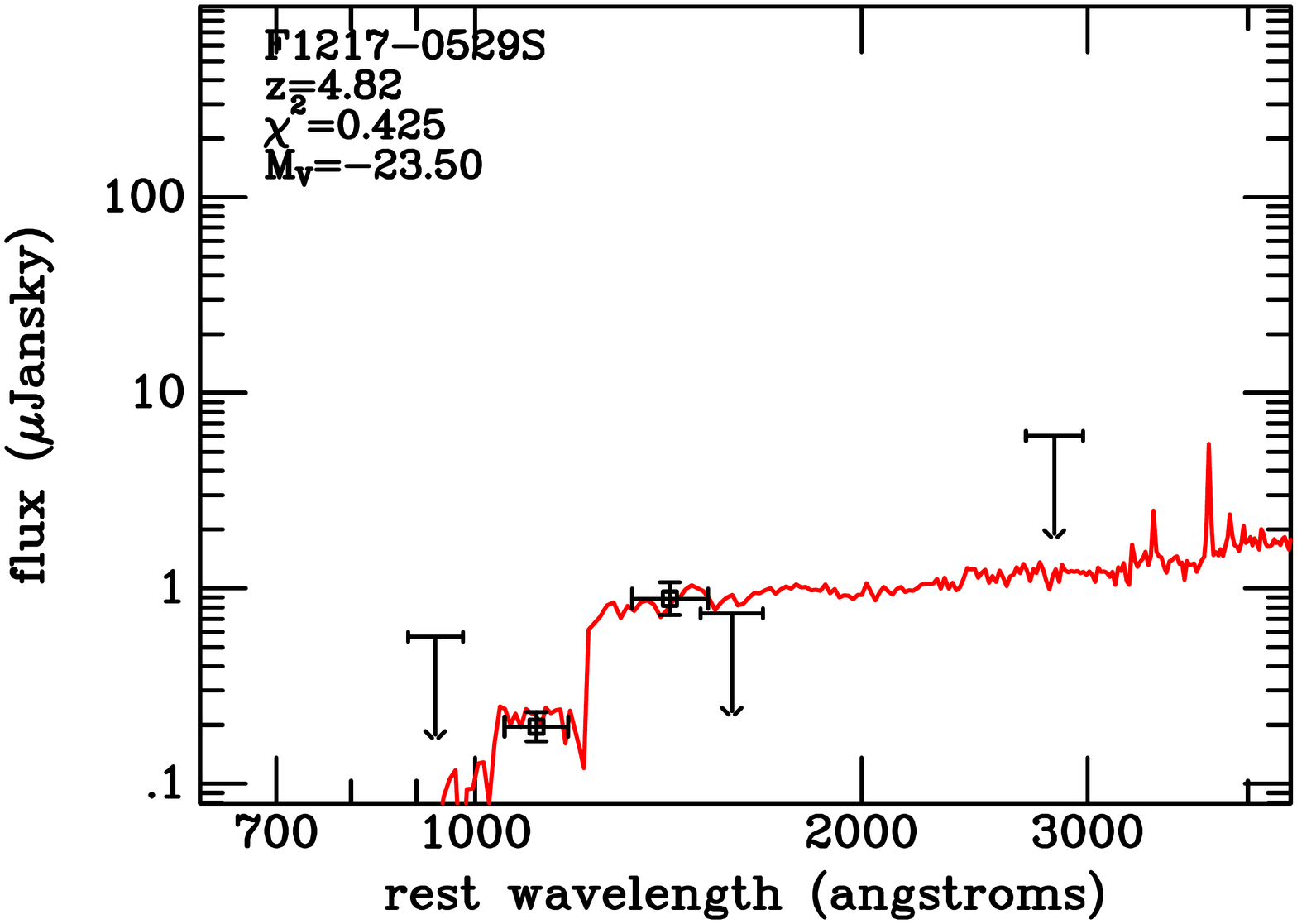}}
\end{center}
\vspace{-1.5in}
\begin{center}
\includegraphics[scale=0.42,angle=0,viewport= 120 50 500 575]{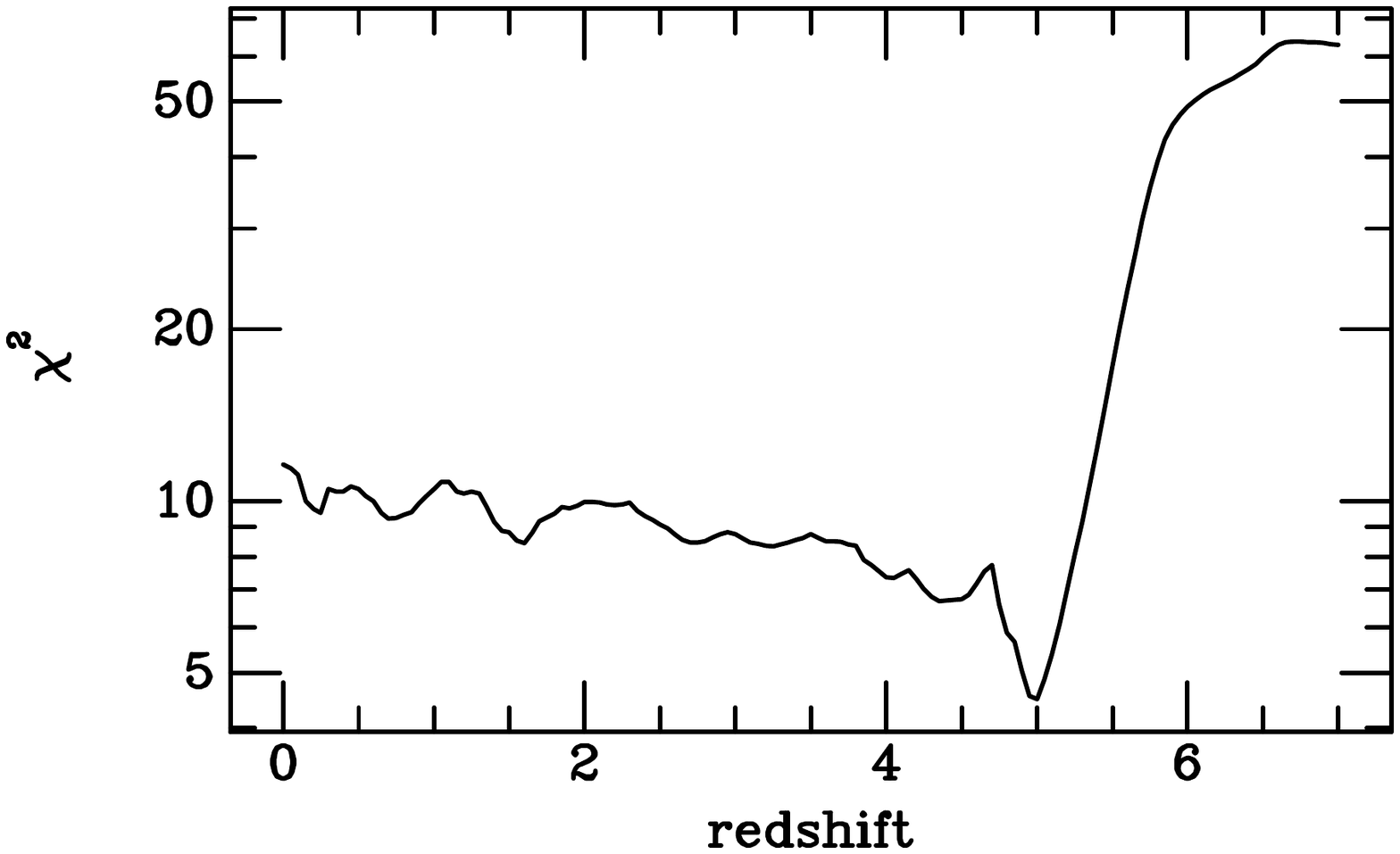}
\includegraphics[scale=0.42,angle=0,viewport= -50 50 500 575]{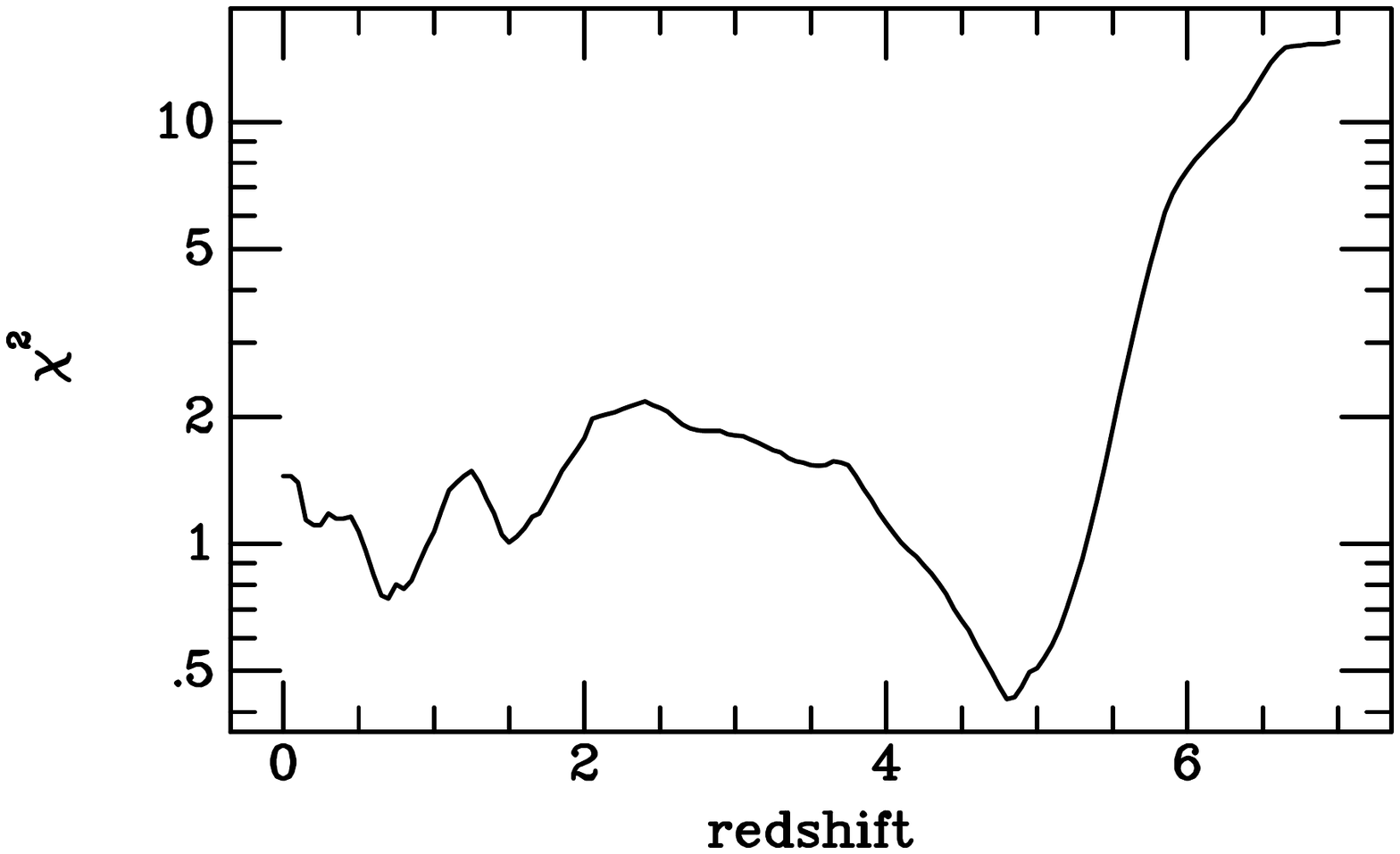}
\end{center}
\vspace{-0.75in}
\begin{center}
\rotatebox{270}{\includegraphics[width=2.0in,height=3.2in]{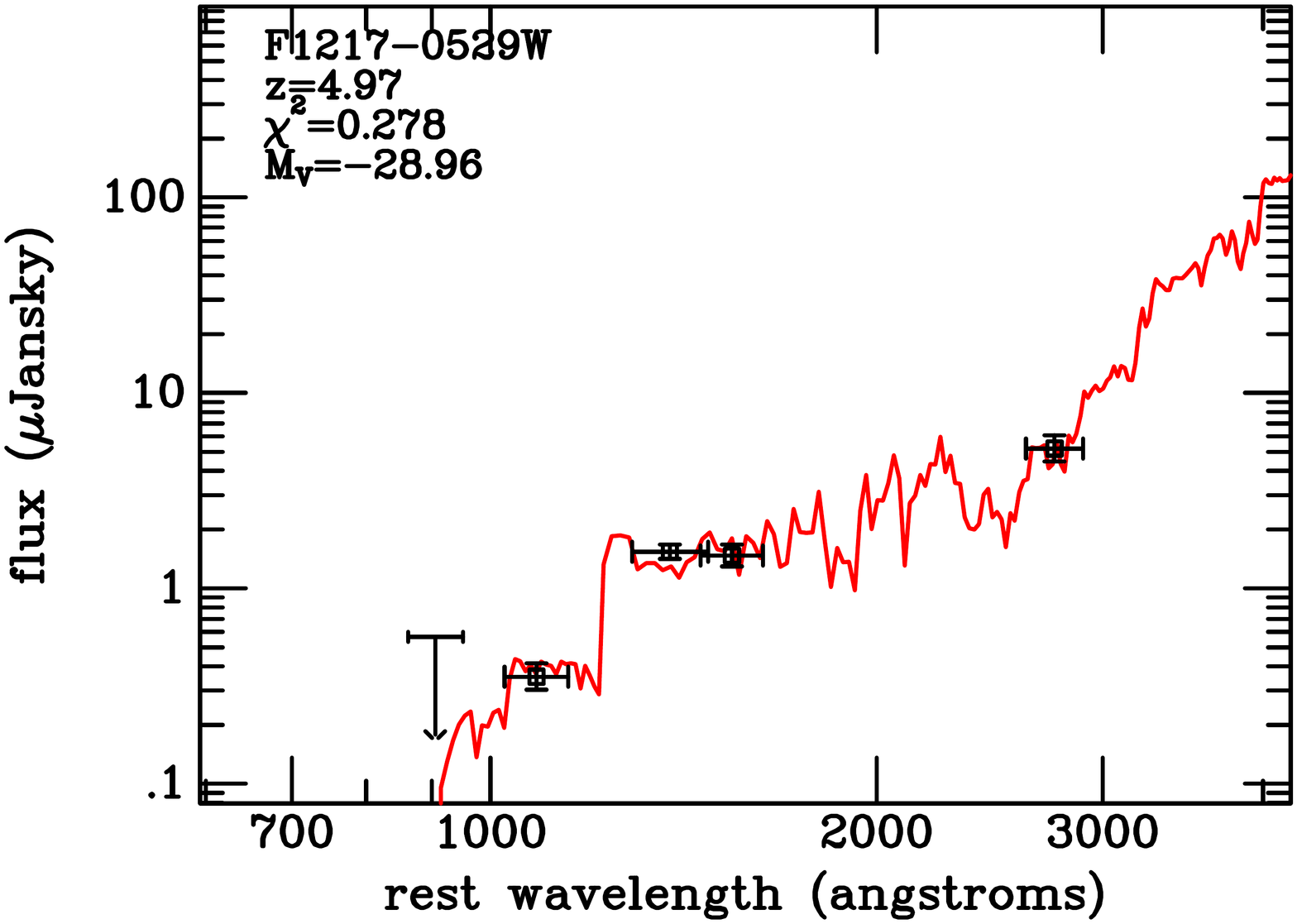}}
\rotatebox{270}{\includegraphics[width=2.0in,height=3.2in]{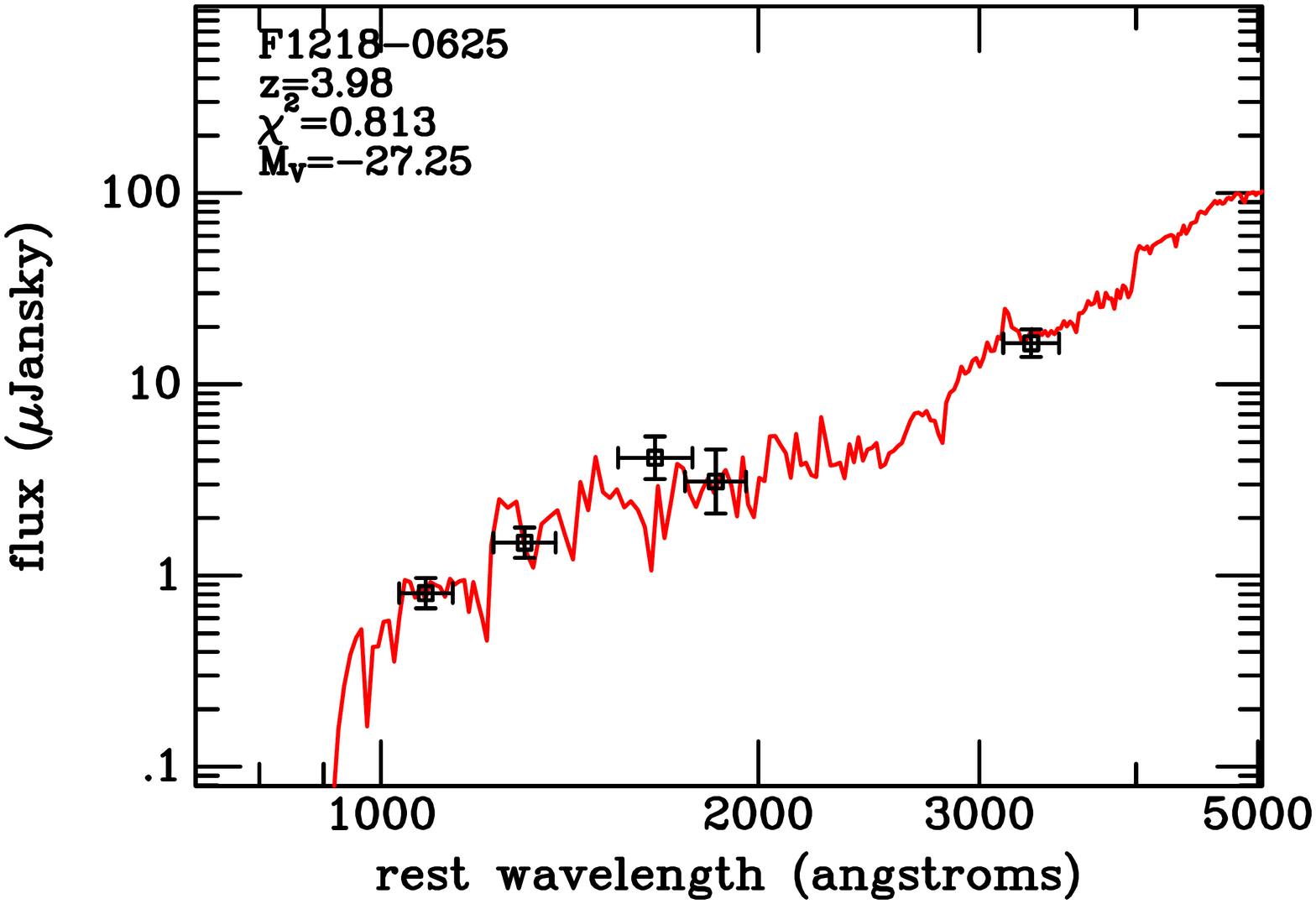}}
\end{center}
\vspace{-1.5in}
\begin{center}
\includegraphics[scale=0.42,angle=0,viewport= 120 50 500 575]{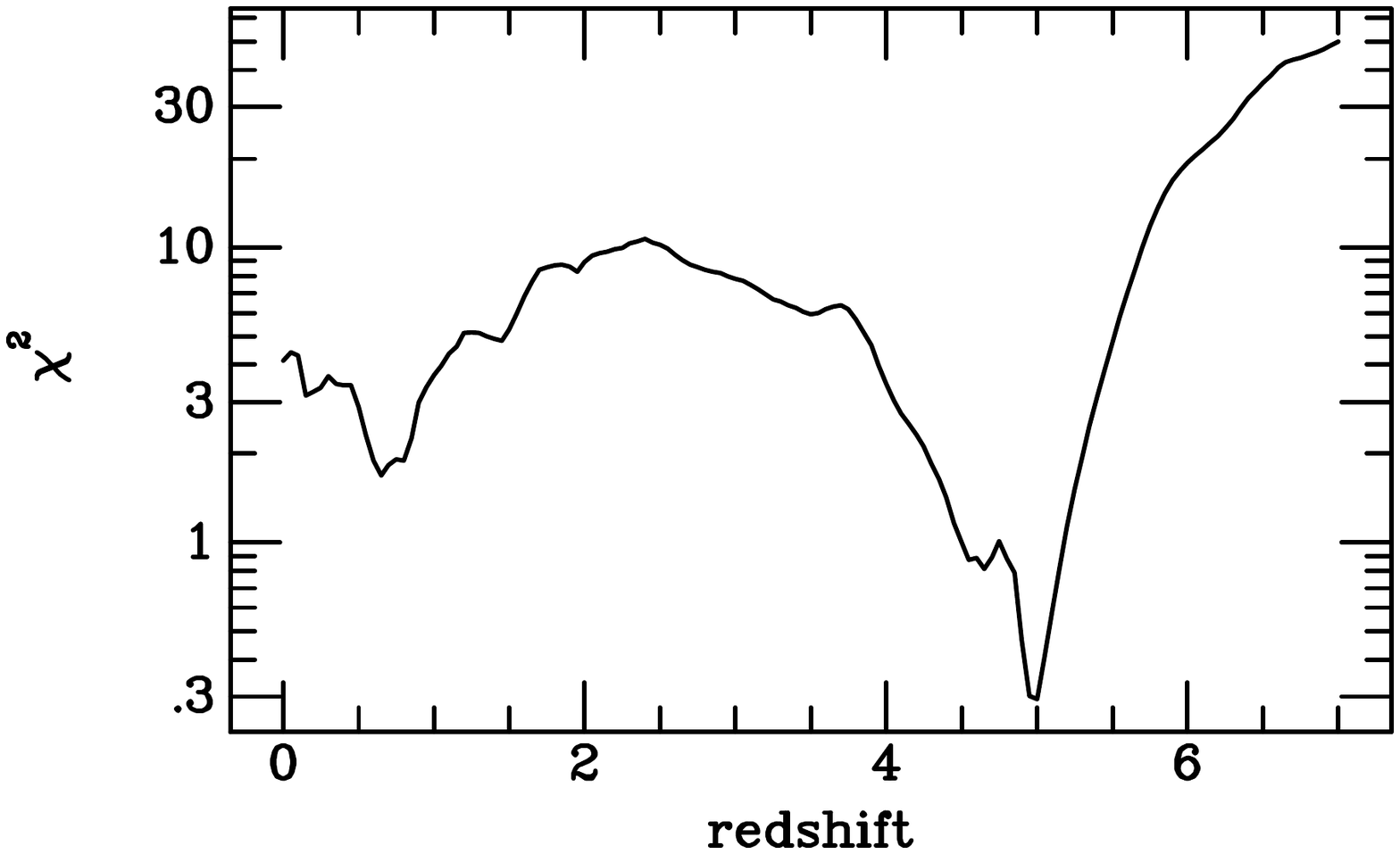}
\includegraphics[scale=0.42,angle=0,viewport= -50 50 500 575]{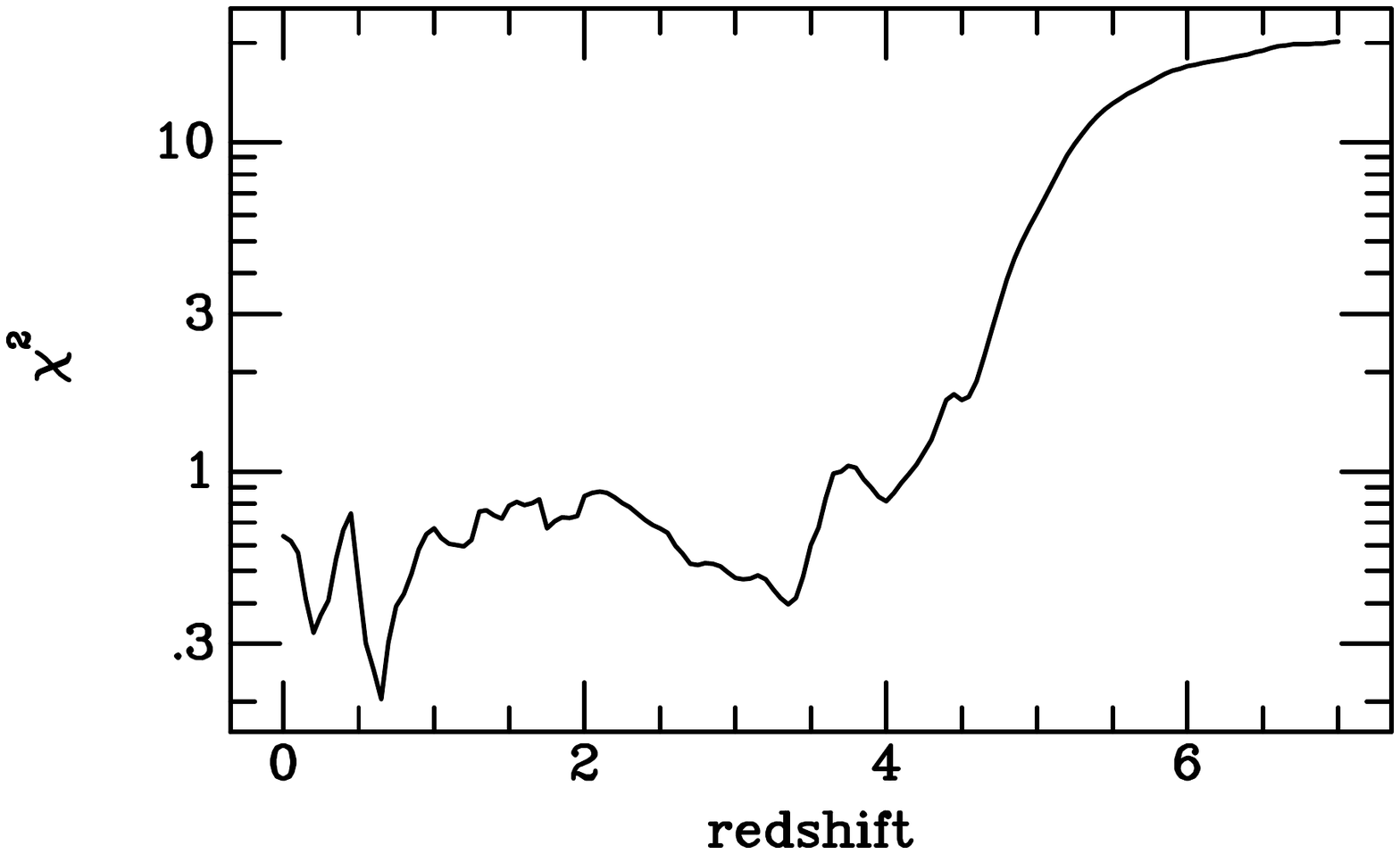}
\end{center}
\vspace{-0.15in}
\caption{\scriptsize
Same as Figure 9 for 4 additional FIRST-BNGS sources.
}
\end{figure*}

\begin{figure*}[htp]
\vspace{+0.4in}
\begin{center}
\rotatebox{270}{\includegraphics[width=2.0in,height=3.2in]{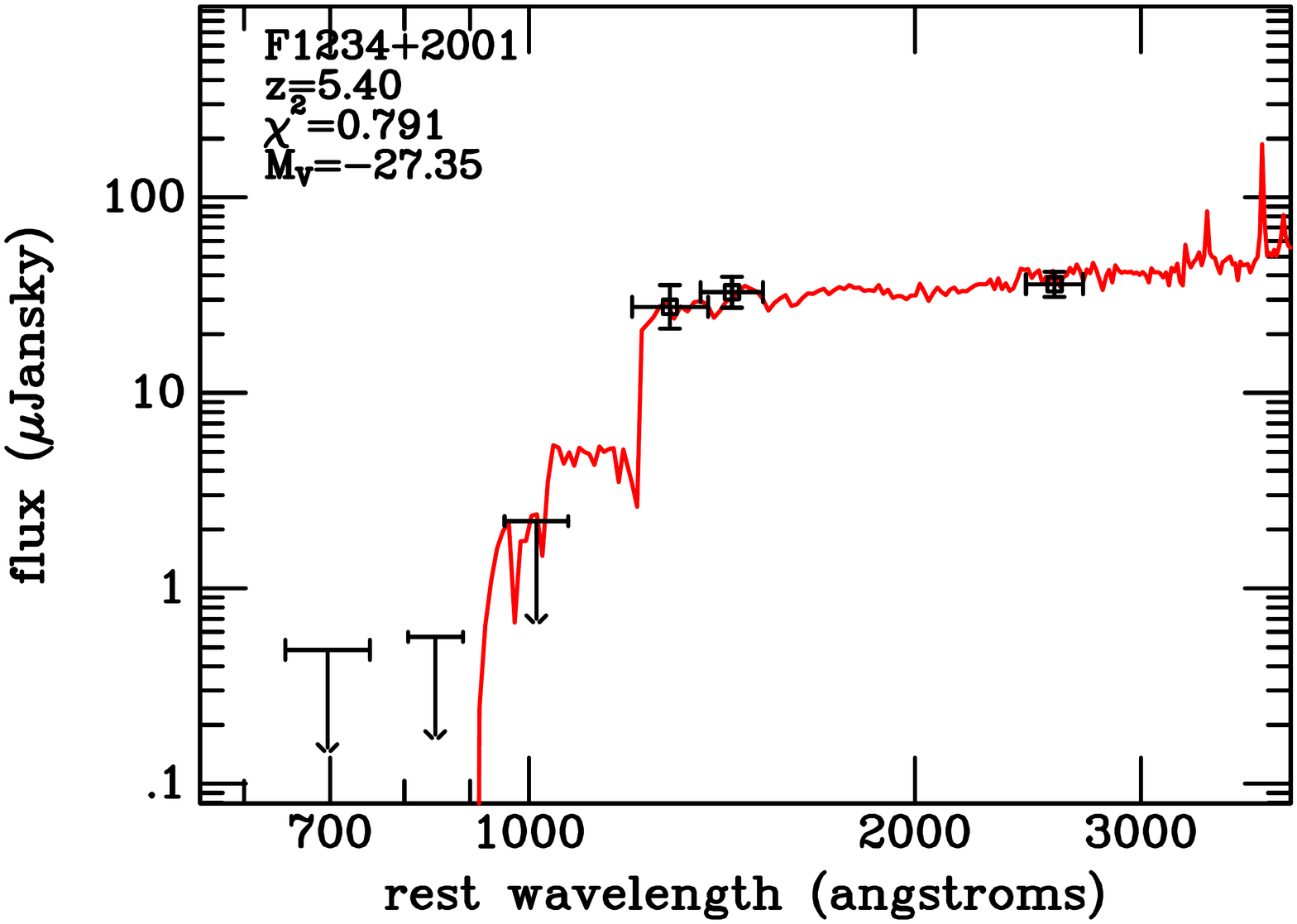}}
\rotatebox{270}{\includegraphics[width=2.0in,height=3.2in]{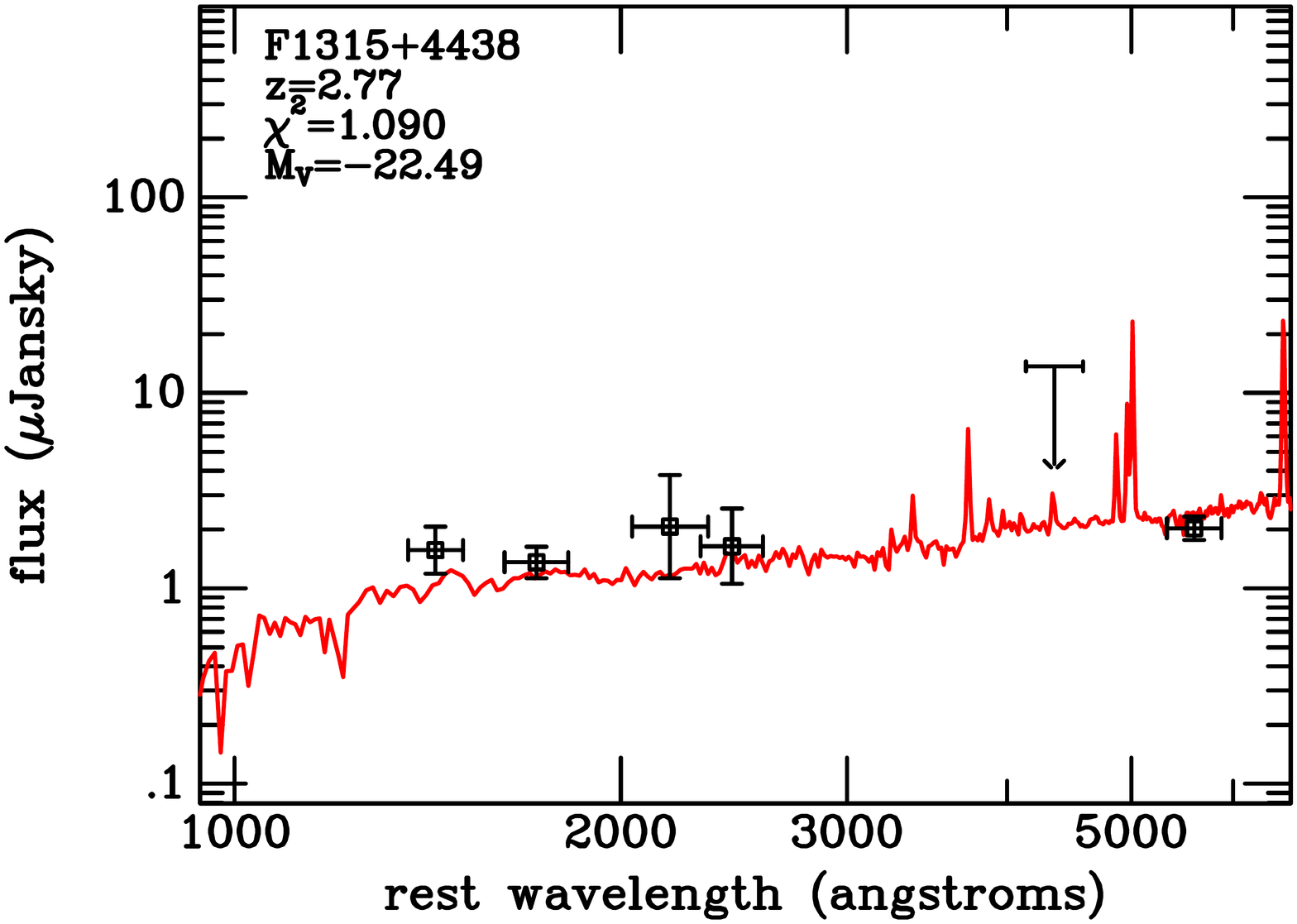}}
\end{center}
\vspace{-1.5in}
\begin{center}
\includegraphics[scale=0.42,angle=0,viewport= 120 50 500 575]{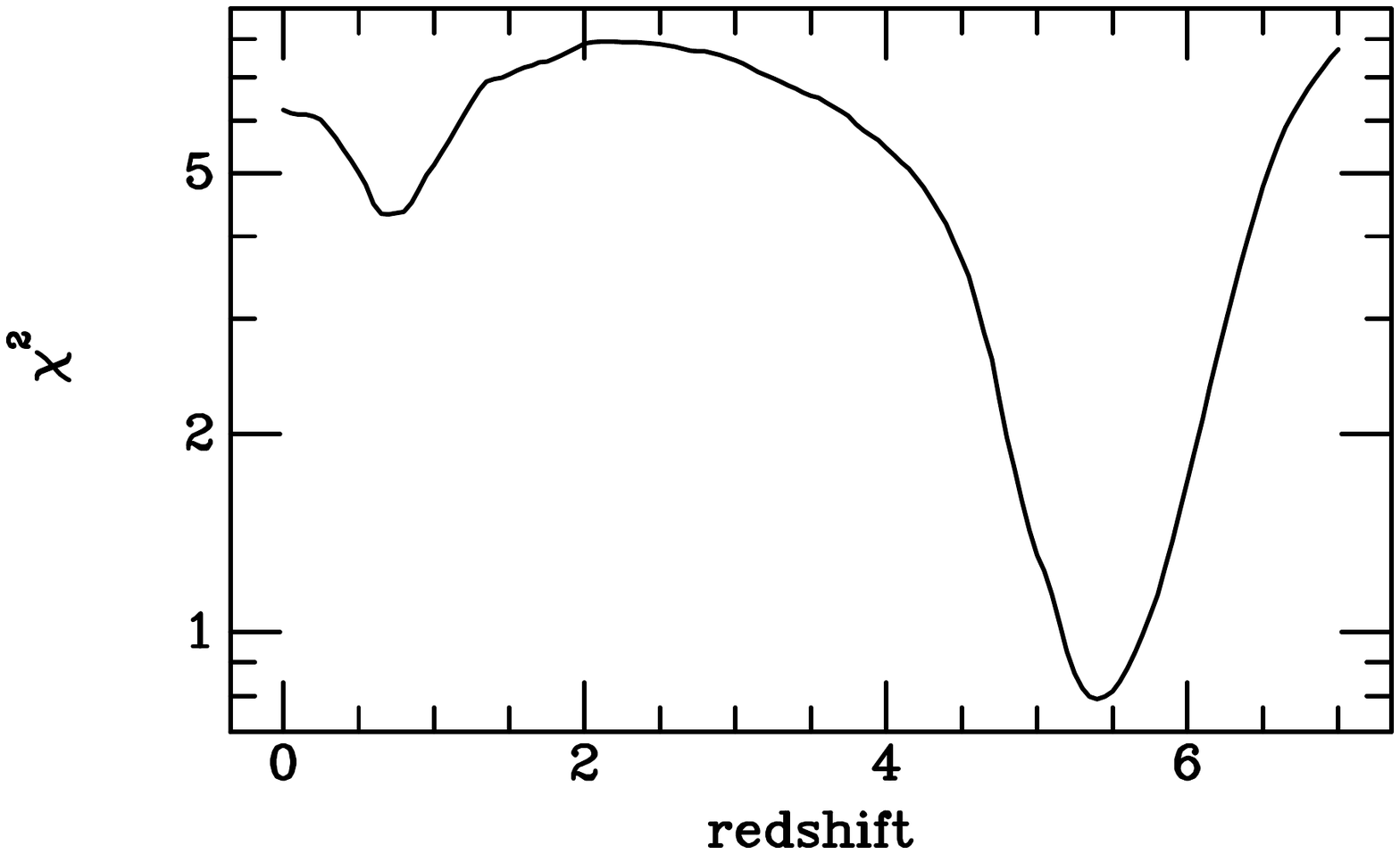}
\includegraphics[scale=0.42,angle=0,viewport= -50 50 500 575]{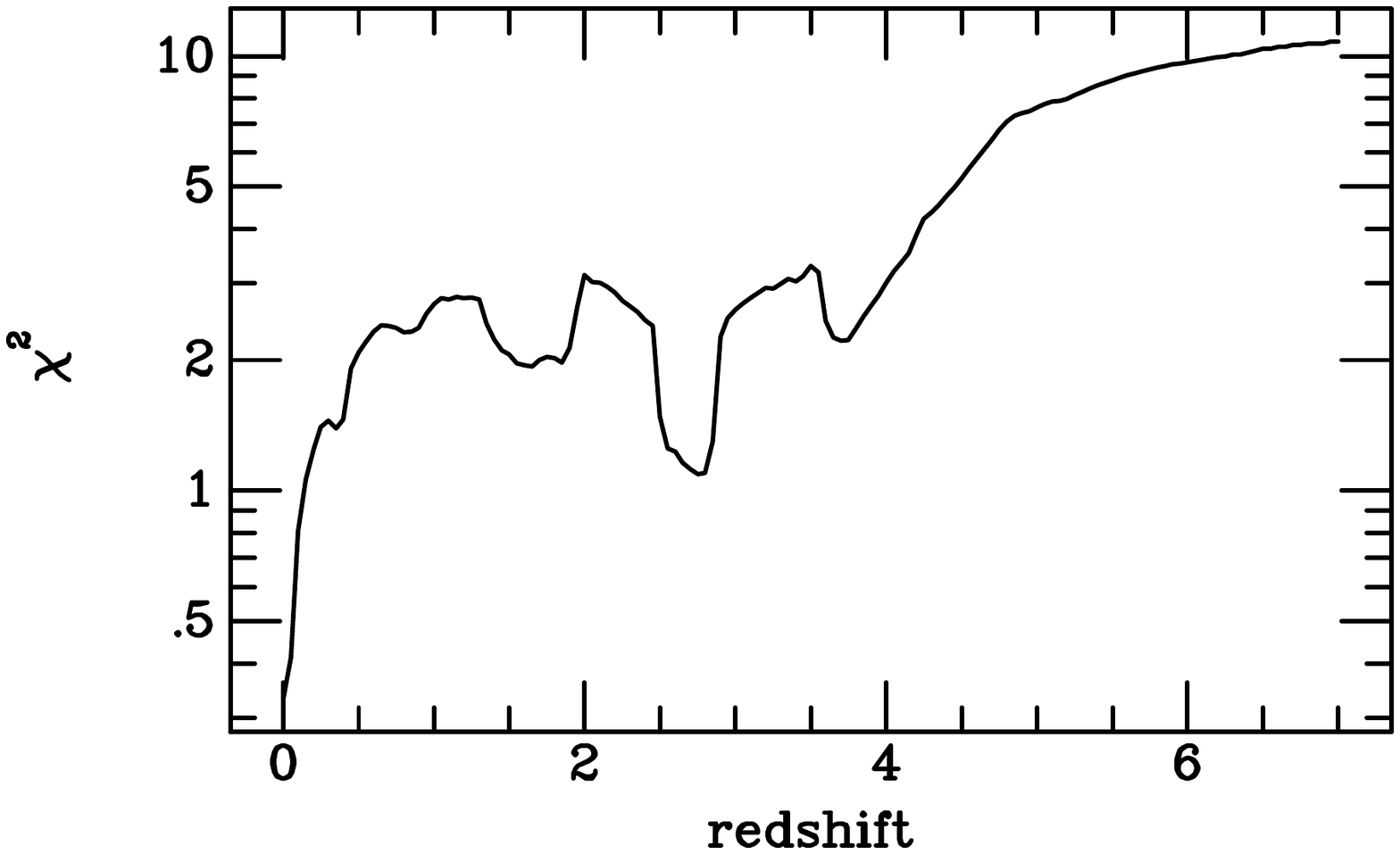}
\end{center}
\vspace{-0.75in}
\begin{center}
\rotatebox{270}{\includegraphics[width=2.0in,height=3.2in]{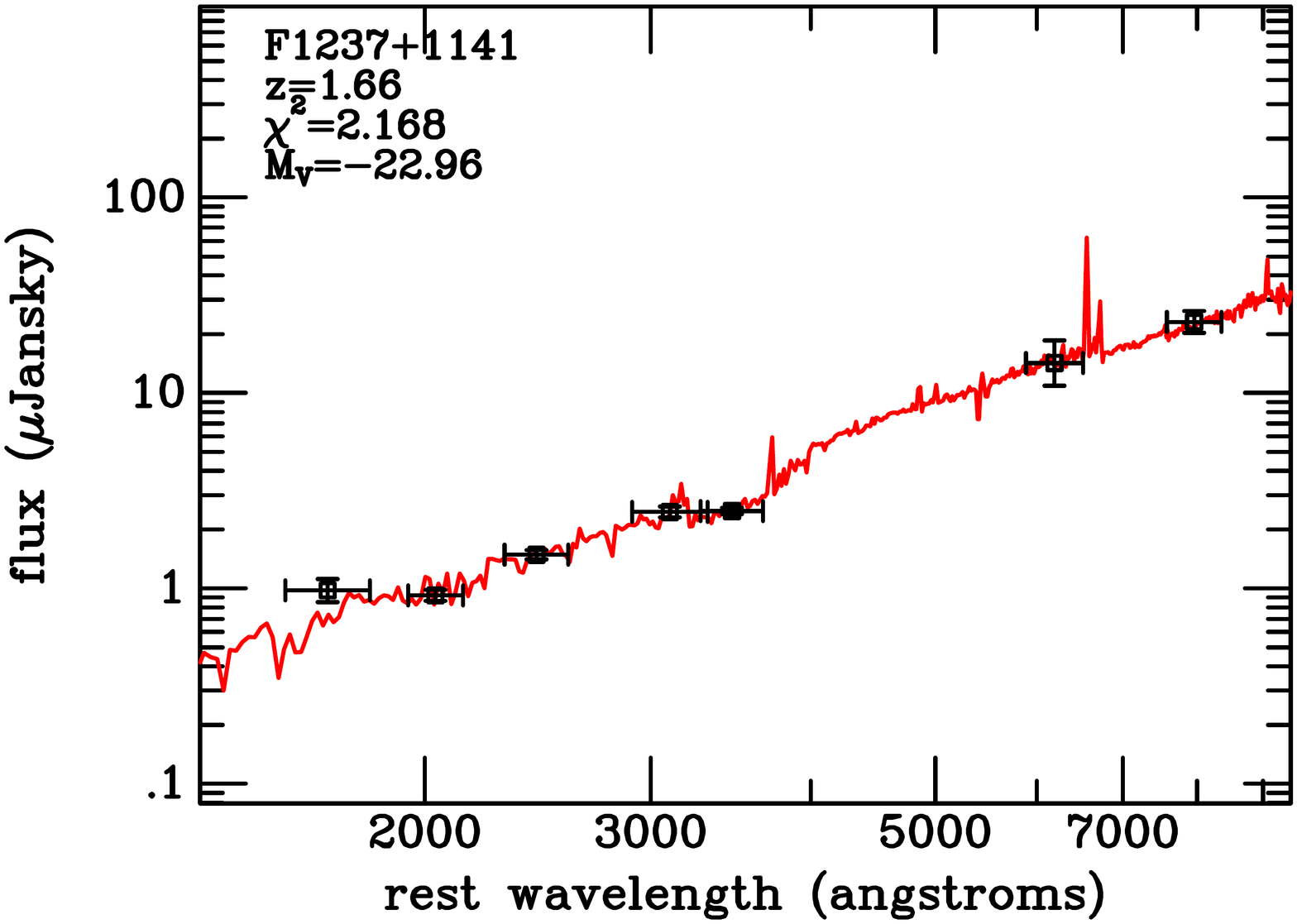}}
\rotatebox{270}{\includegraphics[width=2.0in,height=3.2in]{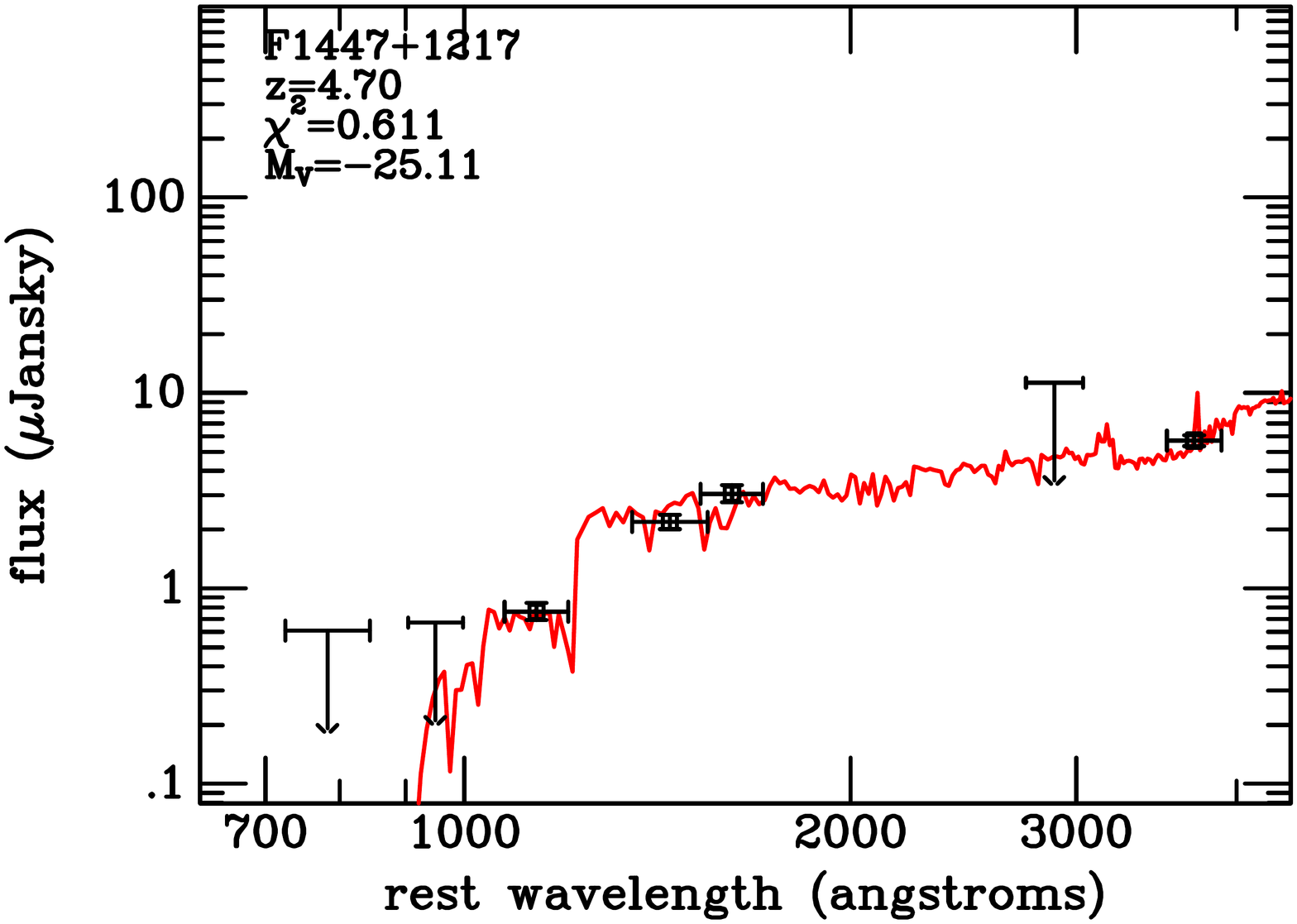}}
\end{center}
\vspace{-1.5in}
\begin{center}
\includegraphics[scale=0.42,angle=0,viewport= 120 50 500 575]{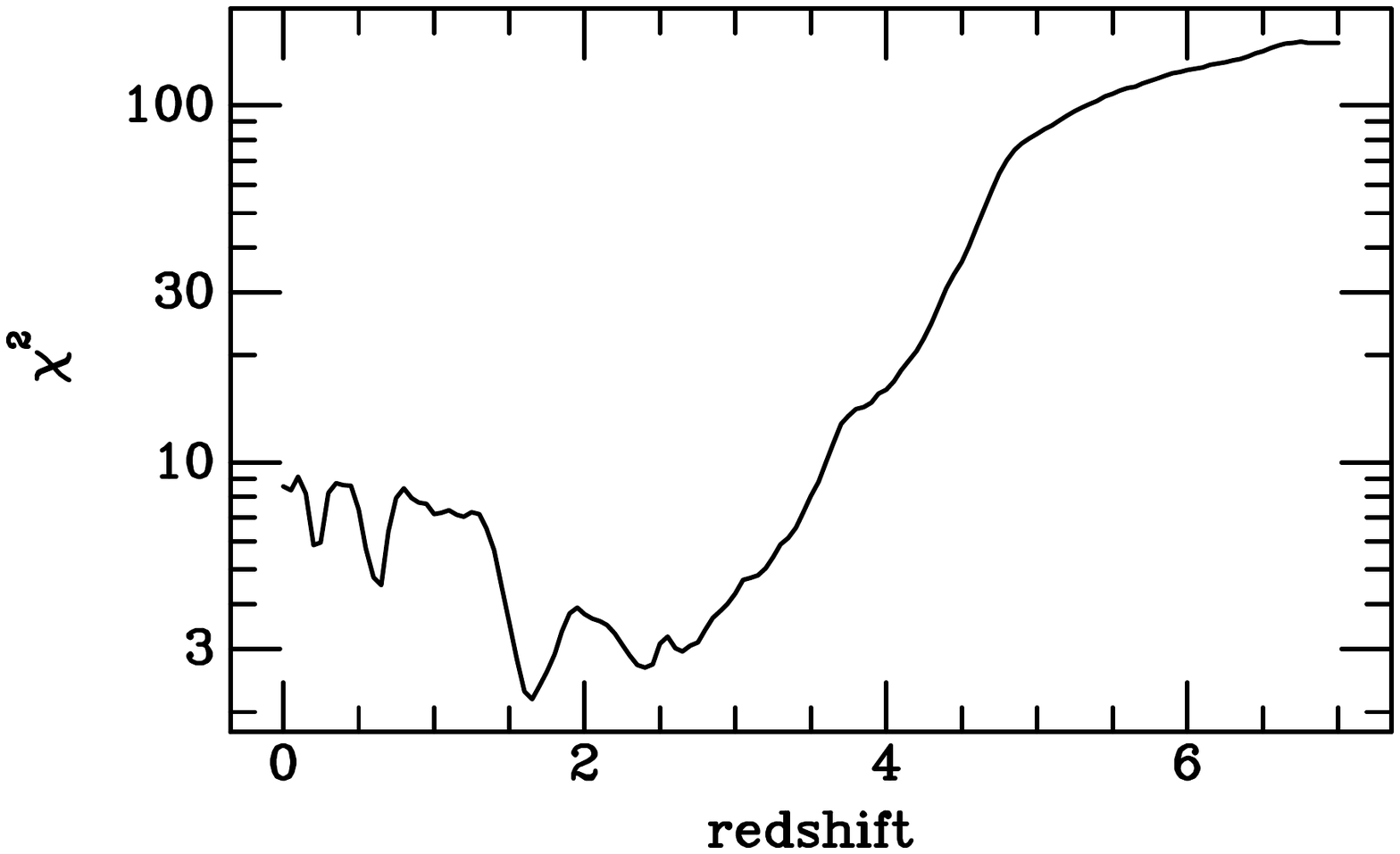}
\includegraphics[scale=0.42,angle=0,viewport= -50 50 500 575]{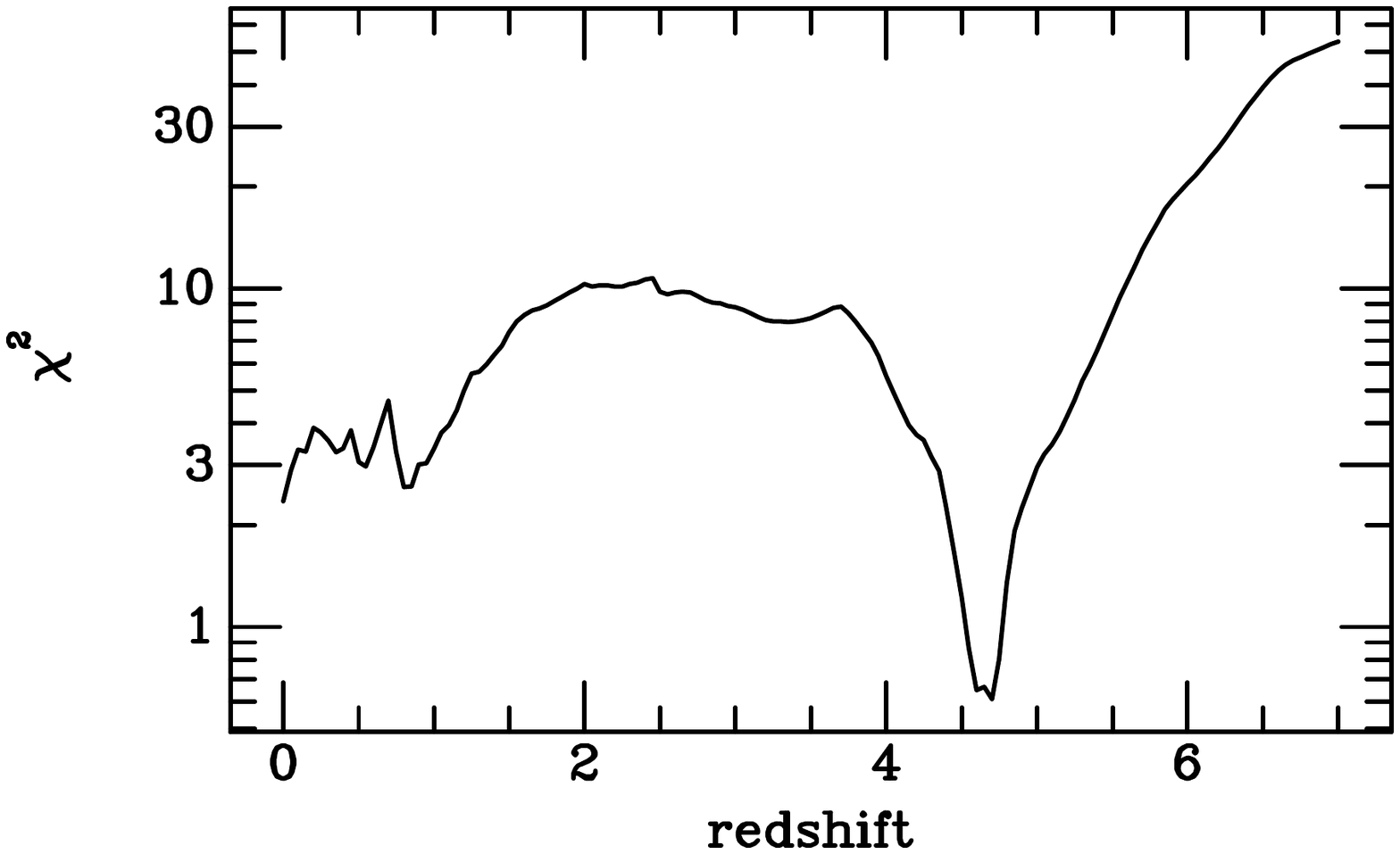}
\end{center}
\vspace{-0.15in}
\caption{\scriptsize
Same as Figure 9 for 4 additional FIRST-BNGS sources.
}
\end{figure*}

\begin{figure*}[htp]
\vspace{+0.4in}
\begin{center}
\rotatebox{270}{\includegraphics[width=2.0in,height=3.2in]{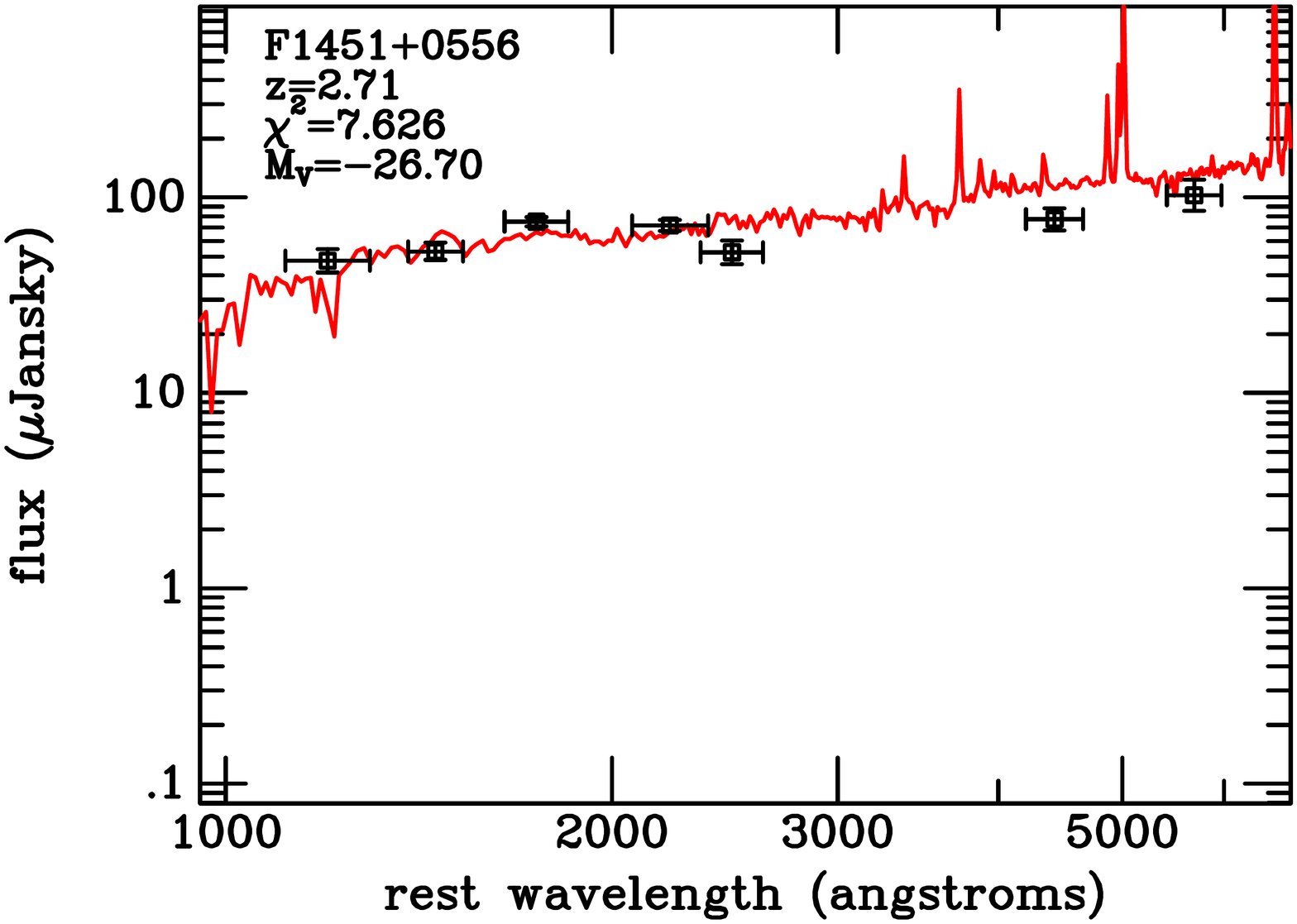}}
\rotatebox{270}{\includegraphics[width=2.0in,height=3.2in]{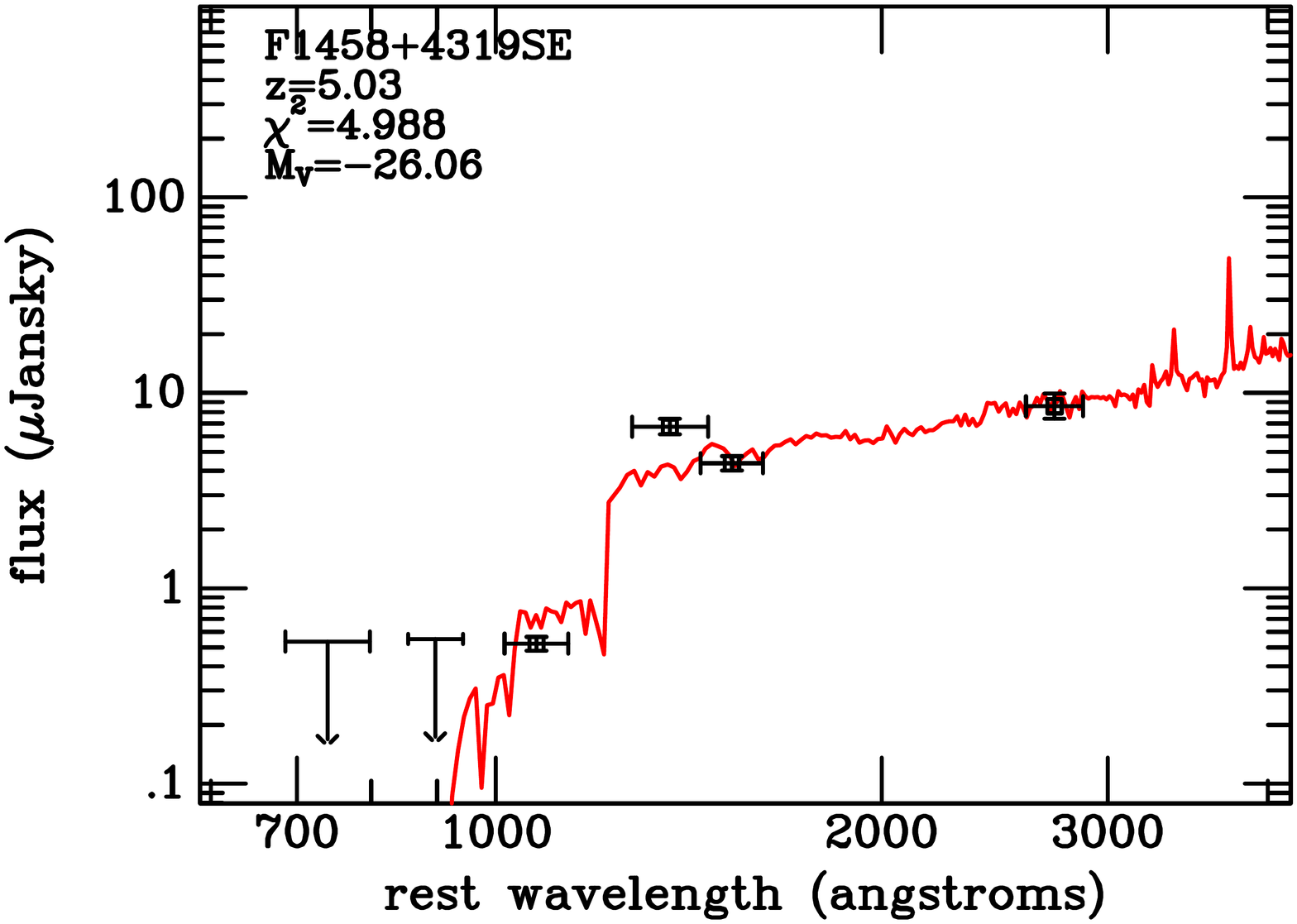}}
\end{center}
\vspace{-1.5in}
\begin{center}
\includegraphics[scale=0.42,angle=0,viewport= 120 50 500 575]{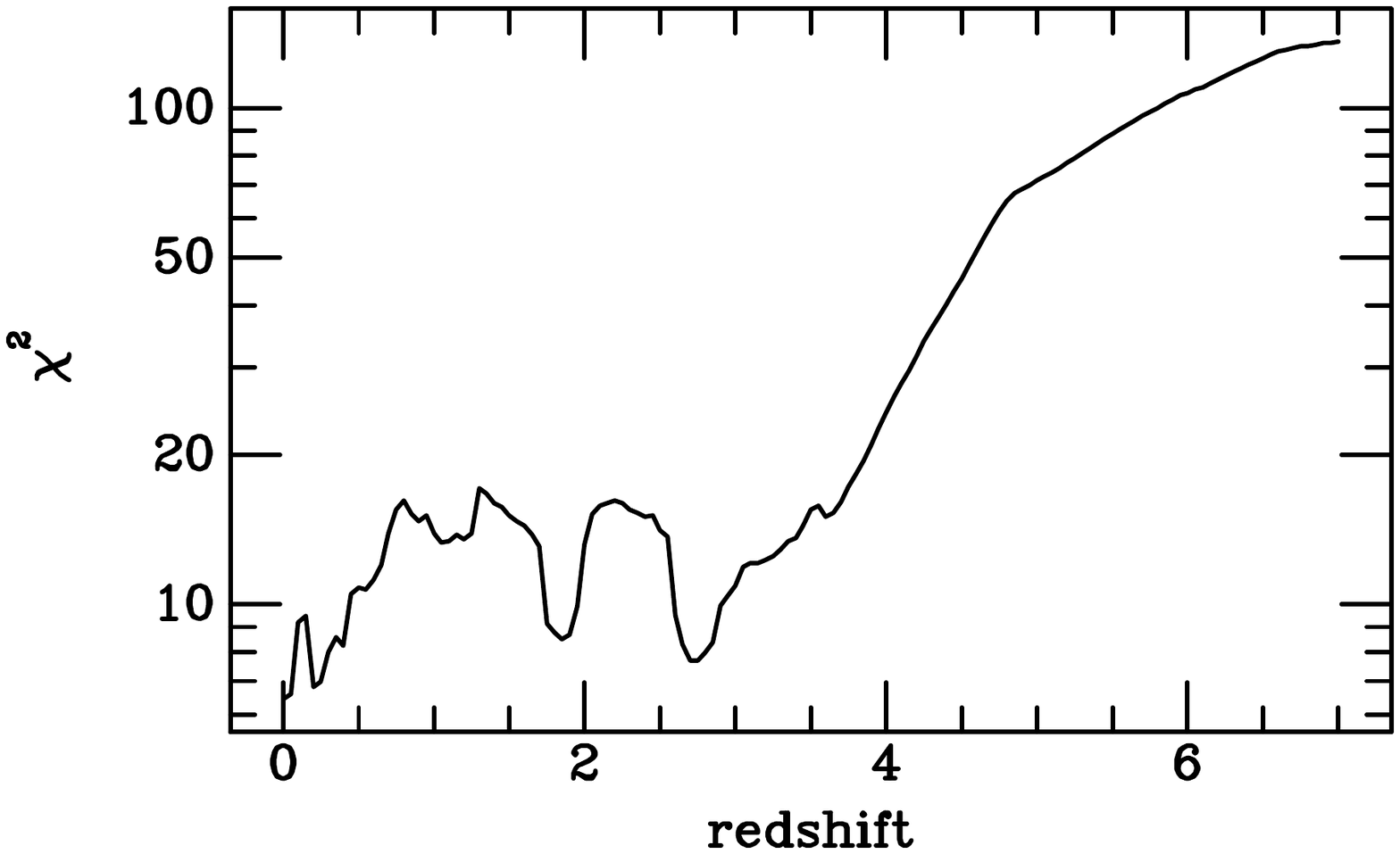}
\includegraphics[scale=0.42,angle=0,viewport= -50 50 500 575]{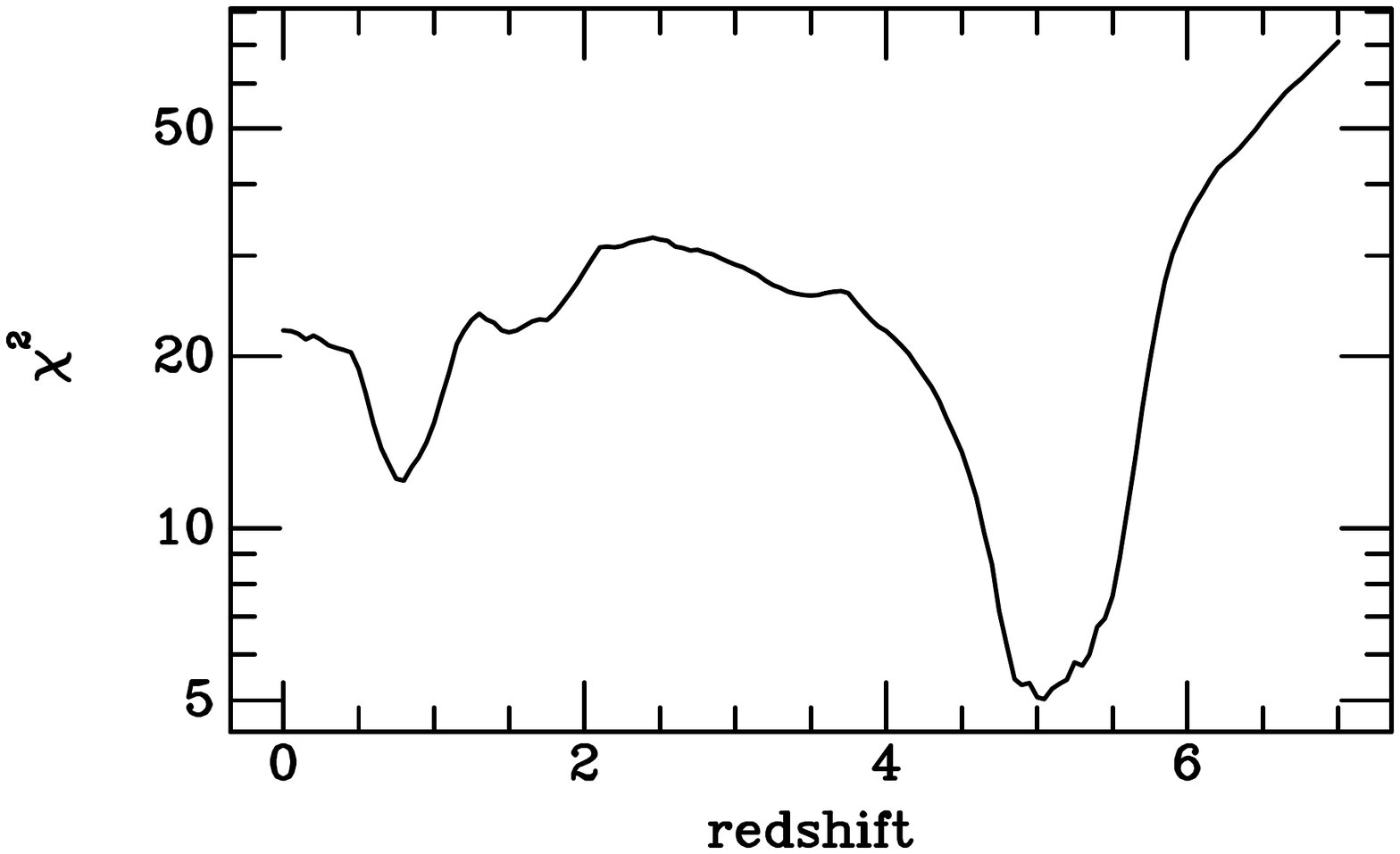}
\end{center}
\vspace{-0.75in}
\begin{center}
\rotatebox{270}{\includegraphics[width=2.0in,height=3.2in]{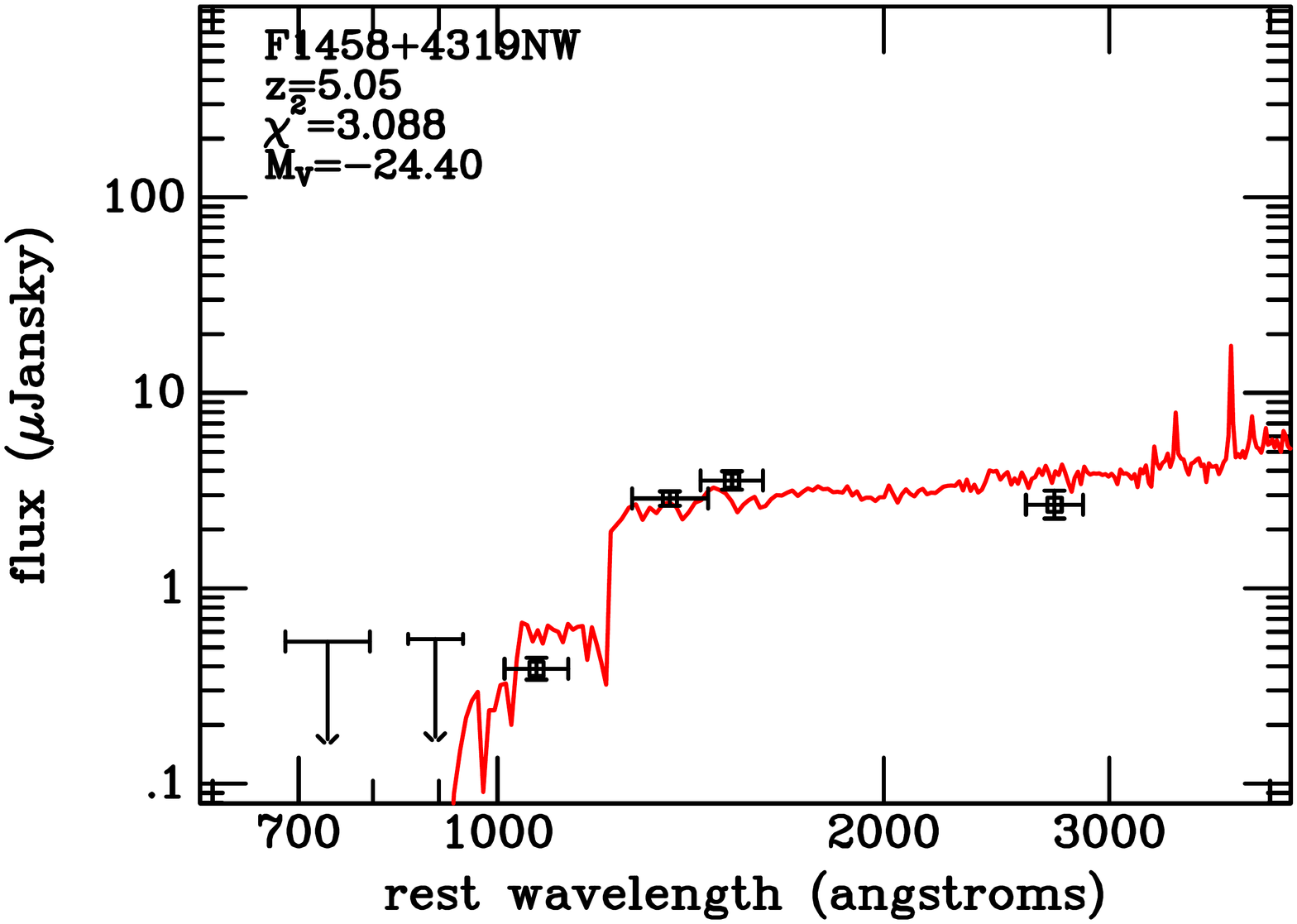}}
\rotatebox{270}{\includegraphics[width=2.0in,height=3.2in]{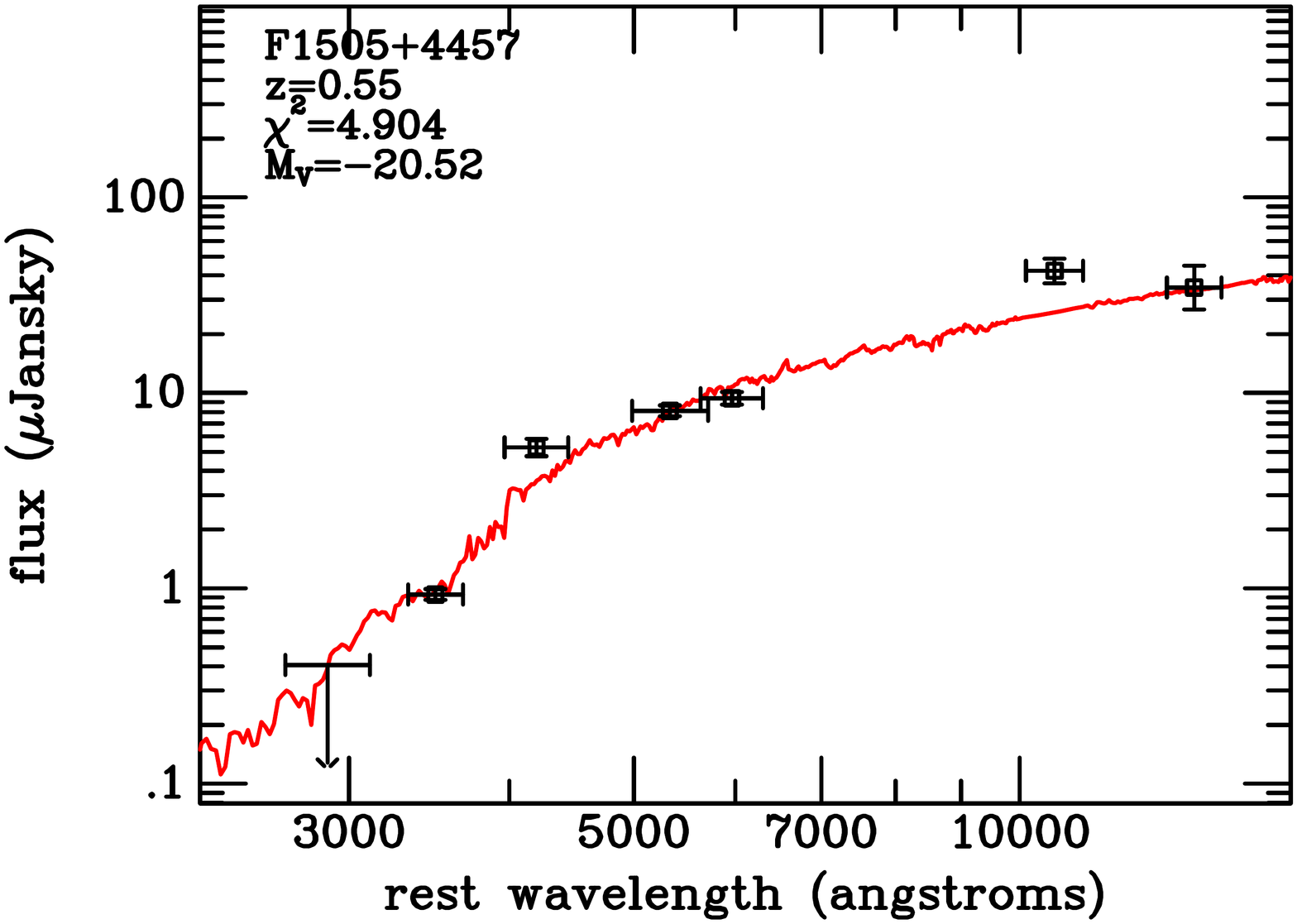}}
\end{center}
\vspace{-1.5in}
\begin{center}
\includegraphics[scale=0.42,angle=0,viewport= 120 50 500 575]{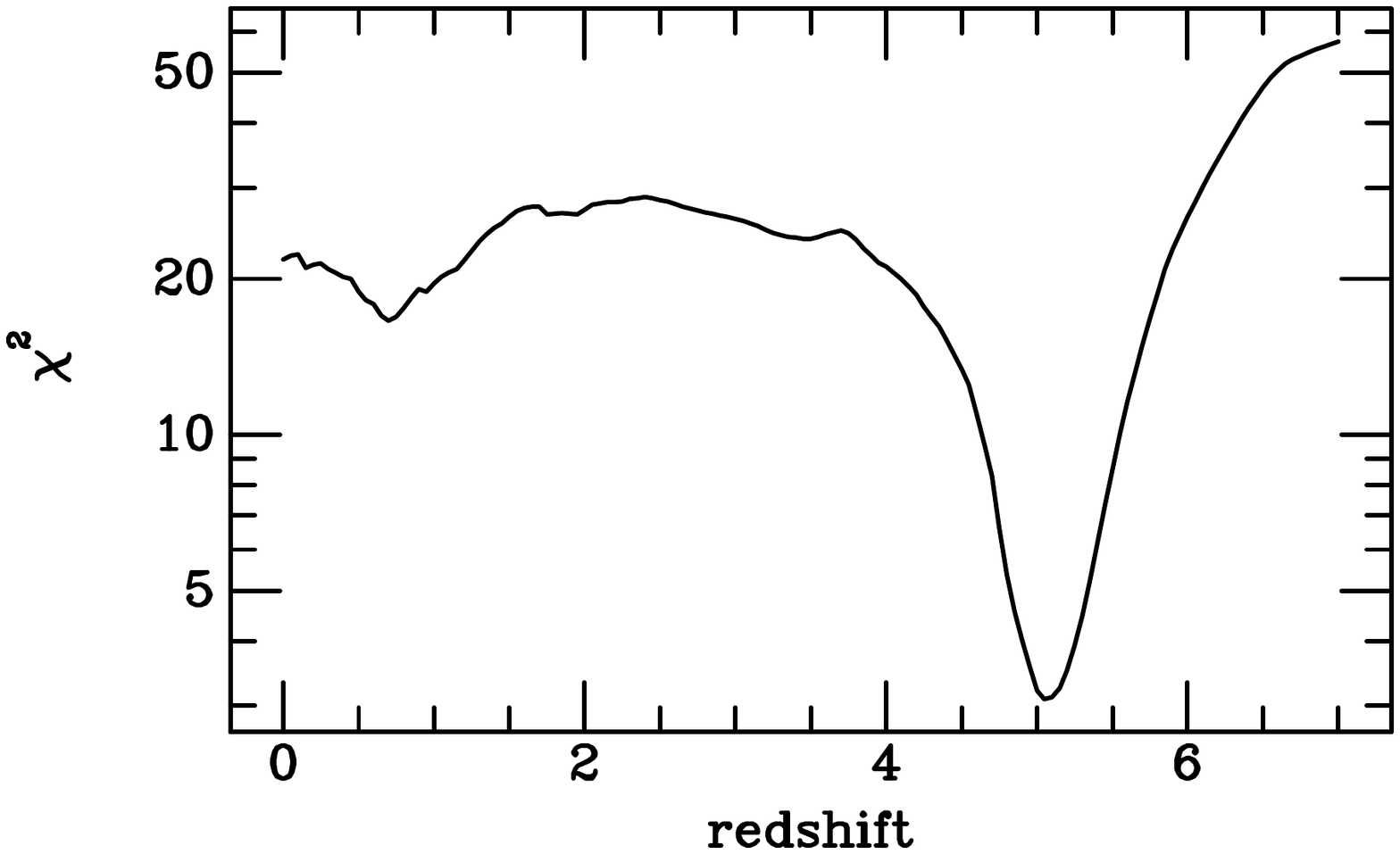}
\includegraphics[scale=0.42,angle=0,viewport= -50 50 500 575]{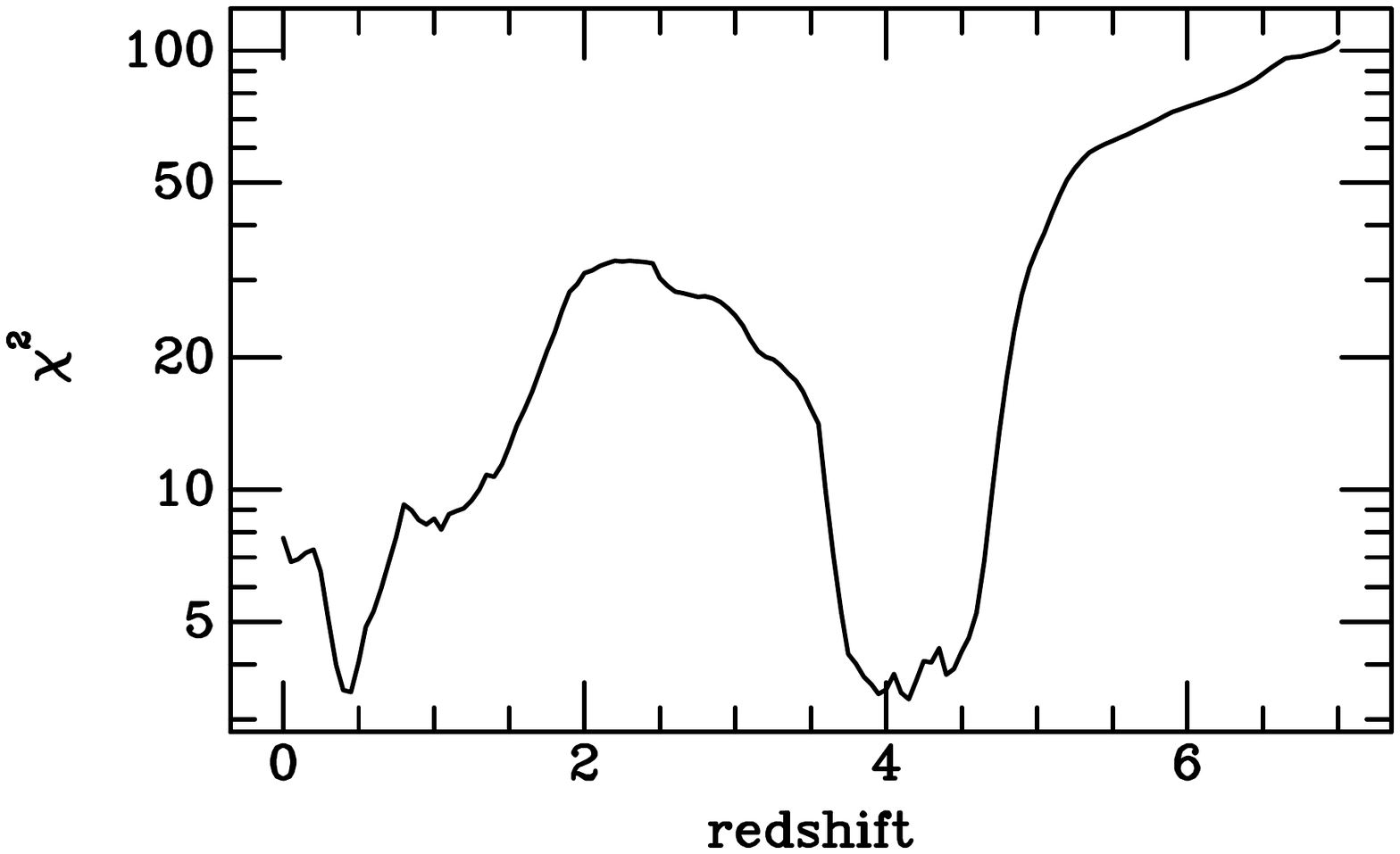}
\end{center}
\vspace{-0.15in}
\caption{\scriptsize
Same as Figure 9 for 4 additional FIRST-BNGS sources.
}
\end{figure*}

\begin{figure*}[htp]
\vspace{+0.4in}
\begin{center}
\rotatebox{270}{\includegraphics[width=2.0in,height=3.2in]{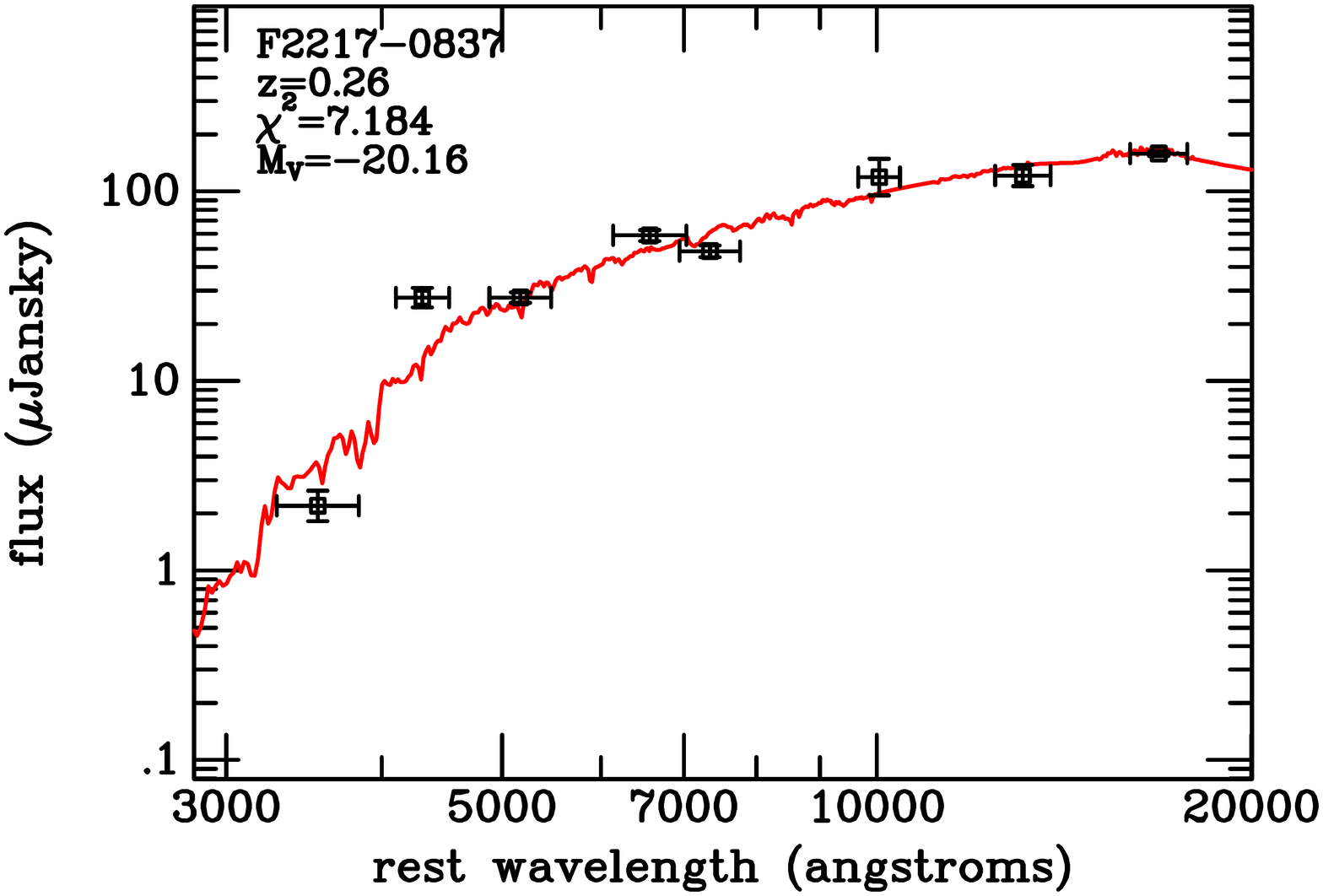}}
\rotatebox{270}{\includegraphics[width=2.0in,height=3.2in]{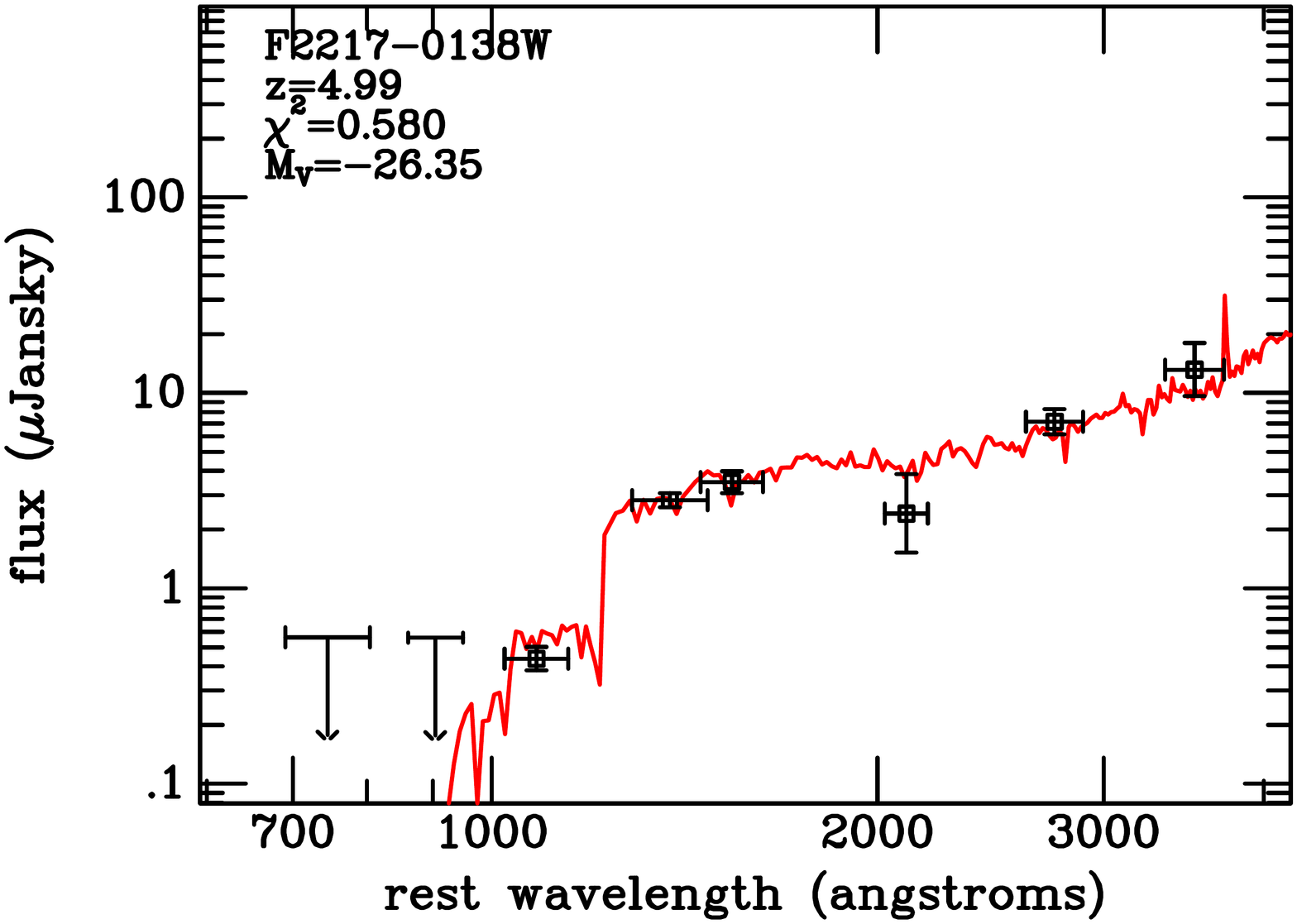}}
\end{center}
\vspace{-1.5in}
\begin{center}
\includegraphics[scale=0.42,angle=0,viewport= 120 50 500 575]{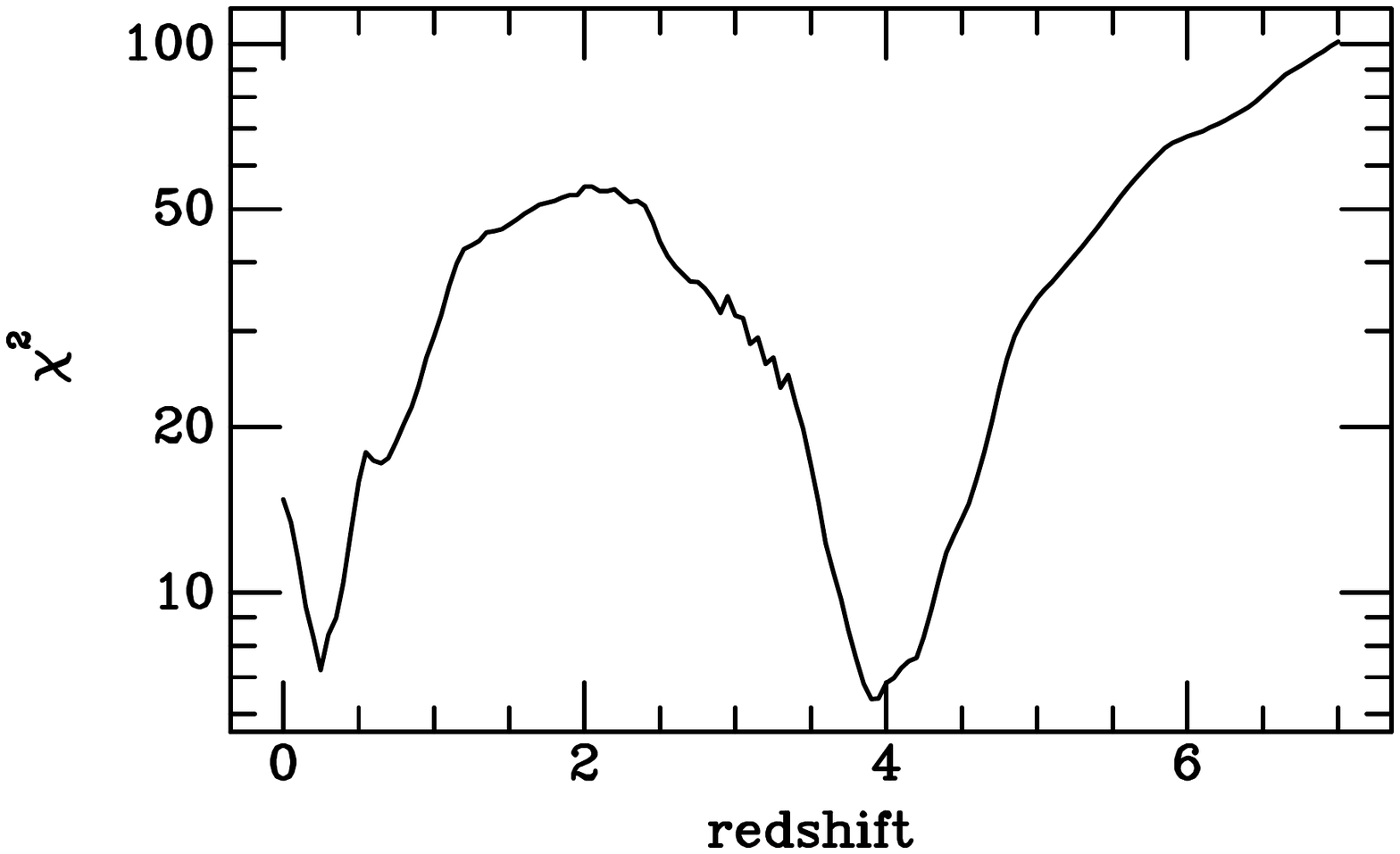}
\includegraphics[scale=0.42,angle=0,viewport= -50 50 500 575]{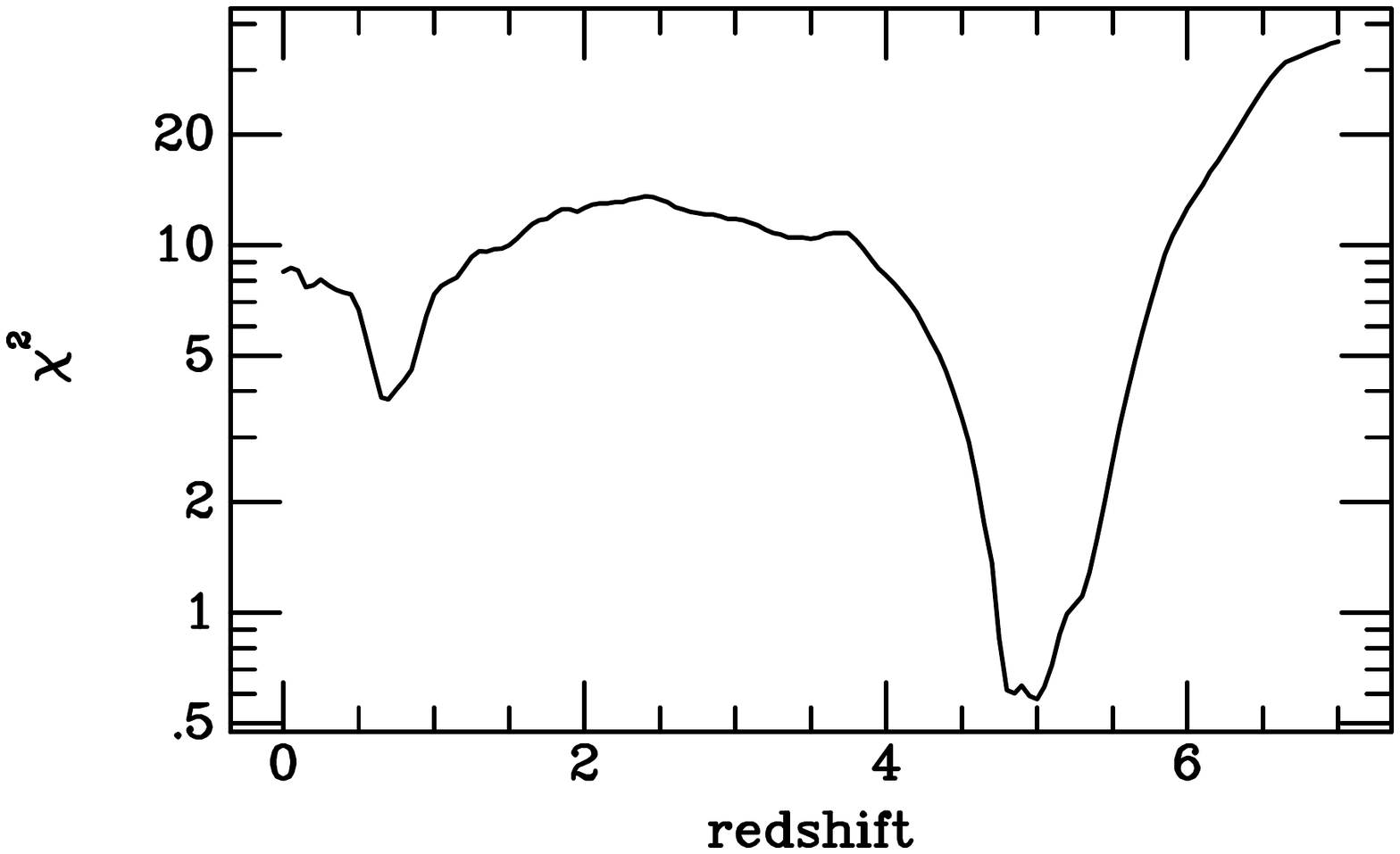}
\end{center}
\vspace{-0.75in}
\begin{center}
\rotatebox{270}{\includegraphics[width=2.0in,height=3.2in]{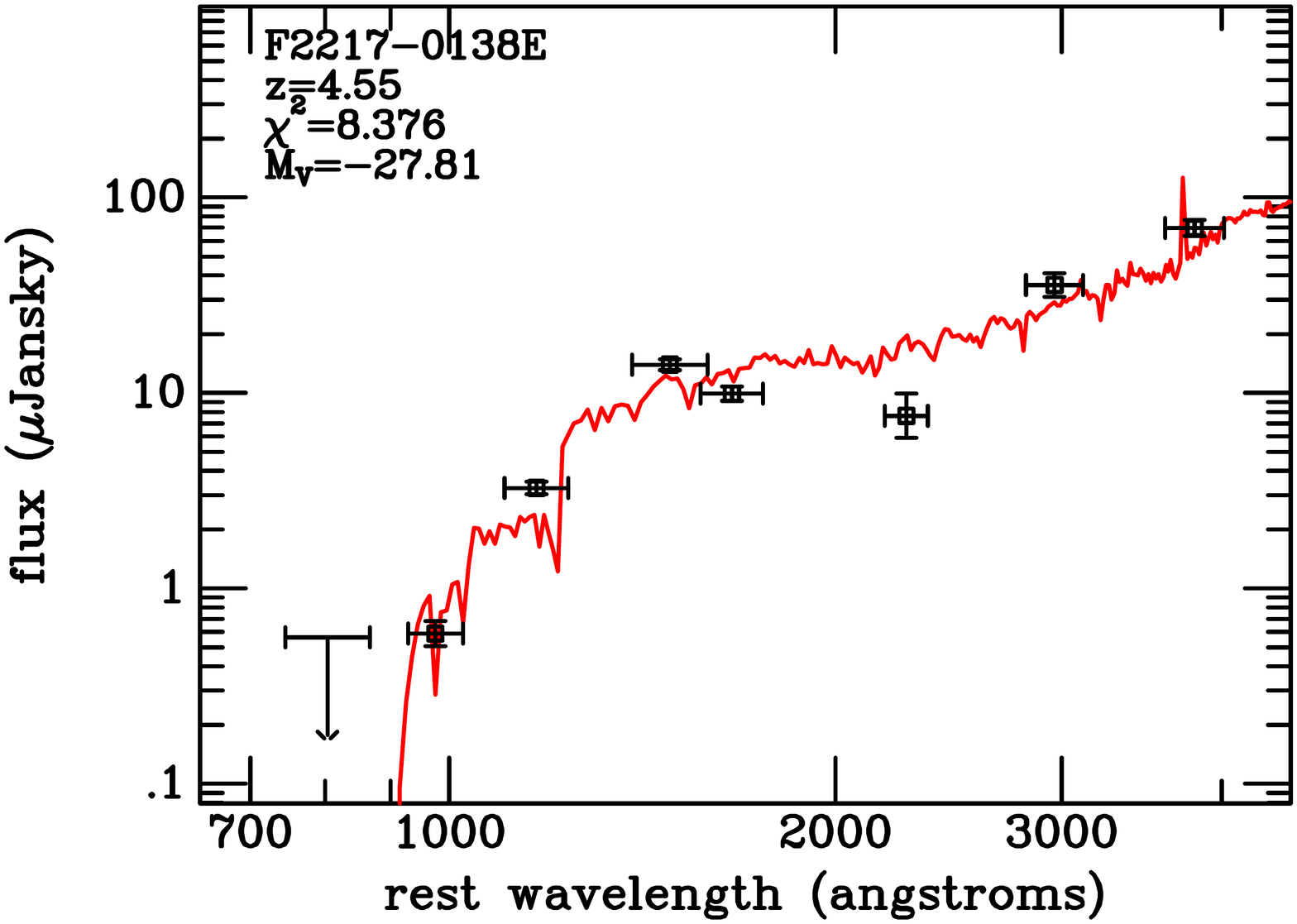}}
\end{center}
\vspace{-1.5in}
\begin{center}
\includegraphics[scale=0.42,angle=0,viewport= 120 50 500 575]{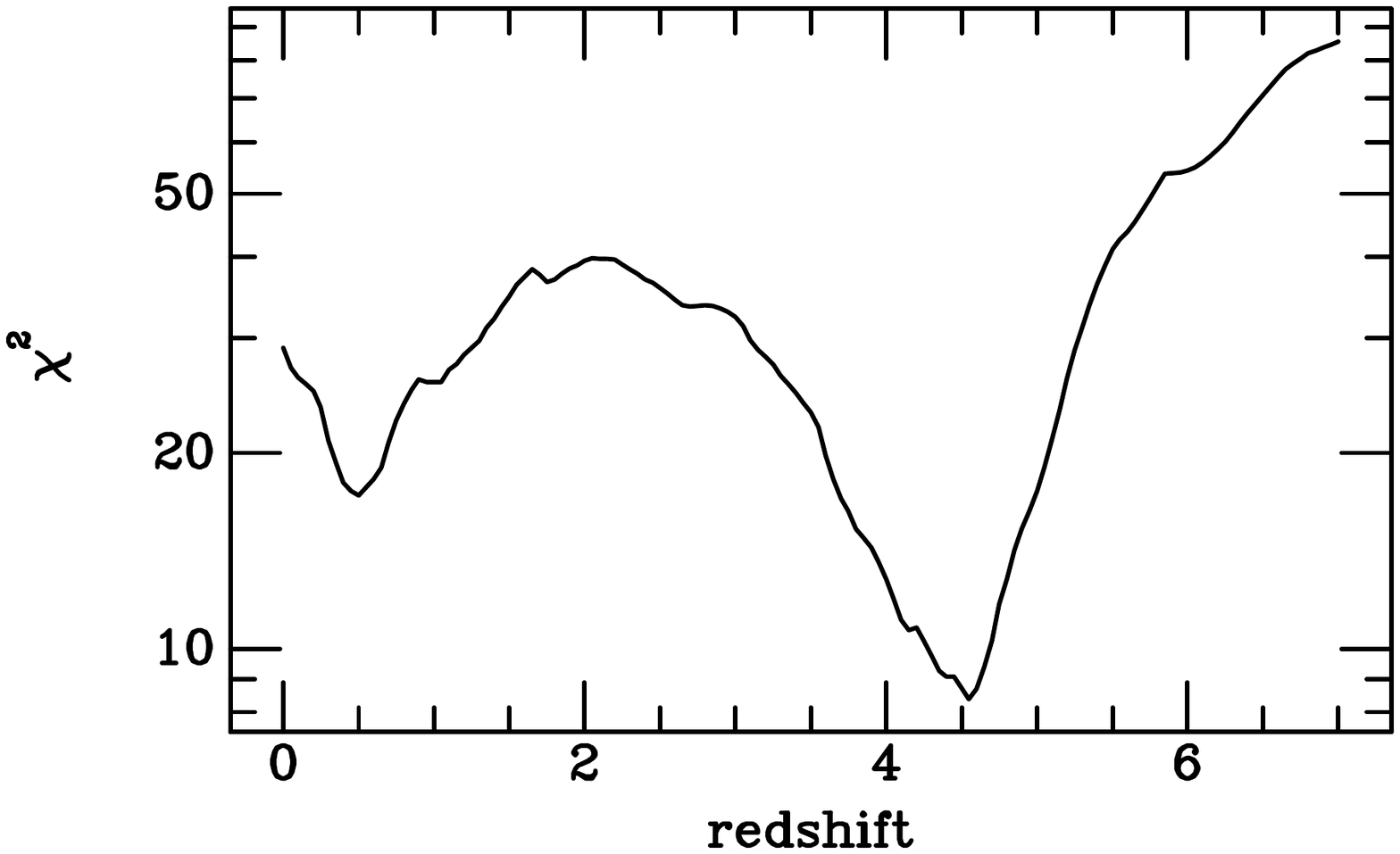}
\end{center}
\vspace{-0.15in}
\caption{\scriptsize
Same as Figure 9 for 3 additional FIRST-BNGS sources.
}
\end{figure*}

\subsection{FIR-Submillimeter}

The photometric measurements from the long-wavelength data were added
to the previous optical and NIR photometry.  The SED templates used
previously cover the rest-frame wavelengths from 100\AA\  to 3$\mu$m,
so were further extended to the submillimeter with 4 long-wavelength
templates (see Appendix A for a more detailed description of the
procedure).  The templates used were Arp 220
\citep{bressan2002} representing a ULIRG SED, M82 \citep{bressan2002}
representing a starburst SED, and 2 templates in the synthetic library
from \citet{dale2001} representing LIRGs ($\alpha$=1.06) and quiescent
($\alpha$=2.5) SEDs.  The best fit SEDs with each long-wavelength SED
template is shown in Figures 18-21.   

For the four objects with FIR flux measurements, all seem to exclude an Arp-220 (ULIRG).
Also the three at high redshift ($z>1$) seem to exclude a LIRG-type SED.  The only 
constrained model (from multiple Spitzer detections) at z$=$2.90 quite definitely shows a 
M-82 (starburst) type SED.

\begin{figure*}[htp]
\vspace{0.5in}
\begin{center}
\includegraphics[scale=0.9,angle=0,viewport= 20 50 600 700]{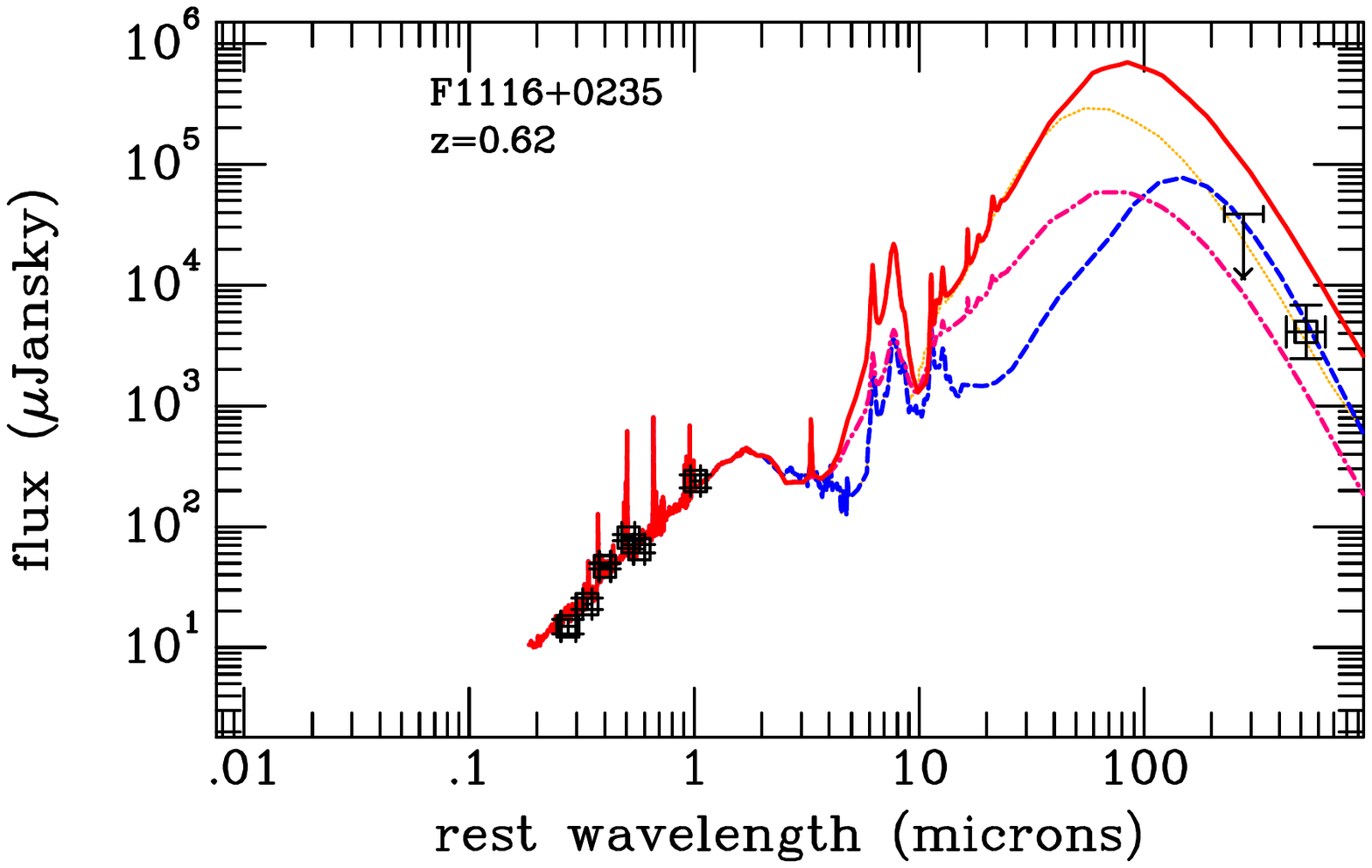}
\end{center}
\vspace{-4.55in}
\caption{\scriptsize
Best-fit SEDs for F1116+0235 with ULIRG (solid), LIRG (dotted),
starburst (dot-dashed), and quiescent (dashed) long wavelength templates. 
}
\end{figure*}

\begin{figure*}[htp]
\begin{center}
\includegraphics[scale=0.3,angle=0,viewport= 0 0 220 200]{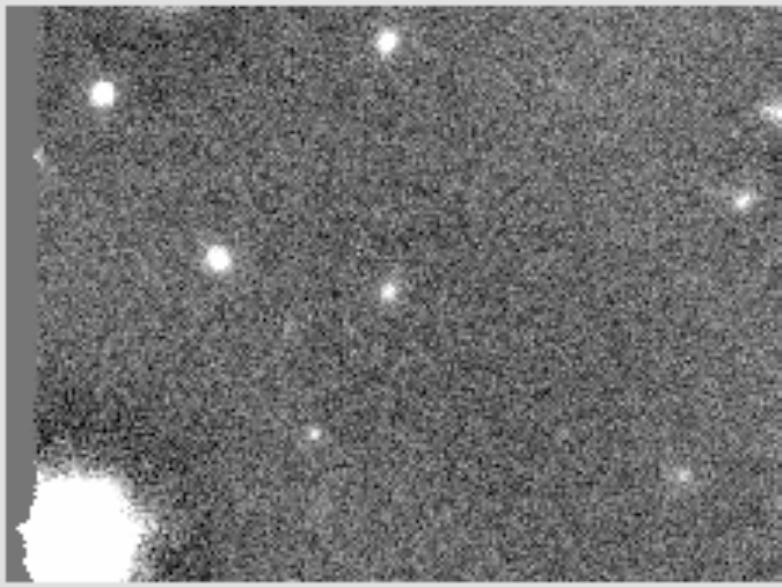}
\includegraphics[scale=0.3,angle=0,viewport= 0 0 220 200]{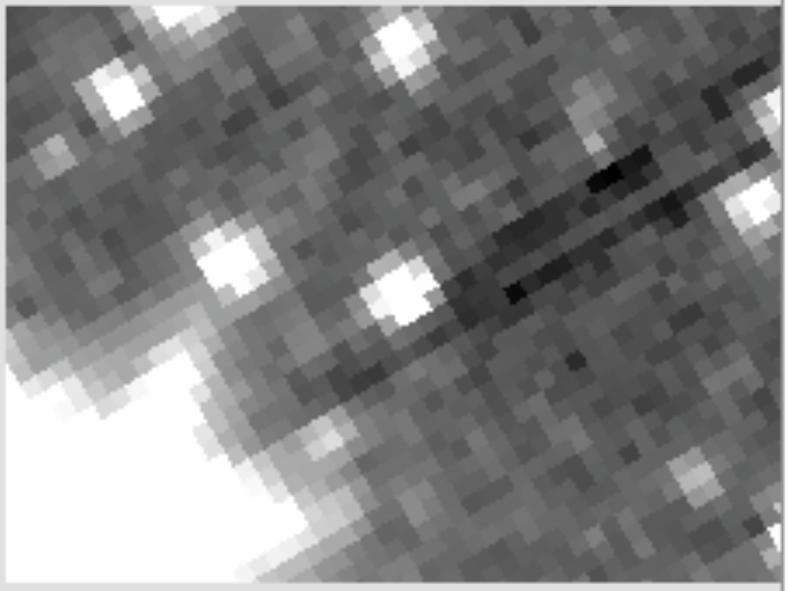}
\includegraphics[scale=0.3,angle=0,viewport= 0 0 220 200]{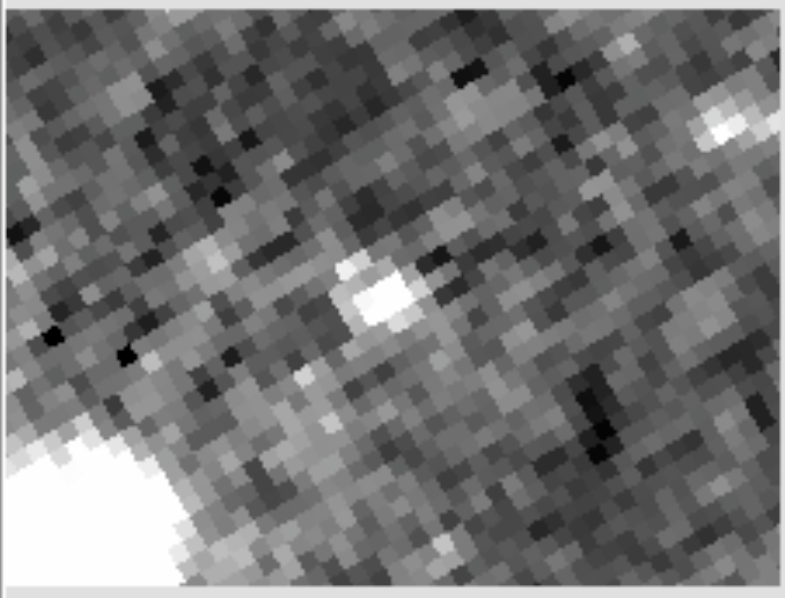}
\includegraphics[scale=0.3,angle=0,viewport= 0 0 220 200]{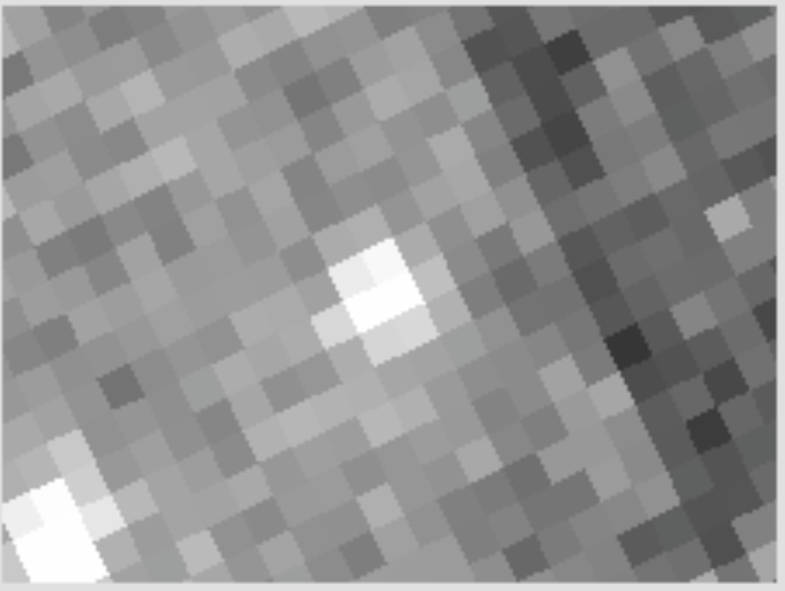}
\includegraphics[scale=0.3,angle=0,viewport= 0 0 220 200]{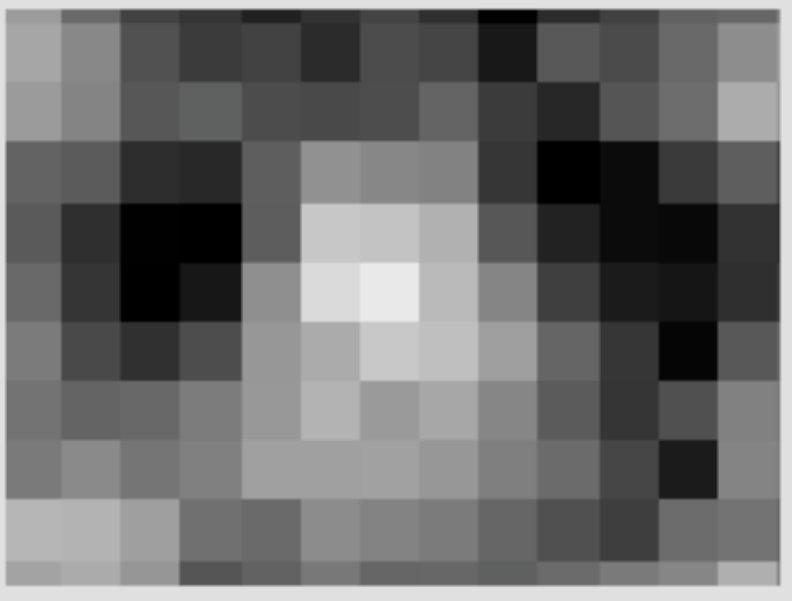}
\includegraphics[scale=0.3,angle=0,viewport= 0 0 220 200]{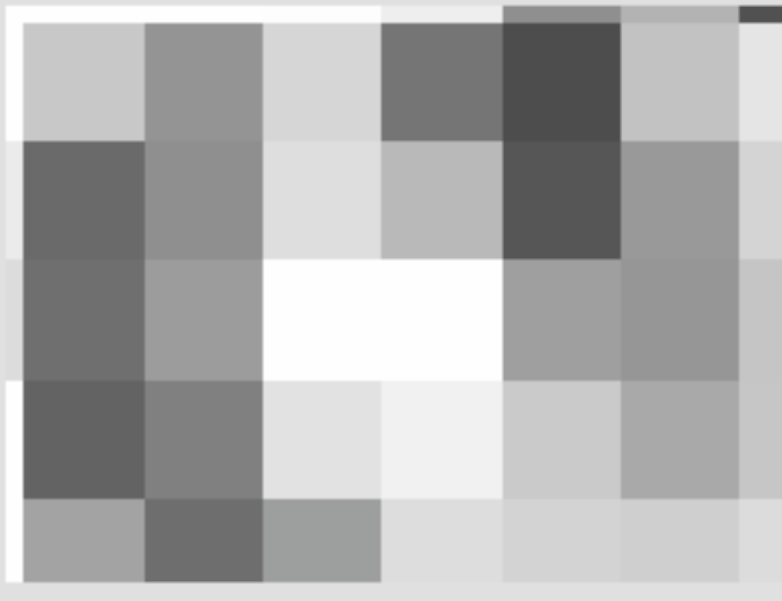}
\end{center}
\vspace{+0.0in}
\begin{center}
\includegraphics[scale=0.9,angle=0,viewport= 20 50 600 700]{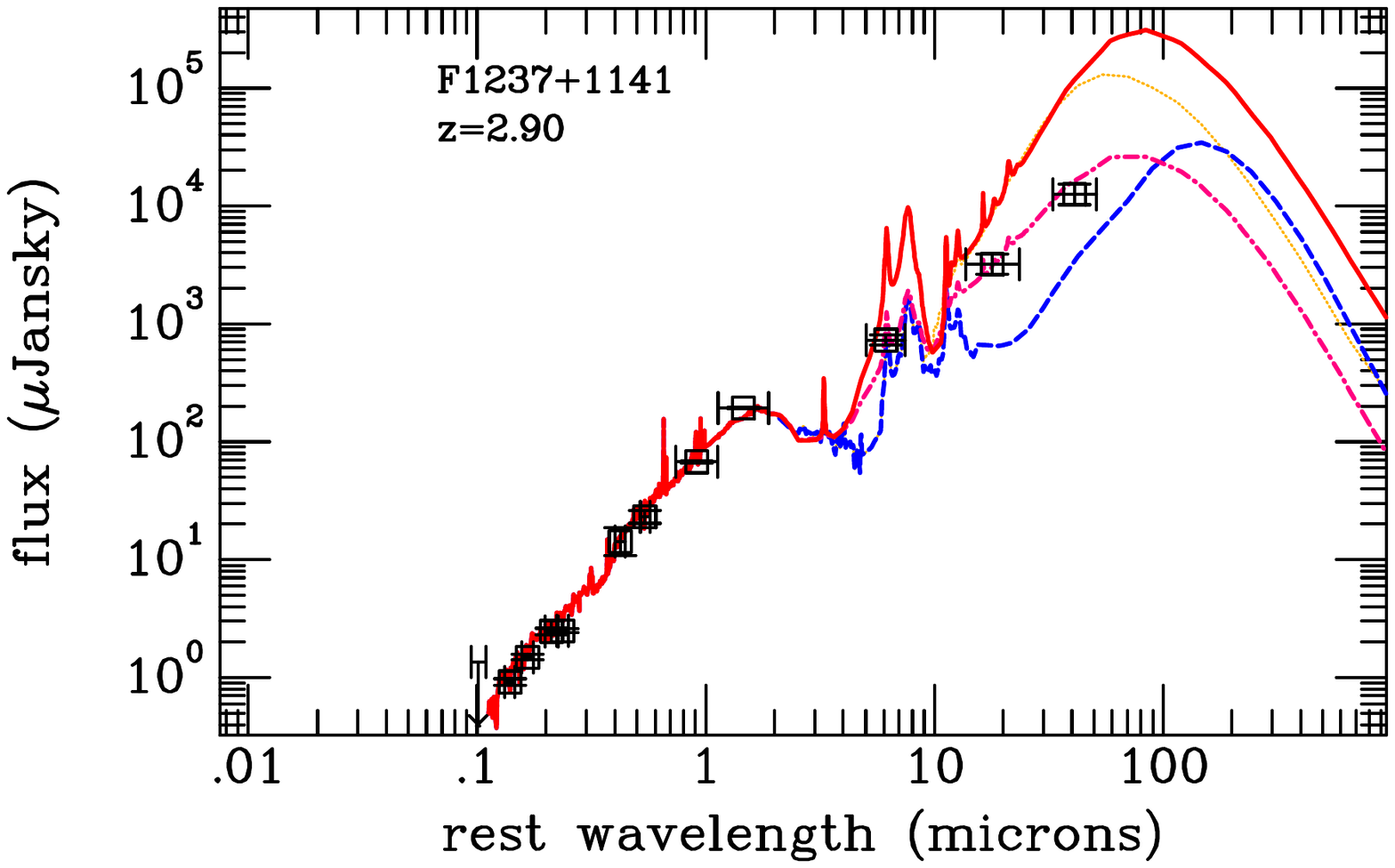}
\end{center}
\vspace{-4.55in}
\caption{\scriptsize
Thumbnail images (approximately 45$''$x35$''$, in K', 3.6$\mu$m,
5.8$\mu$m, 24$\mu$m, 70$\mu$m, 160$\mu$m) and best-fit SEDs for
F1237+1141 with ULIRG (solid), LIRG (dotted),
starburst (dot-dashed), and quiescent (dashed) long wavelength
templates. 
}
\end{figure*}

\begin{figure*}[htp]
\begin{center}
\includegraphics[scale=0.3,angle=0,viewport= 0 0 220 200]{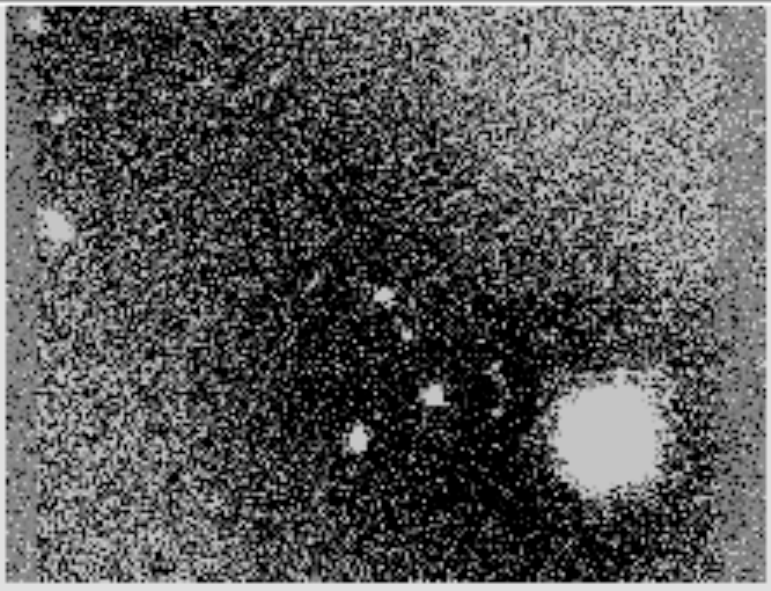}
\includegraphics[scale=0.3,angle=0,viewport= 0 0 220 200]{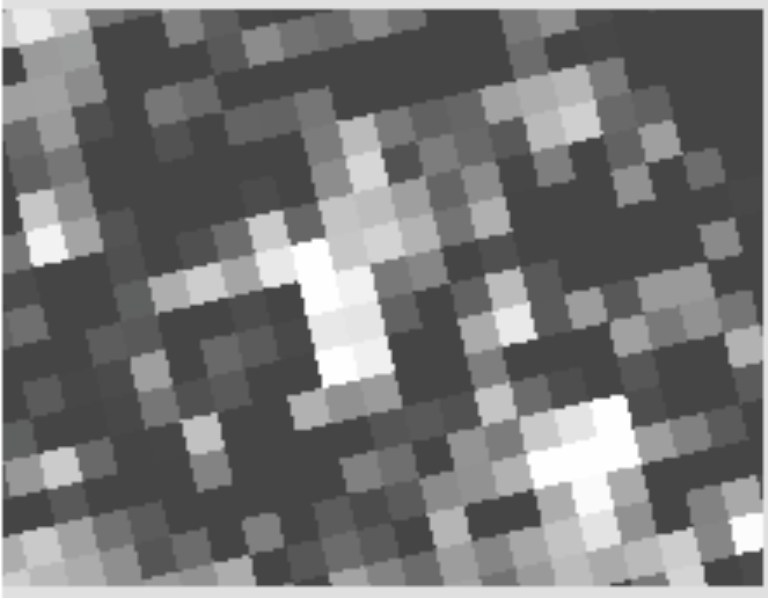}
\includegraphics[scale=0.3,angle=0,viewport= 0 0 220 200]{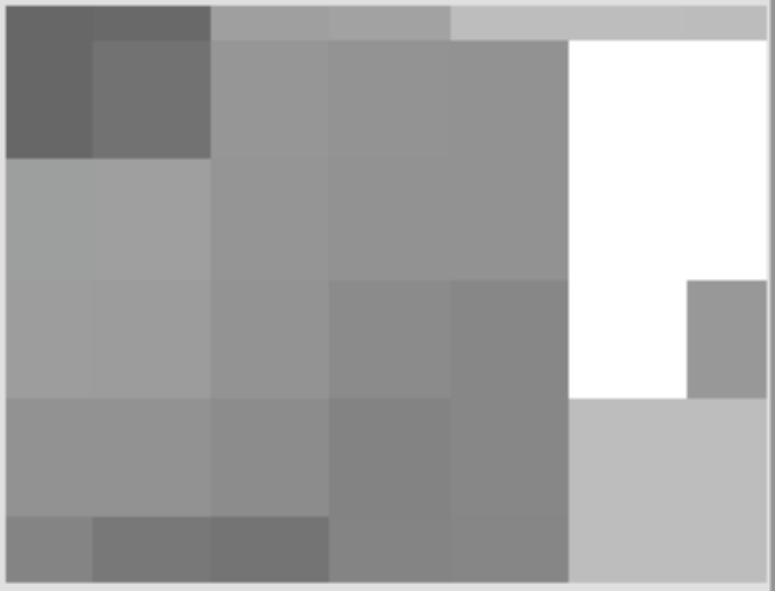}
\end{center}
\vspace{+0.1in}
\begin{center}
\includegraphics[scale=0.9,angle=0,viewport= 20 50 600 700]{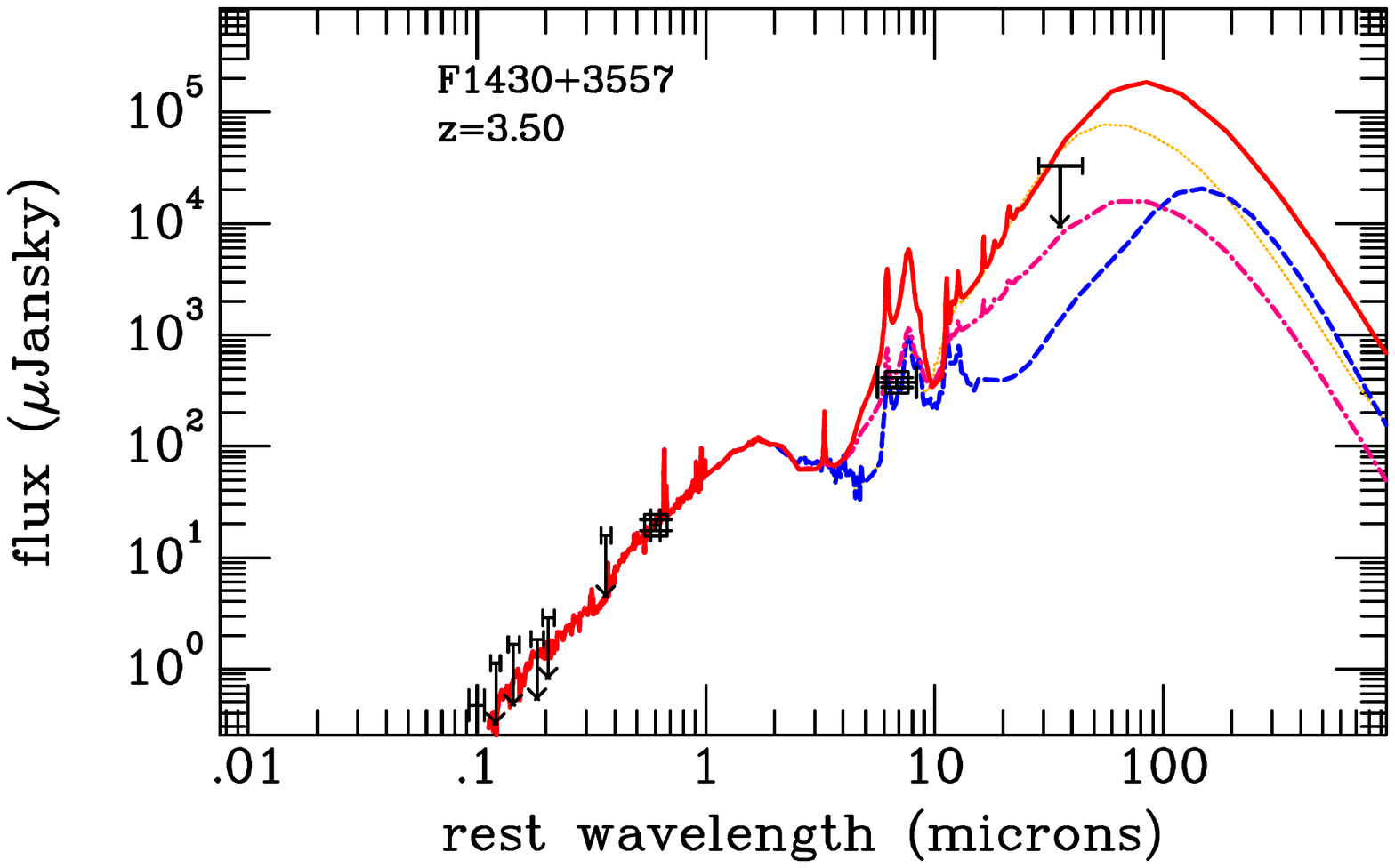}
\end{center}
\vspace{-4.55in}
\caption{\scriptsize
Thumbnail images (approximately 45$''$x35$''$, in K', 24$\mu$m,
160$\mu$m) and best-fit SEDs for F1430+3557 with ULIRG (solid), LIRG (dotted),
starburst (dot-dashed), and quiescent (dashed) long wavelength
templates. 
}
\end{figure*}

\begin{figure*}[htp]
\begin{center}
\includegraphics[scale=0.9,angle=0,viewport= 20 50 600 700]{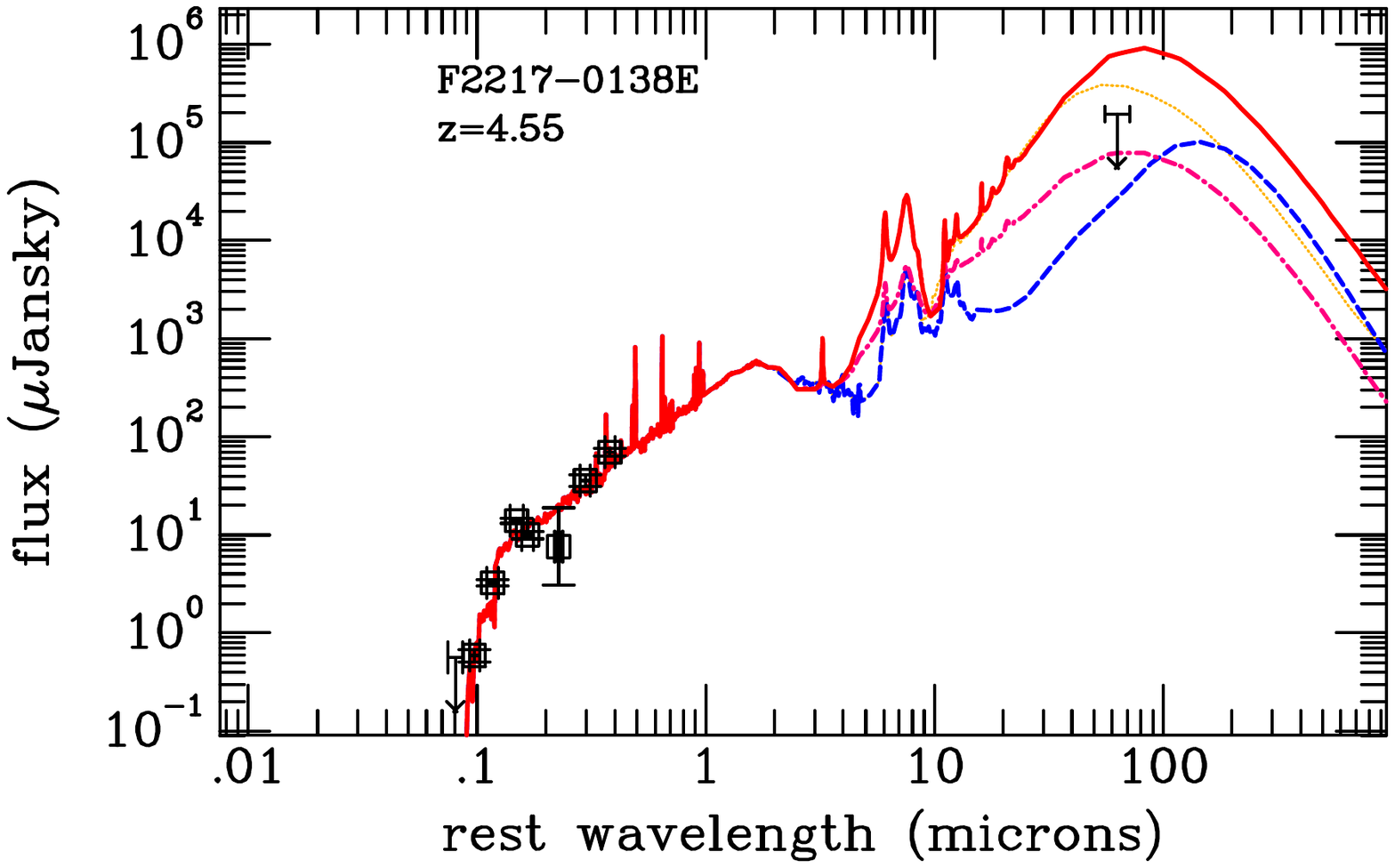}
\end{center}
\vspace{-4.55in}
\caption{\scriptsize
Best-fit SEDs for F2217-0138E with ULIRG (solid), LIRG (dotted),
starburst (dot-dashed), and quiescent (dashed) long wavelength
templates. 
}
\end{figure*}

\subsection{The IR Hubble Diagram}

In Figure 22 we plot the K-band Hubble Diagram, with literature data
compiled by \citet{willott2003}.  These data sets
include more powerful radio galaxies and quasars with spectroscopic redshifts
compared to the FIRST-BNGS sample with photometric redshifts
(including the long wavelength SED solutions from section 4.2 and H-band fluxes
converted to K-band when necessary according to the best-fit SED).  The literature 
magnitudes have been corrected to a common physical aperture (63.9 kpc) 
and also accounted for expected strong emission line
flux according to a prescription by \citet{jarvis2001}.
Our K-band fluxes were corrected to the same physical aperture according to the 
GALFIT models.  The best-fit second-order polynomial of the literature data traces
closely a passively evolving stellar population that formed at z$\sim$10 from \citet{bruzual2003}
as noted by \citet{willott2003}.
The agreement between the literature and the FIRST-BNGS sample is reasonable
considering the accuracy of photometric redshifts relative to the
spectroscopic redshifts of the literature.  Also it is probable that our data 
is incomplete at faint flux levels (there are about 20 FIRST-BNGS sources not detected in H or K) since we were only surveying in H-band to a particular depth (H$\sim$21).  The FIRST-BNGS sample seems to follow the passively evolving population out to at least $z=5$.


\begin{figure*}[htp]
\vspace{-0.8in}
\begin{center}
\includegraphics[scale=0.8]{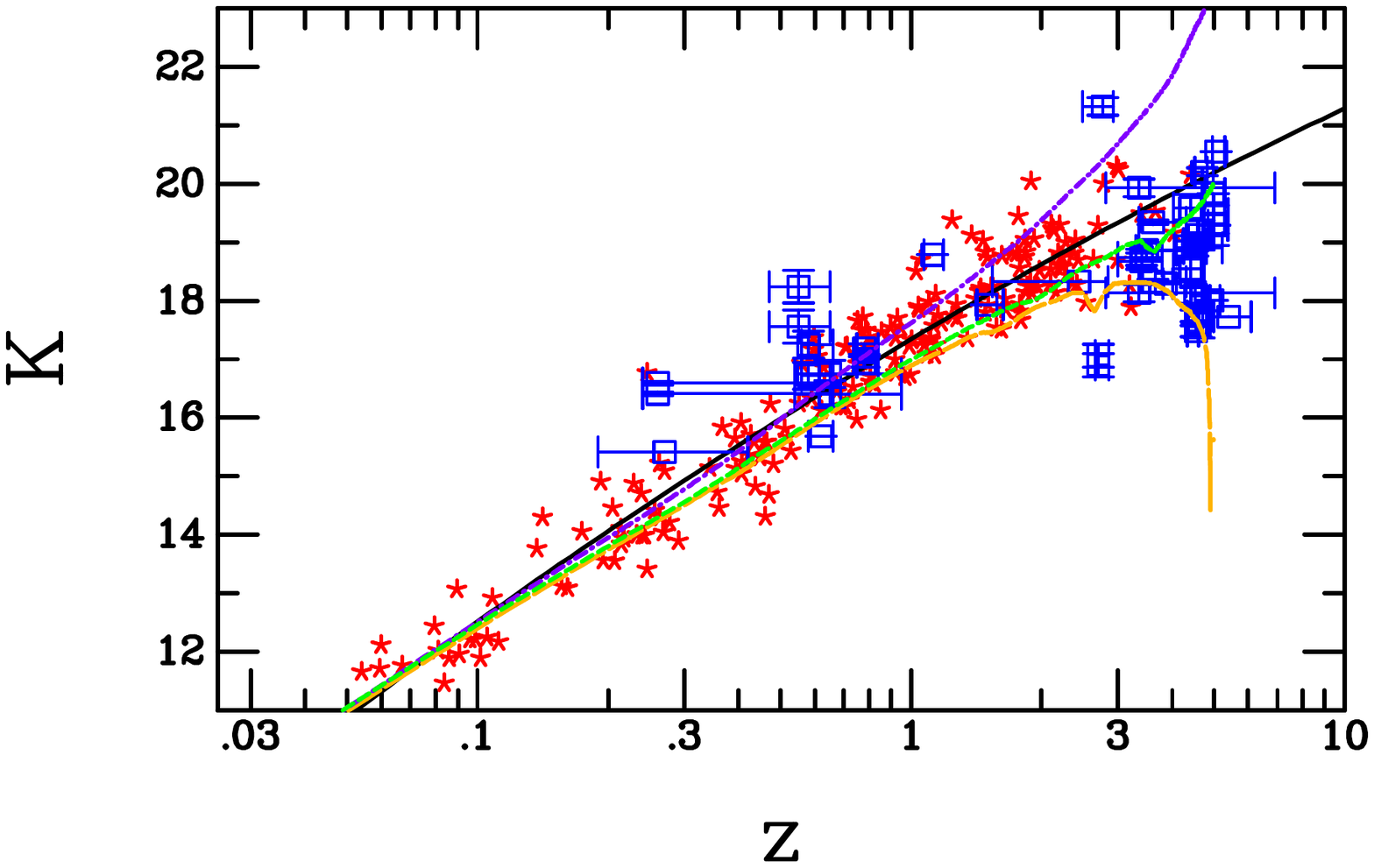}
\end{center}
\vspace{-4.4in}
\caption{\scriptsize
The K-z relation for 3CRR, 6CE, 6C*, and 7CRS radio galaxies from \citet{willott2003}
(stars) with spectroscopic redshifts and the FIRST-BNGS sample with photometric redshifts (boxes), 
all corrected to a physical aperture of 63.9 kpc.  A polynomial fit (solid line) to the \citet{willott2003} data
is K = 17.34 + $4.385 * log(z) - 0.4439 * log(z)^2$, and single stellar population models from \citet{bruzual2003} is shown.
The models are for a non-evolving population (dot-dashed) and passively evolving populations from $z_f=10$ 
(short dashed) and $z_f=5$ (long dashed).  
}
\end{figure*}

\section{Spectroscopy}

Throughout the initial and follow-up phases of this project, we had
the opportunity to observe these objects with various spectrographs on
Mauna Kea to confirm the photometric redshifts.  These observations
were challenging because of the faintness of the science
targets, particularly for slit alignment and guiding over long
integration times.  However, the acquisition problems were largely
solved because of the accurate offsets (from the Pueo-KIR AO
observations) from the nearby bright AO guide star.  This also helped
with tracking using the AO star on the guider.  Table 7 lists the
observed targets, dates, and observing mode used. 

Several of the brightest NIR sources in the sample were observed with
SpeX \citep{rayner2003} on IRTF for several hours each in
low-resolution prism mode to get the widest spectral coverage (0.8-2.5
$\mu$m) and best sensitivity.  Reductions were done with Spextool
\citep{cushing2004,vacca2003} including wavelength, flux, and telluric
calibrations. 

Some of the optically brightest objects were observed with the
Gemini-North Multiobject Spectrograph (GMOS; \citet{hook2003}) in
long-slit, nod and shuffle mode.  Reductions were done using the GMOS
IRAF reduction package with the nod and shuffle routines to carry out
wavelength and flux calibrations.  No telluric correction was done. 

A few of the most interesting sources were observed with the
OH-suppressing IR Imaging Spectrograph (OSIRIS; \citet{larkin2006}) on
Keck II.  This AO-fed integral field spectrograph utilizes a lenslet
array to obtain full near-IR broadband spectral coverage (z, J, H or
K) at R$\sim$4000 resolution at approximately 1000 well-sampled
spatial locations.  It is more efficient than a traditional longslit
spectrograph as no light from extended objects is blocked out by the
slit.  In addition, the high resolution enables observations between
the bright OH sky lines, resulting in lower background noise.  The
data reductions were done with the OSIRIS data reduction pipeline
written in IDL, which provided a wavelength rectified data cube for
each object observed.  Flux calibration, telluric correction, and
spectral extraction were accomplished with our own IDL scripts.   

For each object observed with each instrument, we fit the flux and
wavelength calibrated 1-D spectra to the same SED library of 7
templates we used for the photometric redshifts, allowing redshift and
luminosity to vary, and matching the spectral resolution to the data
via convolution.  We used IDL-based routines to perform a
least-squares fit of the templates to each spectrum. 
Figure 23 shows examples of the best fits of template models with the data
and Table 8 lists the results from these fits.  A number of real and
spurious features were found by examining the individual beam
exposures.  We discuss each object briefly below.

\emph{F0023-0904} with GMOS has a confirmed [OII] 3727 detection at
7280\AA\  and forbidden [NeIII] 3869 at 7550\AA.

\emph{F0916+1134} with GMOS has a confirmed [OII] 3727 detection at
6640\AA\ , and contamination from a nearby source is present at
8160\AA.  

\emph{F1014+1438} with GMOS shows a steep break at about
6100\AA\  which was well fit to the Lyman break of a SB2 SED.
This is consistent with the Hyperz fit. 

\emph{F1234+2001} with GMOS has 
a steep break at 7900\AA.  This was successfully fit to
the Lyman break, but with no other features is not conclusive.  The
two OSIRIS spectra show faint continuum with no large features, and the SpeX
data shows a faint continuum and possibly the Lyman break just at the
edge of sensitivity.  All four fits roughly agree are also consistent
with the photometric Hyperz fit of a z$\sim$5.5 starburst. 

\emph{F1237+1141} with GMOS is featureless and flat except some bad
sky subtraction at 7900\AA\  to 8000\AA.  The OSIRIS data shows a
break at 1.6$\mu$m and perhaps [OII] 3727 at 1.5$\mu$m. 

\emph{F1451+0556} with GMOS shows the edge of [CIV] 1550 at
5400\AA\  and [CIII] 1909 at 6800\AA.  SpeX shows [MgII] 2798 at
1.0$\mu$m, [OII] 3727 at 1.35$\mu$m, H$\beta$/[OIII] 5007 at
1.73$\mu$m, and H$\alpha$ at 2.32$\mu$m.  The bad fit to H$\beta$ is
probably because in the templates there is a sharp transition in
H$\beta$ flux between SB2 and SB3 templates.  OSIRIS also shows
H$\beta$ at 1.73$\mu$m, [OIII] at 1.79$\mu$m, and H$\alpha$ at
2.32$\mu$m.  The equivalent widths of H$\alpha$ and H$\beta$ were
measured by Gaussian fitting and given in Table 9.  If we adopt the
best S/N (SpeX) measurements, the derived reddening is E(B-V)=1.73. 

\emph{F1505+4457} with GMOS shows 
the Balmer break fit at about 8000\AA.  The SpeX data is consistent with the 
Balmer break and a smooth continuum. 

\emph{F2217-0138} with GMOS shows a break which was fit to the Lyman
break.

The rest of the spectroscopy did not produce any well-constrained redshifts.  
A few objects were detected in continuum at relatively low S/N, but without a definitive
spectral break or emission lines, the fits could only confirm that the photometric 
redshifts were consistent with the spectra.  We present these less-constrained 
spectroscopic results in Table 10.

\begin{figure*}[htp]
\vspace{-1.7in}
\begin{center}
\includegraphics[scale=0.45,angle=0,viewport= 0 50 500 700]{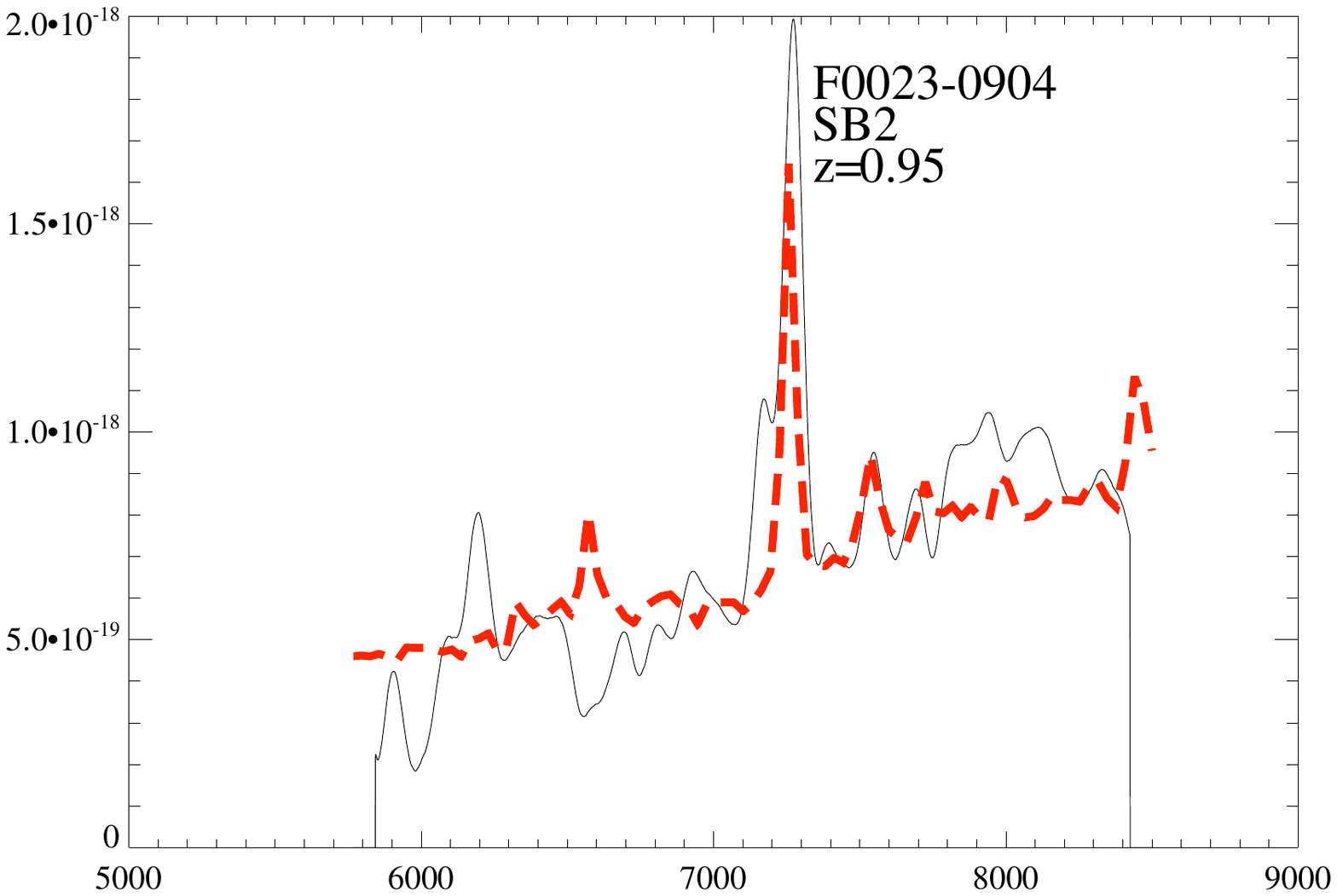}
\includegraphics[scale=0.45,angle=0,viewport= 0 50 500 700]{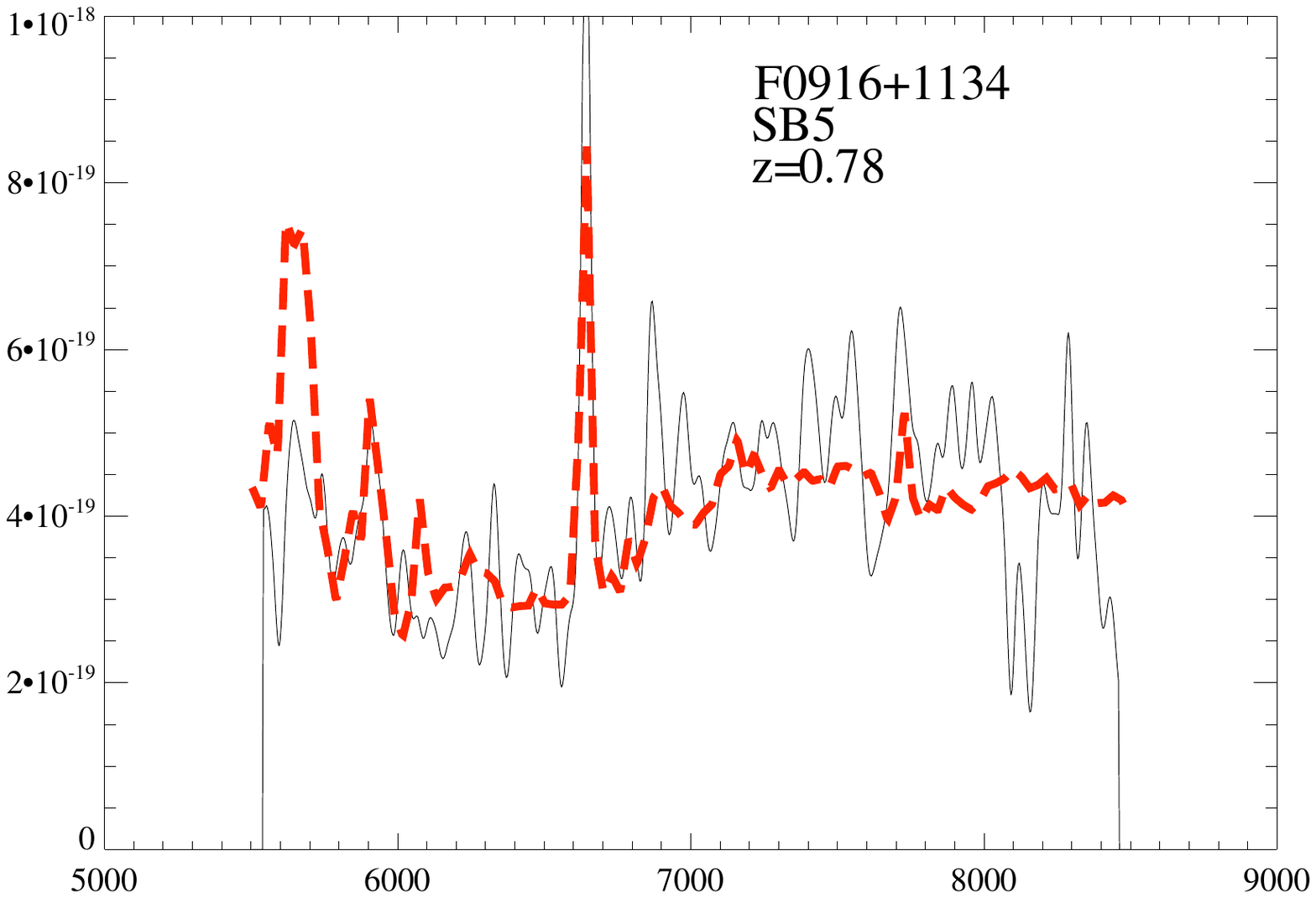}
\end{center}
\vspace{-2.1in}
\begin{center}
\includegraphics[scale=0.45,angle=0,viewport= 0 50 500 700]{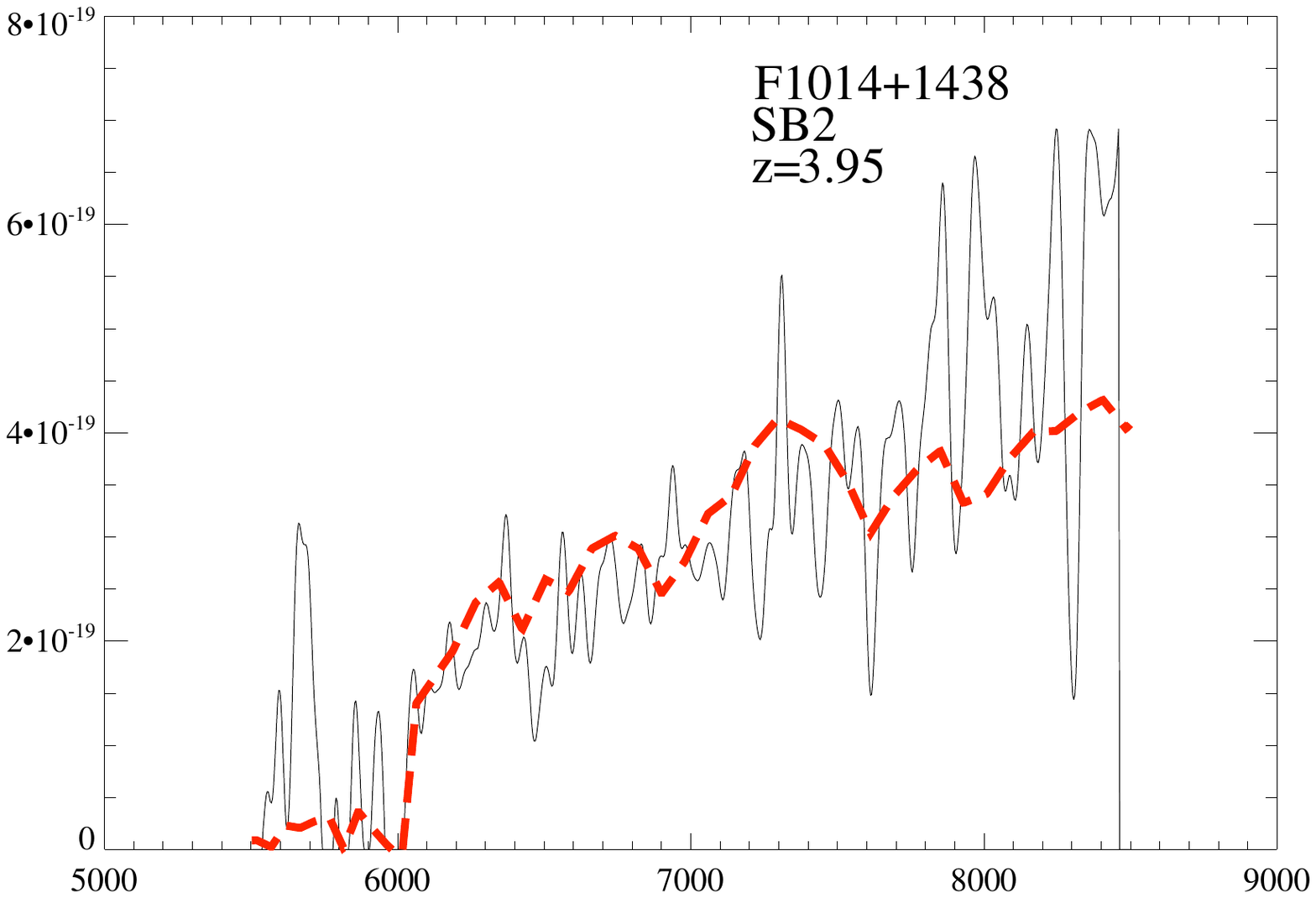}
\includegraphics[scale=0.45,angle=0,viewport= 0 50 500 700]{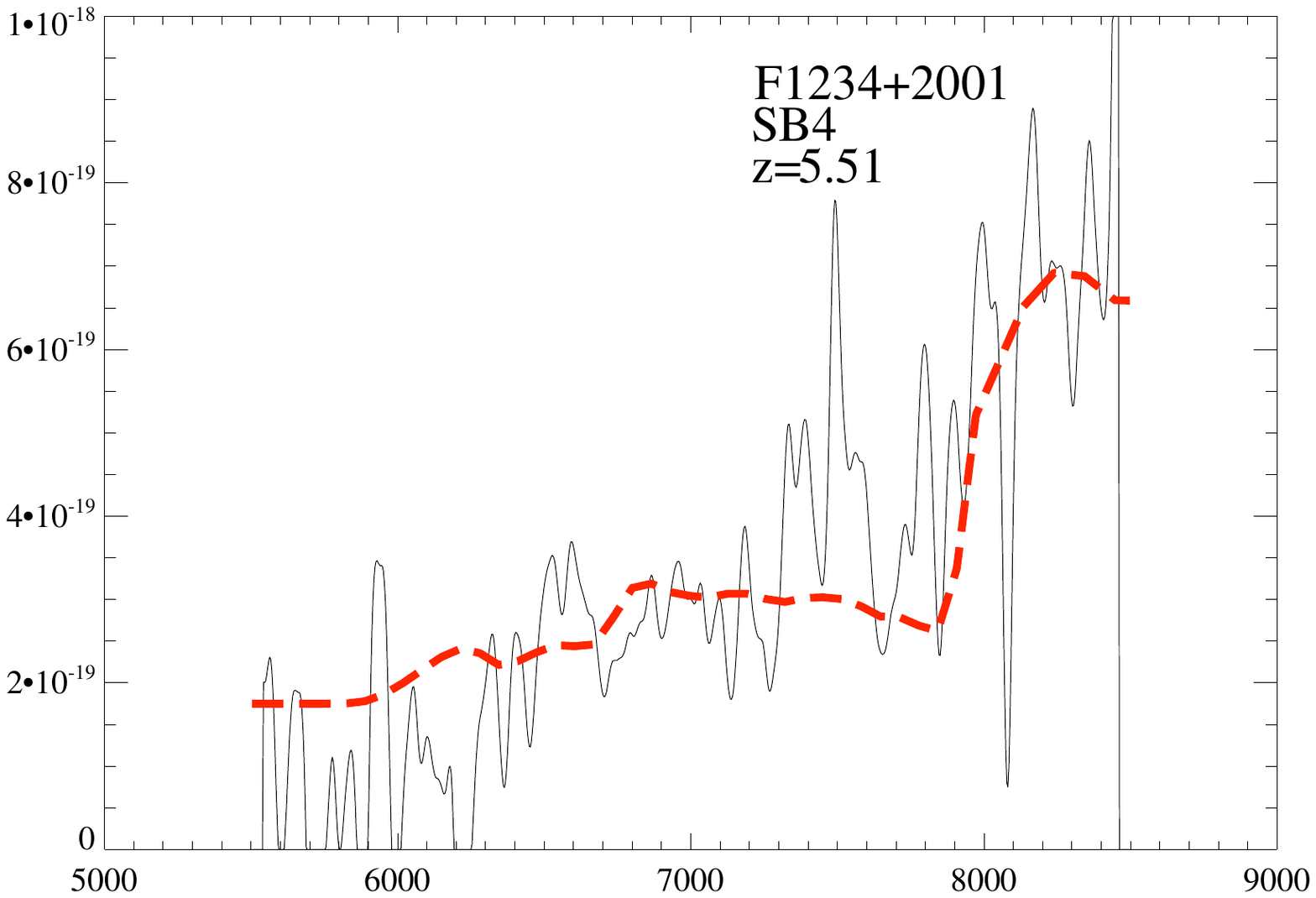}
\end{center}
\vspace{-2.1in}
\begin{center}
\includegraphics[scale=0.45,angle=0,viewport= 0 50 500 700]{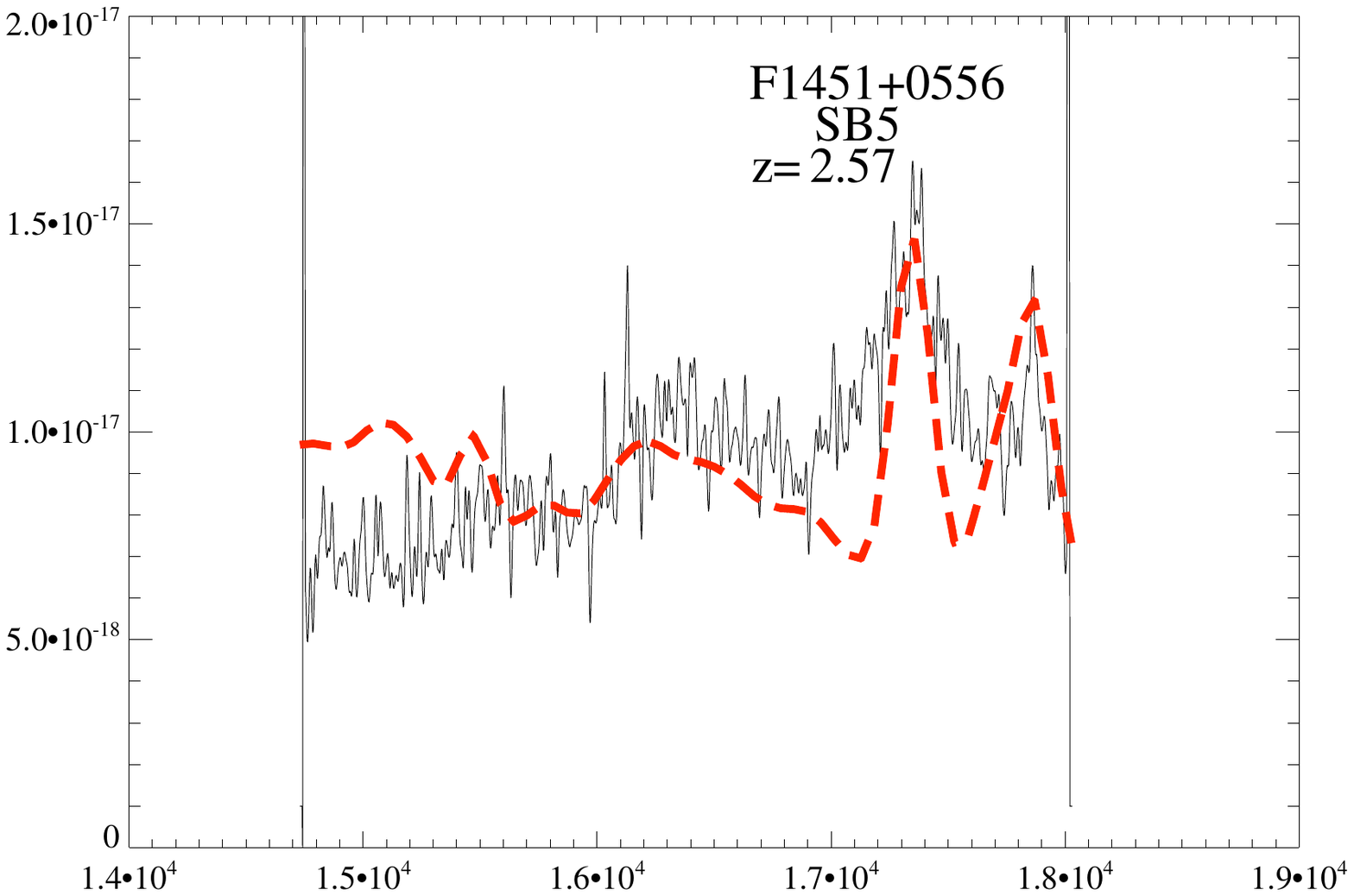}
\includegraphics[scale=0.45,angle=0,viewport= 0 50 500 700]{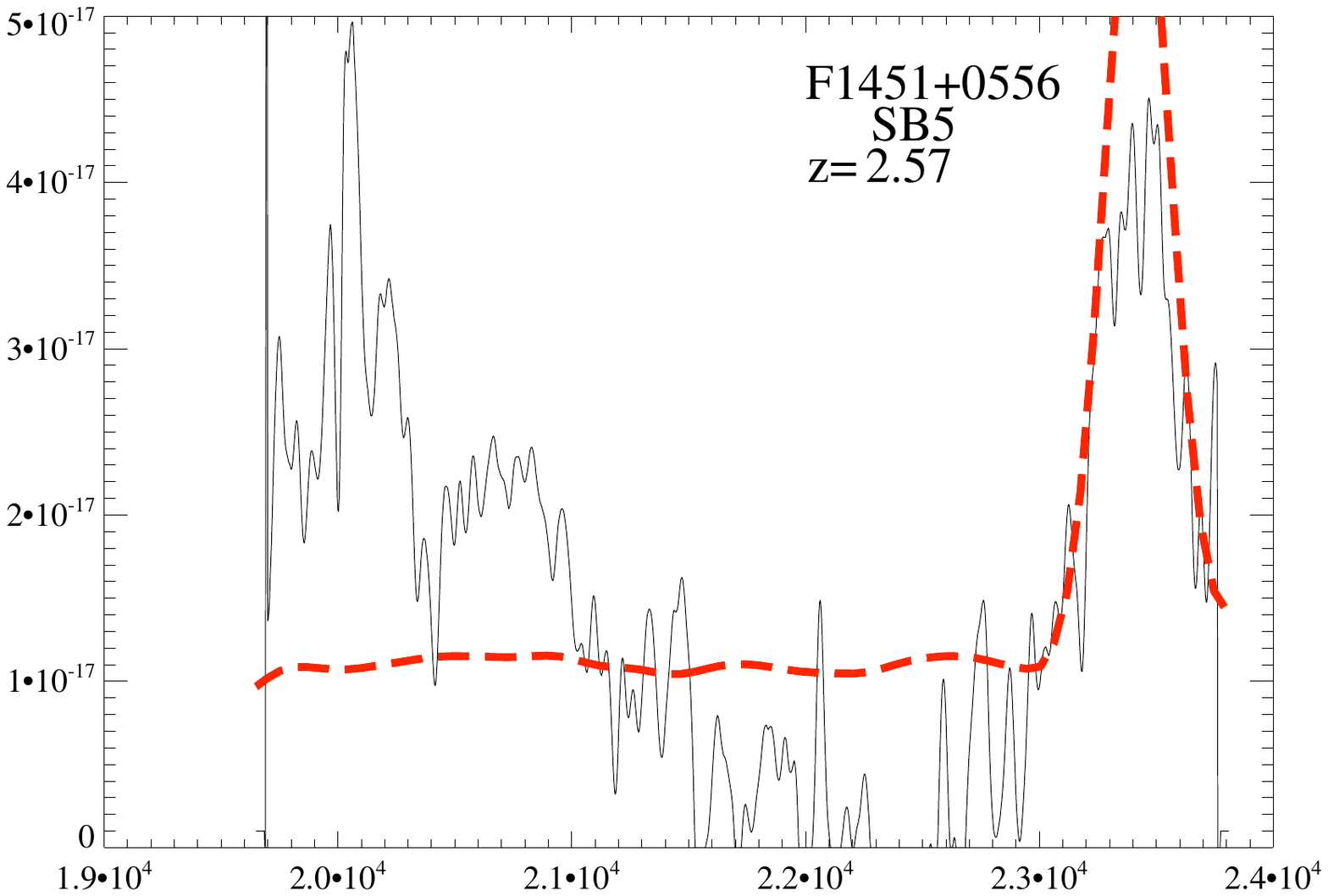}
\end{center}
\vspace{+0.3in}
\caption{\scriptsize
Examples of spectroscopic fitting of 5 FIRST-BNGS objects taken with
various instruments (solid) with best fit template model (dashed).  
The object name, template model, and redshift are indicated in each
panel.  First row: F0023-0904 with GMOS (left) and F0916+0038 with
GMOS (right).  Second row: F1014+1438 with GMOS (left) and F1234+2001
with GMOS (right).  Forth row: F1451+0556 with OSIRIS
in 2 bands (left and right). 
}
\end{figure*}

\begin{figure*}[htp]
\vspace{-0.8in}
\begin{center}
\includegraphics[scale=0.7]{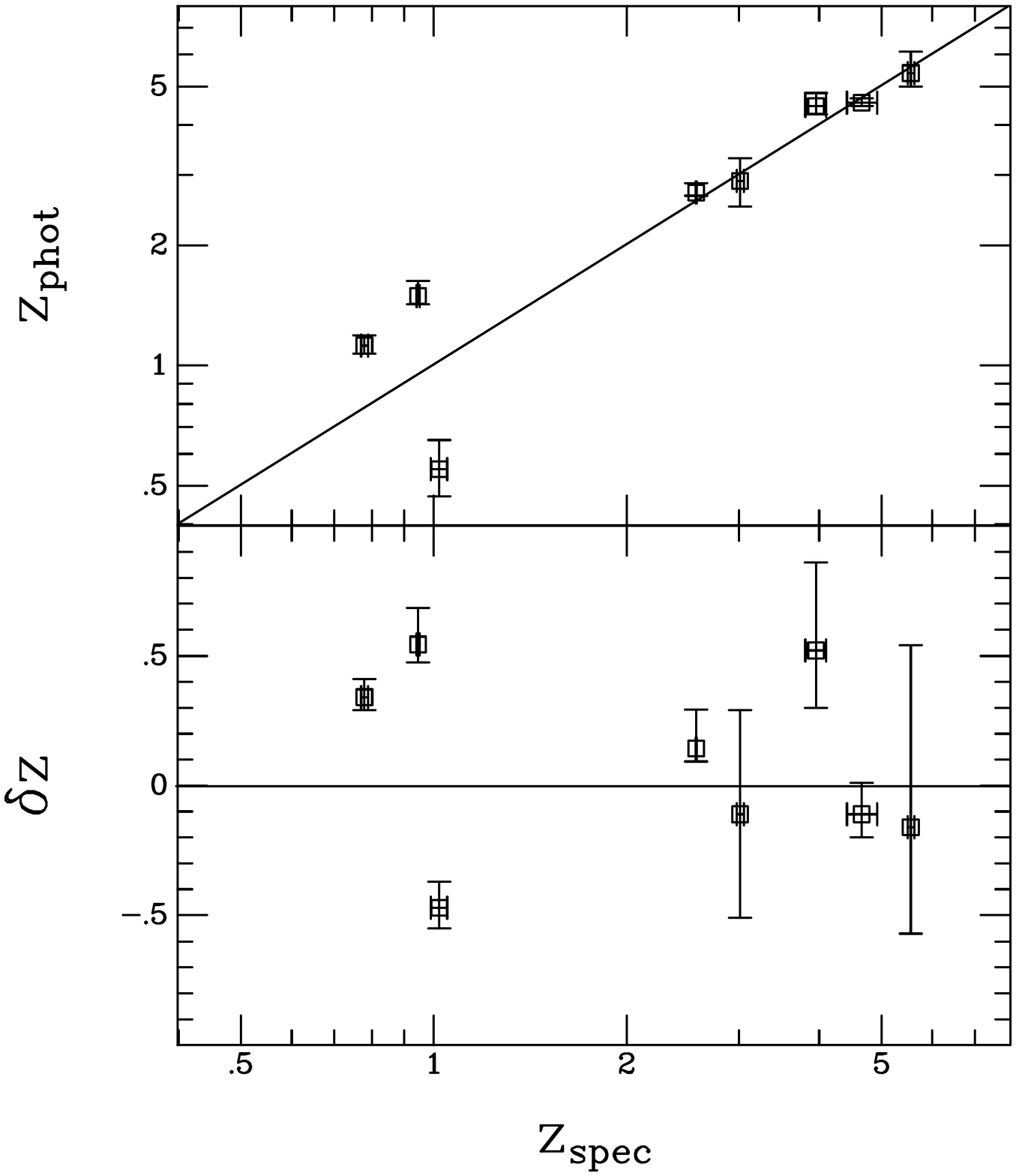}
\end{center}
\vspace{-0.5in}
\caption{\scriptsize
Comparison of redshifts derived from spectroscopic fitting and
photometric fitting.  Bottom panel is the residuals between
spectroscopic and photometric redshifts.  Calculated RMS
$\delta$z/(1+z)=0.146. 
}
\end{figure*}

The well-constrained spectroscopic redshifts may be compared to the
photometric redshifts derived in the previous section.  Figure 24
shows the comparison between the two.

The photometric redshifts match the spectroscopic ones well with a RMS of
$\delta$z/(1+z)=0.146.  This confirms that the photometric redshifts,
though less accurate than spectroscopic ones, are still valid overall.
Historically, Monte Carlo simulations for photometric redshifts from
optical+NIR surveys have determined a typical error of
$\delta$z/(1+z)=0.08 \citep{labbe2003}, but more conservatively
estimate 0.1.  This is an encouraging result considering our spectral
template library may not be an accurate representation of the high
redshift stellar populations we observed.  It also adds confidence to
our photometry measurements that were made in difficult regions (near
a bright star with diffraction spikes and near the edge of the
detector).

\begin{deluxetable*}{llllll}
\tabletypesize{\scriptsize}
\tablecaption{Spectroscopy Results}
\tablehead{
\multicolumn{1}{c}{Name} & \multicolumn{1}{c}{Instruments} &
\multicolumn{1}{c}{$z_{spec}$} & \multicolumn{1}{c}{$\sigma_{z}$} & \multicolumn{1}{c}{$M_V$} & 
\multicolumn{1}{l}{SED} }
\startdata
F0023-0904\tablenotemark{a}   	 & GMOS             & 0.95 & 0.01 & -19.41 & SB2  \\
F0916+1134\tablenotemark{a}  	 & GMOS             & 0.78 & 0.01 & -20.53 & SB5   \\
F1014+1438\tablenotemark{b}  	 & GMOS             & 3.95 & 0.15 & -23.80 & SB2   \\
F1234+2001\tablenotemark{b}  	 & GMOS,SpeX,OSIRIS & 5.56 & 0.07 & -27.59 & SB4   \\
F1237+1141\tablenotemark{b}  	 & GMOS,OSIRIS      & 3.01 & 0.04 & -23.42 & SB3   \\
F1451+0556\tablenotemark{c}  	 & GMOS,SpeX,OSIRIS & 2.57 & 0.01 & -26.76 & SB5  \\
F1505+4457\tablenotemark{b}  	 & GMOS,SpeX        & 1.02 & 0.03 & -20.45 & Sa   \\
F2217-0138E\tablenotemark{b}  	 & GMOS             & 4.66 & 0.26 & -27.99 & SB2    \\
\enddata
\tablenotetext{a}{ [$OII$]3727 detected}
\tablenotetext{b}{ break detected}
\tablenotetext{c}{ [$MgII$]2799, [$OII$]3727, $H_\beta$, [$OIII$]5009, and $H_\alpha$ detected (possibly [$CIII$1909)}
\end{deluxetable*}
\normalsize

\section{VLA observations}

VLA observations were carried out on Nov 28th, 2004, Feb 28th, 2006, and April 20th, 2006 with the VLA in A-array configuration at 3.6cm.  In addition, a VLA archive search resulted in 2 observations of F1445+2702 (known as IRAS14434+2714) on April 13th 1998 and January 14th, 2001 with the same VLA configuration.  We used calibrators 0016-002, 0137+331, 0219+013, 0954+177, 0956+252, 1331+305, 1335+457, 1415+133, 1436+233, 1500+478, 2229-085 for flux and phase calibration.  All data were reduced using standard packages within the Astronomical Image Processing System (AIPS).  Some sources were observed multiple times and were coadded to produce the final map.

\begin{deluxetable}{lrrrr}
\tabletypesize{\scriptsize}
\tablecaption{F1451+0556 Equivalent Widths}
\tablehead{
\multicolumn{1}{c}{} & \multicolumn{2}{c}{$H_\alpha$} & \multicolumn{2}{c}{$H_\beta$}\\[0.5ex]  \\[-1.8ex]
\multicolumn{1}{c}{Instrument} &
\multicolumn{1}{c}{EW} & \multicolumn{1}{c}{$\sigma_{EW}$} & \multicolumn{1}{c}{EW} & \multicolumn{1}{c}{$\sigma_{EW}$} }
\startdata
SpeX & 1515\AA & 519\AA  & 74.9\AA  & 53.6\AA \\
OSIRIS & 2266\AA & 3374\AA & 150\AA  & 36\AA   \\
\enddata
\end{deluxetable}
\normalsize

\begin{deluxetable*}{llllll}
\tabletypesize{\scriptsize}
\tablecaption{Other Spectroscopy}
\tablehead{
\multicolumn{1}{c}{Name} & \multicolumn{1}{c}{Instruments} &
\multicolumn{1}{c}{$z_{spec}$} & \multicolumn{1}{c}{$\sigma_{z}$} & \multicolumn{1}{c}{$M_V$} & 
\multicolumn{1}{l}{SED} }
\startdata
F0129-0140       & GMOS             & 1.71 & 0.40 & -22.80 & SB5    \\
F0202-0021  	 & GMOS,SpeX        & 0.60 & 0.30 & -22.15 & E     \\
F0216+0038  	 & GMOS,SpeX,OSIRIS & 0.56 & 0.30 & -21.49 & S0    \\
F0919+1007	 & OSIRIS           & 0.79 & 0.30 & -23.08 & SB5   \\
F1116+0235  	 & SpeX             & 0.31 & 0.20 & -20.65 & Sc   \\
F1447+1217	 & GMOS             & 4.60 & 0.50 & -24.79 & SB5   \\
F2217-0837  	 & SpeX             & 0.31 & 0.20 & -20.16 & E    \\
\enddata
\end{deluxetable*}
\normalsize

With the VLA A-array at 3.6cm, the largest angular scale detectable is
7.0 arcseconds and the primary beam is about 0.24 arcseconds.  Flux
densities were measured using the IRAF ``phot'' package and converted
to flux units.  Background noise was also measured to derive flux
errors and upper-limits for non-detection.  For multiple sources
(double, triple, distorted morphologies), the fluxes were added
together.  Figure 25 shows example radio maps from FIRST and from our
VLA A-array observations.  Table 11 shows our measurements along with
the best-fit photometric or spectroscopic redshifts plus spectral
indices.  The spectral index, $\alpha$, where
$S_\nu\propto\nu^\alpha$, was calculated for each discrete radio
source in the field as well as the total flux of all the sources
coadded. 

\begin{figure*}[htp]
\begin{center}
\includegraphics[scale=0.85,angle=0,viewport= 30 0 600 700]{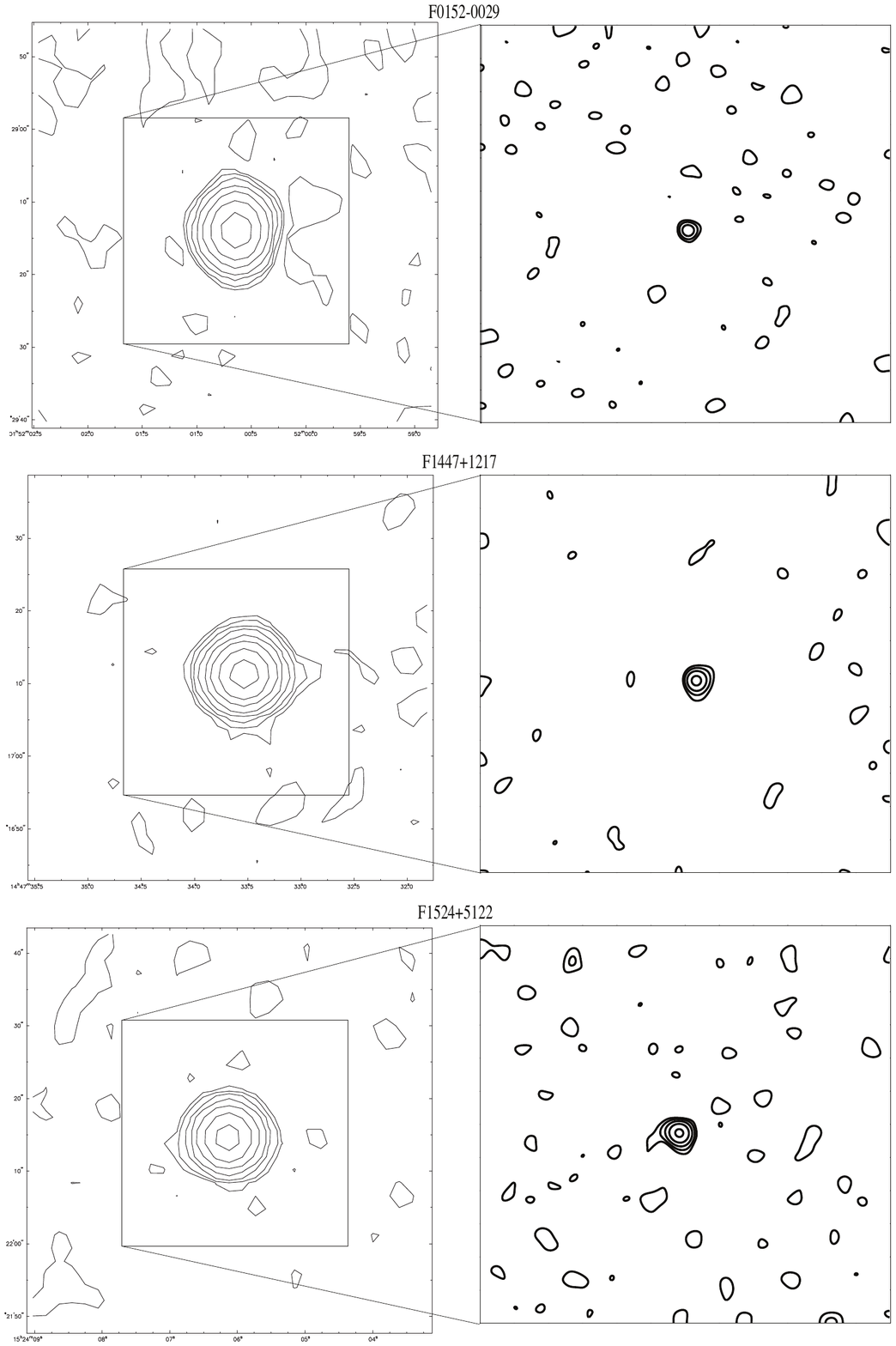}
\end{center}
\vspace{-0.2in}
\caption{\scriptsize
Radio contour maps of 3 FIRST-BNGS sources at 20cm B-array (left) and 3.6cm A-array (right) taken with VLA.
}
\end{figure*}

\begin{deluxetable*}{llllcrrrrrl}
\tabletypesize{\scriptsize}
\tablecaption{VLA 3.6cm and 20cm Fluxes and 3-$\sigma$ upper-limits.}
\tablehead{
\multicolumn{1}{c}{Name} & 
\multicolumn{1}{c}{$z$} & \multicolumn{1}{c}{$\sigma_{z}$} & \multicolumn{1}{c}{$M_V$} & 
\multicolumn{1}{c}{SED} &
\multicolumn{1}{c}{$S_{8.3}$} & 
\multicolumn{1}{c}{$\sigma_{8.3}$} & 
\multicolumn{1}{c}{$S_{1.4}$} &
\multicolumn{1}{c}{$\sigma_{1.4}$} & 
\multicolumn{1}{c}{$\alpha$} &
\multicolumn{1}{l}{morph$^a$}
}
\startdata
F0023-0904 & 0.946 & $^{+0.005}_{-0.005}$ & -19.41 & SB2  & 2.56 & 0.09 & 15.28 & 0.13 & -1.04  &  elong \\
F0129-0140 & 2.76 & $^{+0.27}_{-0.44}$ & -25.07 & SB6 & 0.83 & 0.08 & 3.70 & 0.14 & -0.87  &  elong \\
F0152-0029 &   &                  &  & & 3.79 & 0.09 & 24.69 & 0.17 & -1.09  &  unres \\
F0152+0052E   &      &                  &        &     & 2.17 & 0.09 & 9.88 & 0.17 & -0.88  &  tpl \\
F0152+0052W   &      &                  &        &     & 1.92 & 0.09 & 6.97 & 0.17 & -0.75  &  tpl \\
F0152+0052C   &      &                  &        &     & 0.73 & 0.09 & $<$0.6 &  & $>$0.11  &  tpl \\
F0152+0052tot &      &                  &        &     & 4.83 & 0.09 & 16.85 & 0.17 & -0.73  &  tpl \\
F0202-0021 & 0.58 & $^{+0.08}_{-0.02}$ & -22.13 & E   & $<$0.27 &   & 2.79 & 0.11 & $<$-1.36  &  elong \\
F0216+0038E & 0.65 & $^{+0.30}_{-0.11}$ & -22.09 & S0  & 2.32 & 0.08 & 17.84 & 0.10 & -1.19  &  dbl \\
F0216+0038W & 0.65 & $^{+0.30}_{-0.11}$ & -22.09 & S0  & 0.42 & 0.08 & 12.10 & 0.10 & -1.96  &  dbl \\
F0216+0038tot & 0.65 & $^{+0.30}_{-0.11}$ & -22.09 & S0  & 2.74 & 0.08 & 29.94 & 0.10 & -1.39  &  dbl \\
F0916+1134E & 0.78 & $^{+0.01}_{-0.01}$ & -20.53 & SB5 & 1.38 & 0.20 & 3.45 & 0.15 & -0.53  &  elong \\
F0916+1134W & 0.78 & $^{+0.01}_{-0.01}$ & -20.53 & SB5 & $<$0.6 &  & 1.09 & 0.15 & $<$-0.35  &  elong \\
F0916+1134tot & 0.78 & $^{+0.01}_{-0.01}$ & -20.53 & SB5 & 1.38 & 0.20 & 4.54 & 0.15 & -0.69  &  elong \\
F0919+1007 & 0.79 & $^{+0.03}_{-0.03}$ & -23.08 & SB5  & 9.71 & 0.25 & 8.97 & 0.15 & 0.05  &  elong \\
F0942+1520E & 3.36  & $^{+3.56}_{-0.54}$   &        &     & 6.17 & 0.16 & 11.70 & 0.17 & -0.37  &  dbl \\
F0942+1520W & 3.36  & $^{+3.56}_{-0.54}$   &        &     & 0.72 & 0.16 & 4.00 & 0.17 & -1.00  &  dbl \\
F0942+1520tot & 3.36  &  $^{+3.56}_{-0.54}$  &        &     & 6.89 & 0.16 & 15.70 & 0.17 & -0.48  &  dbl \\
F1010+2527 & 4.63 & $^{+0.08}_{-0.06}$ & -27.32 & SB2 & $<$0.33 &   & 1.31 & 0.15 & $<$-0.80  &  unres \\
F1010+2727N & 4.53 & $^{+0.20}_{-0.11}$ & -27.87 & Sb  & $<$0.33 &  & 2.36 & 0.15 & $<$-1.15   &  dbl \\
F1010+2727S & 4.53 & $^{+0.20}_{-0.11}$ & -27.87 & Sb  & $<$0.33 &  & 3.58 & 0.15 & $<$-1.39   &  dbl \\
F1010+2727tot & 4.53 & $^{+0.20}_{-0.11}$ & -27.87 & Sb  & $<$0.33 &  & 5.94 & 0.15 & $<$-1.69   &  dbl \\
F1014+1438E & 3.95 & $^{+0.15}_{-0.15}$ & -23.80 & SB2 & $<$0.33 &  & 17.14 & 0.17 & $<$-2.30  &  tpl \\
F1014+1438W & 3.95 & $^{+0.15}_{-0.15}$ & -23.80 & SB2 & $<$0.33 &  & 15.60 & 0.17 & $<$-2.25  &  tpl \\
F1014+1438C & 3.95 & $^{+0.15}_{-0.15}$ & -23.80 & SB2 & 4.11 & 0.11 & $<$0.6 &  & $>$1.12  &  tpl \\
F1014+1438tot & 3.95 & $^{+0.15}_{-0.15}$ & -23.80 & SB2 & 4.11 & 0.11 & 32.74 & 0.17 & -1.21  &  tpl \\
F1237+1141N & 3.01  & $^{+0.04}_{-0.04}$   & -23.42 & SB3 & 3.92 & 0.05 & 11.91 & 0.17 & -0.65  &  tpl \\
F1237+1141S & 3.01  & $^{+0.04}_{-0.04}$   & -23.42 & SB3 & 0.45 & 0.05 & 4.56 & 0.17 & -1.35  &  tpl \\
F1237+1141C & 3.01  & $^{+0.04}_{-0.04}$   & -23.42 & SB3 & 1.06 & 0.05 & 1.36 & 0.17 & -0.15  &  tpl \\
F1237+1141tot & 3.01  & $^{+0.04}_{-0.04}$   & -23.42 & SB3 & 5.43 & 0.05 & 17.83 & 0.17 & -0.69  &  tpl \\
F1315+4438 & 2.77 & $^{+0.16}_{-0.28}$ & -22.49 & SB2 & $<$0.08 &   & 0.89 & 0.17 & $<$-1.40  &  dist \\
F1355+3607 &      &                  &        &     & $<$0.23 &   & 3.48 & 0.17 & $<$-1.58  &  dist \\
F1430+3557 & 3.5  & $^{+0.6}_{-0.2}$   & -24.82 & SB3 & 2.13 & 0.08 & 8.54 & 0.17 & -0.81   &  unres \\
F1445+2702 &      &                  &        &     & 13.14 & 0.04 & 31.07 & 0.17 & -0.50  &  unres \\
F1447+1217 & 4.70 & $^{+0.13}_{-0.18}$ & -25.11 & SB3  & 9.15 & 0.14 & 48.90 & 0.17 & -0.98  &  unres \\
F1451+0556 & 2.567 & $^{+0.005}_{-0.005}$ & -26.76 & SB5 & 2.57 & 0.10 & 2.35 & 0.17 & 0.05   &  unres \\
F1505+4457N & 1.02 & $^{+0.03}_{-0.03}$ & -20.45 & Sa & 1.76 & 0.14 & 7.84 & 0.17 & -0.87  &  dbl \\
F1505+4457S & 1.02 & $^{+0.03}_{-0.03}$ & -20.45 & Sa & 8.04 & 0.14 & 9.24 & 0.17 & -0.08  &  dbl \\
F1505+4457tot & 1.02 & $^{+0.03}_{-0.03}$ & -20.45 & Sa & 9.80 & 0.14 & 17.08 & 0.17 & -0.32  &  dbl \\
F1524+5122 &      &                  &        &     & 8.23 & 0.14 & 22.45 & 0.17 & -0.59  &  unres \\
F2217-0138E & 0.48 & $^{+0.12}_{-0.09}$ & -20.78 & S0  & 25.2 & 0.15 & 26.37 & 0.17 & -0.03  &  unres \\
F2354-0055 &      &                  &        &     & 0.31 & 0.04 & 1.38 & 0.17 & -0.87   &  unres \\
\enddata
\tablenotetext{a}{ morphology from 20-cm FIRST survey: unres=unresolved, elong=elongated, dbl=double,tpl=triple, dist=distorted.}
\end{deluxetable*}
\normalsize

Plotting spectral index of each object's total fluxes, $\alpha$ versus
20-cm flux, redshift, absolute magnitude, there is no clear indication
of any trends with $\alpha$ in this data set (Figure 26) though there
are more steep-spectrum ($\alpha<-0.5$) sources observed than
flat-spectrum ($\alpha>-0.5$) sources (9 vs. 3). 

\begin{figure*}[htp]
\vspace{-4.8in}
\begin{center}
\includegraphics[scale=0.85,angle=0,viewport= 0 50 600 700]{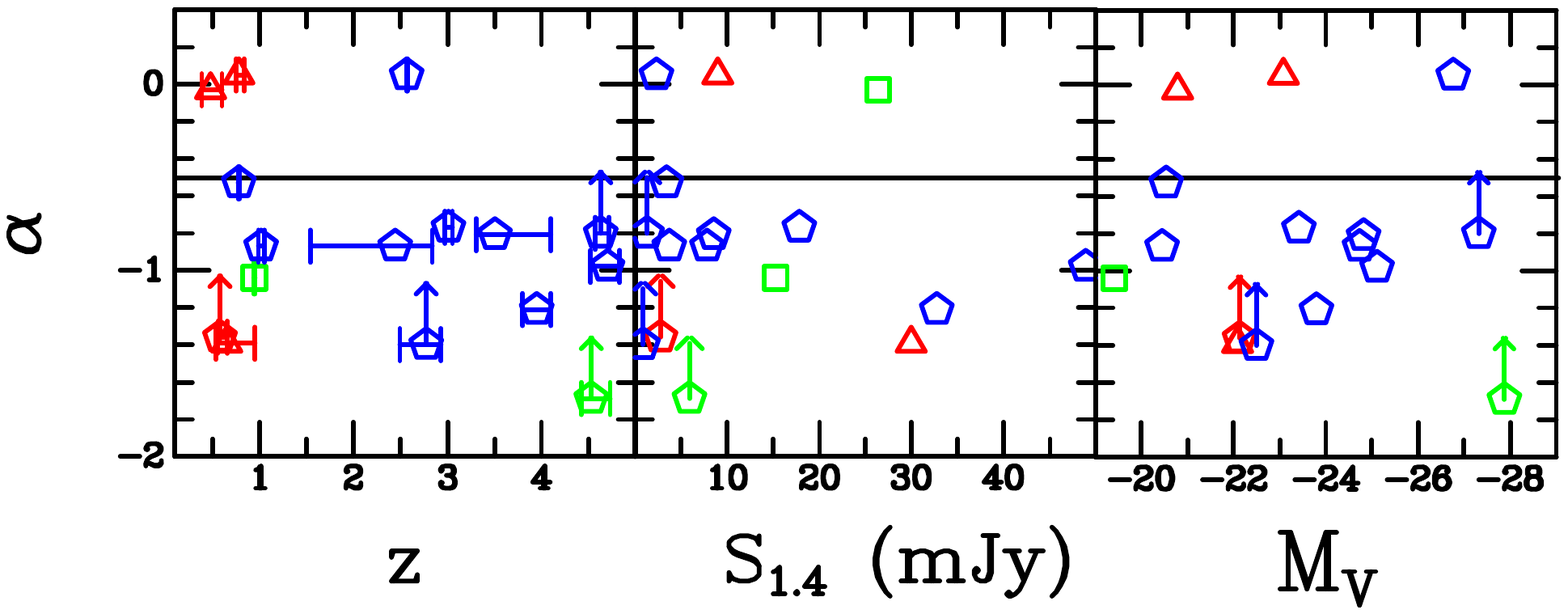}
\end{center}
\vspace{-0.2in}
\caption{\scriptsize
Distribution of $\alpha$ with redshift (left), radio flux (center), and
V-band absolute magnitude (right).  Shapes represent best-fit
optical/NIR SEDs from Hyperz, triangles are early-type (E and S0),
boxes are late-type (Sa,Sb,Sc), and pentagons are starbursts SB1-SB6.
The horizontal line show the division between steep ($\alpha<$-0.5) and flat ($\alpha>$-0.5) spectrum types.
}
\end{figure*}

\section{Discussion of Individual Objects}

Below is a discussion the properties of the 58 individual sources in the FIRST-BNGS sample.

\emph{F0023-0904} was detected in all bands, giving a photometric
redshift of 1.49 with an Sc SED.  Spectroscopy revised the redshift to
0.946 with an [OII] emission detection.  The FIRST radio peak is
offset about 3$''$ from the galaxy, but the outer contours are
elongated in the direction of the galaxy, suggesting a jet structure.
VLA X-band morphology was inconclusive, but provided a spectral slope
of -1.04. 

\emph{F0129-0140} had 2 solutions for photometric redshifts at 0.65
with an SB2 SED and 2.44 with an SB4 SED.  Without any discernible
break, it was not obvious which is the better solution.  Additionally,
spectroscopy gives a rough redshift of 1.7 from 
the spectral slope in the 5500-8000 Angstrom range which does not
break the degeneracy.  A compact
moderate radio flux with faint optical flux suggests a higher redshift
as the low redshift would give such a low luminosity
($9*10^9L_{sun}$), it would be difficult to reconcile with an AGN
host.  We will use the high photometric redshift
solution because it was more constrained than the spectroscopic fit.

\emph{F0152-0029} was undetected in all bands except K.
It does not have a constrained redshift so is either
at z$>$8.0 where the Lyman break is redshifted out of z-band (see
figure 20) or a highly reddened object. 
The radio source is compact with a relatively high radio 20cm
luminosity (25mJy) and steep spectral index (-1.09), suggesting an AGN
source.   

\emph{F0152+0052} was not detected in any band (B, V, R, I, z$'$, J,
and H).  A triple morphology in X-band, and high 20cm flux (18mJy) and
steep spectral index (-0.73) suggests an high redshift AGN.   

\emph{F0202-0021} appears to be well fit by a
low redshift elliptical SED ($z=0.58$), but also has a secondary
solution at $z=4.2$.  The brightness and redshift inferred from the
K-z relation derived in section 2.4 ($z\sim0.9$) supported the first
solution and was roughly confirmed with spectroscopy ($z=0.6$).  An
elongated radio morphology and 
lower radio 20cm flux (3mJy) with a steep spectrum also is consistent
with a giant elliptical with an AGN at low redshift. 

\emph{F0216+0038} has a SED consistent with the S0 template at $z=0.65$
which is lower than that inferred from the K-z relation derived in
section 2.4 ($z\sim0.9$), but was roughly confirmed with spectroscopy.
A moderate 20cm flux (12mJy) and double radio lobe morphology with
steep spectral index (-1.39) suggests a relatively powerful AGN,
perhaps near the end of an active accretion stage. 

\emph{F0916+1134} appears to be a barred-spiral from the residuals in
surface brightness fitting (Stalder \& Chambers\nocite{stalder2009a}, in prep.).  
K-band photometry was excluded from photo-z fitting (due to the poor
residuals in the profile fit),  
which found a best redshift of 1.12 with a
starburst SED.  A confirmed [OIII] 3727\AA\  detection with
spectroscopy gives a final redshift of 0.78.   

\emph{F0919+1007} appears to have a low redshift
($z=0.79$) S0 SED.  Moderate radio 20cm flux (9mJy) with elongated
morphology is also consistent with a low-redshift giant elliptical
with an AGN. 

\emph{F0938+2326} has 2 solutions for photometric redshifts at 0.8
with a SB2 SED and 3.88 with SB3 SED.  The moderate to high 20-cm flux
(8.1mJy) with compact morphology, along with the relatively faint
optical flux, favors the higher redshift solution.   

\emph{F0939-0128} was not detected in V, R, z$'$, or H bands.

\emph{F0942+1520} was undetected in all bands except H and K.
Because of the large H-K color, Hyperz could fit the break with a
redshift of 3 to 6.5.  The high radio 20cm flux (12mJy) and
double-lobe morphology is consistent with a 
high-redshift AGN.

\emph{F0943-0327} was undetected in B, V, R, and H, but barely
detected in I ($< 5 \sigma$) and has a very high radio 20cm flux
(99mJy) and double-lobe morphology.   

\emph{F0950+1619} was detected in R and I though the ID is slightly
offset (about 5$''$) from the radio peak so it is possibly another
blank field.  There were also no detections in z and H bands. 

\emph{F0952+2405} was detected in both I and H bands.  It is bright in
H band, and a faint radio source (1.3mJy).  It is probably not at high
redshift (the K-z relation gives likely redshift of z=0.76). 

\emph{F0955+2951} has a distorted radio morphology with the radio peak
offset from the optical ID by about 5$''$.  The optical and IR
morphologies appeared extended.  The best fit photometric redshift is
a z=4.4 starburst.  Since it is not likely the true optical ID for
this radio source, we will not use it in the subsequent analysis. 

\emph{F0955+0113} was not detected in any observed filter (V, R, I,
z$'$ and H).   

\emph{F0956-0533} was not detected in V, R, I, or H bands.

\emph{F0958+2721} similar to F0952+2405, detected in I and H and
relatively bright with faint 20cm radio source (2.4mJy), so probably
not at high redshift. 

\emph{F1000-0636} was not detected in I or H filters.

\emph{F1008-0605} was detected and relatively bright in all observed
filters (B, V, R, I, and H).  A break was found and fit by Hyperz to
either a $z=0.27$ elliptical SED or $z=3.9$ SB4 SED with little
extinction.  If the high redshift solution were correct, it would be
hard to reconcile this particular object with a hierarchical merging
scenario especially since it would have an extremely luminous stellar
population ($>10^{13}L_{\sun}$) dominating its rest frame UV light,
which has to be assembled in less than 2.3Gyr (age of the universe at
that redshift).  We therefore adopt the low redshift solution.  

\emph{F1010+2527} has a close double morphology in most bands. 
There was not any observed evidence of interaction, but the
photometric redshifts of the two galaxies 
were similar ($z=4.41$,$z=4.63$), which suggests they are associated.
Both are consistent with starburst SEDs.  The south
source's parameters were used in the subsequent analyses because the
fit was better constrained.   
The unresolved, low 20cm flux (1mJy) radio source also supports that
this is not an aligned object. 

\emph{F1010+2727} has an SED consistent with a high redshift Sb
model ($z=4.53$).  A moderate 20cm radio source (6mJy total) and
double-lobe morphology also suggests a high redshift. 

\emph{F1014+1438} appears to be a starburst at high redshift
($z=3.95$), but a low redshift ($z=0.5$) solution was also found.
Because of the faintness of the source and the presence of a steep 
break observed with spectroscopy, we chose the high redshift solution.  
Double-lobe moderate 20cm flux (33mJy integrated) radio source is also
consistent with a high redshift AGN. 

\emph{F1016+1513} was undetected in R, I, z$'$, and H bands.

\emph{F1024-0031} was detected in H, and with the most powerful radio
20cm flux (158mJy) in the sample, is almost certainly an AGN.  Based
on the H-band brightness, it is probably at a redshift around 0.8. 

\emph{F1027+0520} was fit to a z=0.6 starburst SED template.  It is a
bright galaxy, so is potentially at lower redshift due to our low
confidence in our photometric measurements for bright extended objects
and determining a good sky level around them.  It has a compact radio
source with relatively high 20cm flux (23mJy). 

\emph{F1039+2602} has an SED best fit by a $z=3.62$
young starburst SB3.  Its moderate radio 20cm flux (12mJy) and double
morphology is also consistent with high redshift.

\emph{F1040+2323} has a weak (1.6mJy) radio source blank in R and I,
but faintly detected in z$'$ and H.   

\emph{F1116+0235} has a best-fit photometric redshift at 0.62 which is
consistent with a continuum fit from spectroscopy.  
It also has a 2$\sigma$ detection at 850$\mu$m from SCUBA, which can
only exclude a Arp 220-type (ULIRG) SED at the same redshift as the
optical/NIR photometric redshift.  

\emph{F1133+0312} was undetected at R, I, z, and H bands and is at the
lower threshold of the FIRST survey (0.8mJy). 

\emph{F1140+1316} was not detected at B, V, R, I, and z$'$.  A
double-lobed morphology suggests an AGN. 

\emph{F1147+2647} has a possible faint optical ID about 6 arcseconds
to the west of the weak radio peak (0.7mJy), so it is probably not the
proper ID and will not be used in later analyses, though the best fit
SED is a z=4.75 SB1 type SED. 

\emph{F1155+2620} has a 1mJy radio source with an optical counterpart
about 3 arcseconds to the southeast of the radio peak.  Hyperz gives
the best-fit photometric redshift at 4.5 with a starburst SED. 

\emph{F1158+1716} undetected in H.

\emph{F1202+0654} undetected in H.

\emph{F1211+3616} undetected in V, R, I, and H.

\emph{F1215+4342} appears to be a bright S0 type galaxy at z=0.1 with
a relatively high 20cm flux (45mJy) radio source.  Although there was a
solution at high redshift, the extreme optical
luminosity ($>10^{13}L_{sun}$) estimated for that redshift and
extended radio morphology suggest the low redshift solution is more
likely. 

\emph{F1217-0529} is another optical double (possibly triple) source
which also has similar photometric redshifts (z=4.95 and z=4.97 for
the east and west sources respectively).  However, the radio peak is
offset slightly (about 3$''$) to the south, closer to the faint
(R=25.5) south optical source. 

\emph{F1217+3810} was detected at H-band and is relatively bright, so
it is probably at low redshift ($z\sim0.3$) with a weak (0.8mJy) radio
source. 

\emph{F1218-0625} was detected in V, R, I, z$'$ and H.  Two
photometric redshift solutions were found though any redshift between
0 and 4.5 gives reasonable fits.  The compact moderate 20-cm flux (4.3mJy)
and faint optical flux suggest the higher redshift is more likely, but
with a large possible range. 

\emph{F1218-0716} was barely detected in V and H bands.  It also is a
weak (0.8mJy) and distorted radio source. 

\emph{F1234+2001} was identified about 4$''$ to the southwest from the
moderate radio peak (5mJy) of the FIRST source J123432.9+200134.  This
seems a bit far given the position accuracy of FIRST (about 1$''$) and
our imaging data (about 0.3$''$) derived from the USNO-A2.0 catalog.
Since the only optical/IR source in the field around the radio source
is F1234+2001 there are 3 possibilities: 1) The source we have
identified is unrelated or a companion to the radio source
J123432.9+200134 in which case the radio source host is below the
detection threshold of H=19.74 (3$\sigma$); 2) The optical emission is
offset from the center of the host galaxy due to the alignment effect
\citep{chambers1987} though the radio source is too weak to be
regarded as a powerful radio source at any epoch which makes this
scenario unlikely; or 3) The radio source is intrinsically asymmetric
due to relativistic beaming, so the radio centroid of J123432.9+200134
is not centered on the host galaxy F1234+2001.  For the remainder of
the paper we accept that optical/IR 
source, F1234+2001, to be the host of J123432.9+200134.  The
confidence in this ID is strengthened from the subsequently
derived redshifts.   

It should be noted that a diffraction spike passes through the optical
ID in V, R, I, and z$'$.  GALFIT successfully modeled this and the good 
residuals raise our confidence in the photometric measurements.
The best fit SED to the broadband photometry is a $z=5.4$ starburst
mainly from fitting the Lyman break between R and I bands.  Though the
R-band image is relatively shallow, the V-band imaging is deeper which
did confirm that this object was at least $z>3$.  At $z=5.40$, the
imaging data spans the UV wavelength range in the galaxy's rest-frame
from 800\AA\  (V-band) to 2800\AA\  (H-band).  The galaxy is
unresolved in all bands, including the AO H-band (0.29$''$ FWHM)
making its physical extent on order or less than 1.7 kpc (1.0$''$
corresponds to 6.0 kpc at $z\sim5.5$).   

Unfortunately if the observed spectral break is the Lyman break, there
are few powerful emission lines to observe spectrally in this
wavelength range.  The Lyman break would be observed at about
8000\AA\  which will make it the primary spectroscopic feature for our
fitting routine.  Because it is so bright (I$=$19.85), absorption features 
may be also detectable with a 8-10 meter class red-sensitive spectrograph.

The GMOS spectroscopy does show a strong break around 7900\AA\ and
confirms our photometric redshift fit that the Lyman break is between R and
I bands.  The signal to noise ratio of the spectrum was insufficient to identify 
any absorption features.

The SpeX spectroscopy also shows the Lyman break though just at
the edge of sensitivity.  Both spectra are also consistent with the
photometric data, give similar redshifts and are within the error of
the photometry-derived redshift (Tables 5 and 8).  A weighted average
of the three redshifts gives a best estimate at $z=5.53\pm0.06$.
These three independent measures suggest that if the optical ID,
F1234+2001, is associated with the radio source, J123432.9+200134, it
would make it the most distant known radio galaxy.  However, this
redshift should be confirmed with deeper spectroscopy.

\emph{F1237+1141} is consistent with a moderately high redshift
($z\sim2.5$) starburst SED.  The radio morphology is a 
triple system at low radio 20cm flux (1mJy) 
with the ID corresponding with the center source consistent with this redshift.
It was also observed with both IRAC and MIPS, which changed the best
fit redshift to 2.9 to reflect the 5.8$\mu$m data point, but with
large error bars.  The SED beautifully fits a $z=2.9$ starburst with a
long wavelength starburst SED (M82) from a rest wavelength of 110 nm
to 40 microns.  This is a superb illustration of the potential of
Spitzer for both photometric redshifts and studies of the stellar
environment of these objects.  Spectroscopy better constrained the
redshift to 3.01.

\emph{F1315+4438} is the faintest detected object in the HzRG
candidate sample ($K=21.32$).  Hyperz found 2 photometric redshifts, a
SB1 SED at $z=1.1$ and a SB2 
SED at $z=2.8$.  The extremely faint optical flux favors the high
redshift solution though a low radio 20cm flux (0.9mJy) with a
slightly distorted morphology may indicate that the radio emission is
from star formation.  However, this interpretation would support an
even lower redshift than $z\sim1$. 

\emph{F1329+1748} is bright in V, R, and I with a weak (1.3mJy) radio
source. 

\emph{F1355+3607} is undetected with $5 \sigma$ limit at $K>20.86$.  A
distorted, faint (3mJy) radio source is consistent with a low
redshift starburst, perhaps highly reddened.

\emph{F1430+3557} was undetected in all bands 
except K so it does not have a photometric redshift until adding the
MIPS 24$\mu$m and 160$\mu$m Spitzer 
archive data, the best-fit SED is a starburst at $z=3.5$ with
starburst or quiescent long wavelength SED (not LIRG or ULIRG). 
The radio source 20cm flux is moderate (9mJy), together with the lack
of resolution of the radio source  
is also consistent with an AGN at high redshift.

\emph{F1435-0029} was undetected in B, V, R, and I, but detected in
z$'$ and H.  It has a moderate radio source (10mJy), and an apparent
significant break between I and z$'$, so probably at high redshift
($z\sim3.6$). 

\emph{F1447+1217} is probably at very high redshift
($z=4.70$).  It is one of the most powerful radio source (49mJy) in the
sample, but a good SED fit, suggests little QSO contamination, so we
might consider objects at lower radio 20cm flux also safe from
contamination. 

\emph{F1451+0556} is unresolved and has a peculiar color ($R-K=0.27$),
which most resembles a young starburst SB2 SED at z=2.71, though not a very good fit.  
Spectroscopy confirms a redshift of 2.567 with several
emission lines.  This suggests significant AGN influence on the SED of
this galaxy explaining the peculiar color and compactness.  

\emph{F1458+4319} has a distorted radio source making it difficult to
identify an optical counterpart.  The brighter two sources have
photometric redshifts at z=5.1 and 5.0 with starburst SEDs.  The
extreme I-R break seems consistent with the SED fit.   

\emph{F1505+4457} has an SED consistent with an Sa galaxy at low
redshift ($z=0.55$) though spectroscopy data suggests a higher
redshift (z=1.02).  The optical ID lies between the 2 sources (which
are at different fluxes.  This seems to suggest a radio jet with the
axis almost perpendicular to the plane of the sky with the north lobe
pointed away.  With a total 20-cm radio of about 17mJy, with
double-lobe morphology, it is consistent with this moderate redshift. 

\emph{F1524+5122} is undetected with $5 \sigma$ limit at $K>20.96$.
The relatively high radio luminosity (22mJy) suggests an AGN.  It is
potentially a very high redshift object, possibly 
the host galaxy could be highly reddened.

\emph{F1644+2554} was undetected at R, I, z$'$, and H also was not
detected with SCUBA so is not constrained as it has not been detected
in any band. 

\emph{F2217-0837} has 2 Hyperz solutions, a $z=0.3$ elliptical and
$z=4.1$ SB2 SED.  The low redshift solution seems more likely as it is
would otherwise be an extremely bright ($>10^{13}L_{sun}$) galaxy at
high redshift.  A weak (1mJy), elongated radio source also supports
the low redshift solution. 

\emph{F2217-0138} has two possible optical IDs, a brighter source
about 1'' to the east of the radio peak
with a photometric redshift of $z=4.6$ (confirmed with spectroscopy),
and a fainter source 1'' to the west with a photometric redshift
$z=5.0$.  An upper limit from
SHARC2 does not help the optical ID nor constrain any of
SED fitting.  A powerful (26mJy) radio source
suggests an AGN. 

\emph{F2354-0055} was undetected in B, V, R, I, z$'$ and H filters,
and a compact, faint, steep spectrum radio source is probably from an
AGN, perhaps at high redshift.

\section{Summary}

A set of 58 VLA FIRST survey sources that lie within the isoplanatic patch of
a bright natural guide star (BNGS) was constructed to search for high redshift radio 
galaxies able to be used with NIR adaptive optics.  These 58 objects were observed in B, V, R, I, 
z$'$, J, H, and K bands and their redshifts were estimated using SED fitting and 
generally confirmed as accurate with spectroscopy.  

It was found that the FIRST-BNGS sample objects generally follow the IR Hubble
diagram for radio galaxies.   \citet{el-bouchefry2007} did a study with FIRST galaxies in the 
NOAO Deep-Wide Field Survey Bo\"{o}tes field and found similar results.  They had a large 
spread at low redshift ($z<0.7$) that we did not observe probably due to our selection 
preference for high redshift objects.  The low redshift sources in our sample would be the 
brightest; we chose not to complete these in all bands in favor of fainter objects and as a 
result we would not have photometric redshifts for them.  Several objects at high redshift (z$>$1)
have best-fit SEDs consistent with young stellar populations (SB1-6).
The few long-wavelength observations tend to favor either a quiescent or M82-like star-forming SED 
(rather than LIRG or ULIRG type SEDs).  This 
may suggest that at least some of these high redshift galaxies may be in an active star-forming phase 
which is not what is seen in the K-z Hubble diagram which shows something like a passively evolving
population from very high redshift (z$>$5).  
No trend with radio spectral index was found though there were more steep-spectrum sources 
observed than flat-spectrum.

Many of these objects are at significant redshift, and this sample provides a unique
tool to study galaxies at high redshift.  Today's ground-based instrumentation even allows 
sufficient resolution to measure the fundamental parameters of the host galaxies; imaging 
provides resolved morphology and color gradients and spectroscopy allows the dynamics of 
each galaxy to be studied.   A recently completed study of the morphologies of 11 
FIRST-BNGS galaxies from a subsample observed with the Subaru Telescope suggests that 
these objects tend to be compact, blue, dynamically-relaxed galaxies (Stalder \& 
Chambers\nocite{stalder2009a}, in prep.).
These intriguing objects provide a glimpse of the detailed picture of the first few Gyr 
of the history of the universe that future projects such as TMT and JWST will provide. 
They also hold great potential in studying the high redshift universe and deserve further 
attention.

\medskip

We thank Michael Connelley, Steve Howell, Elizabeth McGrath, and Barry Rothberg for 
assisting in some of the imaging and spectroscopy observations for this huge data set.  
We also acknowledge the telescope support staffs at the University of Hawai`i 
2.2-meter as well as the Infrared Telescope Facility, which is operated by the 
University of Hawai`i under Cooperative Agreement no. NCC 5-538 with the National 
Aeronautics and Space Administration, Science Mission Directorate, Planetary 
Astronomy Program.  This research was partially supported by the
Extragalactic and Cosmology division of NSF under grant AST 0098349 and also partially 
supported by the Pan-STARRS Camera Group. 
This research has made use of the NASA/IPAC Extragalactic Database
(NED), which is operated by the Jet Propulsion Laboratory, California
Institute of Technology, under contract to the National Aeronautics
and Space Administration.  This work is partly based on observations 
made with the Spitzer Space Telescope, which is operated by the Jet 
Propulsion Laboratory, California Institute of Technology under a 
contract with NASA.  

\appendix

\section{SED Template Library}

For objects with more than 4 bands of photometric data, photometric redshifts 
were derived using the public SED fitting code, Hyperz \citep{bolzonella2000} which 
requires a set of templates to fit to the photometric data.  
See section 4 for a detailed descriptions of the overall procedure.  

The SED template library used were originally from \citet{kinney1996}, E, S0, Sa, Sb, 
Sc and SB1 to SB6 (starburst models over a range of reddening) built using IUE and 
optical data with a range of wavelength from 1200\AA\  to 1.0$\mu$m.  This library 
was later extended by \citet{mannucci2001} into the NIR (2.4$\mu$m) using averaged 
ground-based NIR spectra of local prototypical galaxies which closely matched the 
templates from \citet{kinney1996}.  

Our high redshift SED fitting procedure requires even shorter wavelength data, and 
the Spitzer and submillimeter data require longer wavelength data than provided by 
these libraries.  Therefore further UV and mid-IR extensions were accomplished using 
a procedure similar to that used by \citet{bolzonella2000} to extend \citet{coleman1980} 
templates. An appropriate GISSEL98 \citep{bruzual1993} synthetic blue spectrum (constant 
SFR, and age=0.1Gyr) was chosen for Sa, Sb, Sc and SB1-6 or a red spectrum (with 
delta-function starburst, age=19Gyr) for E and S0.  These models were chosen to 
match the overall slope of the continuum where it overlaps with the template.  These 
spectra were then grafted onto the Kinney/Mannucci templates by matching the average 
continuum levels.  The matching wavelength ranges were chosen to be relatively smooth 
(few spectral lines) and flat in $F_\lambda$  (1230 to 1500\AA\  in the UV and 2.30 to 
2.36$\mu$m in the Near-IR).  Figures A.1 and A.2 show the full wavelength-range SEDs.

In order to fit SEDs to Spitzer or submillimeter data, a further extension was applied 
to these templates using 4 additional FIR-submm template SEDs chosen to reflect an 
assortment of possible prototypical spectra.  The templates used were Arp 220 
\citep{bressan2002} representing a ULIRG SED, M82 \citep{bressan2002} representing a 
starburst SED, and 2 templates in the synthetic library from \citet{dale2001} 
representing LIRGs ($\alpha=1.06$) and quiescent ($\alpha=2.5$) SEDs.  The extension 
was done almost identically to the UV/IR extensions except using the continuum levels 
from 1.95 $\mu$m to 2.05 $\mu$m to scale the spectra.  All 4 templates were applied to 
each of the SED templates so it is possible to fit the SEDs continuously from 100 \AA\ 
to 1 mm.  Figure A.3 shows the 4 FIR-submm extensions to the SB3 optical-IR template.

\begin{figure*}[htp]
\vspace{+0.2in}
\begin{center}
\includegraphics[scale=0.85,angle=0,viewport= 0 50 1000 700]{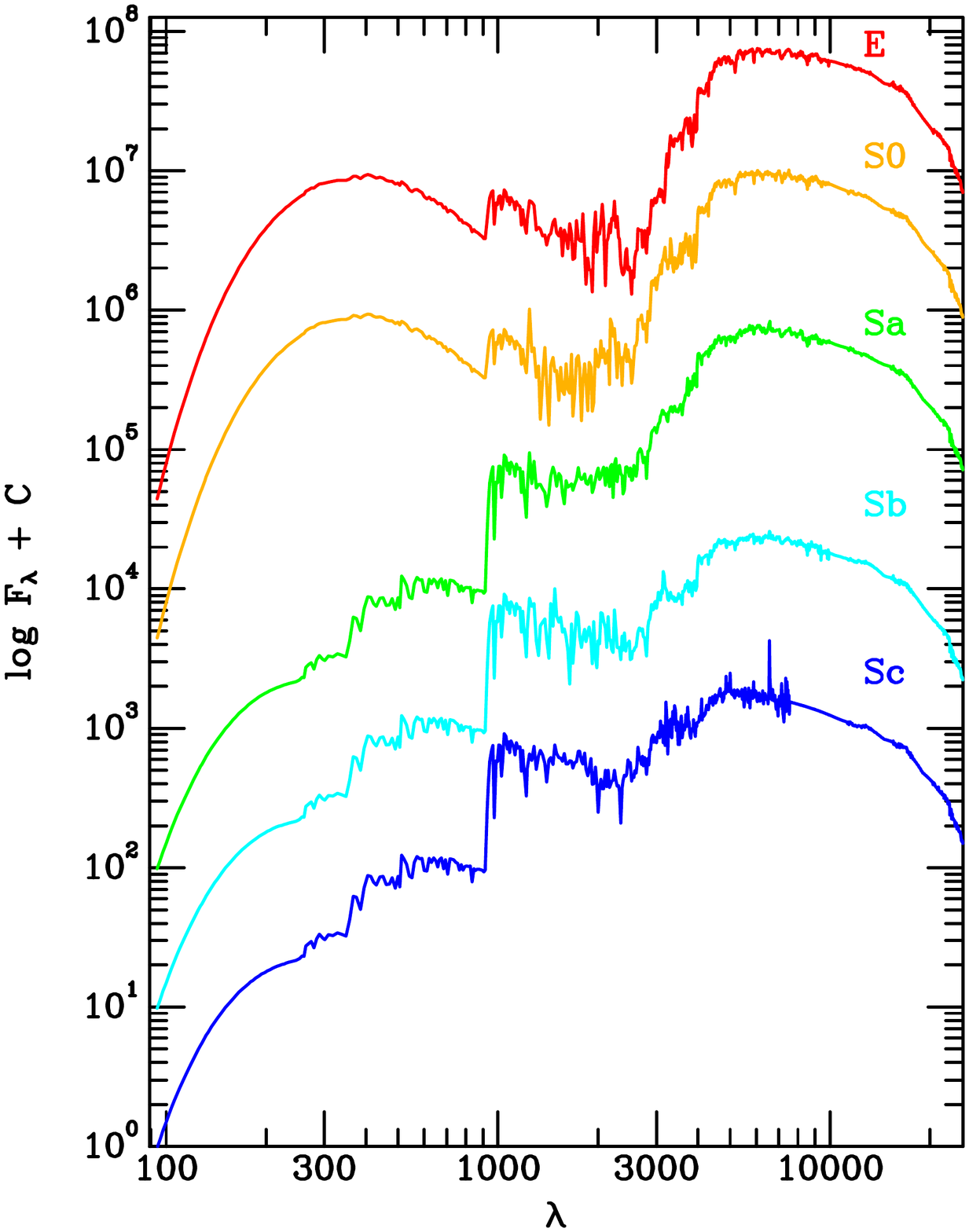}
\end{center}
\vspace{-0.2in}
{\scriptsize
FIG. A.1--- Template Spectra used in Hyperz from \citet{mannucci2001} and extended with GISSEL98 \citep{bruzual1993}.
}
\end{figure*}

\begin{figure*}[htp]
\vspace{+0.2in}
\begin{center}
\includegraphics[scale=0.85,angle=0,viewport= 0 50 1000 700]{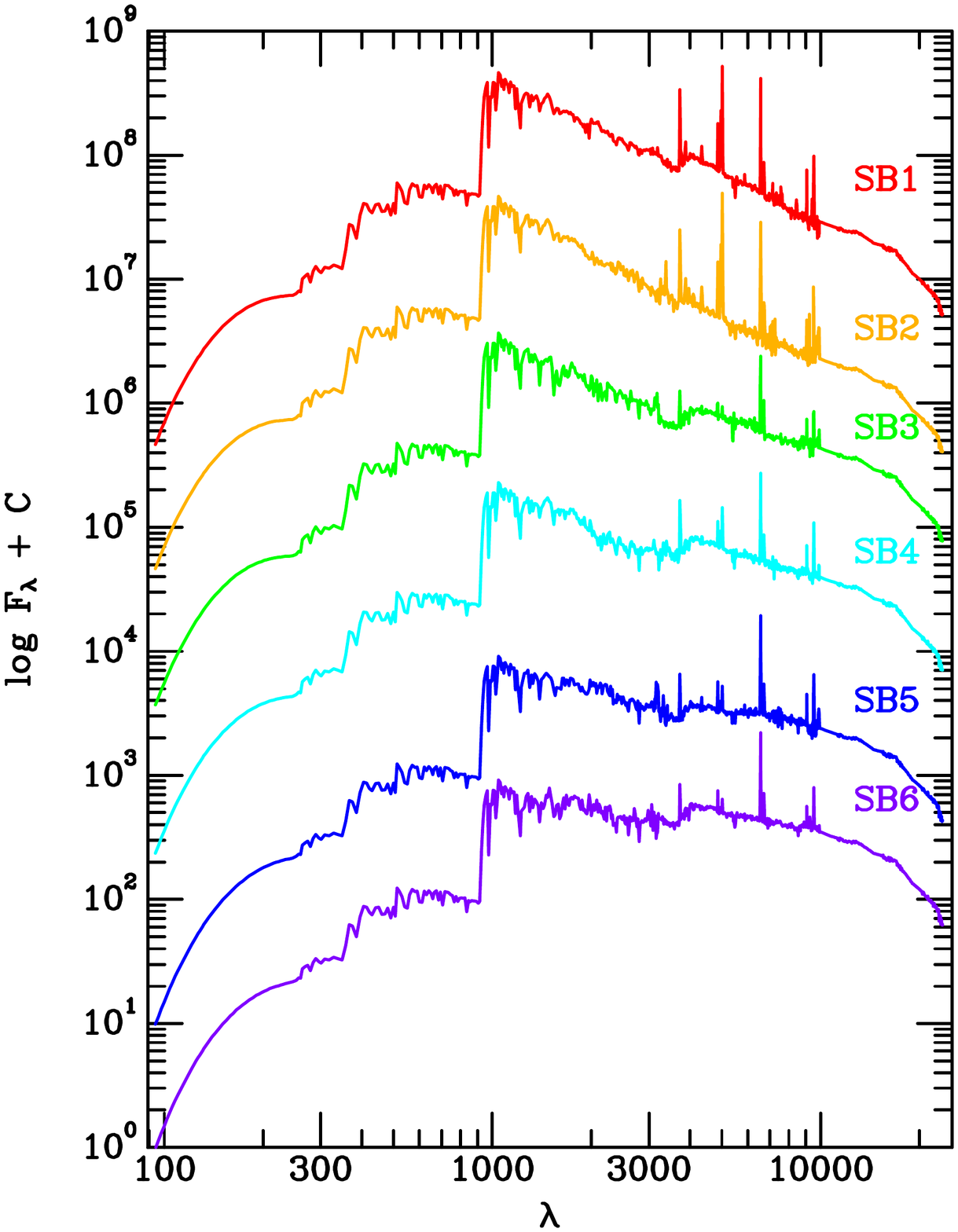}
\end{center}
\vspace{-0.2in}
{\scriptsize
FIG. A.2--- Template Spectra used in Hyperz from \citet{kinney1996} and extended with GISSEL98 \citep{bruzual1993}.
}
\end{figure*}

\begin{figure*}[htp]
\vspace{-0.1in}
\begin{center}
\includegraphics[scale=0.85,angle=0,viewport= 0 50 1000 700]{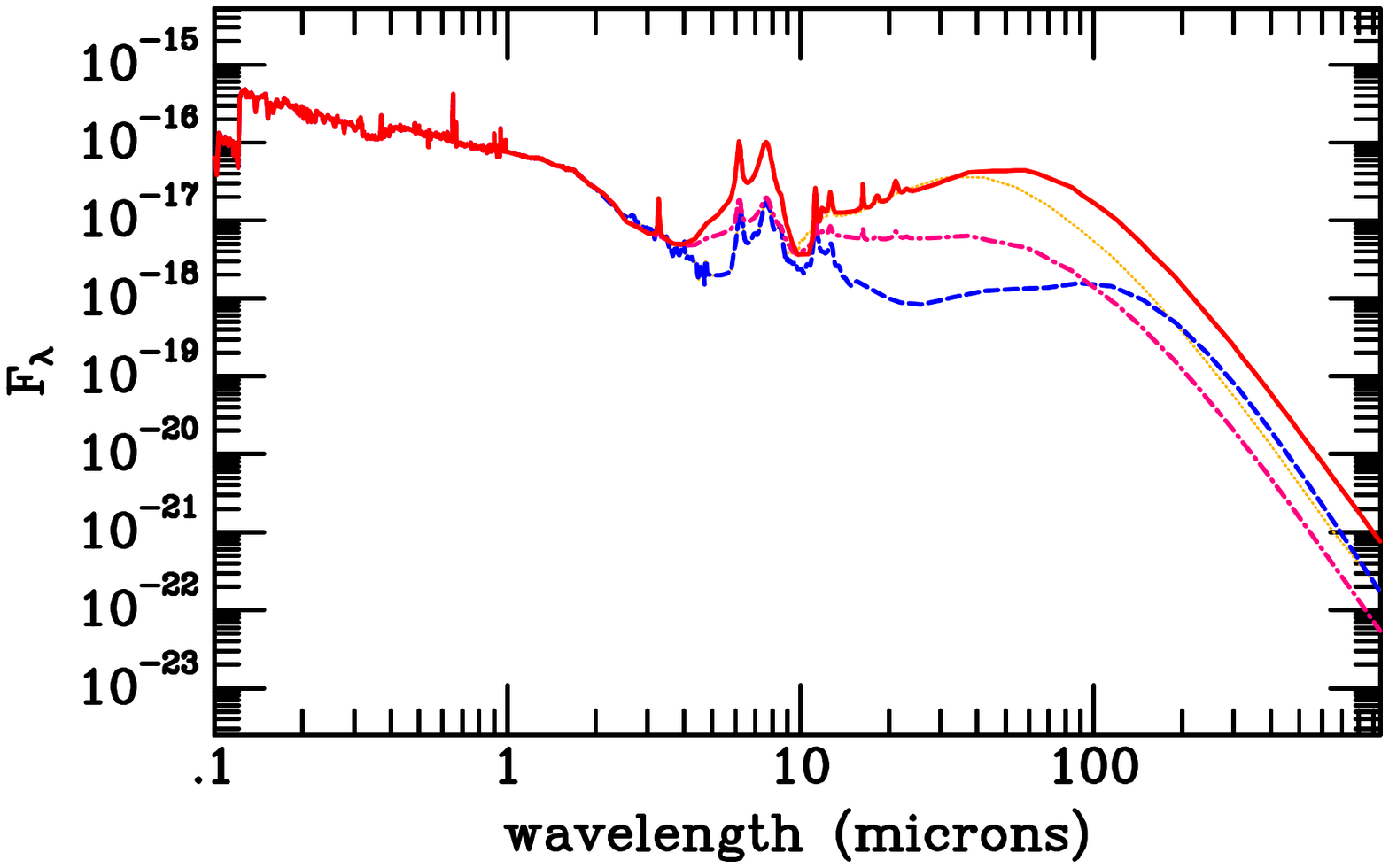}
\end{center}
\vspace{-4.2in}
{\scriptsize
FIG. A.3--- Example template spectra used in fitting long wavelength data.  The templates are ULIRG (solid), LIRG (dotted), starburst (dot-dashed), and quiescent (dashed) long wavelength templates with a Kinney SB3 short wavelength template.
}
\end{figure*}

{\bibliography{References}}

\end{document}